\pdfoutput=1
\documentclass[12pt, a4paper]{article}
\usepackage{jheppub}
\usepackage[usenames, dvipsnames]{xcolor}
\usepackage{graphicx}
\usepackage{microtype}
\usepackage{amssymb} 
\usepackage{amsmath}
\usepackage{framed}
\usepackage{mathtools}
\usepackage{amsfonts}    
\usepackage{dsfont}
\usepackage{pdfpages}
\usepackage{verbatim}
\usepackage{braket}
\usepackage{bm}
\usepackage{tensor}
\usepackage{upgreek}
\usepackage{subcaption}
\usepackage{simpler-wick}
\usepackage{enumitem}
\usepackage[hang, flushmargin]{footmisc}
\usepackage{multirow}
\usepackage[table,xcdraw]{xcolor}
\usepackage{float}

\usepackage{tikz}
\usetikzlibrary{decorations.markings}
\usetikzlibrary{shapes.misc}
\usetikzlibrary{arrows}
\usetikzlibrary{fit, chains}
\usetikzlibrary{patterns}

\usepackage{listings}
\usepackage{color}

\usepackage[framemethod=default]{mdframed}
\usepackage{etoolbox}

\DeclareMathOperator{\arccosh}{arccosh}

\newenvironment{FramedBox}[1][]{%
    \begin{mdframed}[
        skipabove=7pt,
        skipbelow=7pt,
        rightline=true,
        leftline=true,
        topline=true,
        bottomline=true,
        backgroundcolor=gray!10,
        linecolor=gray!10,
        innerleftmargin=10pt,
        innerrightmargin=10pt,
        innertopmargin=10pt,
        innerbottommargin=10pt,
        leftmargin=0cm,
        rightmargin=0cm,
        linewidth=1pt,
        #1
    ]%
    \ignorespaces%
}{%
    \end{mdframed}%
}

\usepackage[framemethod=default]{mdframed}
\newmdenv[skipabove=7pt,
skipbelow=7pt,
rightline=false,
leftline=false,
topline=false,
bottomline=false,
backgroundcolor=gray!10,
linecolor=gray,
innerleftmargin=5pt,
innerrightmargin=5pt,
innertopmargin=5pt,
innerbottommargin=5pt,
leftmargin=0cm,
rightmargin=0cm,
linewidth=4pt]{eBox}

\usepackage{tikz}
\usetikzlibrary{decorations.pathreplacing, decorations.markings,calc,shapes.misc,decorations.pathmorphing,patterns.meta, math, shadings, backgrounds}

\definecolor{rosewood}{rgb}{0.4, 0.0, 0.04}
\definecolor{pyblue}{RGB}{31, 119, 180}
\colorlet{lightpyblue}{pyblue!30!white}
\definecolor{pyorange}{RGB}{255, 127, 14}
\colorlet{lightpyorange}{pyorange!30!white}
\definecolor{pygreen}{RGB}{44, 160, 44}
\colorlet{lightpygreen}{pygreen!30!white}
\definecolor{pyred}{RGB}{214, 39, 40}
\colorlet{lightpyred}{pyred!30!white}
\definecolor{lightgray}{gray}{0.9}

\usepackage{hyperref}
\hypersetup{
    pdfencoding=unicode,
    colorlinks=true,
    urlcolor=rosewood,
    linkcolor=pyblue,
    citecolor=pygreen,
    pdftitle={This is the title},
    pdfauthor={Denis Werth},
    pdfdisplaydoctitle=true,
    pdfstartview=FitH,
    linktocpage=true
}

\newcommand{\kappaone}{\kappa_1}    
\newcommand{\kappatwo}{\kappa_2}    
\newcommand{\kappathree}{\kappa_3}  
\newcommand{\kappafour}{\kappa_4}   

\def \d {\mathrm{d}}

\def \x {\bm{x}}

\def \k {\bm{k}}

\def \a {{\sf{a}}}
\def \b {{\sf{b}}}

\def \C {\mathcal{C}}
\def \D {\mathcal{D}}
\def \E {\mathcal{E}}
\def \F {\mathcal{F}}
\def \K {\mathcal{K}}
\def \L {\mathcal{L}}
\def \G {\mathcal{G}}
\def \I {\mathcal{I}} 

\def \N {\mathcal{N}}

\def \aa {{\sf{a}}}

\def \Im {\mathrm{Im}}
\def \Re {\mathrm{Re}}

\def \dS {\mathrm{dS}}

\def \EAdS {\mathrm{EAdS}}

\def \KLF {\mathrm{KLF}}

\title{Cosmological Correlators in KLF and the Double-Exchange}

\author{Nathan Belrhali,}
\author{Arthur Poisson,}
\author{S\'ebastien Renaux-Petel}

\affiliation{Institut d'Astrophysique de Paris, CNRS, Sorbonne Universit\'e, 98 bis bd Arago, 75014 Paris, France}

\abstract{In this work, we present the procedure to find series representations of tree-level cosmological correlators using the Kontorovich-Lebedev-Fourier (KLF) space formalism. This framework allows us to trade the in-in nested time integrals for frequency integrals over rational propagators and vertex functions, which encode interactions among quantum fields on a de Sitter background. Because these functions are the key objects to understand in order to perform a diagrammatic computation, we derive their relevant analytic properties by using both their integral representation and series representation in terms of Lauricella functions. For a vertex involving any number of fields, we obtain the location of singularities, the corresponding residues and the large-frequency asymptotic behaviour. Gathering these properties at each frequency integration allows us to compute a tree-level correlator directly, without relying on the differential equations it satisfies. To illustrate this procedure, we provide a complete treatment of the double-exchange diagram. The computation naturally distinguishes the different physical contributions, whether to the background or to the cosmological collider signal. The newly derived result is expressed at most in terms of a double series over hypergeometric functions, which simplifies the analytical expression of the correlator.}

\begin{document}

\setcounter{tocdepth}{3}
\maketitle
\setcounter{page}{2}

\newpage

\section{Introduction}

Cosmological correlators are the correlation functions of primordial quantum fluctuations at the end of inflation. They constitute the key observables of primordial cosmology, since their observation from their imprint on the cosmic microwave background and the large-scale structure of the universe allow us to probe the particle content and interactions of the early universe \cite{Achucarro:2022qrl}. Improving the analytical understanding of these objects presents a significant theoretical challenge, because the expansion of the background space-time induces a breaking of time-translation invariance. As a consequence, the free propagation of fields has a complicated time-dependence (e.g. given by Hankel functions for massive scalar fields on a pure $\dS$ background) and nested multi-layer time-integrals have to be carried out, requiring tailored integration techniques even at tree-level.

To this end, a rich toolbox of analytical techniques has been developed to compute cosmological correlators. In this context, the word "compute" refers to turning the integral expressions that arise from quantum field theory (QFT) and perturbation theory principles into series representations, motivated in particular by practical numerical evaluations. A significant part of these methods is based on writing and solving systems of differential equations that are satisfied by a correlator in terms of its kinematic variables: the bootstrap techniques \cite{Arkani-Hamed:2018kmz,Baumann:2019oyu,Baumann:2020dch,Pajer:2020wxk,Pimentel:2022fsc,Jazayeri:2022kjy,Wang:2022eop,Qin:2022fbv,Qin:2023ejc}, the massive family-tree \cite{Xianyu:2023ytd,Liu:2024str,Fan:2025scu,Xianyu:2025lbk} or the kinematic flow \cite{Arkani-Hamed:2023kig,Arkani-Hamed:2023bsv,De:2023xue,Hang:2024xas,Baumann:2024mvm,Westerdijk:2026msm,Ke:2026laa,Baumann:2026atn}, often using energy singularities and discontinuities to fully determine the correlator \cite{Maldacena:2011nz,Raju:2012zr,Baumann:2020dch,Goodhew:2020hob,Melville:2021lst,Goodhew:2021oqg,Jazayeri:2021fvk}. Other tools rely on the use of integral transforms to rewrite the correlator in a more suitable form for analytical computations: these include methods based on the Mellin-Barnes transform \cite{Qin:2022fbv,Xianyu:2023ytd,Sleight:2019hfp,Sleight:2019mgd,Sleight:2020obc,Sleight:2021plv,Qin:2023bjk}, the Laplace transform \cite{Belrhali:2026ygh,Belrhali:2026jqe} or on the de Sitter momentum space that uses the Kontorovich-Lebedev-Fourier (KLF) transform \cite{Belrhali:2026ktb,Belrhali:2026rkn}.

The latter is the perspective we adopt in this work. Its motivation stems from the observation that, because the background expansion breaks energy conservation, the physical meaning of a frequency as a variable dual to conformal time has to be reconsidered. As a result, and unlike in Minkowski space-time for scattering amplitudes, there is no proper definition of a momentum space, and perturbative computations cannot be recast in terms of a frequency Fourier variable. The KLF framework circumvents this by exploiting the maximal symmetry of de Sitter space-time, the relevant background for inflationary fluctuations: it reformulates QFT in de Sitter by labelling the quantum states with eigenvalues of well-chosen symmetry generators. In particular, the quadratic Casimir operator defines a new frequency $\mu$, while the spatial translation generators yield the spatial momenta, as in the Minkowski case.
Equipped with this tool, we can rewrite perturbative computation in these variables, exchanging time integration for integrals over an off-shell frequency for each internal line of a correlator. In this language, the time-ordered propagators have a rational form, like the flat-space ones, differing only by the $i\epsilon$-prescription proper to de Sitter. The crucial difference with flat-space spectral integrals is the presence of the so-called \textit{vertex functions}, which encode the properties of the interaction at a given vertex and have a highly non-trivial analytic dependence on frequency and momentum variables. The single-exchange diagram, studied in \cite{Belrhali:2026ktb,Belrhali:2026rkn,Werth:2024mjg}, is the first non-trivial example treated with the KLF formalism. This was made possible by the relative simplicity of its vertex functions, which in turn stems from the fact that a single massive leg is attached to each vertex of the diagram. The complexity of the vertex functions, however, increases with the number of massive legs connected to the corresponding vertex. Computing generic diagrams therefore requires understanding their analytical properties in these more involved cases. This is precisely the scope of the present work, which we illustrate through the explicit example of a diagram involving the exchange of two massive fields.

This diagram has already been studied, as part of a broader effort on massive cosmological correlators of increasing complexity \cite{Xianyu:2023ytd,Aoki:2024uyi,Liu:2024str,Fan:2025scu,Xianyu:2025lbk}, from the perspective of solving a system of differential equations satisfied by the object of interest or by using the Mellin-Barnes transform. In this work, we choose a different route and show how the integrals for a given correlator can be directly computed from their expression in frequency-momentum variables, knowing the analytic properties of the integrand and using complex analysis. Indeed, the frequency integrals are computed using contour integration in the complex frequency plane. This requires the knowledge of the following properties of the vertex functions, which we derive systematically: the location and nature of their singularities, which appear to be simple poles, their corresponding residues, and the large-frequency asymptotic behaviour that is used to provide contour closure prescriptions. The derivation of these properties relies on the interplay between the two representations of a vertex function. The first is the integral representation, which arises from the construction of KLF momentum space; the second is the series representation in terms of Lauricella $F_C$ functions, obtained from the former through Mellin-Barnes transforms. Used together, these two representations make the analytic structure of the vertex functions explicit and provide a systematic route to series representations of massive tree-level correlators.

These general statements are exemplified by the explicit calculation of the double-exchange diagram. In that case, the integral that has to be carried out is a double-layer integral, with one integration for each massive internal line. The vertex functions are given in terms of Legendre and Appell $F_4$ functions, and for each integration layer two cases have to be distinguished depending on a condition on momenta flowing in the diagram. Then, for each of the four kinematic cases, we end up with a series representation written in terms of residues of the integrand. Among several components, the most complicated is given in terms of a double series involving Gauss hypergeometric and Appell $F_4$ functions, which has to be compared to previous representations that were given in terms of series with up to four summation layers.
These results allow for an explicit evaluation of the correlator that accounts for a process with the exchange of two massive particles.

\paragraph{Outline.} The outline of this work is the following: section \ref{section_cc_with_KLF} begins with an introduction to the computation of cosmological correlators using the KLF formulation of QFT in de Sitter, and then provides the analytical properties of the vertex functions that allow for the computation of frequency integrals entering tree-level correlators. As an illustration, section \ref{section_the_double_exchange_diagram} displays the essential features of the computation of the double-exchange diagram, collecting all the different pieces for a given kinematic configuration. Finally, section \ref{sec:conclusion} contains the concluding remarks and a discussion of future perspectives.
This material is complemented by appendices that provide the following: useful relations and formulas both used in the main text and in appendices, details on the general analytic structure of vertex functions discussed in the first part of the paper, and the completion of the double-exchange calculation by adding the remaining kinematic configurations not displayed in the main text, as well as additional details worth mentioning.

\section{Cosmological Correlators in KLF}\label{section_cc_with_KLF}

In this section, we start by reviewing the KLF momentum space for cosmological correlators in \ref{subsec_intro_KLF}, as formalised in \cite{Belrhali:2026ktb,Belrhali:2026rkn}. Then, in \ref{vertex_fct_N} we provide details on how to perform perturbative computations, i.e. finding series representations of the KLF integrals by using the meromorphic properties of the integrands. To this end, analytic properties of the so-called vertex functions, that appear in KLF integrands in perturbative computations when one encounters a vertex in a Feynman diagram, are derived. In particular, we show how their representation with Lauricella $F_C$ functions makes their pole structure and residues apparent, and derive their asymptotic behaviour.

\subsection{KLF Momentum Space}\label{subsec_intro_KLF}

Cosmological correlators are in-in correlation functions of quantum fields evolving over a fixed curved background, that is de Sitter spacetime ($\mathrm{dS}_{d+1}$, of dimension $d+1$). $\mathrm{dS}_{d+1}$ can be defined as a hypersurface in an embedding Minkowski space $\mathbb{M}^{1,d+1}$, as the set of points of $\mathbb{M}^{1,d+1}$ that satisfy
\begin{equation}
    \eta_{AB} X^A X^B = -(X^0)^2 + (X^1)^2 + \ldots + (X^{d+1})^2 = H^{-2} \,,\label{dS_def}
\end{equation}
where $H^{-1}$ is the de Sitter radius (here the inverse of the Hubble parameter $H$).

In a cosmological context, the convenient coordinates to parametrise $\mathrm{dS}_{d+1}$ spacetime are the Poincar\'e coordinates $(\tau,\mathbf{x})$, where $\tau<0$ is the conformal time and $\mathbf{x}$ are the spatial coordinates. This choice leads to the following FLRW metric
\begin{equation}
    \mathrm{d}s^2=\frac{-\d\tau^2+\d\mathbf{x}^2}{(H\tau)^2}\,,
\end{equation}
with a constant Hubble parameter $H$.

On this fixed background spacetime, we then consider the correlation functions of quantum operators $\mathcal{\hat{O}}_i$ (which for instance depend on the quantum field $\varphi$, the scalar inflaton fluctuation) in the Bunch-Davies vacuum state $\ket\Omega$, evaluated at the time boundary $\tau\rightarrow 0$ corresponding to the end of inflation:
\begin{equation}
    \mathcal{C}_B\left(\{\mathbf{x}_i\}\right)=\braket{\Omega\vert\prod_i \mathcal{\hat{O}}_i\left(\mathbf{x}_i\right)\vert\Omega}\,,
    \label{bdy_correlator_def}
\end{equation}
or written in $d$-dimensional spatial Fourier space as
\begin{equation}
    \mathcal{C}_B\left(\{\mathbf{k}_i\}\right)=\braket{\Omega\vert\prod_i \mathcal{\hat{O}}_i\left(\mathbf{k}_i\right)\vert\Omega}=\int\prod_j \d^d \mathbf{x}_j\,e^{i \sum_j \mathbf{x}_j . \mathbf{k}_j} \braket{\Omega\vert\prod_j \mathcal{\hat{O}}_j\left(\mathbf{x}_j\right)\vert\Omega}\,.
    \label{bdy_correlator_Fourier}
\end{equation}

To compute these observables in perturbation theory, it is possible to use the Schwinger-Keldysh (or in-in) formalism \cite{Weinberg:2005vy,Chen:2017ryl} to express each correlator at a given interaction coupling order as a sum of Feynman diagrams. Now, to perform these perturbative computations in a well-defined momentum space, we want to make use of the $\mathrm{dS}_{d+1}$ symmetries.
 
\paragraph{KLF space.} To identify the Hilbert space $\mathcal{H}$ of a quantum field theory in de Sitter, we use its decomposition over unitary irreducible representations of the $\mathrm{dS}_{d+1}$ symmetry group.
From the definition \eqref{dS_def}, we see that $\mathrm{dS}_{d+1}$ is symmetric under the Lorentz group $\mathrm{SO}(1,d+1)$. Following \cite{Belrhali:2026ktb,Belrhali:2026rkn}, we consider the corresponding Lie algebra and its generators to find a basis of $\mathcal{H}$ by diagonalising a maximal set\footnote{In this context we choose to work with the quadratic Casimir operator and the momentum generators, however this largest set is not necessarily unique, for instance the construction of the CFT basis is made by diagonalising the quadratic Casimir and the dilatation generator \cite{Hogervorst:2021uvp}.} of commuting operators among the generators of this algebra. This set of operators consists of the quadratic Casimir operator and the generators of spatial translations. Then, the eigenfunctions of these operators, which are the harmonic functions, can be found to be:
\begin{equation}
\label{def_harmonic_function}
    \Phi_{\k}^{\mu}(z, \x) = \frac{H^{\frac{d+1}{2}}}{\sqrt{\pi}} \, e^{-i \k \cdot \x} z^{\frac{d}{2}} K_{i\mu}(kz)\,,
\end{equation}
where $K_{i\mu}$ is the modified Bessel function of the second kind, $\mu$ and $k_i$, $i=1,\ldots,d$ are related respectively to the Casimir operator and translation generators eigenvalues, and $z$ $\in[0,+\infty[$ is the coordinate obtained by Wick-rotating the conformal time to the Euclidean anti-de Sitter ($\mathrm{EAdS}_{d+1}$) geometry:
\begin{equation}\label{z_Lorentzian}
    z=- \tau e^{i\, \a \left(\frac{\pi}{2}-\epsilon\right)}\;,
\end{equation}
where the $\epsilon$ factor comes from the in-in $i\epsilon$-prescription and $\a=\pm$ is the Schwinger-Keldysh index.

Then, it is possible to show that the harmonic functions \eqref{def_harmonic_function} form a basis of the square-integrable functions on $\mathrm{EAdS}_{d+1}$ (i.e. of $L^2\left[\mathrm{EAdS}_{d+1},\d z\,\d^d \mathbf{x}\sqrt{-g}\right]$), allowing for the definition of the KLF integral transform:
\begin{equation}
\label{KLF_transform}
\begin{aligned}
    f(z,\x) &= \int_{\KLF} \;\Phi^{\mu}_{\k}(z,\x)\;f^{\mu}_{\k}\;,\\
    f^{\mu}_{\k} &= \int_{\EAdS}\left[\Phi^{\mu}_{\k}(z,\x)\right]^*\, f(z,\x) \,,
\end{aligned}
\end{equation}
where $f\in L^2\left[\mathrm{EAdS}_{d+1},\d z\d^d x\sqrt{-g}\right]$ and the integration measure is given by
\begin{equation}
    \begin{aligned}
        \int_{\KLF} &\equiv \int \displaylimits_{-\infty}^{+\infty}\d\mu \, \N_{\mu}\int_{\mathbb{R}^d}\frac{\d^d\k}{(2\pi)^d} \,, \\
        \int_{\EAdS} &\equiv \int \displaylimits_0^\infty \frac{\d z}{(Hz)^{d+1}}\int_{\mathbb{R}^d}\d^d\x \,,
    \end{aligned}
    \label{eq_KLF_measure}
\end{equation}
where $\N_{\mu}=\frac{\mu}{\pi}\sinh{(\pi\mu)}$ is the de Sitter density of states.

\paragraph{Correlators in KLF.}
Correlators in KLF space are defined using the usual path integral formulation:
\begin{equation}
\label{def_KLF_correlators}
    \mathcal{G}_{\aa_n \ldots \aa_n} \left(^{\mu_1...\mu_n}_{\k_1...\k_n}\right) = \prod_{i=1}^n\frac{(2\pi)^d}{\N_{\mu_i}}\frac{\delta Z[J_+,J_-]}{\delta J_{\aa_i,\k_i}^{\mu_i}}  \Bigg\rvert_{J_\pm=0} \,,
\end{equation}
where the functional generator $Z$ is expressed in KLF space as
\begin{equation}
\label{def_functional_generator_KLF}
    Z[J_+,J_-] \equiv \int\D\varphi_\pm e^{i S_+ - i S_- + \int_{\KLF}(\varphi_+J_+ +\varphi_-J_-)} \,.
\end{equation}
From \eqref{def_KLF_correlators}, the $\EAdS_{d+1}$ correlators (and then the $\dS_{d+1}$ correlators \eqref{bdy_correlator_def}) can be recovered using the integral transform \eqref{KLF_transform}. Before going back to the boundary correlators, let us see how diagrammatic rules work in KLF space, in particular free propagators and interactions at vertices.

To find the in-in propagators in KLF space, which are the building blocks of any perturbative computation, one can compute the two-point functions that arise from the functional generator given by the quadratic action of the free theory using \eqref{def_KLF_correlators}. For principal series fields (i.e. with parameter $\mu_\varphi$ real, on which we will focus in this work\footnote{For propagators of fields in the other unitary irreducible representations, see sec 3.4 and 4.2 of \cite{Belrhali:2026rkn}}), we then have:
\begin{equation}
\label{eq_propagator_def}
    \boldsymbol{\mathcal{G}}^{\mu_\varphi}\left(^{\mu\,\mu'}_{\k\,\k'}\right) 
    \equiv \left[\G_{\a\b}^{\mu_\varphi}\right]_{\a,\b=\pm,\pm}\left(^{\mu\,\mu'}_{\k\,\k'}\right)
    = \frac{1}{H^2}\delta\left(^{\mu \,\mu'}_{\k\,\k'}\right)\,\Pi^{\mu_\varphi}_{\a\b}(\mu)\,,
\end{equation}
where $\Pi^{\mu_\varphi}_{\a\b}$ is the reduced propagator:
\begin{equation}
    \left[\Pi^{\mu_\varphi}_{\a\b}(\mu)\right]_{\a,\b=\pm,\pm}\equiv\left(\begin{array}{cc}
    \frac{e^{-\frac{i\pi(d-1)}{2}}}{(\mu^2-\mu_{\varphi}^2)_{i\epsilon}} & \frac{\delta\left(\mu-\mu_{\varphi}\right)}{\N_{\mu}} \\
    \frac{\delta\left(\mu-\mu_{\varphi}\right)}{\N_{\mu}} & \frac{e^{+\frac{i\pi(d-1)}{2}}}{(\mu^2-\mu_{\varphi}^2)_{-i\epsilon}}
    \end{array}\right)\,.
    \label{eq_def_reduced_propagator}
\end{equation}
Each matrix element is a propagator with Schwinger-Keldysh indices $\a,\b=\pm,\pm$. The $i\epsilon$-prescription, which enters the diagonal propagator expressions and encodes particle propagation \cite{Melville:2024ove,Werth:2024mjg,Belrhali:2026rkn}, is given by
\begin{equation}
     \frac{1}{(\mu^2-\mu_{\varphi}^2)_{i\epsilon}} \equiv \frac{1}{2\sinh(\pi\mu_{\varphi})}\left[\frac{e^{+\pi\mu_{\varphi}}}{\mu^2-\mu_{\varphi}^2+i\epsilon}-\frac{e^{-\pi\mu_{\varphi}}}{\mu^2-\mu_\varphi^2-i\epsilon}\right] \,.
     \label{ieps_precription}
\end{equation}
Then, any line in a Feynman diagram carries a propagator:
\begin{equation}
    \begin{aligned}
        &\begin{tikzpicture}[baseline={(0,0)}]
            \draw[thick, black] (-1,0) to (1,0);
            \filldraw[color=black,fill=black] (-1,0) circle (3pt);
            \filldraw[color=black,fill=black] (1,0) circle (3pt);
            \node[black] at (-1,0.4) {$\a$};
            \node[black] at (-1,-0.4) {$\mu\;,\k$};
            \node[black] at (1,-0.4) {$\mu'\;,\k'$};
            \node[black] at (1,0.4) {$\b$};
            \node[black] at (0.,+0.4) {$\varphi,\;\mu_\varphi$};
            \end{tikzpicture}
            & \hspace*{-0.8cm}= \G_{\a\b}^{\mu_\varphi}\left(^{\mu\,\mu'}_{\k\,-\k'}\right)\;.
    \end{aligned}
\end{equation}

Interacting theories under consideration in this work contain polynomial interactions:
\begin{equation}
    \L \supset \lambda\,\prod_{i} \sigma_i^{\alpha_i}\,,
\end{equation}
where the fields $\sigma_i$ have arbitrary mass in the principal series (and $\alpha_i\in\mathbb{N}$). Then, for a given interaction involving $N$ massive fields, each with mass $\mu_i$ and momentum $\k_i$ that flows in the corresponding line, the vertex contribution to the diagram is given by
\begin{equation}\label{eq_feynman_rule_vertex}
\begin{aligned}
    &\begin{tikzpicture}[baseline={(0,0)}]
    \draw[thick, black] (0, 0) to (-1,1) node[above] {\scriptsize $\mu_1,\k_1$};
    \draw[thick, black] (0, 0) to (1,1) node[above] {\scriptsize $\mu_N,\k_N$};
    \draw[thick, black] (0,0) to (-0.33,1) node[above, xshift=1mm] {\scriptsize $\mu_2,\k_2$};
    \node[black] at (0.2,0.7) {$\ldots$};
    \draw[color=black] (0,0) circle (3pt) node[below] {\scriptsize$\aa=\pm$};
    \filldraw[color=black,fill=black] (0,0) circle (3pt);
    \end{tikzpicture}
    &\hspace*{-0.8cm}= - i\,\aa\,\C\,\lambda(2\pi)^d\delta^{(d)}\left(\sum_{j=1}^N\k_j\right)\frac{H^{\frac{(N-2)(d+1)}{2}}e^{\aa\frac{i \pi d}{2}}}{\pi^\frac{N}{2}}
    \mathcal{I}_d\begin{bmatrix}
        \mu_1 &\ldots &\mu_N \\ k_1 &\ldots & k_N \\ K &\ldots & K
    \end{bmatrix}
    \,,
    \end{aligned}
\end{equation}
where $\a$ is the Schwinger-Keldysh index of the vertex and $\C$ is a combinatorial factor depending on the theory under consideration. The crucial feature in KLF space diagrammatics is the presence of the quantity $\mathcal{I}_d\left[\{\mu_j\},\{k_j\},\{K\}\right]$
\footnote{In \cite{Belrhali:2026rkn,Belrhali:2026ktb}, the notation $\mathcal{I}^{\mu_1 \ldots \mu_N}_{k_1 \ldots k_N}$ was used for the vertex function that appears in \eqref{eq_feynman_rule_vertex}. In this work we use the convention \eqref{def_vertex_function_K} corresponding to the compact notation $\mathcal{I}_d\left[\{\mu_j\},\{k_j\},\{K\}\right]$.},
called a \textit{vertex function}, and defined by the following integral representation:\footnote{These integrals have been extensively studied for $N=3$ in \cite{Bzowski:2012ih,Bzowski:2013sza,Bzowski:2015yxv}}
\begin{equation}
    \mathcal{I}_d\begin{bmatrix}
        \mu_1 &\ldots &\mu_N \\ k_1 &\ldots & k_N \\ K &\ldots & K
    \end{bmatrix}
    = \int \limits^\infty_0\d z\; z^{\frac{d(N-2)}{2}-1}\prod_{j=1}^N K_{i\mu_j}(k_j z) \,,
    \label{def_vertex_function_K}
\end{equation}
The subscript $d$ accounts for the spatial dimension. The letters $K$ in the third row of indices denote that the Euclidean time dependence attached to leg $j$ is given by the modified Bessel function of the second kind $K_{i\mu_j}$. The presence of these indices arises from the definition of the following generalisation of \eqref{def_vertex_function_K}:
\begin{equation}
    \mathcal{I}_d\begin{bmatrix}
        \mu_1 &\ldots &\mu_N \\ k_1 &\ldots & k_N \\ A_1 &\ldots & A_N
    \end{bmatrix}
    = \int \limits^\infty_0\d z\; z^{\frac{d(N-2)}{2}-1}\prod_{j=1}^N \left(A_j\right)_{i\mu_j}(k_j z) \,,
    \label{def_vertex_function_general}
\end{equation}
where the letter $A_j$ can be $I$ or $K$, respectively for the modified Bessel function of the first or second kind \eqref{def_bessel_I}-\eqref{connection_formula_bessel}. This last definition will prove useful for reasons that will become clear in the next subsection. For instance, the vertex function \eqref{def_vertex_function_K} that enters a vertex in \eqref{eq_feynman_rule_vertex} has $A_j=K$, for every $j\in\{1,\ldots,N\}$.

These vertex functions can also be represented in terms of Lauricella hypergeometric series \cite{Lauricella1893,Matsumoto_2020}. This hypergeometric representation will be given and used in the next subsection to investigate the analytical properties of vertex functions that we will use in the rest of this work.

Let us add for completeness that frequencies and momenta attached to internal lines must be integrated over, using \eqref{eq_KLF_measure}. For a tree-level diagram, momentum integrals straightforwardly give momentum conservation relations. However, integrals over KLF frequencies are non-trivial, as an imprint of broken time translation symmetry.

\paragraph{Late-time $\dS$ correlators.} Finally, gathering propagators and interaction terms, performing the inverse KLF transform and the Wick rotation back, the full contribution to a $\dS_{d+1}$ boundary correlator \eqref{bdy_correlator_Fourier} with $n$ fields of momentum $\mathbf{k}_i$ and mass $\mu_i$ is given by
\begin{equation}\label{eq_general_late_time_correlator}
     \mathcal{C}_B\left(\{\mathbf{k}_i\}\right)= (2\pi)^d\delta^{(d)}\left(\sum_{j=1}^n\k_j\right)
     \sum_{\{\a^V_E\}}\E_{\{\a_E^V\}}(\tau_0)\G^A_{\{\a^V_E\}}\left(^{\mu_{1}\ldots\mu_{_n}}_{\k_1\;\ldots\;\k_n}\right)\;.
\end{equation}
To recover the full correlator $\C_B$, the sum is taken over in-in indices of external vertices $\{\a_E^V\}$, i.e. vertices attached to at least one external leg (the sum over internal indices is implicit in the amputated correlator). The factor $\E_{\{\a_E^V\}}(\tau_0)$, that encodes the bulk-to-boundary propagator dependence, is given in general by equation (4.67) of \cite{Belrhali:2026rkn}, however here we will focus on the case where all external legs carry conformally coupled scalar fields. Then, it reduces to
\begin{equation}\label{eq_Eps_contribution_external_legs_cc}
    \lim_{\tau_0\to 0}\E_{\{\a_E^V\}}(\tau_0) = \frac{H^{\frac{n(d-3)}{2}}}{2^{\frac{n}{2}}}\frac{(-\tau_0)^{\frac{n(d-1)}{2}}}{\sqrt{k_1\ldots k_n}}e^{-\frac{i\pi(d-1)}{4}\sum_{j=1}^{n_E} n_j\a_j}\,,
\end{equation}
where $\a_j\in\{\a^V_E\}$, $n_j$ is the number of external legs attached to the external vertex labelled by $\a_j$, $n_E$ is the number of external vertices and we then have $\sum_{j=1}^{n_E}n_j=n$\footnote{The sum of external legs attached to each external vertex is the number of external legs of the correlator.}.
The amputated KLF correlator $\G^A_{\{\a^V_E\}}$ in the expression \eqref{eq_general_late_time_correlator}, which is the non-trivial part in perturbative computations, is given by removing the external legs from the original diagram and sending the corresponding KLF frequencies appearing in vertex functions to their on-shell value. It is possible to write its expression using the rules summarised in this subsection (propagators, vertices, and integration over KLF variables). In section \ref{section_the_double_exchange_diagram}, we will apply these rules to the case of the double-exchange diagram.

\subsection{Analytical Properties of Vertex Functions}\label{vertex_fct_N}

As seen previously, in any perturbative computation performed in KLF variables, each vertex contributes a vertex function \eqref{def_vertex_function_K}. Since at tree-level, integration over momentum is immediate and gives momentum conservation, only integrals over internal KLF frequencies remain and form the non-trivial piece to compute. To perform these integrals, our aim is to use complex analysis, in particular the Cauchy residue theorem. Therefore, we have to integrate over closed contours in the complex frequency plane and collect residues of the integrand, whose non-trivial part is given by the vertex functions \eqref{def_vertex_function_K}. That is why we need to know the two following analytical properties of vertex functions: the location of their singularities with the corresponding value of the residue, and their asymptotic behaviour when one frequency is taken to be large (in order to choose the right path to close the integration contour), in any direction in the complex plane.

\paragraph{Series representation of the vertex function.} The integral representation that defines the vertex functions \eqref{def_vertex_function_K} can be divergent for some complex values of the different frequencies $\{\mu_j\}$.
Indeed, this is due to the $z\sim 0$ leading behaviour of modified Bessel functions \eqref{small_z_bessel_K}\footnote{The integral is always well defined at $z\rightarrow+\infty$ because of the decreasing exponential asymptotic behaviour of $K_{i\mu}$ \eqref{large_z_bessel_K}.}.

Then, to perform the frequency integrals we need to analytically continue the function $\mu_k\mapsto \mathcal{I}_d\left[\{\mu_j\},\{k_j\},\{K\}\right]$ into the regions of the frequency complex plane where the integral \eqref{def_vertex_function_K} diverges and then investigate the location of singularities. Such an analytic continuation can be obtained by writing the modified Bessel functions $K_{i\mu_j}$ in terms of their Mellin-Barnes integral representation \eqref{eq_MellinB_K}, exchanging the order of integration by performing the Euclidean time integral first and then the Mellin-Barnes integrals using the Cauchy residue theorem. The detailed derivation is given in Appendix  \ref{appendix_derivation_vertex_fct} (see also \cite{Liu:2024str}). It leads to the following representation of $\mathcal{I}_d\left[\{\mu_j\},\{k_j\},\{K\}\right]$:  
\begin{multline}
    \mathcal{I}_d\begin{bmatrix}
        \mu_1 &\ldots &\mu_N \\ k_1 &\ldots & k_N \\ K &\ldots & K
    \end{bmatrix} = 
    \frac{(i\pi)^{N-1}}{2^{N+1}}\left(\frac{2}{k_N}\right)^{\frac{d(N-2)}{2}}\sum_{c_1,\dots,c_{N-1}=\pm} \prod_{j=1}^{N-1}\Biggl[\frac{c_j}{\sinh(\pi\mu_j)}\left(\frac{k_j}{k_N}\right)^{i c_j\mu_j}\Biggr]\\
    \times
    \Gamma\biggl(\frac{d(N-2)}{4}+\sum_{j=1}^{N-1} \frac{i c_j \mu_j}{2}\pm \frac{i\mu_N}{2}\biggr)
    \\\times\F_C^{(N-1)}\Biggl(\begin{matrix}
    \frac{d(N-2)}{4}+\sum\limits_{j=1}^{N-1} \frac{i c_j \mu_j}{2}+ \frac{i\mu_N}{2}\,,\;  \frac{d(N-2)}{4}+\sum\limits_{j=1}^{N-1} \frac{i c_j \mu_j}{2}- \frac{i\mu_N}{2}\\ 1+i c_1\mu_1\,,\;\ldots\,,\;1+i c_{N-1}\mu_{N-1}
    \end{matrix};\left(\frac{k_1}{k_N}\right)^2,\ldots,\left(\frac{k_{N-1}}{k_N}\right)^2\Biggr)\,.
    \label{eq_vertex_function_vf}
    \end{multline}
written in terms of the dressed Lauricella function $\F_C^{(N-1)}$ \eqref{eq_def_Lauricella_C}-\eqref{eq_def_dressed_Lauricella_C} \cite{Lauricella1893,Matsumoto_2020}, where
\begin{equation}
    \Gamma(a\pm b)\equiv \Gamma(a+b)\Gamma(a-b)\,.
    \label{eq_convention_Gamma_fct}
\end{equation}
In terms of momentum variables, the original integral representation \eqref{def_vertex_function_K} is convergent for $\sum_{j=1}^N k_j>0$, i.e. in every physical configuration. However, the convergence domain of the Lauricella function $F_C^{(N-1)}$ \eqref{eq_cvg_radius_Lauricella_FC} implies the following condition on kinematic variables for the use of the representation \eqref{eq_def_Lauricella_C}:
\begin{equation}
    \sum_{j=1}^{N-1} k_j<k_N\,.
    \label{eq_cvg_radius_kinematics}
\end{equation}
Since the series representation \eqref{eq_def_Lauricella_C}-\eqref{eq_def_dressed_Lauricella_C} is very useful to identify singularities and compute residues as we will see in the following, in the rest of this work, we will suppose \eqref{eq_cvg_radius_kinematics} to be satisfied when integrating over frequencies.

Before going into the singularity structure, let us introduce useful conventions that are the connection formulas for the vertex functions.

\paragraph{Connection formulas.} Let us see how connection formulas for vertex functions emerge from the series representation \eqref{eq_vertex_function_vf}. These formulas will prove useful when studying the singularities and deriving the large-frequency asymptotic behaviours. The connection formulas for vertex functions are defined from the integral representation \eqref{def_vertex_function_K} by using the connection formula between modified Bessel functions \eqref{connection_formula_bessel} and writing it at the integral level. For instance, considering \eqref{connection_formula_bessel} for $\mu_1$ leads to
\begin{equation}
    \mathcal{I}_d\begin{bmatrix}
        \mu_1 &\mu_2&\ldots &\mu_N \\ k_1 &k_2&\ldots & k_N \\ K & K&\ldots & K
    \end{bmatrix}=\frac{i\pi}{2}\frac{1}{\sinh(\pi\mu_1)}\left(
    \mathcal{I}_d\begin{bmatrix}
        \mu_1 &\mu_2&\ldots &\mu_N \\ k_1 &k_2&\ldots & k_N \\ I & K&\ldots & K
    \end{bmatrix}-
    \mathcal{I}_d\begin{bmatrix}
        -\mu_1 &\mu_2&\ldots &\mu_N \\ k_1 &k_2&\ldots & k_N \\ I & K&\ldots & K
    \end{bmatrix}\right)
    \label{eq_connection_formula_bdy}
\end{equation}
where the vertex functions that appear in the right-hand side have the integral representation \eqref{def_vertex_function_general} with $A_1=I$ and $A_j=K$, $J\in\{2,\ldots,N\}$\footnote{Concerning the convergence properties of the integrals \eqref{def_vertex_function_general}, the large $z$ asymptotic behaviour of $I_{i\mu_j}(z)$/$K_{i\mu_j}(z)$ \eqref{large_z_bessel_I}-\eqref{large_z_bessel_K} imposes a kinematic condition depending on each $A_j=I/K$. However, since we assume \eqref{eq_cvg_radius_kinematics} in KLF computations, these kinematic conditions are always satisfied as long as $A_N=K$.}.
Their series representation can be straightforwardly deduced from that of $\mathcal{I}_d\left[\{\mu_j\},\{k_j\},\{K\}\right]$, since the sum of the two terms and the $1/\sinh(\pi\mu_1)$ factor are explicit in \eqref{eq_vertex_function_vf}, we thus have, defining $\alpha_\pm\equiv\frac{d(N-2)}{4}+\frac{i \mu_1}{2}+\sum_{j=2}^{N-1} \frac{i c_j \mu_j}{2}\pm \frac{i\mu_N}{2}$,
\begin{equation}
\begin{aligned}
    &\mathcal{I}_d\begin{bmatrix}
        \mu_1 & \mu_2&\ldots &\mu_N \\ k_1 &k_2&\ldots & k_N \\ I &K&\ldots & K
    \end{bmatrix} = 
    \frac{(i\pi)^{N-2}}{2^{N}}\left(\frac{2}{k_N}\right)^{\frac{d(N-2)}{2}}
    \left(\frac{k_1}{k_N}\right)^{i \mu_1} \sum_{c_2,\dots,c_{N-1}=\pm} \prod_{j=2}^{N-1}\Biggl[\frac{c_j}{\sinh(\pi\mu_j)}\left(\frac{k_j}{k_N}\right)^{i c_j\mu_j}\Biggr]\\
    &\times \Gamma\!\left(\alpha_\pm\right) \F_C^{(N-1)}\Biggl(\begin{matrix}
    \alpha_+\,,\;\alpha_-\\ 1+i\mu_1\,,\,1+i c_2\mu_2\,,\;\ldots\,,\;1+i c_{N-1}\mu_{N-1}
    \end{matrix};\left(\frac{k_1}{k_N}\right)^2,\ldots,\left(\frac{k_{N-1}}{k_N}\right)^2\Biggr)\,.
    \end{aligned}
    \label{eq_vertex_function_IK...K}
\end{equation}
This procedure can be iterated for $j=2,\ldots,N-1$ to derive connection formulas with respect to every frequency. The general connection formula among vertex functions is derived in Appendix \ref{appendix_connection_formula}.

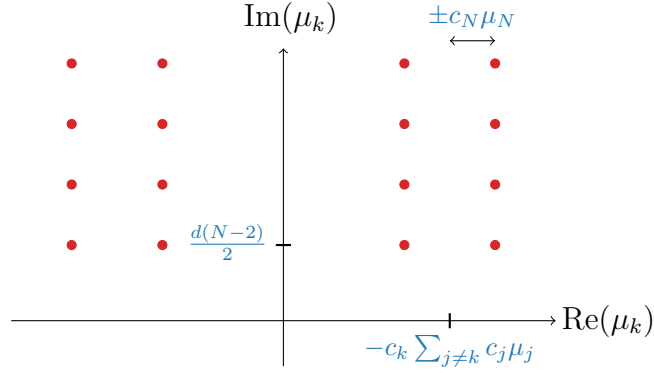
\begin{figure}\centering
\hspace{0cm}
    	\begin{tikzpicture}[scale = 2]

        \draw[black, ->] (-1.8,0) -- (1.8,0) coordinate (xaxis);
		\draw[black, ->] (0,-0.3) -- (0,1.8) coordinate (yaxis);
		\node at (2.15, 0) {$\text{Re}(\mu_k)$};
		\node at (0.05, 2) {$\text{Im}(\mu_k)$};
        
        \draw[pyred, fill = pyred] (1.4, 0.5) circle (.03cm);
		\draw[pyred, fill = pyred] (1.4, 0.9) circle (.03cm);
		\draw[pyred, fill = pyred] (1.4, 1.3) circle (.03cm);
		\draw[pyred, fill = pyred] (1.4, 1.7) circle (.03cm);
        \draw[pyred, fill = pyred] (.8, 0.5) circle (.03cm);
		\draw[pyred, fill = pyred] (.8, 0.9) circle (.03cm);
		\draw[pyred, fill = pyred] (.8, 1.3) circle (.03cm);
		\draw[pyred, fill = pyred] (.8, 1.7) circle (.03cm);

        \draw[pyred, fill = pyred] (-1.4, 0.5) circle (.03cm);
		\draw[pyred, fill = pyred] (-1.4, 0.9) circle (.03cm);
		\draw[pyred, fill = pyred] (-1.4, 1.3) circle (.03cm);
		\draw[pyred, fill = pyred] (-1.4, 1.7) circle (.03cm);
        \draw[pyred, fill = pyred] (-.8, 0.5) circle (.03cm);
		\draw[pyred, fill = pyred] (-.8, 0.9) circle (.03cm);
		\draw[pyred, fill = pyred] (-.8, 1.3) circle (.03cm);
		\draw[pyred, fill = pyred] (-.8, 1.7) circle (.03cm);
        \node at (1.1, -.2) {\textcolor{pyblue}{\footnotesize$-c_k\sum_{j\neq k} c_j\mu_j$}};
        \node at (1.25, 2) {\textcolor{pyblue}{\footnotesize$\pm c_N\mu_N$}};
        \node at (-.37,0.52) {\textcolor{pyblue}{\footnotesize$\frac{d(N-2)}{2}$}};
		\draw[black, <->] (1.1,1.85) -- (1.4,1.85);
		\draw[black,thick, -] (1.1,-.05) -- (1.1,.05);
		\draw[black,thick, -] (-.05,.5) -- (.05,.5);
    \end{tikzpicture}
\caption{Pole structure in the $\mu_k$ complex plane of the function $\mu_k\mapsto\mathcal{I}_d\left[\{\mu_j\},\{k_j\},\{K\}\right]$, which has the expression given by \eqref{eq_vertex_function_vf}.}
\label{fig_poles_of_IKK_N}
\end{figure}

\paragraph{Singularities and residues.}
From the representation \eqref{eq_vertex_function_vf}, it can be shown that singularities of the vertex function $\mathcal{I}_d\left[\{\mu_j\},\{k_j\},\{K\}\right]$ only come from the product of two $\Gamma$-functions. 
Indeed, $\F_C^{(N-1)}$ is analytic with respect to $\mu_1,\ldots,\mu_N$ (see Appendix \ref{appendix_def_special_functions}) and similarly for the power prefactors. Moreover, there are no singularities coming from the divergences of $\mu_k\mapsto1/\sinh(\pi\mu_k)$, as proven in Appendix \ref{subappendix_poles_residues_vertex_fct_general}. Consequently, the only singularities that appear in \eqref{eq_vertex_function_vf} are simple poles with location and corresponding residues obtained from the well-known properties of the $\Gamma$-function \eqref{eq_poles_residue_gamma}. To summarise, the function $\mu_k\mapsto\mathcal{I}_d\left[\{\mu_j\},\{k_j\},\{K\}\right]$ has poles at
\begin{equation}
    \mu_k=c_k\left(\tilde{\mu}_{k,N}+i\left(2 p_k+\frac{d(N-2)}{2}\right)\right)\;,\; p_k\in\mathbb{N}\,,
    \label{eq_pole_vertex_fct_N_body}
\end{equation}
where
\begin{equation}
    \tilde{\mu}_{k,N}\equiv -\sum_{j=1,j\neq k}^{N-1} c_j\mu_j - c_N \mu_N \,.
    \label{eq_tilde_mu_k_N}
\end{equation}
Here we have $c_j=\pm$ for every $j\in\{1,\ldots,N\}$. However, note that compared to previous expressions we additionally introduced the index $c_N$ that selects the factor in the $\Gamma$-function product where the singularity comes from, and that the choice of $c_k$ in \eqref{eq_pole_vertex_fct_N_body} similarly fixes one term in the corresponding sum in \eqref{eq_vertex_function_vf}.
The distribution of these singularities is represented in Fig. \ref{fig_poles_of_IKK_N}. 
The corresponding residues are
\begin{multline}
\text{Res}\Biggl(\mu_k \mapsto\mathcal{I}_d\left[\{\mu_j\},\{k_j\},\{K\}\right], \mu_k=c_k\biggl(\tilde{\mu}_{k,N}+i\Bigl(2 p_k+\frac{d(N-2)}{2}\Bigl)\biggl)\;,\; p_k\in\mathbb{N}\Biggr)\\
    = 
    \frac{(i\pi)^{N-1}}{2^{N+1}}\left(\frac{2}{k_N}\right)^{\frac{d(N-2)}{2}}\sum_{\substack{c_1,\dots,c_{k-1},\\c_{k+1},\ldots,c_{N-1}=\pm}} \prod_{j=1,j\neq k}^{N-1} \Biggl[\frac{c_j}{\sinh(\pi\mu_j)}\left(\frac{k_j}{k_N}\right)^{i c_j\mu_j}\Biggr]
    \\
    \times(-2i c_k) \frac{(-1)^{p_k}}{p_k !}
    \frac{ \Gamma\left(-p_k-i c_N \mu_N\right)}{\sinh\left(\pi\left(\tilde{\mu}_{k,N}+i\Bigl(2 p_k+\frac{d(N-2)}{2}\Bigl)\right)\right)}\left(\frac{k_k}{k_N}\right)^{i \tilde{\mu}_{k,N}-\Bigl(2 p_k+\frac{d(N-2)}{2}\Bigl)}\\
    \times\F_C^{(N-1)}\Biggl(\begin{matrix}
    -p_k\,,\;  -p_k-i c_N \mu_N \\ 1+i c_1\mu_1\,,\;\ldots\,,\;1+i\tilde{\mu}_{k,N}-2 p_k -\frac{d(N-2)}{2}\,,\;\ldots
    \end{matrix};\left(\frac{k_1}{k_N}\right)^2,\ldots,\left(\frac{k_{N-1}}{k_N}\right)^2\Biggr)\,,
    \label{eq_general_N_residue_vertex_fct}
    \end{multline}
In this last expression, $c_j=\pm$ for every $j\in\{1,\ldots,N\}$, but recall that $c_k$ and $c_N$ are fixed, and we still sum over $c_j=\pm$ for $j\in\{1,\ldots,N\}/\{k,N\}$.

In practical computations, after performing an integration layer over $\mu_k$, we want to integrate again on other frequencies. To this end, we need to use the vertex function analytic structure discussed here for the other internal lines that carry a different frequency $\mu_j$, $j\neq k$, but we must also investigate the singularity structure in the $\mu_j$ plane of the residues \eqref{eq_general_N_residue_vertex_fct} coming from integration over $\mu_k$. In fact, this structure is simple, since it has only simple poles arising from the $1/\sinh\left(\pi\left(\tilde{\mu}_{k,N}+i\left(2 p_k+d(N-2)/2\right)\right)\right)$ factor\footnote{In some particular cases, there can be degeneracies such that the singularities of $1/\sinh\left(\pi\left(\tilde{\mu}_{k,N}+i\left(2 p_k+d(N-2)/2\right)\right)\right)$ in the $\mu_j$ plane are cancelled by zeros of the dressed Lauricella function. It will be the case for instance for the double-exchange diagram studied in sec. \ref{section_the_double_exchange_diagram}.} (recall that $\tilde{\mu}_{k,N}$ is given by \eqref{eq_tilde_mu_k_N}), since $\mathcal{F}_C^{(N-1)}$ is analytic with respect to all its parameters $a,b,c_1,\ldots,c_{N-1}$ (see \eqref{eq_def_dressed_Lauricella_C}). The asymptotic behaviour of the residues at large $\mu_j$ is given in Appendix \ref{subappendix_asymptotic_behavior_N}.

Finally, let us add that for certain conditions on the parameters entering the vertex functions (i.e. $d$,$N$, frequencies $\mu_j$ entering \eqref{eq_vertex_function_vf} from external legs), a singularity \eqref{eq_pole_vertex_fct_N_body} can be cancelled by a zero of the dressed Lauricella function\footnote{See for instance sec. 5.2 and Appendix F of \cite{Belrhali:2026rkn} for $d$ even, involving the dressed hypergeometric function $_2\mathcal{F}_1$ \eqref{eq_2F1_dressed_def}}. Details about this feature can be found in Appendix \ref{subappendix_poles_residues_vertex_fct_general}.

\paragraph{Large-frequency asymptotic behaviour.}
The last property of the vertex functions that we need in order to compute KLF frequency integrals is their large $\mu_k$ asymptotic behaviour, which provides a suitable contour prescription. This asymptotic behaviour is derived from the integral representation of vertex functions \eqref{def_vertex_function_general}. Details about this derivation can be found in Appendix \ref{subappendix_asymptotic_behavior_N}.
To have a convenient asymptotic behaviour at large $\mu_k$ for contour prescription, it turns out that we need to consider the vertex functions with $A_k=I$. The other indices $A_j$ can be $K$ or $I$, except for $j=N$ where we stick to $A_N=K$.
In this case, the expression of the leading-order behaviour is
\begin{equation}
\begin{aligned}
\mathcal{I}_d&\begin{bmatrix}
        \mu_1 &\ldots &\mu_k&\ldots&\mu_N \\ k_1 &\ldots&k_k&\ldots  & k_N \\ A_1 &\ldots & I &\ldots & K
    \end{bmatrix}\underset{\mu_k\rightarrow\infty}{\sim}
    \frac{\pi^{\frac{1}{2}\left(\lvert\K\rvert-\lvert\I\rvert\right)}}{2^{\frac{N-1}{2}} k_k^{\frac{(d-1)(N-2)-1}{2}}
    \prod\limits_{j=1,j\neq k}^N \sqrt{k_j}}
    \left(i\mu_k\right)^{\frac{(d-1)(N-2)-3}{2}} \\
    &\times 
    \sum_{d_l=0,1;l\in\I} \left(i e^{-\pi\mu_l}\right)^{d_l}\left(\left(\frac{k_t\left(\{d_l\}\right)}{k_k}\right)^2-1\right)^{\frac{1-(d-1)(N-2)}{4}} e^{-i\mu_k \arccosh\left(\frac{k_t\left(\{d_l\}\right)}{k_k}\right)}
    \,,
\end{aligned}
    \label{eq_vertex_function_app_large_mu_vf}
\end{equation}
where the sets $\I$ and $\K$ are defined in \eqref{eq_asymptotics_sets_IK} and $k_t\left(\{d_l\}\right)$ in \eqref{eq_kt_dl}.
The asymptotic behaviour of the vertex function with $A_k=K$ can be deduced from this last expansion by using the connection formulas with respect to $\mu_k$ (see Appendix \ref{subappendix_poles_residues_vertex_fct_general}).

Before turning to an explicit computation in the next section, let us also mention that the asymptotic behaviour of residues \eqref{eq_general_N_residue_vertex_fct} at any large $\mu_j$ can be deduced from the Lauricella function properties (see Appendix \ref{subappendix_asymptotic_behavior_N}).

\begin{FramedBox}
\vspace{-.4cm}
\paragraph{Summary.} This section gathers the information needed to compute a tree-level correlator with the KLF formalism. 
\begin{itemize}
    \item A correlator can be written according to the diagrammatic rules presented in \ref{subsec_intro_KLF}, the non-trivial part being given in terms of frequency integrals over vertex functions \eqref{def_vertex_function_K}.
    \item To perform these integrals, one can use complex analysis and the residue theorem, given that the integrands have only simple poles \eqref{eq_pole_vertex_fct_N_body} in the complex KLF frequency plane.
    \item Each frequency integral can be rewritten using connection formulas for the vertex functions (e.g. \eqref{eq_connection_formula_bdy}, see also \ref{appendix_connection_formula}), in order to have tractable large-frequency asymptotic behaviour.
    \item The knowledge of these asymptotic behaviours \eqref{eq_vertex_function_app_large_mu_vf} provides a contour closure prescription. The latter can be used to collect residues of the integrand components, originating from vertex functions \eqref{eq_general_N_residue_vertex_fct} and from the $i\epsilon$-prescription in KLF propagators (see \eqref{eq_i_eps_residues}) to find hypergeometric representations of a correlator.
\end{itemize}
\end{FramedBox}

To be more concrete, in the next section we turn to the complete study of an explicit non-trivial example of a double-exchange diagram.

\section{The Double-Exchange Diagram}\label{section_the_double_exchange_diagram}

In this section, we illustrate the KLF formalism presented previously by performing the computation of the following diagram, which we will refer to as the double-exchange diagram:
\begin{equation}
    \begin{aligned}
        \mathcal{G}_{\a_1\a_2\a_3}^{\mu_1,\mu_2} \left(\{k_i,s_j\}\right)\equiv&\begin{tikzpicture}[baseline={(0,0.3)}]
            \draw[thick,black] (-1.9,1) to (-1.5,0);
            \draw[thick,black] (-1.1,1) to (-1.5,0);
            \draw[thick,black] (0.1,1) to (0.5,0);
            \draw[thick,black] (.9,1) to (0.5,0);
            \draw[thick,black] (2.1,1) to (2.5,0);
            \draw[thick,black] (2.9,1) to (2.5,0);
            \draw[thick,pyblue] (-1.5,0) to (0.5,0);
            \draw[thick,pyblue] (.5,0) to (2.5,0);
            \node[black] at (-1.5,-.3) {$\a_1$};
            \node[black] at (0.5,-.3) {$\a_2$};
            \node[black] at (2.5,-.3) {$\a_3$};           \filldraw[color=black,fill=black] (-1.5,0) circle (2pt);
            \filldraw[color=black,fill=black] (0.5,0) circle (2pt);
            \filldraw[color=black,fill=black] (2.5,0) circle (2pt);
            \filldraw[color=black,fill=white] (-1.95,1) rectangle ++(4pt,4pt);
            \filldraw[color=black,fill=white] (-1.15,1) rectangle ++(4pt,4pt);
            \filldraw[color=black,fill=white] (.05,1) rectangle ++(4pt,4pt);
            \filldraw[color=black,fill=white] (.85,1) rectangle ++(4pt,4pt);
            \filldraw[color=black,fill=white] (2.05,1) rectangle ++(4pt,4pt);
            \filldraw[color=black,fill=white] (2.85,1) rectangle ++(4pt,4pt);
            \node[black] at  (-1.95,1.4) {$k_1$};
            \node[black] at  (-1.15,1.4) {$k_2$};
            \node[black] at  (.05,1.4) {$k_3$};
            \node[black] at  (.85,1.4) {$k_4$};
            \node[black] at  (2.05,1.4) {$k_5$};
            \node[black] at  (2.85,1.4) {$k_6$};
            \node[pyblue] at  (-0.5,.2) {$\nu,s_1,\mu_1$};
            \node[pyblue] at  (1.5,.2) {$\mu,s_2,\mu_2$};
        \end{tikzpicture}\,.
\end{aligned}\label{draw_double_exch_diagram_general}
\end{equation}
In the following, we consider that the spatial dimension $d$ is not an even integer, that case being reachable by analytic continuation at the end of the calculation.

Let us first define the notations that are used in this section for this diagrammatic computation, and that are sketched in \eqref{draw_double_exch_diagram_general}. The variables $\mu$ and $\nu$ are the KLF frequencies over which we will integrate, following the method described in the previous section. Concerning kinematic variables, $k_i$ ($i\in\{1,\ldots,6\}$) are the momenta flowing in the external lines, each carrying a conformally coupled scalar field. The momenta $s_1$ and $s_2$ flow in the first and second internal lines, respectively, and $\mu_1$, $\mu_2$ are the on-shell frequencies of the exchanged fields (which are related to their mass). The indices $\a_1,\a_2,\a_3$ attached to each vertex are the Schwinger-Keldysh indices, that can be either $+$ or $-$.
This diagram can arise from the following interactions
\begin{equation}
    \lambda \,\varphi_i \varphi_j \sigma_k\,\subset\, \L_{\text{int}}\,,
\end{equation}
for the left and right vertices, where $\varphi$ denotes a conformally coupled field and $\sigma$ a scalar field with arbitrary mass. The central vertex can come from
\begin{equation}
    g \,\varphi_i \varphi_j \,\sigma_k \sigma_l\,\subset\, \L_{\text{int}}\,.
\end{equation}

Using the KLF diagrammatic rules and the late-time correlator general expression \eqref{eq_general_late_time_correlator} with fixed in-in indices $\a_1$, $\a_2$, $\a_3$, the double-exchange diagram can be expressed as
\begin{equation}
    \mathcal{G}_{\a_1\a_2\a_3}^{\mu_1,\mu_2} \left(\{k_i,s_j\}\right)= (2\pi)^d\delta^{(d)}\left(\sum_{i=1}^6\k_i\right)
    \E_{\a_1,\a_2,\a_3}(\tau_0)\; \G^{A,\mu_1,\mu_2}_{\a_1\a_2\a_3} \left(\{k_i,s_j\}\right)\;,
    \label{eq_relation_G_Gamputated}
\end{equation}
where $i=1,\ldots,6$ and $j=1,2$. The prefactor $\E_{\a_1,\a_2,\a_3}$ is obtained from \eqref{eq_Eps_contribution_external_legs_cc} with $n=6$, $n_E=3$ and $n_j=2$ for $j=1,2,3$:
\begin{equation}
    \E_{\a_1,\a_2,\a_3}(\tau_0)=\frac{H^{3(d-3)}}{2^3}\frac{(-\tau_0)^{3(d-1)}}{\sqrt{k_1\ldots k_6}}e^{-\frac{i\pi(d-1)}{2}(\a_1+\a_2+\a_3)}\,.
    \label{eq_eps_factor}
\end{equation}
Using the diagrammatic rules, the amputated diagram $\G^A$ is given by
\begin{equation}
\begin{aligned}
\G^{A,\mu_1,\mu_2}_{\a_1\a_2\a_3} &\left(\{k_i,s_j\}\right)=
    i\,\a_1\a_2\a_3\,\lambda_1 \lambda_2 \lambda_3\,(2\pi)^{3d} \delta^{(d)}\left(\k_{12}+\boldsymbol{s}_1\right) \delta^{(d)}\left(\k_{34}+\boldsymbol{s}_{12}\right) \delta^{(d)}\left(\boldsymbol{s}_2+\k_{34}\right)\\
    &\times\frac{H^{2(d-1)} e^{(\a_1+\a_2+\a_3)\frac{i \pi d}{2}}}{\pi^5}
    \int\limits_{-\infty}^{+\infty}\d\nu \, \N_{\nu}\;
    \int\limits_{-\infty}^{+\infty}\d\mu \, \N_{\mu}\,
    \mathcal{I}_d\begin{bmatrix}
        \nu& i/2 & i/2 \\ s_1&k_1 & k_2  \\ K & K & K
    \end{bmatrix}
    \Pi^{\mu_1}_{\a_1\a_2}(\nu)\\
    &\times
    \mathcal{I}_d\begin{bmatrix}
        \nu &\mu &i/2 & i/2 \\ s_1 & s_2 & k_3 & k_4  \\ K &K & K & K
    \end{bmatrix}
    \Pi^{\mu_2}_{\a_2\a_3}(\mu)\,
    \mathcal{I}_d\begin{bmatrix}
        \mu &i/2 & i/2  \\ s_2 &k_5 & k_6 \\ K & K & K
    \end{bmatrix}\,,
    \end{aligned}
    \label{eq_amputated_double_exchange}
\end{equation}
where $\k_{ij}\equiv \k_i+\k_j$ (and similarly $\boldsymbol{s}_{ij}=\boldsymbol{s}_i+\boldsymbol{s}_j$) and we set the frequencies of external legs to $\mu_{c.c}=i/2$. Recall that $\Pi_{\mathrm{\a_1\a_2}}^{\mu_1}$ and $\Pi_{\mathrm{\a_2\a_3}}^{\mu_2}$ are massive internal propagators \eqref{eq_def_reduced_propagator}, and $\mathcal{I}_d$ are the vertex functions. From \eqref{eq_relation_G_Gamputated}, the expression for the full diagram $\mathcal{G}_{\a_1\a_2\a_3}^{\mu_1,\mu_2}$ \eqref{draw_double_exch_diagram_general} then becomes
\begin{equation}
\begin{aligned}
    \mathcal{G}_{\a_1\a_2\a_3}^{\mu_1,\mu_2} &\left(\{k_i,s_j\}\right)= i^{1+\a_1+\a_2+\a_3} \a_1\a_2\a_3\,\lambda_1 \lambda_2 \lambda_3\,\frac{H^{5d-11}}{8 \pi^5}\frac{(-\tau_0)^{3(d-1)}}{\sqrt{k_1\ldots k_6}}\\
    &\times(2\pi)^{4d} \delta^{(d)}\left(\sum_{i=1}^6\k_i\right)
    \delta^{(d)}\left(\k_{12}+\boldsymbol{s}_1\right) \delta^{(d)}\left(\k_{34}+\boldsymbol{s}_{12}\right) \delta^{(d)}\left(\boldsymbol{s}_2+\k_{34}\right)\\
    &\times
    G_{\a_1,\a_2,\a_3} \left(\{k_i,s_j\}\right)\;,
    \end{aligned}
    \label{eq_G_a1a2a3}
\end{equation}
where we also used the expression for the $\E_{\a_1,\a_2,\a_3}(\tau_0)$ factor \eqref{eq_eps_factor}. The quantity $G^{\mu_1,\mu_2}_{\a_1\a_2\a_3}$, which is the non-trivial piece that we will compute in the following, is given by
\begin{equation}
\begin{aligned}
    G_{\a_1\a_2\a_3} &\left(\{k_i,s_j\}\right)
    \equiv\int\limits_{-\infty}^{+\infty}\d\nu \, \N_{\nu}
    \int\limits_{-\infty}^{+\infty}\d\mu \, \N_{\mu}
    \;
    \mathcal{I}_d\begin{bmatrix}
        \nu &i/2 & i/2 \\ s_1 &k_1 & k_2 \\ K & K & K
    \end{bmatrix} \Pi^{\mu_1}_{\a_1\a_2}(\nu)\\
    &\times
    \mathcal{I}_d\begin{bmatrix}
        \nu &\mu &i/2 & i/2 \\ s_1 & s_2 &k_3 & k_4 \\ K &K & K & K
    \end{bmatrix}\Pi^{\mu_2}_{\a_2\a_3}(\mu)\;
    \mathcal{I}_d\begin{bmatrix}
        \mu &i/2 & i/2 \\ s_2 &k_5 & k_6 \\ K & K & K
    \end{bmatrix}\,,
    \end{aligned}
    \label{eq_G_integrals_a1a2a3}
\end{equation}
where in the label $G_{\a_1,\a_2,\a_3}$ we omit the reference to the on-shell frequencies of the exchanged fields $\mu_1$ and $\mu_2$ or to any momentum variable for clarity.

We have shown that computing the diagram \eqref{draw_double_exch_diagram_general} amounts now to performing the integral \eqref{eq_G_integrals_a1a2a3}. To this end, we start by gathering all the relevant analytical properties of the KLF integrand (in subsection \ref{subsection_analytics_properties}). Then, we focus on the computation of the most involved piece of the diagram, given by the component $\mathcal{G}_{+++}$ (in \ref{subsection_+++}). Then we turn to the computation of the partially-factorised component $\mathcal{G}_{-++}$ (\ref{subsection_-++}) and of the totally factorised one $\mathcal{G}_{+-+}$ (\ref{subsection_+-+}). The remaining components of the diagram \eqref{draw_double_exch_diagram_general} can be obtained by complex conjugation or symmetry by exchanging kinematic parameters associated with vertices labelled by $\a_1$ and $\a_3$. In subsection \ref{subsection_full_result}, we gather all the results to obtain an expression for the full correlator.

\subsection{Analytical Properties of the KLF Integrand}\label{subsection_analytics_properties}

Before turning to the computation of the correlator components, let us gather in this subsection the relevant analytical properties of the different vertex functions that appear in \eqref{eq_G_integrals_a1a2a3}.

\subsubsection{Vertex function with one massive line}

The analytical properties of the vertex function that corresponds to left and right vertices in \eqref{draw_double_exch_diagram_general} have already been studied in \cite{Belrhali:2026rkn} in the case of a single-exchange diagram. We recall here the relevant ones for the present computation. From \eqref{def_vertex_function_K} for $N=3$, the integral representation gives
\begin{equation}
    \mathcal{I}_d\begin{bmatrix}
        \nu &i/2 & i/2 \\ s_1 &k_1 & k_2 \\ K & K & K
    \end{bmatrix}=
    \int \limits^\infty_0\d z\; z^{\frac{d}{2}-1} K_{i\nu}(s_1 z) K_{\frac{1}{2}}(k_1 z) K_{\frac{1}{2}}(k_2 z) \,.
\end{equation}
Noting that the modified Bessel function $K_{1/2}$ has the explicit simple form \eqref{eq_besselK_cc}, using the connection formula for the modified Bessel functions \eqref{connection_formula_bessel} and the integral representation for the Legendre $Q$ function \eqref{eq_Q_integral_I}, the previous expression can be written as
\begin{equation}
\begin{aligned}
    \mathcal{I}_d\begin{bmatrix}
        \nu &i/2 & i/2 \\ s_1 &k_1 & k_2 \\ K & K & K
    \end{bmatrix}
    &=\left(\frac{\pi}{2}\right)^{\frac{3}{2}}\frac{e^{-i\pi\frac{d}{2}}}{\sqrt{k_1 k_2}s_1^{\frac{d-2}{2}}}
    \frac{1}{\sinh(\pi\nu)}
    \left(\left(\frac{k_{12}}{s_1}\right)^2-1\right)^{\frac{3-d}{4}}\\
    &\times\left(Q_{i\nu-\frac{1}{2}}^{\frac{d-3}{2}}\left(\frac{k_{12}}{s_1}\right)-Q_{-i\nu-\frac{1}{2}}^{\frac{d-3}{2}}\left(\frac{k_{12}}{s_1}\right)\right)\,.
    \end{aligned}
    \label{eq_vertex_fct_12_Q}
\end{equation}
Note that $k_{12}>s_1$ from momentum conservation at the vertex. This last expression gives a connection formula of the form \eqref{eq_connection_formula_bdy}. Similarly, we can obtain the following
\begin{equation}
\begin{aligned}
    \mathcal{I}_d\begin{bmatrix}
        \mu &i/2 & i/2 \\ s_2 &k_5 & k_6 \\ K & K & K
    \end{bmatrix}
    &=\left(\frac{\pi}{2}\right)^{\frac{3}{2}}\frac{e^{-i\pi\frac{d}{2}}}{\sqrt{k_5 k_6}s_2^{\frac{d-2}{2}}}
    \frac{1}{\sinh(\pi\mu)}
    \left(\left(\frac{k_{56}}{s_2}\right)^2-1\right)^{\frac{3-d}{4}}\\
    &\times\left(Q_{i\mu-\frac{1}{2}}^{\frac{d-3}{2}}\left(\frac{k_{56}}{s_2}\right)-Q_{-i\mu-\frac{1}{2}}^{\frac{d-3}{2}}\left(\frac{k_{56}}{s_2}\right)\right)\,.
    \end{aligned}
    \label{eq_vertex_fct_56_Q}
\end{equation}
For these two vertex functions, singularities with residues and large-frequency asymptotic behaviour are related to the properties of the Legendre $Q$ function \eqref{eq_large_mu_Q}-\eqref{eq_residues_legendre_Q}.

\subsubsection{Vertex function with two massive lines}

We now want to study the vertex function that comes from the central vertex of the diagram. Its integral representation is (for $N=4$ in \eqref{def_vertex_function_K}):
\begin{equation}
    \mathcal{I}_d\begin{bmatrix}
        \nu &\mu &i/2 & i/2 \\ s_1 & s_2 & k_3 & k_4  \\ K &K & K & K
    \end{bmatrix}
    =\int \limits^\infty_0\d z\; z^{d-1}
    K_{i\nu}(s_1 z) K_{i\mu}(s_2 z) K_{\frac{1}{2}}(k_3 z)
    K_{\frac{1}{2}}(k_4 z)\,.
    \label{eq_int_rep_vertex_F4}
\end{equation}
Using the explicit expression of $K_{\frac{1}{2}}$ \eqref{eq_besselK_cc}, we can simplify this integral representation by using the identity \eqref{eq_prod_K_identity} with $M=2$ for the two conformally coupled external legs, to get
\begin{equation}
\begin{aligned}
    \mathcal{I}_d\begin{bmatrix}
        \nu &\mu &i/2 & i/2 \\ s_1 & s_2 & k_3 & k_4  \\ K &K & K & K
    \end{bmatrix}
    &=\sqrt{\frac{\pi\, k_{34}}{2 k_3 k_4}}
    \mathcal{I}_{d'}\begin{bmatrix}
    \nu &\mu &i/2 \\ s_1 & s_2 & k_{34}  \\ K &K & K
    \end{bmatrix}\\
    &=\sqrt{\frac{\pi\, k_{34}}{2 k_3 k_4}}\int \limits^\infty_0\d z\; z^{\frac{d'}{2}-1}
    K_{i\nu}(s_1 z) K_{i\mu}(s_2 z) K_{\frac{1}{2}}(k_{34} z)\,,
    \end{aligned}
    \label{eq_simplification_to_F4}
\end{equation}
where the parameter $d'$ satisfies $d'/2-1=d-3/2$. This simplification generally occurs when dealing with a vertex attached to two or more legs carrying conformally coupled fields.

\paragraph{Series representation.} To obtain the series representation of this vertex function, we combine the simplification \eqref{eq_simplification_to_F4} and the general formula \eqref{eq_vertex_function_vf} that we derived in the previous section with $N=3$ \footnote{For the other parameters than $N=3$, we perform the following replacements in \eqref{eq_vertex_function_vf}: $\mu_1\rightarrow\nu$, $\mu_2\rightarrow\mu$, $\mu_3\rightarrow i/2$, $k_1\rightarrow s_1$, $k_2\rightarrow s_2$, $k_3\rightarrow k_{34}$ and $d\rightarrow d'$. Then, we obtain \eqref{eq_series_rep_vertex_F4_IKK}}. This gives
\begin{equation}
\begin{aligned}
    \mathcal{I}_d&\begin{bmatrix}
        \nu &\mu &i/2 & i/2 \\ s_1 & s_2 & k_3 & k_4  \\ K &K & K & K
    \end{bmatrix}
    =\frac{\pi}{2\sqrt{k_3 k_4}k_{34}^{d-1}}
    \sum_{c_1,c_2=\pm}\left(\frac{i\pi}{2}\right)^2\frac{1}{\sinh(\pi c_1\nu)\sinh(\pi c_2 \mu)}\\
    &\times\left(\frac{s_1}{2 k_{34}}\right)^{i c_1 \nu} \left(\frac{s_2}{2 k_{34}}\right)^{i c_2 \mu}
    \Gamma\left(d-1+i\left(c_1\nu+c_2\mu\right)\right)\\
    &\times\F_4\Biggl(\begin{matrix}
    \frac{d-1}{2}+\frac{i}{2}\left(c_1\nu+c_2\mu\right)\,,\, \frac{d}{2}+\frac{i}{2}\left(c_1\nu+c_2\mu\right)\\
    1+i c_1\nu\,,\,1+i c_2\mu\end{matrix}\,;
    \left(\frac{s_1}{k_{34}}\right)^2,\left(\frac{s_2}{k_{34}}\right)^2\Biggr)\,,
    \end{aligned}
    \label{eq_series_rep_vertex_F4_IKK}
\end{equation}
where we simplified the products of $\Gamma$-functions that appears in \eqref{eq_vertex_function_vf} using the duplication formula \eqref{eq_gamma_duplication}.
This representation is given in terms of the dressed Appell $F_4$ function, which has the definition \eqref{eq_def_dressed_F4} and convergence radius \eqref{eq_cvg_radius_Lauricella_FC} (for $N=2$). The kinematic condition \eqref{eq_cvg_radius_kinematics} that comes out of the convergence properties becomes in the present case
\begin{equation}
    s_1+s_2<k_{34}\,.
    \label{eq_kinematic_cond_double_exchange}
\end{equation}
As discussed in the general case, we stick to this condition during the computation and let the analytic continuation be performed afterwards. Indeed, \eqref{eq_kinematic_cond_double_exchange} covers only part of the physically relevant domain. While a systematic continuation over the whole kinematic domain of a general diagram is beyond the scope of this work, as we will see at the end of the computation, for the double-exchange diagram studied in this section, some significant simplifications occur in the evaluation of the series representation. Note that in general, analytic continuation of the central vertex function can be reached through Mellin-Barnes representations, see Eq.~\eqref{eq_one_fold_MB_N3} for the vertex function relevant here.

\paragraph{Connection formulas.} As noted in general in the series \eqref{eq_vertex_function_vf}, this representation \eqref{eq_series_rep_vertex_F4_IKK} immediately gives us the connection formulas for the vertex function. These formulas are all detailed in Appendix \ref{subapp_vertex_fct_double_exch}.

Depending on the component of $\mathcal{G}_{\a_1,\a_2,\a_3}^{\mu_1,\mu_2}$, we may or may not make use of these connection formulas to rewrite the integral $G_{\a_1,\a_2,\a_3}$ \eqref{eq_G_integrals_a1a2a3} that we need to compute. Therefore, let us focus first on the totally nested component of the diagram, i.e. $G_{+++}$.

\subsection{Totally Nested Component}\label{subsection_+++}

The totally nested component $\mathcal{G}_{+++}$ is given by the diagram \eqref{draw_double_exch_diagram_general} where Schwinger-Keldysh indices $a_i$ are all taken to be $a_1=a_2=a_3=+$. Therefore, the internal massive propagators are given by (see \eqref{eq_def_reduced_propagator})
\begin{equation}
\Pi^{\mu_1}_{++}(\nu)=\frac{e^{-\frac{i\pi(d-1)}{2}}}{(\nu^2-\mu_{1}^2)_{i\epsilon}}\,,
\label{eq_propagator_++_line1}
\end{equation}
for the left internal line and
\begin{equation}
\Pi^{\mu_2}_{++}(\mu)=\frac{e^{-\frac{i\pi(d-1)}{2}}}{(\mu^2-\mu_2^2)_{i\epsilon}}\,,
\label{eq_propagator_++_line2}
\end{equation}
for the right one. Also, the integral \eqref{eq_G_integrals_a1a2a3} with $\a_1=\a_2=\a_3=+$ takes the expression
\begin{equation}
\begin{aligned}
    G_{+++} &\left(\{k_i,s_j\}\right)
    \equiv e^{-i\pi(d-1)}
    \int\limits_{-\infty}^{+\infty}\d\nu \, \N_{\nu}\,
    \int\limits_{-\infty}^{+\infty}\d\mu \, \N_{\mu}\;
    \mathcal{I}_d\begin{bmatrix}
        \nu &i/2 & i/2 \\ s_1 &k_1 & k_2 \\ K & K & K
    \end{bmatrix} \frac{1}{(\nu^2-\mu_{1}^2)_{i\epsilon}}\\
    &\times
    \mathcal{I}_d\begin{bmatrix}
        \nu &\mu &i/2 & i/2 \\ s_1 & s_2 &k_3 & k_4 \\ K &K & K & K
    \end{bmatrix}
    \frac{1}{(\mu^2-\mu_2^2)_{i\epsilon}}\;
    \mathcal{I}_d\begin{bmatrix}
        \mu &i/2 & i/2 \\ s_2 &k_5 & k_6 \\ K & K & K
    \end{bmatrix}\,.
    \end{aligned}
    \label{eq_G_integrals_+++}
\end{equation}
The aim of this subsection is to write this integral as a series over residues. We will perform the two layers of integration one by one, starting with the integral over $\mu$ and then doing the one over $\nu$.

\subsubsection{First integration layer}\label{subsubsec_1st_layer}

The first step of the computation is to focus on the integral
\begin{equation}
\begin{aligned}
    F_{+++} \left(\{k_i,s_j\},\nu\right)
    \equiv
    \int\limits_{-\infty}^{+\infty}\d\mu \, \N_{\mu}\,
    \mathcal{I}_d\begin{bmatrix}
        \nu &\mu &i/2 & i/2 \\ s_1 & s_2 &k_3 & k_4 \\ K &K & K & K
    \end{bmatrix}
    \frac{1}{(\mu^2-\mu_2^2)_{i\epsilon}}\;
    \mathcal{I}_d\begin{bmatrix}
        \mu &i/2 & i/2 \\ s_2 &k_5 & k_6 \\ K & K & K
    \end{bmatrix}\,,
    \end{aligned}
    \label{eq_def_F_+++}
\end{equation}
such that
\begin{equation}
\begin{aligned}
    G_{+++} &\left(\{k_i,s_j\}\right)
    = e^{-i\pi(d-1)}
    \int\limits_{-\infty}^{+\infty}\d\nu \, \N_{\nu}\;
    \mathcal{I}_d\begin{bmatrix}
        \nu &i/2 & i/2 \\ s_1 &k_1 & k_2 \\ K & K & K
    \end{bmatrix} \frac{1}{(\nu^2-\mu_{1}^2)_{i\epsilon}}\,
    F_{+++} \left(\{k_i,s_j\},\nu\right)\,.
    \end{aligned}
    \label{eq_G_+++_wrt_F_+++}
\end{equation}
In order to rewrite $F_{+++}$ in a convenient way to have a clear closing contour prescription, we use the connection formulas \eqref{eq_vertex_fct_56_Q} and \eqref{eq_connection_IKK_IKI} for the two vertex functions that enter \eqref{eq_def_F_+++}. It yields
\begin{equation}
    F_{+++} \left(\{k_i,s_j\},\nu\right)=i\left(\frac{\pi}{2}\right)^{\frac{3}{2}} \frac{e^{-i\pi\frac{d}{2}}}{\sqrt{k_5 k_6} \,s_2^{\frac{d-2}{2}}}\,\left(\left(\frac{k_{56}}{s_2}\right)^2-1\right)^{\frac{3-d}{4}}
    \left[F_{+++}^0+F_{+++}^1\right]\,,
    \label{eq_F+++_F0_F1}
\end{equation}
where $K_{56}\equiv k_5+k_6$ and
\begin{equation}
\begin{aligned}
    F_{+++}^0\left(\{k_i,s_j\},\nu\right)
    \equiv\int\limits_{-\infty}^{+\infty}\d\mu \, \frac{\mu}{\sinh(\pi\mu)}\,
    \mathcal{I}_d\begin{bmatrix}
        \nu &\mu &i/2 & i/2 \\ s_1 & s_2 &k_3 & k_4 \\ K & I & K & K
    \end{bmatrix}
    \frac{1}{(\mu^2-\mu_2^2)_{i\epsilon}}\;
    Q_{i\mu-\frac{1}{2}}^{\frac{d-3}{2}}\left(\frac{k_{56}}{s_2}\right)\,,
    \end{aligned}
    \label{eq_F_+++_0}
\end{equation}
\begin{equation}
\begin{aligned}
    F_{+++}^1\left(\{k_i,s_j\},\nu\right)
    \equiv-\int\limits_{-\infty}^{+\infty}\d\mu \, \frac{\mu}{\sinh(\pi\mu)}\,
    \mathcal{I}_d\begin{bmatrix}
        \nu &\mu &i/2 & i/2 \\ s_1 & s_2 &k_3 & k_4 \\ K &I & K & K
    \end{bmatrix}
    \frac{1}{(\mu^2-\mu_2^2)_{i\epsilon}}\;
    Q_{-i\mu-\frac{1}{2}}^{\frac{d-3}{2}}\left(\frac{k_{56}}{s_2}\right)\,,
    \end{aligned}
    \label{eq_F_+++_1}
\end{equation}
and we used the change of variable $\mu'=-\mu$ twice to reduce the expression of $F_{+++}$ \eqref{eq_F+++_F0_F1} to two terms.

For convenience, we will use the notation $\I_{AB}$ in the text for the vertex function of the central vertex where $A,B=K,I$ are the lower indices related to frequencies $\nu$ and $\mu$. In particular, $\I_{KI}$ is the function that appears in \eqref{eq_F_+++_0}-\eqref{eq_F_+++_1} and $\I_{KK}$ is the original vertex function coming from the diagrammatic rules \eqref{eq_int_rep_vertex_F4}.

\paragraph{Background and signal residues.}
The different terms in the integrands of $F_{+++}^0$ and $F_{+++}^1$ have poles that contribute to $F_{+++}$. In all the rest of this computation and for every integral that we will encounter, we separate the different poles and corresponding residues into two categories, according to which term they come from. First, the residues coming from the $i\epsilon$-prescription \eqref{ieps_precription} (see \eqref{eq_i_eps_residues}) in any $++$ propagator are called \textit{signal residues}, and we will use the letter $S$ for reference. Secondly, we refer to all other residues, in particular those coming from the vertex functions, as \textit{background residues}. For instance, in $F_{+++}$ we denote $F_{+++}^S$ the contribution coming from signal residues and $F_{+++}^B$ the contribution coming from background ones, therefore we have
\begin{equation}
    F_{+++}=F_{+++}^B+F_{+++}^S\,.
    \label{eq_F+++BS}
\end{equation}
In addition, each of the two components in the decomposition \eqref{eq_F+++_F0_F1} of $F_{+++}$ will have background and signal residues. For instance, we also write
\begin{equation}
    F_{+++}^0=F_{+++}^{0,B}+F_{+++}^{0,S}\,.
    \label{eq_F+++0_BS}
\end{equation}
and similarly for $F_{+++}^1$.
Let us now start the computation by finding the contour prescription for $F_{+++}^0$ and $F_{+++}^1$, and then gather background and signal residues.

\begin{figure}[h!]
\hspace{-1cm}
\begin{subfigure}[h!]{0.4\textwidth}
        \hspace{-0.25cm}
    	\begin{tikzpicture}[scale = 2]

        \draw[black, ->] (-1.8,0) -- (1.8,0) coordinate (xaxis);
		\draw[black, ->] (0,-1.8) -- (0,1.8) coordinate (yaxis);
		\node at (2.1, 0) {$\text{Re}(\mu)$};
		\node at (0, 1.95) {$\text{Im}(\mu)$};
        
        \draw[pyred, fill = pyred] (0, 0.25) circle (.03cm);
		\draw[pyred, fill = pyred] (0, 0.75) circle (.03cm);
		\draw[pyred, fill = pyred] (0, 1.25) circle (.03cm);
        \node at (.7, 0.25) {\textcolor{pyred}{\footnotesize$Q_{i\mu-\frac{1}{2}}^{\frac{d-3}{2}}\left(k_{56}/s_2\right)$}};

        \draw[pyorange, fill = pyorange] (1.2, 1) circle (.03cm);
		\draw[pyorange, fill = pyorange] (1.2, 1.5) circle (.03cm);
        \draw[pyorange, fill = pyorange] (-1.2, 1) circle (.03cm);
		\draw[pyorange, fill = pyorange] (-1.2, 1.5) circle (.03cm);        \node at (-.9, 1) {\textcolor{pyorange}{\footnotesize$\I_{KI}$}};

        \draw[pyblue, fill = pyblue] (0, -0.5) circle (.03cm);
		\draw[pyblue, fill = pyblue] (0, -1) circle (.03cm);
		\draw[pyblue, fill = pyblue] (0, -1.5) circle (.03cm);
        \draw[pyblue, fill = pyblue] (0, 0.5) circle (.03cm);
		\draw[pyblue, fill = pyblue] (0, 1) circle (.03cm);
		\draw[pyblue, fill = pyblue] (0, 1.5) circle (.03cm);
        \node at (0.4, -1.03) {\textcolor{pyblue}{$\frac{\mu}{\sinh\left(\pi\mu\right)}$}};
        
        \draw[xshift=0,pyblue!80!black,decoration={markings,mark=between positions 0.1 and 1 step 0.2 with \arrow{>}},postaction={decorate}] (-1.70,0) -- (1.70,0) arc (0:-180:1.70);
    \end{tikzpicture}
\end{subfigure}
\hfill
\begin{subfigure}[h!]{0.4\textwidth}
    \hspace{-1.5cm}
    	\begin{tikzpicture}[scale = 2]

        \draw[black, ->] (-1.8,0) -- (1.8,0) coordinate (xaxis);
		\draw[black, ->] (0,-1.8) -- (0,1.8) coordinate (yaxis);
		\node at (2.1, 0) {$\text{Re}(\mu)$};
		\node at (0, 1.95) {$\text{Im}(\mu)$};
        
        \draw[pyorange, fill = pyorange] (1.2, 1) circle (.03cm);
		\draw[pyorange, fill = pyorange] (1.2, 1.5) circle (.03cm);
        \draw[pyorange, fill = pyorange] (-1.2, 1) circle (.03cm);
		\draw[pyorange, fill = pyorange] (-1.2, 1.5) circle (.03cm);        \node at (-.9, 1) {\textcolor{pyorange}{\footnotesize$\I_{KI}$}};

        \draw[pyred, fill = pyred] (0, -0.25) circle (.03cm);
		\draw[pyred, fill = pyred] (0, -0.75) circle (.03cm);
		\draw[pyred, fill = pyred] (0, -1.25) circle (.03cm);
        \node at (0.75, -0.75) {\textcolor{pyred}{\footnotesize$Q_{-i\mu-\frac{1}{2}}^{\frac{d-3}{2}}\left(k_{56}/s_2\right)$}};

        \draw[pyblue, fill = pyblue] (0, -0.5) circle (.03cm);
		\draw[pyblue, fill = pyblue] (0, -1) circle (.03cm);
		\draw[pyblue, fill = pyblue] (0, -1.5) circle (.03cm);
        \draw[pyblue, fill = pyblue] (0, 0.5) circle (.03cm);
		\draw[pyblue, fill = pyblue] (0, 1) circle (.03cm);
		\draw[pyblue, fill = pyblue] (0, 1.5) circle (.03cm);
        \node at (-0.45, -1.03) {\textcolor{pyblue}{$\frac{\mu}{\sinh\left(\pi\mu\right)}$}};
    \end{tikzpicture}
\end{subfigure}
\caption{Background residues in the $\mu$ complex plane of $F_{+++}^{0}$ \eqref{eq_F_+++_0} (left) and $F_{+++}^{1}$ \eqref{eq_F_+++_1} (right), for $d=3$. Recall that $\I_{KI}$ denotes the vertex function in the right-hand side of these integrals. To compute $F_{+++}^{B,0}$, we always close the contour with a half-circle in the lower half-plane. For $F_{+++}^{B,1}$, the contour is closed in the lower (resp. upper) half-plane for $\kappaone<1$ \eqref{eq_condition_C2} (resp. $\kappaone>1$).}
\label{fig_analytic_structure_F+++B}
\end{figure}
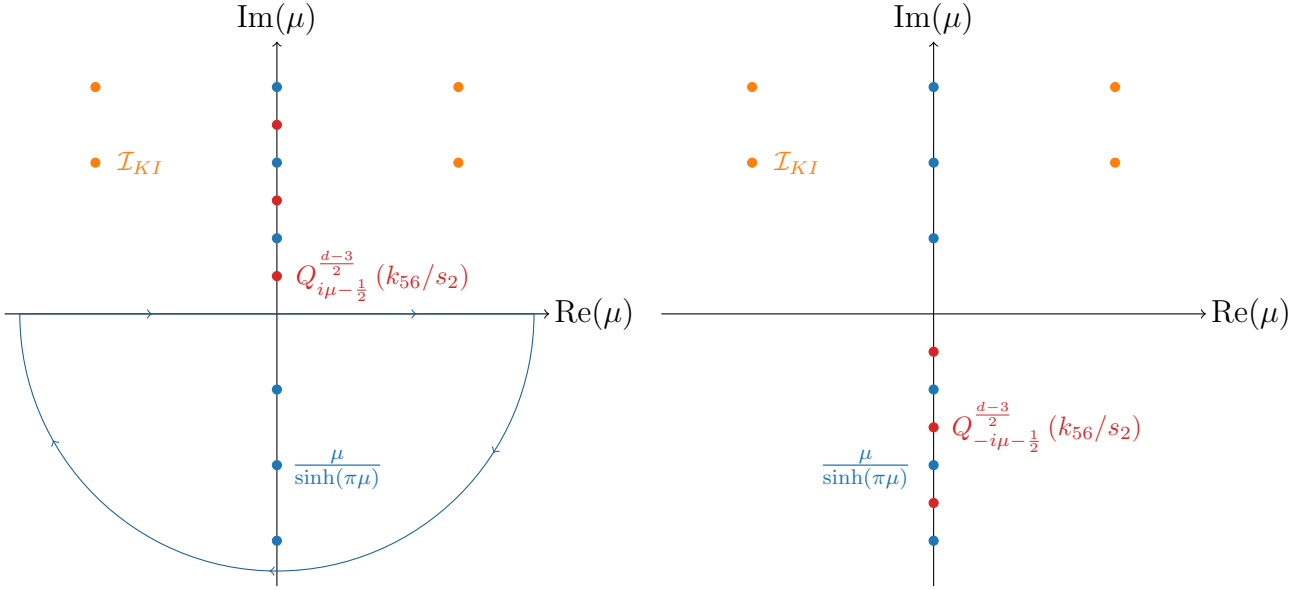

\paragraph{Contour prescription.}
The integrals $F_{+++}^0$ \eqref{eq_F_+++_0} and $F_{+++}^1$ \eqref{eq_F_+++_1} have a singularity structure with signal poles (see Fig. \ref{fig_poles_ieps}) and background poles displayed in Fig. \ref{fig_analytic_structure_F+++B}. The latter come from different pieces: the factor $\textcolor{pyblue}{\mu/\sinh(\pi\mu)}$, the central vertex function $\textcolor{pyorange}{\I_{KI}}$ and the Legendre function $\textcolor{pyred}{Q_{\pm i\mu-1/2}^{(d-3)/2}(k_{56}/s_2)}$ arising from the right vertex function. The residues that will contribute to each integral depend on their specific contour prescription.

In the case of $F_{+++}^0$, this contour prescription arises from the leading terms in the large $\mu$ asymptotic behaviour of $\textcolor{pyred}{Q_{i\mu-1/2}^{(d-3)/2}(k_{56}/s_2)}$ \eqref{eq_large_mu_Q} and of $\textcolor{pyorange}{\I_{KI}}$ \eqref{eq_large_mu_IKI}, which are\footnote{The factor $\mu/\sinh(\pi\mu)$ is at most an oscillating function in the complex $\mu$ plane (when $\mu\in i\mathbb{R}$), and $(\mu^2-\mu_2^2)_{i\epsilon}^{-1}\underset{\mu\to\infty}{\sim}\mu^{-2}$.}:
\begin{equation}
    \mathcal{I}_d\begin{bmatrix}
        \nu &\mu &i/2 & i/2 \\ s_1 & s_2 &k_3 & k_4 \\ K & I & K & K
    \end{bmatrix}
    Q_{i\mu-\frac{1}{2}}^{\frac{d-3}{2}}\left(\frac{k_{56}}{s_2}\right)
    \underset{\mu\to\infty}{\propto}
    e^{-i\mu\left[\arccosh\left(\frac{k_{34}+s_1}{s_2}\right)+\arccosh\left(\frac{k_{56}}{s_2}\right)\right]}\,.
    \label{eq_asymptotics_F+++0}
\end{equation}
Then, because $k_{56}>s_2$ and from the kinematic condition \eqref{eq_kinematic_cond_double_exchange} we have $k_{34}+s_1>s_2$, we can close the integration contour in $F_{+++}^0$ in the lower-half complex plane.

Similarly, for $F_{+++}^1$, the leading terms come from the large $\mu$ asymptotic behaviour of $\textcolor{pyred}{Q_{-i\mu-1/2}^{(d-3)/2}(k_{56}/s_2)}$ \eqref{eq_large_mu_Q} and of $\textcolor{pyorange}{\I_{KI}}$ \eqref{eq_large_mu_IKI}, but now we have
\begin{equation}
    \mathcal{I}_d\begin{bmatrix}
        \nu &\mu &i/2 & i/2 \\ s_1 & s_2 &k_3 & k_4 \\ K & I & K & K
    \end{bmatrix}
    Q_{-i\mu-\frac{1}{2}}^{\frac{d-3}{2}}\left(\frac{k_{56}}{s_2}\right)
    \underset{\mu\to\infty}{\sim}e^{-i\mu\left[\arccosh\left(\frac{k_{34}+s_1}{s_2}\right)-\arccosh\left(\frac{k_{56}}{s_2}\right)\right]}\,.
    \label{eq_asymptotics_F+++1}
\end{equation}
In that case, the coefficient that multiplies $i\mu$ in the exponential can have both signs, depending on the value of the following kinematic ratio:
\begin{equation}
    \kappaone\equiv\frac{k_{56}}{k_{34}+s_1}\,.
    \label{eq_condition_C2}
\end{equation}
If $\kappaone>1$ (resp. $\kappaone<1$), the contour has to be closed in the upper-half (resp. lower-half) complex $\mu$-plane.

\paragraph{Background residues of $F_{+++}^0$.}
Since for the computation of $F_{+++}^0$ we close the contour in the lower half-plane, the only background residues that contribute are then residues of $\mu\mapsto\mu/\sinh(\pi\mu)$ in the negative imaginary axis (see Fig. \ref{fig_analytic_structure_F+++B}). We then have
\begin{equation}
    F_{+++}^{0,B}=-2 \sum_{n=0}^{+\infty} (-1)^n\,
    \mathcal{I}_d\begin{bmatrix}
        \nu &-i(n+1) &i/2 & i/2 \\ s_1 & s_2 &k_3 & k_4 \\ K & I & K & K
    \end{bmatrix}
    \frac{n+1}{(n+1)^2+\mu_2^2}\,
    Q_{n+\frac{1}{2}}^{\frac{d-3}{2}}\left(\frac{k_{56}}{s_2}\right)\,.
    \label{eq_F+++0B_res}
\end{equation}

\paragraph{Background residues of $F_{+++}^1$.}

\subparagraph{Case $\kappaone>1$.} Here, the contour is closed in the upper half-plane and we collect two different contributions:
\begin{itemize}
\item \textbf{\textcolor{pyblue}{Poles from $\frac{\mu}{\sinh(\pi\mu)}$.}}
The contour encircles the poles from $\mu\mapsto\mu/\sinh(\pi\mu)$ located in the positive imaginary axis. The corresponding residues are:
\begin{equation}
    F_{+++}^{1,B}\supset 2 \sum_{n=0}^{+\infty} (-1)^n\,
    \mathcal{I}_d\begin{bmatrix}
        \nu &i(n+1) &i/2 & i/2 \\ s_1 & s_2 &k_3 & k_4 \\ K & I & K & K
    \end{bmatrix}
    \frac{n+1}{(n+1)^2+\mu_2^2}\,
    Q_{n+\frac{1}{2}}^{\frac{d-3}{2}}\left(\frac{k_{56}}{s_2}\right)\,.
    \label{eq_F+++1B_C2>1_res_sh}
\end{equation}
\item \textbf{\textcolor{pyorange}{Poles from $\I_{KI}$.}} The other poles that contribute to $F_{+++}^{1,B}$ come from the vertex function $I_{KI}$ and have residues
\begin{equation}
\begin{aligned}
    F_{+++}^{1,B}&\supset -i\frac{2^{d-2}\pi^3}{\sqrt{k_3 k_4} \,s_2^{d-1}}
    \sum_{c=\pm}\frac{1}{\sinh(\pi c \nu)} \left(\frac{s_1}{s_2}\right)^{i c\nu} \sum_{n=0}^{+\infty} \left(\frac{2 k_{34}}{s_2}\right)^n \frac{1}{n!}\\
    &\times \frac{1}{\sinh(\pi c\nu-i\pi d)}\,\F_4\Biggl(\begin{matrix}
    -\frac{n}{2}\,,\, \frac{1-n}{2}\\
    1+i c\nu\,,\,1-i c \nu-(n+d-1)\end{matrix}\,;
    \left(\frac{s_1}{k_{34}}\right)^2,\left(\frac{s_2}{k_{34}}\right)^2\Biggr)\\
    &\times \frac{-c\nu+i(n+d-1)}{\left(c\nu-i(n+d-1)\right)^2-\mu_2^2}\,
    Q_{i c \nu +n+d-\frac{3}{2}}^{\frac{d-3}{2}}\left(\frac{k_{56}}{s_2}\right)\,.
    \end{aligned}
    \label{eq_F+++1B_C2>1_res_IKI}
\end{equation}
\end{itemize}
Now, $F_{+++}^B$ is found by adding together the background contributions \eqref{eq_F+++0B_res}, \eqref{eq_F+++1B_C2>1_res_sh} and \eqref{eq_F+++1B_C2>1_res_IKI} in the expression \eqref{eq_F+++_F0_F1}. However, it appears that the sum of the first two contributions vanishes, using the connection formula \eqref{eq_connection_IKK_IKI} for $\mu=i(n+1)$, noting that each term in the series is $\propto\sinh(\pi i(n+1))=0$. Then, in the case where $\kappaone>1$, $F_{+++}^B$ can be written as
\begin{equation}
\begin{aligned}
    F_{+++}^{B}&\left(\{k_i,s_j\},\nu\right)=\frac{e^{-i\pi\frac{d}{2}}2^{d-\frac{7}{2}}\pi^\frac{9}{2}}{\sqrt{k_3 k_4 k_5 k_6} \,s_2^{\frac{3d}{2}-2}}\left(\left(\frac{k_{56}}{s_2}\right)^2-1\right)^{\frac{3-d}{4}}\\
    &\times
    \sum_{c=\pm}\frac{1}{\sinh(\pi c \nu)} \left(\frac{s_1}{s_2}\right)^{i c\nu} \sum_{n=0}^{+\infty} \left(\frac{2 k_{34}}{s_2}\right)^n \frac{1}{n!}\\
    &\times \frac{1}{\sinh(\pi c\nu-i\pi d)}\,\F_4\Biggl(\begin{matrix}
    -\frac{n}{2}\,,\, \frac{1-n}{2}\\
    1+i c\nu\,,\,1-i c \nu-(n+d-1)\end{matrix}\,;
    \left(\frac{s_1}{k_{34}}\right)^2,\left(\frac{s_2}{k_{34}}\right)^2\Biggr)\\
    &\times \frac{-c\nu+i(n+d-1)}{\left(c\nu-i(n+d-1)\right)^2-\mu_2^2}\,
    Q_{i c \nu +n+d-\frac{3}{2}}^{\frac{d-3}{2}}\left(\frac{k_{56}}{s_2}\right)\,.
    \end{aligned}
    \label{eq_F+++B_C2>1}
\end{equation}

\subparagraph{Case $\kappaone<1$.} For $\kappaone<1$, we close the contour in the lower-half $\mu$-plane. We will also have two different contributions to background residues of $F_{+++}^1$:
\begin{itemize}
\item \textbf{\textcolor{pyblue}{Poles from $\frac{\mu}{\sinh(\pi\mu)}$.}}
We need to add the residues corresponding to the poles from $\mu\mapsto\mu/\sinh(\pi\mu)$ located in the negative imaginary axis. These residues are:
\begin{equation}
    F_{+++}^{1,B}\supset 2 \sum_{n=0}^{+\infty} (-1)^n\,
    \mathcal{I}_d\begin{bmatrix}
        \nu &-i(n+1) &i/2 & i/2 \\ s_1 & s_2 &k_3 & k_4 \\ K & I & K & K
    \end{bmatrix}
    \frac{n+1}{(n+1)^2+\mu_2^2}\,
    Q_{-n-\frac{3}{2}}^{\frac{d-3}{2}}\left(\frac{k_{56}}{s_2}\right)\,.
    \label{eq_F+++1B_C2<1_res_sh}
\end{equation}
\item \textbf{\textcolor{pyred}{Poles from $Q_{-i\mu-\frac{1}{2}}^{\frac{d-3}{2}}\left(k_{56}/s_2\right)$.}} The other poles that contribute to $F_{+++}^{1,B}$ come from the Legendre $Q$ function, with residues:
\begin{equation}
\begin{aligned}
    F_{+++}^{1,B}&\supset \frac{2^{\frac{d-1}{2}} \pi^\frac{3}{2} e^{i\pi\frac{d-1}{2}}}{\sin\left(\pi\frac{d}{2}\right)}
    \left(\left(\frac{k_{56}}{s_2}\right)^2-1\right)^{\frac{d-3}{4}}
    \sum_{n=0}^{+\infty} \frac{1}{n!}\left(\frac{2 k_{56}}{s_2}\right)^n\\
    &\times
    \mathcal{I}_d\begin{bmatrix}
        \nu &-i\left(n+\frac{d-2}{2}\right) &i/2 & i/2 \\ s_1 & s_2 & k_3 & k_4  \\ K & I & K & K
    \end{bmatrix}
    \frac{n+\frac{d-2}{2}}{\left(n+\frac{d-2}{2}\right)^2+\mu_2^2}
    \,_2\F_1\left(
    \begin{matrix}
        \frac{1-n}{2}\,,\,-\frac{n}{2} \\ -n-\frac{d-4}{2}
    \end{matrix}\,,\left(\frac{s_2}{k_{56}}\right)^2\right)\,.
    \end{aligned}
    \label{eq_F+++1B_C2<1_res_Q}
\end{equation}
\end{itemize}
As in the previous case, the sum of the two contributions \eqref{eq_F+++0B_res} and \eqref{eq_F+++1B_C2<1_res_sh} to $F_{+++}^B$ vanishes due to the connection formula for Legendre $Q$ functions \eqref{connection_formula_P_Q} with $\mu=-i(n+1)$. Finally, only the third term \eqref{eq_F+++1B_C2<1_res_Q} contributes to $F_{+++}^B$:
\begin{equation}
\begin{aligned}
    F_{+++}^{B}&\left(\{k_i,s_j\},\nu\right)=
    \frac{2^{\frac{d}{2}-2} \pi^3 }{\sin\left(\pi\frac{d}{2}\right)\sqrt{k_5 k_6} \,s_2^{\frac{d-2}{2}}}
    \sum_{n=0}^{+\infty} \frac{1}{n!}\left(\frac{2 k_{56}}{s_2}\right)^n\\
    &\times
    \mathcal{I}_d\begin{bmatrix}
        \nu &-i\left(n+\frac{d-2}{2}\right) &i/2 & i/2 \\ s_1 & s_2 & k_3 & k_4  \\ K & I & K & K
    \end{bmatrix}
    \frac{n+\frac{d-2}{2}}{\left(n+\frac{d-2}{2}\right)^2+\mu_2^2}
    \,_2\F_1\left(
    \begin{matrix}
        \frac{1-n}{2}\,,\,-\frac{n}{2} \\ -n-\frac{d-4}{2}
    \end{matrix}\,,\left(\frac{s_2}{k_{56}}\right)^2\right)
    \,.
    \end{aligned}
    \label{eq_F+++B_C2<1}
\end{equation}

\paragraph{Signal residues.} From our contour prescription, the signal residues of $F_{+++}^0$ come from the poles located at $\mu=\pm\mu_2-i\epsilon$, in the lower half complex plane. Then,
\begin{equation}
\begin{aligned}
    F_{+++}^{0,S}=&-\frac{i\pi}{2\sinh^2(\pi\mu_2)}\Biggl(e^{\pi\mu_2}\mathcal{I}_d\begin{bmatrix}
        \nu & \mu_2 &i/2 & i/2 \\ s_1 & s_2 & k_3 & k_4  \\ K & I & K & K
    \end{bmatrix}
    Q_{i\mu_2-\frac{1}{2}}^{\frac{d-3}{2}}\left(\frac{k_{56}}{s_2}\right)
    \\&+
    e^{-\pi\mu_2}\mathcal{I}_d\begin{bmatrix}
        \nu & -\mu_2 &i/2 & i/2 \\ s_1 & s_2 & k_3 & k_4  \\ K & I & K & K
    \end{bmatrix}
    Q_{-i\mu_2-\frac{1}{2}}^{\frac{d-3}{2}}\left(\frac{k_{56}}{s_2}\right)
    \Biggr)\,.
    \end{aligned}
    \label{eq_F+++_0S}
\end{equation}
Now, the expression of $F_{+++}^{1,S}$ depends on the value of $\kappaone$.

\subparagraph{Case $\kappaone>1$.} In this configuration, the relevant signal poles are those located in the upper half-plane, i.e. at $\mu=\pm\mu_2+i\epsilon$. Their residues give
\begin{equation}
\begin{aligned}
    F_{+++}^{1,S}=&\frac{i\pi}{2\sinh^2(\pi\mu_2)}\Biggl(e^{-\pi\mu_2}\mathcal{I}_d\begin{bmatrix}
        \nu & \mu_2 &i/2 & i/2 \\ s_1 & s_2 & k_3 & k_4  \\ K & I & K & K
    \end{bmatrix}
    Q_{-i\mu_2-\frac{1}{2}}^{\frac{d-3}{2}}\left(\frac{k_{56}}{s_2}\right)
    \\&+
    e^{\pi\mu_2}\mathcal{I}_d\begin{bmatrix}
        \nu & -\mu_2 &i/2 & i/2 \\ s_1 & s_2 & k_3 & k_4  \\ K & I & K & K
    \end{bmatrix}
    Q_{i\mu_2-\frac{1}{2}}^{\frac{d-3}{2}}\left(\frac{k_{56}}{s_2}\right)
    \Biggr)\,.
    \end{aligned}
    \label{eq_F+++_1S_C2>1}
\end{equation}
The sum of the two signal contributions \eqref{eq_F+++_0S} and \eqref{eq_F+++_1S_C2>1} can be written in a more compact way using the connection formula between $\I_{KK}$ and $\I_{KI}$ \eqref{eq_connection_IKK_IKI} and the analytic continuation formula \eqref{eq_analytic_cont_P} for Legendre $P$. Using also \eqref{eq_F+++_F0_F1}, we end with
\begin{equation}
\begin{aligned}
    F_{+++}^S \left(\{k_i,s_j\},\nu\right)&=\left(\frac{\pi}{2}\right)^{\frac{3}{2}} \frac{\left(\left(\frac{k_{56}}{s_2}\right)^2-1\right)^{\frac{3-d}{4}}}{\sqrt{k_5 k_6} \,s_2^{\frac{d-2}{2}}}\,
    \Gamma\left(\frac{d-2}{2}\pm i\mu_2\right)\\
    &\times\mathcal{I}_d\begin{bmatrix}
        \nu & \mu_2 &i/2 & i/2 \\ s_1 & s_2 & k_3 & k_4  \\ K & K & K & K
    \end{bmatrix}
    P_{i\mu_2-\frac{1}{2}}^{-\frac{d-3}{2}}\left(e^{+i\pi}\frac{k_{56}}{s_2}\right)
    \,,
    \end{aligned}
    \label{eq_F+++_S_C2>1}
\end{equation}

\subparagraph{Case $\kappaone<1$.} Otherwise, we collect the residues of the poles located at $\mu=\pm\mu_2-i\epsilon$. Therefore we have
\begin{equation}
\begin{aligned}
    F_{+++}^{1,S}=&\frac{i\pi}{2\sinh^2(\pi\mu_2)}\Biggl(e^{\pi\mu_2}\mathcal{I}_d\begin{bmatrix}
        \nu & \mu_2 &i/2 & i/2 \\ s_1 & s_2 & k_3 & k_4  \\ K & I & K & K
    \end{bmatrix}
    Q_{-i\mu_2-\frac{1}{2}}^{\frac{d-3}{2}}\left(\frac{k_{56}}{s_2}\right)
    \\&+
    e^{-\pi\mu_2}\mathcal{I}_d\begin{bmatrix}
        \nu & -\mu_2 &i/2 & i/2 \\ s_1 & s_2 & k_3 & k_4  \\ K & I & K & K
    \end{bmatrix}
    Q_{i\mu_2-\frac{1}{2}}^{\frac{d-3}{2}}\left(\frac{k_{56}}{s_2}\right)
    \Biggr)\,.
    \end{aligned}
    \label{eq_F+++_1S_C2<1}
\end{equation}
The sum of the two contributions \eqref{eq_F+++_0S} and \eqref{eq_F+++_1S_C2<1} can be written using the connection formula for Legendre functions \eqref{connection_formula_P_Q} and the analytic continuation formula \eqref{eq_analytic_continuation_IKK_2} for $\I_{KK}$. It yields
\begin{equation}
\begin{aligned}
    F_{+++}^S \left(\{k_i,s_j\},\nu\right)&=i\left(\frac{\pi}{2}\right)^{\frac{3}{2}} \frac{\left(\left(\frac{k_{56}}{s_2}\right)^2-1\right)^{\frac{3-d}{4}}}{\sqrt{k_5 k_6} \,s_2^{\frac{d-2}{2}}}\,
    \Gamma\left(\frac{d-2}{2}\pm i\mu_2\right)\\
    &\times\mathcal{I}_d\begin{bmatrix}
        \nu & \mu_2 &i/2 & i/2 \\ s_1 & e^{-i\pi}s_2 & k_3 & k_4  \\ K & K & K & K
    \end{bmatrix}
    P_{i\mu_2-\frac{1}{2}}^{-\frac{d-3}{2}}\left(\frac{k_{56}}{s_2}\right)
    \,.
    \end{aligned}
    \label{eq_F+++_S_C2<1}
\end{equation}

\paragraph{Summary.} We computed here the first integral layer $F_{+++}$ \eqref{eq_def_F_+++} of the desired integral $G_{+++}$ \eqref{eq_G_integrals_+++}. Using the decomposition \eqref{eq_F+++0_BS} into a background piece and a signal piece, $F_{+++}$ is given by the sum of \eqref{eq_F+++B_C2>1} and \eqref{eq_F+++_S_C2>1} when $\kappaone>1$ (with the kinematic ratio $\kappaone$ given by \eqref{eq_condition_C2}) and by the sum of \eqref{eq_F+++B_C2<1} and \eqref{eq_F+++_S_C2<1} when $\kappaone<1$.

\subsubsection{Second integration layer}\label{subsec_second_int_layer}

The second integral to perform is given by \eqref{eq_G_+++_wrt_F_+++}. Given the separation made for the first integration layer between the background and signal contributions, here we similarly consider the following four different contributions to $G_{+++}$:
\begin{equation}
    G_{+++}=G_{+++}^{B,B}+G_{+++}^{B,S}+G_{+++}^{S,B}+G_{+++}^{S,S}\,.
    \label{eq_decomposition_G+++_BorS}
\end{equation}
The two superscripts correspond to the type of residues collected (which we denote by $B$ for background or $S$ for signal) at each integration layer\footnote{For example, $G_{+++}^{B,S}$ contains the background residues of the second integral \eqref{eq_G_+++_wrt_F_+++} performed over the signal part of the first integral $F_{+++}$ that is $F_{+++}^S$.}.

\paragraph{Background-Background component.} Let us start by considering the component $G_{+++}^{B,B}$ that contains the background-background residues. For computational convenience, we define from \eqref{eq_G_+++_wrt_F_+++} the following quantity as an integral over $F_{+++}^B$:
\begin{equation}
\begin{aligned}
    &G_{+++}^{.,B}\left(\{k_i,s_j\}\right)\equiv G_{+++}^{B,B}+G_{+++}^{S,B}\\
    &=e^{-i\pi(d-1)}
    \int\limits_{-\infty}^{+\infty}\d\nu \, \N_{\nu}\;
    \mathcal{I}_d\begin{bmatrix}
        \nu &i/2 & i/2 \\ s_1 &k_1 & k_2 \\ K & K & K
    \end{bmatrix} \frac{1}{(\nu^2-\mu_{1}^2)_{i\epsilon}}\,
    F_{+++}^B \left(\{k_i,s_j\},\nu\right)\,.
    \end{aligned}
    \label{eq_G_+++_.B_def_wrt_F+++B}
\end{equation}
This last integral is given by the sum of two terms in the decomposition \eqref{eq_decomposition_G+++_BorS}: its signal residues encoded in $G_{+++}^{S,B}$, which we will compute later, and the sum of its background residues $G_{+++}^{B,B}$. Let us focus on the case $\kappa_1<1$. From the expression of $F_{+++}^B$ in that case \eqref{eq_F+++B_C2<1}, the integral \eqref{eq_G_+++_.B_def_wrt_F+++B} can be rewritten as
\begin{equation}
\begin{aligned}
    G_{+++}^{.,B}\left(\{k_i,s_j\}\right)
    &= \frac{e^{-i\pi\frac{3}{2}(d-1)} 2^{\frac{d-7}{2}} \pi^\frac{9}{2}}{\sqrt{k_1 k_2 k_5 k_6}s_1^{\frac{d-2}{2}}s_2^{\frac{d-2}{2}}\sin\left(\pi\frac{d}{2}\right)}
    \left(\left(\frac{k_{12}}{s_1}\right)^2-1\right)^{\frac{3-d}{4}}\\
    &\times \sum_{n=0}^{+\infty} \frac{n+\frac{d-2}{2}}{\left(n+\frac{d-2}{2}\right)^2+\mu_2^2}\frac{1}{n!}\left(\frac{2 k_{56}}{s_2}\right)^n
    \,_2\F_1\left(
    \begin{matrix}
        \frac{1-n}{2}\,,\,-\frac{n}{2} \\ -n-\frac{d-4}{2}
    \end{matrix}\,,\left(\frac{s_2}{k_{56}}\right)^2\right)\\
    &\times
    \Biggl[G_{+++}^{0,.,B}+G_{+++}^{1,.,B}\Biggr]
    \,.
    \end{aligned}
    \label{eq_G_+++_.B_def_wrt_G_+++_01B_C2<1}
\end{equation}
with
\begin{subequations}
\begin{equation}
\begin{aligned}
    G_{+++}^{0,.,B}&=\int\limits_{-\infty}^{+\infty}\d\nu \, \frac{\nu}{\sinh(\pi\nu)}\,
    Q_{i\nu-\frac{1}{2}}^{\frac{d-3}{2}}\left(\frac{k_{12}}{s_1}\right)
    \frac{1}{(\nu^2-\mu_{1}^2)_{i\epsilon}}
    \mathcal{I}_d\begin{bmatrix}
        \nu &-i\left(n+\frac{d-2}{2}\right) &i/2 & i/2 \\ s_1 & s_2 & k_3 & k_4  \\ I & I & K & K
    \end{bmatrix}
    \,,
    \end{aligned}
    \label{eq_G_+++_0.B_def_C2<1}
\end{equation}
\begin{equation}
\begin{aligned}
G_{+++}^{1,.,B}&=-\int\limits_{-\infty}^{+\infty}\d\nu \, \frac{\nu}{\sinh(\pi\nu)}\,
Q_{-i\nu-\frac{1}{2}}^{\frac{d-3}{2}}\left(\frac{k_{12}}{s_1}\right)
    \frac{1}{(\nu^2-\mu_{1}^2)_{i\epsilon}}
    \mathcal{I}_d\begin{bmatrix}
        \nu &-i\left(n+\frac{d-2}{2}\right) &i/2 & i/2 \\ s_1 & s_2 & k_3 & k_4  \\ I & I & K & K
    \end{bmatrix}
    \,.
    \end{aligned}
    \label{eq_G_+++_1.B_def_C2<1}
\end{equation}
\end{subequations}
To obtain this decomposition, we used the expression of the left vertex function in terms of Legendre $Q$ functions \eqref{eq_vertex_fct_12_Q}, the connection formula between $\I_{KI}$ and $\I_{II}$ \eqref{eq_connection_IKI_III} to write the central vertex function in terms of $\I_{II}$, and the change of variable $\nu'=-\nu$ in the integrals \eqref{eq_G_+++_0.B_def_C2<1}-\eqref{eq_G_+++_1.B_def_C2<1}.

\paragraph{Asymptotic behaviour and contour prescription.}
We now use the large $\nu$ asymptotic behaviours of the integrands to obtain the contour prescription for $G_{+++}^{0,.,B}$ and $G_{+++}^{1,.,B}$.
These asymptotic behaviours can be inferred from those of the Legendre $Q$ function \eqref{eq_large_mu_Q} and of the vertex function $\I_{II}$ \eqref{eq_large_nu_III}. Dropping the prefactors in front of the leading exponential terms that matter in the choice of the contour prescription, we have
\begin{equation}
\begin{aligned}
    Q_{\pm i\nu-\frac{1}{2}}^{\frac{d-3}{2}}\left(\frac{k_{12}}{s_1}\right)
    \mathcal{I}_d\begin{bmatrix}
        \nu &-i\left(n+\frac{d-2}{2}\right) &i/2 & i/2 \\ s_1 & s_2 & k_3 & k_4  \\ I & I & K & K
    \end{bmatrix}
    &\underset{\nu\to +\infty}{\propto}
    e^{-i\nu\left(\pm\arccosh\left(\frac{k_{12}}{s_1}\right)+ \arccosh\left(\frac{k_{34}-s_2}{s_1}\right)\right)}\\
    &+e^{-i\nu\left(\pm\arccosh\left(\frac{k_{12}}{s_1}\right)+ \arccosh\left(\frac{k_{34}+s_2}{s_1}\right)\right)}\,,
    \end{aligned}
    \label{eq_large_nu_integrand_C2<1}
\end{equation}
where the $\pm$ sign in the right-hand side takes the value $+$ for $G_{+++}^{0,.,B}$ (resp. $-$ for $G_{+++}^{1,.,B}$).
For $G_{+++}^{0,.,B}$ \eqref{eq_G_+++_0.B_def_C2<1}, for the asymptotic behaviour to be a decreasing exponential at infinity, we close the contour in the lower-half complex plane (as mentioned in Appendix \ref{subsubapp_contour_G+++01.B}).
For $G_{+++}^{1,.,B}$ \eqref{eq_G_+++_1.B_def_C2<1}, it appears that to close the contour in the upper-half $\nu$-plane, we need to have
\begin{equation}
    \kappatwo\equiv\frac{k_{12}}{k_{34}+s_2}>1\,,
    \label{eq_condition_C11p}
\end{equation}
and to close it on the lower half-plane, we must have
\begin{equation}
    \kappathree\equiv\frac{k_{12}}{k_{34}-s_2}<1\,,
    \label{eq_condition_C12p}
\end{equation}
Note that the asymptotic behaviour \eqref{eq_large_nu_integrand_C2<1} forbids us to study the case where $\kappatwo<1$ and $\kappathree>1$. This can be solved by a further decomposition of $G_{+++}^{1,.,B}$ into two pieces, each having a single leading exponential factor at large frequency, which we leave to Appendix \ref{appendix_kinematic_cases_G+++}.
More generally, in this section, we only focus on the computation when $\kappaone<1$ and $\kappatwo>1$. Because very similar technical tools are used, all the other cases, including $\kappaone>1$, are left to Appendix \ref{appendix_kinematic_cases_G+++}.

These contour prescriptions, together with the residue computations that follow, are derived assuming the condition \eqref{eq_kinematic_cond_double_exchange}, $k_{34}>s_1+s_2$. They in fact remain valid throughout the entire physical domain, and in particular when $|s_1-s_2|<k_{34}<s_1+s_2$, where the central vertex function is defined 
thanks to the analytic continuation theorem. Consequently, the results in terms of series representations in this section and in Appendix~\ref{appendix_kinematic_cases_G+++} hold for all physical kinematics, the only modification being that the Appell $F_4$ function that appears in the results is evaluated through its analytic continuation.

\paragraph{Poles and residues.}
The background singularity structure of $G_{+++}^{0,.,B}$ \eqref{eq_G_+++_0.B_def_C2<1} and $G_{+++}^{1,.,B}$ \eqref{eq_G_+++_1.B_def_C2<1} is represented in Fig. \ref{fig_analytic_structure_G+++B_C2<1}. As discussed before, for the contour closure of $G_{+++}^{1,.,B}$ we have two possibilities, depending on the value of the ratios $\kappatwo$ \eqref{eq_condition_C11p} and $\kappathree$ \eqref{eq_condition_C12p}. This prescription enables us now to compute the integral by collecting the right residues.

\begin{figure}[h!]
\hspace{-1cm}
\begin{subfigure}[h!]{0.4\textwidth}
        \hspace{-0.25cm}
    	\begin{tikzpicture}[scale = 2]

        \draw[black, ->] (-1.8,0) -- (1.8,0) coordinate (xaxis);
		\draw[black, ->] (0,-1.8) -- (0,1.9) coordinate (yaxis);
		\node at (2.1, 0) {$\text{Re}(\nu)$};
		\node at (0, 2) {$\text{Im}(\nu)$};
        
        \draw[pyred, fill = pyred] (0, 0.25) circle (.03cm);
		\draw[pyred, fill = pyred] (0, 0.75) circle (.03cm);
		\draw[pyred, fill = pyred] (0, 1.75) circle (.03cm);        
    \draw[pyred] (0,1.22cm) -- (0,1.28cm) arc(90:270:.03cm);
    \draw[pyorange] (0,1.28cm) -- (0,1.22cm) arc(-90:90:.03cm);
    \begin{scope}
        \clip (0,1.25cm) circle(.03cm);
        \fill[pyred] (0,1.22cm) rectangle (-.03cm,1.28cm);
        \fill[pyorange] (0,1.22cm) rectangle (.03cm,1.28cm);
    \end{scope}
    \draw[pyred] (0,1.72cm) -- (0,1.78cm) arc(90:270:.03cm);
    \draw[pyorange] (0,1.78cm) -- (0,1.72cm) arc(-90:90:.03cm);
    \begin{scope}
        \clip (0,1.75cm) circle(.03cm);
        \fill[pyred] (0,1.72cm) rectangle (-.03cm,1.78cm);
        \fill[pyorange] (0,1.72cm) rectangle (.03cm,1.78cm);
    \end{scope}
        \node at (-.7, 1.25) {\textcolor{pyred}{\footnotesize$Q_{i\nu-\frac{1}{2}}^{\frac{d-3}{2}}\left(k_{12}/s_1\right)$}};
        \node at (.3, 1.25) {\textcolor{pyorange}{\footnotesize$
        I_{II}$}};

        \draw[pyblue, fill = pyblue] (0, -0.5) circle (.03cm);
		\draw[pyblue, fill = pyblue] (0, -1) circle (.03cm);
		\draw[pyblue, fill = pyblue] (0, -1.5) circle (.03cm);
        \draw[pyblue, fill = pyblue] (0, 0.5) circle (.03cm);
		\draw[pyblue, fill = pyblue] (0, 1) circle (.03cm);
		\draw[pyblue, fill = pyblue] (0, 1.5) circle (.03cm);
        \node at (0.4, -.53) {\textcolor{pyblue}{$\frac{\nu}{\sinh\left(\pi\nu\right)}$}};
        
        \draw[xshift=0,pyblue!80!black,decoration={markings,mark=between positions 0.1 and 1 step 0.2 with \arrow{>}},postaction={decorate}] (-1.70,0) -- (1.70,0) arc (0:-180:1.70);
    \end{tikzpicture}
\end{subfigure}
\hfill
\begin{subfigure}[h!]{0.4\textwidth}
    \hspace{-1.5cm}
    	\begin{tikzpicture}[scale = 2]

        \draw[black, ->] (-1.8,0) -- (1.8,0) coordinate (xaxis);
		\draw[black, ->] (0,-1.8) -- (0,1.9) coordinate (yaxis);
		\node at (2.1, 0) {$\text{Re}(\nu)$};
		\node at (0, 2) {$\text{Im}(\nu)$};

        \draw[pyred, fill = pyred] (0, -0.25) circle (.03cm);
		\draw[pyred, fill = pyred] (0, -0.75) circle (.03cm);
		\draw[pyred, fill = pyred] (0, -1.25) circle (.03cm);
		\draw[pyred, fill = pyred] (0, -1.75) circle (.03cm);
        \node at (-.8, -1.25) {\textcolor{pyred}{\footnotesize$Q_{-i\nu-\frac{1}{2}}^{\frac{d-3}{2}}\left(k_{12}/s_1\right)$}};

        \draw[pyorange, fill = pyorange] (0, 1.25) circle (.03cm);
		\draw[pyorange, fill = pyorange] (0, 1.75) circle (.03cm);
        \node at (.3, 1.25) {\textcolor{pyorange}{\footnotesize$I_{II}$}};

        \draw[pyblue, fill = pyblue] (0, -0.5) circle (.03cm);
		\draw[pyblue, fill = pyblue] (0, -1) circle (.03cm);
		\draw[pyblue, fill = pyblue] (0, -1.5) circle (.03cm);
        \draw[pyblue, fill = pyblue] (0, 0.5) circle (.03cm);
		\draw[pyblue, fill = pyblue] (0, 1) circle (.03cm);
		\draw[pyblue, fill = pyblue] (0, 1.5) circle (.03cm);
        \node at (0.4, -.53) {\textcolor{pyblue}{$\frac{\nu}{\sinh\left(\pi\nu\right)}$}};
    \end{tikzpicture}
\end{subfigure}
\caption{Background residues in the $\nu$ complex plane of $G_{+++}^{0,.,B}$ \eqref{eq_G_+++_0.B_def_C2<1} (left) and $G_{+++}^{1,.,B}$ \eqref{eq_G_+++_1.B_def_C2<1} (right) for $\kappaone<1$. The figure is drawn for $d=3$ and $n=0$. Under these conditions, in the integrand of  $G_{+++}^{0,.,B}$ poles coming from the two Legendre $Q$ functions are at the same location in the imaginary axis, however we avoid them by always closing the contour with a half-circle in the lower half-plane.
For $G_{+++}^{1,.,B}$, the contour is closed in the upper (resp. lower) half-plane for $\kappatwo>1$ \eqref{eq_condition_C11p} (resp. $\kappathree<1$ \eqref{eq_condition_C12p}). For $\kappatwo<1$ and $\kappathree>1$, see Appendix \ref{appendix_kinematic_cases_G+++}.}
\label{fig_analytic_structure_G+++B_C2<1}
\end{figure}

As represented in the left panel of Fig. \ref{fig_analytic_structure_G+++B_C2<1}, the only factor that has poles contributing to the background residues of $G_{+++}^{0,.,B}$ is $\nu/\sinh(\pi\nu)$. These poles are located at $\nu=-i(p+1)$, $p\in\mathbb{N}$, and their sum, which we denote $G_{+++}^{0,B,B}$, is then
\begin{equation}
\begin{aligned}
    G_{+++}^{0,B,B}&=
    -2\sum_{p=0}^{+\infty}(-1)^{p}\frac{p+1}{(p+1)^2+\mu_1^2}
    \\
    &\times Q_{p+\frac{1}{2}}^{\frac{d-3}{2}}\left(\frac{k_{12}}{s_1}\right)\,
    \mathcal{I}_d\begin{bmatrix}
        -i(p+1) &-i\left(n+\frac{d-2}{2}\right) &i/2 & i/2 \\ s_1 & s_2 & k_3 & k_4  \\ I & I & K & K
    \end{bmatrix}\,.
\end{aligned}
\label{eq_G+++_0BB_C2<1}
\end{equation}

We similarly define the quantity $G_{+++}^{1,B,B}$, which contains the background residues of $G_{+++}^{1,.,B}$ \eqref{eq_G_+++_1.B_def_C2<1}.
Under the conditions $\kappaone<1$ and $\kappatwo>1$, the contour in the integral $G_{+++}^{1,.,B}$ is closed in the upper half-plane, and according to Fig. \ref{fig_analytic_structure_G+++B_C2<1}, we pick the two following sets of residues:
\begin{itemize}
\item \textbf{\textcolor{pyblue}{Poles from $\frac{\nu}{\sinh(\pi\nu)}$.}}
We first collect residues from the singularities of $\nu\mapsto\nu/\sinh(\pi\nu)$ that are located at $\nu=i(p+1)$, $p\in\mathbb{N}$:
\begin{equation}
\begin{aligned}
    G_{+++}^{1,B,B}&\supset
    2\sum_{p=0}^{+\infty}(-1)^{p}\frac{p+1}{(p+1)^2+\mu_1^2}
    \\
    &\times Q_{p+\frac{1}{2}}^{\frac{d-3}{2}}\left(\frac{k_{12}}{s_1}\right)\,
    \mathcal{I}_d\begin{bmatrix}
        i(p+1) &-i\left(n+\frac{d-2}{2}\right) &i/2 & i/2 \\ s_1 & s_2 & k_3 & k_4  \\ I & I & K & K
    \end{bmatrix}\,.
    \end{aligned}
    \label{eq_G+++_1BB_C2<1_C1p>1_res_sh}
\end{equation}
\item \textbf{\textcolor{pyorange}{Poles from $\I_{II}$.}} The other residues come from the vertex function $\I_{II}$. From \eqref{eq_app_poles_III}, it has poles at
\begin{equation}
    \nu=i\left(p+n+\frac{3d-4}{2}\right)\,,\,p\in\mathbb{N}\,,
    \label{eq_G+++_1BB_C2<1_C1p>1_poles_III}
\end{equation}
and residues given by \eqref{eq_residues_III}. The contribution of these residues to $G_{+++}^{1,B,B}$ is then
\begin{equation}
\begin{aligned}
    G_{+++}^{1,B,B}&\supset
    \frac{\pi^2}{\sqrt{k_3 k_4}k_{34}^{d-1} \sin\left(\frac{3\pi d}{2}\right)} (-1)^{n} \sum_{p=0}^{+\infty} 
    \frac{1}{p!}\left(\frac{2 k_{34}}{s_1}\right)^{\frac{3d-4}{2}+n+p} \left(\frac{s_2}{2 k_{34}}\right)^{\frac{d-2}{2}+n}\\
    &\times
    Q_{p+n+\frac{3d-5}{2}}^{\frac{d-3}{2}}\left(\frac{k_{12}}{s_1}\right)
    \frac{\frac{3d-4}{2}+n+p}{\left(\frac{3d-4}{2}+n+p\right)^2+\mu_1^2}\\
    &\times\F_4\left(
    \begin{matrix}
        -\frac{p}{2}\,,\,\frac{1-p}{2} \\ 3-\frac{3d}{2}-p-n\,,\, n+\frac{d}{2}
    \end{matrix}\,,\left(\frac{s_1}{k_{34}}\right)^2\,,\,\left(\frac{s_2}{k_{34}}\right)^2\right)\,.
    \end{aligned}
    \label{eq_G+++_1BB_C2<1_C1p>1_res_III}
\end{equation}
\end{itemize}

\paragraph{Final result for $\kappaone<1$, $\kappatwo>1$.}
To write the final result for $G_{+++}^{B,B}$, we gather the three components \eqref{eq_G+++_0BB_C2<1}, \eqref{eq_G+++_1BB_C2<1_C1p>1_res_sh} and \eqref{eq_G+++_1BB_C2<1_C1p>1_res_III}. However, in a similar way to the first integration layer, we first note that the sum of the two contributions that originate from poles of $\nu/\sinh(\pi\nu)$ (\eqref{eq_G+++_0BB_C2<1} and \eqref{eq_G+++_1BB_C2<1_C1p>1_res_sh}) vanishes, due to the connection formula between the vertex functions $\I_{KI}$ and $\I_{II}$ \eqref{eq_connection_IKI_III} applied at $\nu=i(p+1)$. Indeed, it implies that each term of this sum is $\propto \cos\left(\pi(p+1/2)\right)=0$. The third contribution \eqref{eq_G+++_1BB_C2<1_C1p>1_res_III} is the only remaining one, hence from \eqref{eq_G_+++_.B_def_wrt_G_+++_01B_C2<1} we have
\begin{framed}
\vspace{-.5cm}
\begin{equation}
\begin{aligned}
    G_{+++}^{B,B}&\left(\{k_i,s_j\}\right)
    = \frac{e^{-i\pi(d-1)} \pi^7}{2^3 \sqrt{k_1 k_2 k_3 k_4 k_5 k_6} k_{34}^{d-1} s_1^{\frac{d-2}{2}}s_2^{\frac{d-2}{2}}\sin\left(\pi\frac{d}{2}\right)\sin\left(\frac{3\pi d}{2}\right)}
    \left(\frac{k_{34}}{k_{12}}\right)^{\frac{3d-4}{2}}
    \\
    &\times \left(\frac{s_1 s_2}{k_{12} k_{34}}\right)^{\frac{d-2}{2}}\sum_{n=0}^{+\infty} \sum_{p=0}^{+\infty} \frac{(-1)^n \Gamma\left(2 d+ p+n-3\right)}{n! p!}
    \left(\frac{k_{56}}{k_{12}}\right)^n
    \left(\frac{k_{34}}{k_{12}}\right)^{p} 
    \\
    &\times
    \,_2\F_1\left(
    \begin{matrix}
        d-\frac{1}{2}+ \frac{p+n}{2}\,,\,d-\frac{3}{2}+ \frac{p+n}{2} \\ p+n+\frac{3d-2}{2}
    \end{matrix}\,,\left(\frac{s_1}{k_{12}}\right)^2\right)
    \frac{\frac{3d-4}{2}+n+p}{\left(\frac{3d-4}{2}+n+p\right)^2+\mu_1^2}
    \\
    &\times\F_4\left(
    \begin{matrix}
        -\frac{p}{2}\,,\,\frac{1-p}{2} \\ 3-\frac{3d}{2}-p-n\,,\, n+\frac{d}{2}
    \end{matrix}\,,\left(\frac{s_1}{k_{34}}\right)^2\,,\,\left(\frac{s_2}{k_{34}}\right)^2\right)\\
    &\times 
    \frac{n+\frac{d-2}{2}}{\left(n+\frac{d-2}{2}\right)^2+\mu_2^2}
    \,_2\F_1\left(
    \begin{matrix}
        \frac{1-n}{2}\,,\,-\frac{n}{2} \\ -n-\frac{d-4}{2}
    \end{matrix}\,,\left(\frac{s_2}{k_{56}}\right)^2\right)
    \,.
    \end{aligned}
    \label{eq_G+++_BB_C2<1_C11p>1_full_res}
\end{equation}
\end{framed}

As mentioned before, we leave the other kinematic cases to Appendix \ref{appendix_kinematic_cases_G+++} for the interested reader. We make a similar choice for the other upcoming components of $G_{+++}$, focusing in the main text on the case $\kappaone<1$ and $\kappatwo>1$.
The next component we study is $G_{+++}^{S,B}$ \eqref{eq_decomposition_G+++_BorS}, which encodes the Signal-Background residues.

\paragraph{Signal-Background residues, $\kappaone<1$, $\kappatwo>1$.}
The computation of $G_{+++}^{S,B}$ when $\kappaone<1$ can be done from the decomposition of the quantity $G_{+++}^{.,B}$ \eqref{eq_G_+++_.B_def_wrt_G_+++_01B_C2<1} into the two pieces $G_{+++}^{0,.,B}$ and $G_{+++}^{1,.,B}$ \eqref{eq_G_+++_0.B_def_C2<1}-\eqref{eq_G_+++_1.B_def_C2<1}. From these integrals, $G_{+++}^{S,B}$ is obtained by taking their signal residues, following the correct contour prescriptions that were derived previously. These signal residues in the integral expression of $G_{+++}^{.,B}$ \eqref{eq_G_+++_.B_def_wrt_G_+++_01B_C2<1} arise from the KLF propagator $\left(\nu^2-\mu_1^2\right)_{i\epsilon}$ (see Appendix \ref{subapp_details_double_exch} and Fig. \ref{fig_poles_ieps}).

For the first component $G_{+++}^{0,S,B}$, the contour has to be closed in the lower-half $\nu$-plane (see Fig. \ref{fig_analytic_structure_G+++B_C2<1}). Then, the two signal residues that are contained in $G_{+++}^{0,S,B}$ are
\begin{equation}
\begin{aligned}
    G_{+++}^{0,S,B}&= \frac{-\pi i}{2\sinh^2(\pi\mu_1)} \sum_{c=\pm}
    e^{\pi c\mu_1}
    Q_{i c\mu_1-\frac{1}{2}}^{\frac{d-3}{2}}\left(\frac{k_{12}}{s_1}\right)
    \mathcal{I}_d\begin{bmatrix}
        c\mu_1 &-i\left(n+\frac{d-2}{2}\right) &i/2 & i/2 \\ s_1 & s_2 & k_3 & k_4  \\ I & I & K & K
    \end{bmatrix}
\end{aligned}
\label{eq_G+++_0SB_C2<1}
\end{equation}
For the second piece $G_{+++}^{1,S,B}$, as for the computation of the background-background residues, we need to distinguish cases depending on the value of the ratios $\kappatwo$ \eqref{eq_condition_C11p} and $\kappathree$ \eqref{eq_condition_C12p}. As for the background-background component we focus here on the case $\kappaone<1$ and $\kappatwo>1$, then the contour closure has to be done in the lower half-plane. It gives
\begin{equation}
\begin{aligned}
    G_{+++}^{1,S,B}&=\frac{\pi i }{2\sinh^2(\pi\mu_1)} \sum_{c=\pm} e^{\pi c\mu_1}
    Q_{i c\mu_1-\frac{1}{2}}^{\frac{d-3}{2}}\left(\frac{k_{12}}{s_1}\right)
    \mathcal{I}_d\begin{bmatrix}
        -c\mu_1 &-i\left(n+\frac{d-2}{2}\right) &i/2 & i/2 \\ s_1 & s_2 & k_3 & k_4  \\ I & I & K & K
    \end{bmatrix}\,.
\end{aligned}
\label{eq_G+++_0SB_C2<1_C1p>1}
\end{equation}
Gathering the two contributions \eqref{eq_G+++_0SB_C2<1} and \eqref{eq_G+++_0SB_C2<1_C1p>1}, we can use the connection formula between $\I_{KI}$ and $\I_{II}$ \eqref{eq_connection_IKI_III} and the analytic continuation formula for the Legendre $P$ function \eqref{eq_analytic_cont_P} to simplify their sum:
\begin{equation}
\begin{aligned}
    G_{+++}^{0,S,B}+G_{+++}^{1,S,B}&=
    e^{i\pi\frac{d-1}{2}}\Gamma\left(\frac{d-2}{2}\pm i\mu_1\right)
    P_{i \mu_1-\frac{1}{2}}^{-\frac{d-3}{2}}\left(e^{+i\pi}\frac{k_{12}}{s_1}\right) \\
    &\quad\times\mathcal{I}_d\begin{bmatrix}
        \mu_1 &-i\left(n+\frac{d-2}{2}\right) &i/2 & i/2 \\ s_1 & s_2 & k_3 & k_4  \\ K & I & K & K
    \end{bmatrix}
    \,.
\end{aligned}
\label{eq_G+++_0SB_+_G+++_1SB_C2<1_C1p>1}
\end{equation}
Inserting this expression in the decomposition \eqref{eq_G_+++_.B_def_wrt_G_+++_01B_C2<1} leads to the following expression for $G_{+++}^{S,B}$:
\begin{framed}
\vspace{-.5cm}
\begin{equation}
\begin{aligned}
    G_{+++}^{S,B}&=
    \frac{e^{-i\pi(d-1)} 2^{\frac{d-7}{2}} \pi^\frac{9}{2}}{\sqrt{k_1 k_2 k_5 k_6}s_1^{\frac{d-2}{2}}s_2^{\frac{d-2}{2}}\sin\left(\pi\frac{d}{2}\right)}
    \left(\left(\frac{k_{12}}{s_1}\right)^2-1\right)^{\frac{3-d}{4}}\\
    &\times \Gamma\left(\frac{d-2}{2}\pm i\mu_1\right)
    P_{i \mu_1-\frac{1}{2}}^{-\frac{d-3}{2}}\left(e^{+i\pi}\frac{k_{12}}{s_1}\right)\\
    &\times\sum_{n=0}^{+\infty}
    \mathcal{I}_d\begin{bmatrix}
        \mu_1 &-i\left(n+\frac{d-2}{2}\right) &i/2 & i/2 \\ s_1 & s_2 & k_3 & k_4  \\ K & I & K & K
    \end{bmatrix}\\
    &\times\frac{n+\frac{d-2}{2}}{\left(n+\frac{d-2}{2}\right)^2+\mu_2^2}
    \frac{1}{n!}\left(\frac{2 k_{56}}{s_2}\right)^n
    \,_2\F_1\left(
    \begin{matrix}
        \frac{1-n}{2}\,,\,-\frac{n}{2} \\ -n-\frac{d-4}{2}
    \end{matrix}\,,\left(\frac{s_2}{k_{56}}\right)^2\right)\,.
\end{aligned}
\label{eq_G+++_SB_C2<1_C11p>1}
\end{equation}
\end{framed}

\paragraph{Background-Signal residues, $\kappaone<1$, $\kappathree>1$.}
To obtain the complete expression for $G_{+++}$, two components remain to be computed, according to \eqref{eq_decomposition_G+++_BorS}, which are $G_{+++}^{B,S}$ and $G_{+++}^{S,S}$. To obtain these components, we need the signal residues of the first integration layer, contained in $F_{+++}^S$ \eqref{eq_F+++_S_C2>1}, and integrate them over the frequency $\nu$. To this end, we can define the quantity $G_{+++}^{.,S}$ from \eqref{eq_G_+++_wrt_F_+++} as
\begin{equation}
\begin{aligned}
    &G_{+++}^{.,S}\left(\{k_i,s_j\}\right)\equiv G_{+++}^{B,S}+G_{+++}^{S,S}\\
    &=e^{-i\pi(d-1)}
    \int\limits_{-\infty}^{+\infty}\d\nu \, \N_{\nu}\;
    \mathcal{I}_d\begin{bmatrix}
        \nu &i/2 & i/2 \\ s_1 &k_1 & k_2 \\ K & K & K
    \end{bmatrix} \frac{1}{(\nu^2-\mu_{1}^2)_{i\epsilon}}\,
    F_{+++}^S \left(\{k_i,s_j\},\nu\right)\,.
    \end{aligned}
    \label{eq_G_+++_.S_def_wrt_F+++S}
\end{equation}
Using the expression of $F_{+++}^S$ found for $\kappaone<1$ \eqref{eq_F+++_S_C2<1}, and the connection formulas \eqref{eq_vertex_fct_12_Q} and \eqref{eq_connection_IKK_IIK} for the vertex functions, it turns out that $G_{+++}^{.,S}$ can be written as
\begin{equation}
\begin{aligned}
    G_{+++}^{.,S}\left(\{k_i,s_j\}\right)
    &=\frac{e^{-i\pi\frac{3}{2}d} \pi^3}{2^4\sqrt{k_1 k_2 k_5 k_6}s_1^{\frac{d-2}{2}}s_2^{\frac{d-2}{2}}}
    \left(\left(\frac{k_{12}}{s_1}\right)^2-1\right)^{\frac{3-d}{4}}
    \left(\left(\frac{k_{56}}{s_2}\right)^2-1\right)^{\frac{3-d}{4}}\\
    &\times\Gamma\left(\frac{d-2}{2}\pm i\mu_2\right)
    P_{i\mu_2-\frac{1}{2}}^{-\frac{d-3}{2}}\left(\frac{k_{56}}{s_2}\right)\\
    &\times\Biggl[G_{+++}^{0,.,S}+G_{+++}^{1,.,S}\Biggr]
    \,.
    \end{aligned}
    \label{eq_G_+++_.S_wrt_G+++_01.S_C2<1}
\end{equation}
with
\begin{subequations}
\begin{equation}
\begin{aligned}
    G_{+++}^{0,.,S}&=
    \int\limits_{-\infty}^{+\infty}\d\nu \, \frac{\nu}{\sinh(\pi\nu)}\, Q_{i\nu-\frac{1}{2}}^{\frac{d-3}{2}}\left(\frac{k_{12}}{s_1}\right) \frac{1}{(\nu^2-\mu_{1}^2)_{i\epsilon}}\,\mathcal{I}_d\begin{bmatrix}
        \nu &\mu_2 &i/2 & i/2 \\ s_1 & e^{-i\pi}s_2 & k_3 & k_4 \\ I &K & K & K
    \end{bmatrix}\,,
    \end{aligned}
    \label{eq_G_+++_0.S_def_C2<1}
\end{equation}
\begin{equation}
\begin{aligned}
G_{+++}^{1,.,S}&=
    -\int\limits_{-\infty}^{+\infty}\d\nu \, \frac{\nu}{\sinh(\pi\nu)}\, Q_{-i\nu-\frac{1}{2}}^{\frac{d-3}{2}}\left(\frac{k_{12}}{s_1}\right) \frac{1}{(\nu^2-\mu_{1}^2)_{i\epsilon}}\,\mathcal{I}_d\begin{bmatrix}
        \nu &\mu_2 &i/2 & i/2 \\ s_1 & e^{-i\pi}s_2 & k_3 & k_4 \\ I &K & K & K
    \end{bmatrix}\,,
    \end{aligned}
    \label{eq_G_+++_1.S_def_C2<1}
\end{equation}
\label{eq_G_+++_0and1.S_def_C2<1}
\end{subequations}
where we used the change of variable $\nu'=-\nu$ to reduce the number of terms in \eqref{eq_G_+++_.S_wrt_G+++_01.S_C2<1} to two.

Now, we observe that the two integrals $G_{+++}^{0,.,S}$ and $G_{+++}^{1,.,S}$ are respectively identical to $F_{+++}^0$ \eqref{eq_F_+++_0} and $F_{+++}^1$ \eqref{eq_F_+++_1}, up to a suitable exchange of parameters\footnote{More precisely, to go from $F_{+++}^0$/$F_{+++}^1$ to $G_{+++}^{0,.,S}$/$G_{+++}^{1,.,S}$ we need to do the following replacements: $\mu\to\nu$, $\mu_2\to\mu_1$, $\nu\to\mu_2$, $k_{56}\to k_{12}$, $s_2\to s_1$ and $s_1\to e^{-i\pi} s_2$.}. 
First, the contour closure prescription for $G_{+++}^{0,.,S}$ and $G_{+++}^{1,.,S}$ can be derived in a similar way to $F_{+++}^0$ and $F_{+++}^1$, by considering the leading exponential terms in the large $\nu$ asymptotic behaviour of the integrands:
\begin{equation}
    \mathcal{I}_d\begin{bmatrix}
        \nu &\mu_2 &i/2 & i/2 \\ s_1 & e^{-i\pi}s_2 &k_3 & k_4 \\ I & K & K & K
    \end{bmatrix}
    Q_{\pm i\nu-\frac{1}{2}}^{\frac{d-3}{2}}\left(\frac{k_{12}}{s_1}\right)
    \underset{\nu\to\infty}{\propto}e^{-i\nu\left[\arccosh\left(\frac{k_{34}-s_2}{s_1}\right)\pm\arccosh\left(\frac{k_{12}}{s_1}\right)\right]}\,.
    \label{eq_asymptotics_G+++01.S_C2<1}
\end{equation}
For $G_{+++}^{0,.,S}$, we pick the $+$ sign in the above expression and since $k_{34}-s_2>s_1$ and $k_{12}>s_1$ in our setup, we can always close the contour in the lower-half $\nu$ plane. For $G_{+++}^{1,.,S}$, the closure will depend on the kinematic ratio $\kappathree$ \eqref{eq_condition_C12p}: if $\kappathree>1$ (resp. $\kappathree<1$), the contour must be closed in the upper-half (resp. lower-half) $\nu$-plane.

These features of $G_{+++}^{.,S}$ will allow us to compute both $G_{+++}^{B,S}$ and $G_{+++}^{S,S}$. Starting with $G_{+++}^{B,S}$, we first compute the component $G_{+++}^{0,B,S}$ by collecting the background residues of $G_{+++}^{0,.,S}$ \eqref{eq_G_+++_0.S_def_C2<1}. Using the expression already found for $F_{+++}^{0,B}$ \eqref{eq_F+++0B_res}, we have
\begin{equation}
    G_{+++}^{0,B,S}=-2 \sum_{n=0}^{+\infty} (-1)^n\frac{n+1}{(n+1)^2+\mu_1^2}\,
    \mathcal{I}_d\begin{bmatrix}
        -i(n+1)& \mu_2 &i/2 & i/2 \\ s_1 & e^{-i\pi}s_2 &k_3 & k_4 \\ I & K & K & K
    \end{bmatrix}
    Q_{n+\frac{1}{2}}^{\frac{d-3}{2}}\left(\frac{k_{12}}{s_1}\right)\,.
    \label{eq_G+++0BS_C2<1_res}
\end{equation}
Since the integral $G_{+++}^{1,.,S}$ for $\kappathree>1$ \eqref{eq_G_+++_1.S_def_C2<1} is identical to $F_{+++}^1$ \eqref{eq_F_+++_1} (up to a change of external kinematic labels), to find the contributions to $G_{+++}^{1,B,S}$ we can follow the computation of $F_{+++}^{1,B}$ already performed previously in the case $\kappaone>1$. It is made of the two contributions \eqref{eq_F+++1B_C2>1_res_sh} and \eqref{eq_F+++1B_C2>1_res_IKI}, which we use here to find the corresponding ones of $G_{+++}^{1,B,S}$.
\begin{itemize}
\item \textbf{\textcolor{pyblue}{Poles from $\frac{\mu}{\sinh(\pi\mu)}$.}}
The corresponding residues are:
\begin{equation}
    G_{+++}^{1,B,S}\supset 2 \sum_{n=0}^{+\infty} (-1)^n\frac{n+1}{(n+1)^2+\mu_1^2}\,
    \mathcal{I}_d\begin{bmatrix}
        i(n+1)&\mu_2 &i/2 & i/2 \\ s_1 & e^{-i\pi}s_2 &k_3 & k_4 \\ I & K & K & K
    \end{bmatrix}
    Q_{n+\frac{1}{2}}^{\frac{d-3}{2}}\left(\frac{k_{12}}{s_1}\right)\,.
    \label{eq_G+++1BS_C2<1_C1p>1_res_sh}
\end{equation}
\item \textbf{\textcolor{pyorange}{Poles from $\I_{IK}$.}} The other poles that contribute come from $\I_{IK}$, with residues
\begin{equation}
\begin{aligned}
    G_{+++}^{1,B,S}&\supset -i\frac{2^{d-2}\pi^3}{\sqrt{k_3 k_4} \,s_1^{d-1}}
    \sum_{c=\pm}\frac{1}{\sinh(\pi c \mu_2)} \left(-\frac{s_2}{s_1}\right)^{i c\mu_2} \sum_{n=0}^{+\infty} \left(\frac{2 k_{34}}{s_1}\right)^n \frac{1}{n!}\\
    &\times \frac{1}{\sinh(\pi c\mu_2-i\pi d)}\,\F_4\Biggl(\begin{matrix}
    -\frac{n}{2}\,,\, \frac{1-n}{2}\\
    2-i c \mu_2-n-d\,,\,1+i c\mu_2\end{matrix}\,;
    \left(\frac{s_1}{k_{34}}\right)^2,\left(\frac{s_2}{k_{34}}\right)^2\Biggr)\\
    &\times Q_{i c \mu_2 +n+d-\frac{3}{2}}^{\frac{d-3}{2}}\left(\frac{k_{12}}{s_1}\right)\,\frac{-c\mu_2+i(n+d-1)}{\left(c\mu_2-i(n+d-1)\right)^2-\mu_1^2}\,.
    \end{aligned}
    \label{eq_G+++1BS_C2<1_C1p>1_res_IKI}
\end{equation}
\end{itemize}
The cancellation between the first two terms \eqref{eq_G+++0BS_C2<1_res} and \eqref{eq_G+++1BS_C2<1_C1p>1_res_sh} still occurs thanks to the connection formula \eqref{eq_connection_IKK_IIK} between $\I_{KK}$ and $\I_{IK}$, and we finally obtain from \eqref{eq_G_+++_.S_wrt_G+++_01.S_C2<1}:
\begin{framed}
\vspace{-.5cm}
\begin{equation}
\begin{aligned}
    G_{+++}^{B,S}&\left(\{k_i,s_j\}\right)
    =\frac{-e^{-i\pi\frac{3d}{2}}2^{d-6}\pi^6}{\sqrt{k_1 k_2 k_3 k_4 k_5 k_6}s_1^{\frac{3d-4}{2}} s_2^{\frac{d-2}{2}}}
    \left(\left(\frac{k_{12}}{s_1}\right)^2-1\right)^{\frac{3-d}{4}}
    \left(\left(\frac{k_{56}}{s_2}\right)^2-1\right)^{\frac{3-d}{4}}\\
    &\times
    \Gamma\left(\frac{d-2}{2}\pm i\mu_2\right)
    P_{i\mu_2-\frac{1}{2}}^{-\frac{d-3}{2}}\left(\frac{k_{56}}{s_2}\right)
    \sum_{c=\pm}\frac{1}{\sinh(\pi c \mu_2)} \left(-\frac{s_2}{s_1}\right)^{i c\mu_2}\\
    &\times\sum_{n=0}^{+\infty} \left(\frac{2 k_{34}}{s_1}\right)^n \frac{1}{n!}
    Q_{i c \mu_2 +n+d-\frac{3}{2}}^{\frac{d-3}{2}}\left(\frac{k_{12}}{s_1}\right)\,\frac{-c\mu_2+i(n+d-1)}{\left(c\mu_2-i(n+d-1)\right)^2-\mu_1^2}\\
    &\times\frac{1}{\sinh(\pi c\mu_2-i\pi d)}\,\F_4\Biggl(\begin{matrix}
    -\frac{n}{2}\,,\, \frac{1-n}{2}\\
    2-i c \mu_2-n-d\,,\,1+i c\mu_2\end{matrix}\,;
    \left(\frac{s_1}{k_{34}}\right)^2,\left(\frac{s_2}{k_{34}}\right)^2\Biggr).
    \end{aligned}
    \label{eq_G_+++_BS_C2<1_C11p>1}
\end{equation}
\end{framed}

\paragraph{Signal-Signal residues, $\kappaone<1$, $\kappathree>1$.}
The last component that we need to compute to obtain the complete solution for $G_{+++}$ is the one given by signal-signal residues, i.e. $G_{+++}^{S,S}$, given by the signal residues of the quantity $G_{+++}^{.,S}$ written for $\kappaone<1$ \eqref{eq_G_+++_.S_wrt_G+++_01.S_C2<1}. These residues simply come from the $i\epsilon$-prescription carried by the propagator $\left(\nu^2-\mu_1^2\right)_{i\epsilon}^{-1}$ (see Appendix \ref{subapp_details_double_exch} and Fig. \ref{fig_poles_ieps}) in the integrals $G_{+++}^{0,.,S}$ \eqref{eq_G_+++_0.S_def_C2<1} and $G_{+++}^{1,.,S}$ \eqref{eq_G_+++_1.S_def_C2<1}.

Let us start by computing $G_{+++}^{0,S,S}$, i.e. the signal residues of $G_{+++}^{0,.,S}$ \eqref{eq_G_+++_0.S_def_C2<1}. For this quantity, the contour has to be closed in the lower half-plane. Therefore, we collect the residues corresponding to the poles located at $\nu=\pm\mu_1-i\epsilon$:
\begin{equation}
\begin{aligned}
    G_{+++}^{0,S,S}=&-\frac{i\pi}{2\sinh^2(\pi\mu_1)}
    \sum_{c=\pm}e^{\pi c\mu_1}\mathcal{I}_d\begin{bmatrix}
        c\mu_1 & \mu_2 &i/2 & i/2 \\ s_1 & e^{-i\pi}s_2 & k_3 & k_4  \\ I & K & K & K
    \end{bmatrix}
    Q_{i c\mu_1-\frac{1}{2}}^{\frac{d-3}{2}}\left(\frac{k_{12}}{s_1}\right)
    \,.
    \end{aligned}
    \label{eq_G+++_0SS_C2<1}
\end{equation}
For $\kappathree>1$, $G_{+++}^{1,S,S}$ is obtained from $G_{+++}^{1,.,S}$ \eqref{eq_G_+++_1.S_def_C2<1} by collecting the signal residues corresponding to the poles at $\nu=\pm\mu_1+i\epsilon$:
\begin{equation}
\begin{aligned}
    G_{+++}^{1,S,S}=&\frac{i\pi}{2\sinh^2(\pi\mu_1)}
    \sum_{c=\pm}e^{-\pi c\mu_1}\mathcal{I}_d\begin{bmatrix}
        c\mu_1 &\mu_2 & i/2 & i/2 \\ s_1 &e^{-i\pi}s_2 & k_3 & k_4  \\ I & K & K & K
    \end{bmatrix}
    Q_{-i c\mu_1-\frac{1}{2}}^{\frac{d-3}{2}}\left(\frac{k_{12}}{s_1}\right)
    \,.
    \end{aligned}
    \label{eq_G+++_1SS_C2<1_C1p>1}
\end{equation}
The sum of the two terms \eqref{eq_G+++_0SS_C2<1} and \eqref{eq_G+++_1SS_C2<1_C1p>1} is rewritten using the connection formula between $\I_{KK}$ and $\I_{IK}$ \eqref{eq_connection_IKK_IIK} and the analytical continuation formula \eqref{eq_analytic_cont_P} for Legendre $P$. Moreover, using the decomposition \eqref{eq_G_+++_.S_wrt_G+++_01.S_C2<1}, we end with
\begin{framed}
\vspace{-.5cm}
\begin{equation}
\begin{aligned}
    G_{+++}^{S,S}
    &=\frac{e^{-i\pi\left(d+\frac{1}{2}\right)} \pi^3}{2^4\sqrt{k_1 k_2 k_5 k_6}s_1^{\frac{d-2}{2}}s_2^{\frac{d-2}{2}}}
    \left(\left(\frac{k_{12}}{s_1}\right)^2-1\right)^{\frac{3-d}{4}}
    \left(\left(\frac{k_{56}}{s_2}\right)^2-1\right)^{\frac{3-d}{4}}\\
    &\times\Gamma\left(\frac{d-2}{2}\pm i\mu_1\right)\Gamma\left(\frac{d-2}{2}\pm i\mu_2\right)
    P_{i\mu_1-\frac{1}{2}}^{-\frac{d-3}{2}}\left(e^{+i\pi}\frac{k_{12}}{s_1}\right)
    \\
    &\times
    \mathcal{I}_d\begin{bmatrix}
        \mu_1 &\mu_2 & i/2 & i/2 \\ s_1 &e^{-i\pi}s_2 & k_3 & k_4  \\ K & K & K & K
    \end{bmatrix}
    P_{i\mu_2-\frac{1}{2}}^{-\frac{d-3}{2}}\left(\frac{k_{56}}{s_2}\right)
    \,.
    \end{aligned}
    \label{eq_G+++_SS_C2<1_C11p>1}
\end{equation}
\end{framed}

\paragraph{Summary.} In this subsection, series expressions of the component $\G_{+++}$ of the double-exchange diagram \eqref{draw_double_exch_diagram_general} have been derived. $\G_{+++}$ is related to the quantity $G_{+++}$ by \eqref{eq_G_a1a2a3}, the latter being given in the kinematic case $\kappaone<1$ and $\kappathree>1$ by the sum of \eqref{eq_G+++_BB_C2<1_C11p>1_full_res}, \eqref{eq_G+++_SB_C2<1_C11p>1}, \eqref{eq_G_+++_BS_C2<1_C11p>1} and \eqref{eq_G+++_SS_C2<1_C11p>1} (since for the two last ones $\kappathree>\kappatwo>1$), according to the decomposition \eqref{eq_decomposition_G+++_BorS} into background and signal pieces.

A complete summary of the whole double-exchange computation is written at the end of the current section. To this end, we now turn to the computation of the components $\G_{-++}$ and $\G_{+-+}$.

\subsection{Partially Nested Component}\label{subsection_-++}

The component $\mathcal{G}_{-++}$ of the double-exchange diagram \eqref{draw_double_exch_diagram_general} is obtained by setting the Schwinger-Keldysh indices $a_i$ to $a_1=-$ and $a_2=a_3=+$. The non-trivial piece to compute here is the integral $G_{-++}$, from $G_{\a_1\a_2\a_3}$ given by  \eqref{eq_G_integrals_a1a2a3} and related to $\G_{-++}$ by \eqref{eq_G_a1a2a3}. Because we still have $\a_2=\a_3=+$, the internal  propagator for the internal line between the central and right vertices is still given by \eqref{eq_propagator_++_line2}, as in the case of $\mathcal{G}_{+++}$ studied in the previous subsection. However, now the KLF propagator for the left internal line is given by
\begin{equation}
    \Pi^{\mu_1}_{-+}(\nu)=\frac{\delta\left(\nu-\mu_1\right)}{\N_{\nu}}\,.
    \label{eq_propagator_-+_line1}
\end{equation}
Consequently, using the expressions of propagators \eqref{eq_propagator_-+_line1} and \eqref{eq_propagator_++_line2} in the integral $G_{\a_1\a_2\a_3}$ \eqref{eq_G_integrals_a1a2a3} with $a_1=-$ and $a_2=a_3=+$, the integral $G_{-++}$ becomes
\begin{equation}
\begin{aligned}
    G_{-++} \left(\{k_i,s_j\}\right)
    &\equiv e^{-\frac{i\pi(d-1)}{2}}
    \int\limits_{-\infty}^{+\infty}\d\nu \,
    \int\limits_{-\infty}^{+\infty}\d\mu \, \N_{\mu}\,
    \mathcal{I}_d\begin{bmatrix}
        \nu &i/2 & i/2 \\ s_1 &k_1 & k_2 \\ K & K & K
    \end{bmatrix} \delta\left(\nu-\mu_1\right)
    \\
    &\times
    \mathcal{I}_d\begin{bmatrix}
        \nu &\mu &i/2 & i/2 \\ s_1 & s_2 &k_3 & k_4 \\ K &K & K & K
    \end{bmatrix}
    \frac{1}{(\mu^2-\mu_2^2)_{i\epsilon}}\,
    \mathcal{I}_d\begin{bmatrix}
        \mu &i/2 & i/2 \\ s_2 &k_5 & k_6 \\ K & K & K
    \end{bmatrix}\,.
    \end{aligned}
    \label{eq_G_integrals_-++}
\end{equation}
The first integration over $\nu$ can be done immediately, due to the presence of the $\delta$-function. We get
\begin{equation}
\begin{aligned}
    G_{-++} &\left(\{k_i,s_j\}\right)
    \equiv e^{-\frac{i\pi(d-1)}{2}}
    \mathcal{I}_d\begin{bmatrix}
        \mu_1 &i/2 & i/2 \\ s_1 &k_1 & k_2 \\ K & K & K
    \end{bmatrix}
    \\
    &\times
    \int\limits_{-\infty}^{+\infty}\d\mu \, \N_{\mu}\,
    \mathcal{I}_d\begin{bmatrix}
        \mu_1 &\mu &i/2 & i/2 \\ s_1 & s_2 &k_3 & k_4 \\ K &K & K & K
    \end{bmatrix}
    \frac{1}{(\mu^2-\mu_2^2)_{i\epsilon}}\,
    \mathcal{I}_d\begin{bmatrix}
        \mu &i/2 & i/2 \\ s_2 &k_5 & k_6 \\ K & K & K
    \end{bmatrix}\,.
    \end{aligned}
    \label{eq_G_integrals_-++_2}
\end{equation}
The remaining integral on $\mu$ is exactly $F_{+++} \left(\{k_i,s_j\},\mu_1\right)$ \eqref{eq_def_F_+++}, which is the first integration layer of the integral $G_{+++}$ in sec. \ref{subsection_+++}. Therefore, we have
\begin{equation}
\begin{aligned}
    G_{-++} &\left(\{k_i,s_j\}\right)
    \equiv e^{-\frac{i\pi(d-1)}{2}}
    \mathcal{I}_d\begin{bmatrix}
        \mu_1 &i/2 & i/2 \\ s_1 &k_1 & k_2 \\ K & K & K
    \end{bmatrix}\,
    F_{+++} \left(\{k_i,s_j\},\mu_1\right)
    \,.
    \end{aligned}
    \label{eq_G_integrals_-++_3}
\end{equation}
Finally, using what has been computed previously in sec. \ref{subsection_+++}, we can immediately obtain the integral $G_{-++}$, depending on the value of the kinematic ratio $\kappaone$ \eqref{eq_condition_C2}.
Recall first the decomposition $F_{+++}=F_{+++}^B+F_{+++}^S$, according to which we can split $G_{-++}$ into
\begin{equation}
    G_{-++}=G_{-++}^B+G_{-++}^S\,.
    \label{eq_G-++_BorS}
\end{equation}

\paragraph{$\kappaone<1$.} Using the background piece \eqref{eq_F+++B_C2<1} and the signal piece for $\kappaone<1$ into \eqref{eq_G_integrals_-++_3}, we have
\begin{framed}
\vspace{-.5cm}
\begin{subequations}
\begin{equation}
\begin{aligned}
    G_{-++}^B
    &= \frac{e^{-\frac{i\pi(d-1)}{2}}2^{\frac{d}{2}-2} \pi^3 }{\sin\left(\pi\frac{d}{2}\right)\sqrt{k_5 k_6} \,s_2^{\frac{d-2}{2}}}
    \mathcal{I}_d\begin{bmatrix}
        \mu_1 &i/2 & i/2 \\ s_1 &k_1 & k_2 \\ K & K & K
    \end{bmatrix}\,
    \sum_{n=0}^{+\infty} \mathcal{I}_d\begin{bmatrix}
        \mu_1 &-i\left(n+\frac{d-2}{2}\right) &i/2 & i/2 \\ s_1 & s_2 & k_3 & k_4  \\ K & I & K & K
    \end{bmatrix}\\
    &\times\frac{n+\frac{d-2}{2}}{\left(n+\frac{d-2}{2}\right)^2+\mu_2^2}\frac{1}{n!}\left(\frac{2 k_{56}}{s_2}\right)^n
    \,_2\F_1\left(
    \begin{matrix}
        \frac{1-n}{2}\,,\,-\frac{n}{2} \\ -n-\frac{d-4}{2}
    \end{matrix}\,,\left(\frac{s_2}{k_{56}}\right)^2\right)
    \,,
    \end{aligned}
    \label{eq_G-++_B_C2<1}
\end{equation}
\begin{equation}
\begin{aligned}
    G_{-++}^S &\left(\{k_i,s_j\}\right)
    \equiv e^{-\frac{i\pi(d-2)}{2}}
    \left(\frac{\pi}{2}\right)^{\frac{3}{2}} \frac{\left(\left(\frac{k_{56}}{s_2}\right)^2-1\right)^{\frac{3-d}{4}}}{\sqrt{k_5 k_6} \,s_2^{\frac{d-2}{2}}}\,
    \Gamma\left(\frac{d-2}{2}\pm i\mu_2\right)\\
    &\times\mathcal{I}_d\begin{bmatrix}
        \mu_1 &i/2 & i/2 \\ s_1 &k_1 & k_2 \\ K & K & K
    \end{bmatrix}\,
    \mathcal{I}_d\begin{bmatrix}
        \mu_1 & \mu_2 &i/2 & i/2 \\ s_1 & e^{-i\pi}s_2 & k_3 & k_4  \\ K & K & K & K
    \end{bmatrix}
    P_{i\mu_2-\frac{1}{2}}^{-\frac{d-3}{2}}\left(\frac{k_{56}}{s_2}\right)
    \,.
    \end{aligned}
    \label{eq_G-++_S_C2<1}
\end{equation}
\label{eq_G-++_C2<1}
\end{subequations}
\end{framed}

\paragraph{$\kappaone>1$.} Here, we insert the background piece \eqref{eq_F+++B_C2>1} and the signal piece \eqref{eq_F+++_S_C2>1} into \eqref{eq_G_integrals_-++_3}, to obtain
\begin{framed}
\vspace{-.5cm}
\begin{subequations}
\begin{equation}
\begin{aligned}
    G_{-++}^B&=
    \frac{e^{-i\pi\left(d-\frac{1}{2}\right)}2^{d-\frac{7}{2}}\pi^\frac{9}{2}}{\sqrt{k_3 k_4 k_5 k_6} \,s_2^{\frac{3d}{2}-2}}\left(\left(\frac{k_{56}}{s_2}\right)^2-1\right)^{\frac{3-d}{4}}\
    \mathcal{I}_d\begin{bmatrix}
        \mu_1 &i/2 & i/2 \\ s_1 &k_1 & k_2 \\ K & K & K
    \end{bmatrix}\\
    &\times
    \sum_{c=\pm}\frac{1}{\sinh(\pi c \mu_1)} \left(\frac{s_1}{s_2}\right)^{i c\mu_1} \sum_{n=0}^{+\infty} \left(\frac{2 k_{34}}{s_2}\right)^n \frac{1}{n!}\\
    &\times \frac{1}{\sinh(\pi c\mu_1-i\pi d)}\,\F_4\Biggl(\begin{matrix}
    -\frac{n}{2}\,,\, \frac{1-n}{2}\\
    1+i c\mu_1\,,\,2-i c \mu_1-n-d\end{matrix}\,;
    \left(\frac{s_1}{k_{34}}\right)^2,\left(\frac{s_2}{k_{34}}\right)^2\Biggr)\\
    &\times Q_{i c \mu_1 +n+d-\frac{3}{2}}^{\frac{d-3}{2}}\left(\frac{k_{56}}{s_2}\right)\,\frac{-c\mu_1+i(n+d-1)}{\left(c\mu_1-i(n+d-1)\right)^2-\mu_2^2}
    \,,
    \end{aligned}
    \label{eq_G-++_B_C2>1}
\end{equation}
\begin{equation}
\begin{aligned}
    G_{-++}^S &
    =\left(\frac{\pi}{2}\right)^{\frac{3}{2}} \frac{e^{-\frac{i\pi(d-1)}{2}}}{\sqrt{k_5 k_6} \,s_2^{\frac{d-2}{2}}}
    \left(\left(\frac{k_{56}}{s_2}\right)^2-1\right)^{\frac{3-d}{4}}
    \Gamma\left(\frac{d-2}{2}\pm i\mu_2\right)\\
    &\times
    \mathcal{I}_d\begin{bmatrix}
        \mu_1 &i/2 & i/2 \\ s_1 &k_1 & k_2 \\ K & K & K
    \end{bmatrix}
    \mathcal{I}_d\begin{bmatrix}
        \mu_1 & \mu_2 &i/2 & i/2 \\ s_1 & s_2 & k_3 & k_4  \\ K & K & K & K
    \end{bmatrix}
    P_{i\mu_2-\frac{1}{2}}^{-\frac{d-3}{2}}\left(e^{+i\pi}\frac{k_{56}}{s_2}\right)
    \,.
    \end{aligned}
    \label{eq_G-++_S_C2>1}
\end{equation}
\label{eq_G-++_C2>1}
\end{subequations}
\end{framed}

The original Schwinger-Keldysh component $\G_{-++}$ can be obtained by its relation \eqref{eq_G_a1a2a3} to the quantity $G_{-++}$, the latter being written by gathering its background and signal pieces, i.e. using \eqref{eq_G-++_BorS}.

\subsection{Factorised Component}\label{subsection_+-+}

The last component that is crucial to compute to obtain all the information needed on the double-exchange diagram is the factorised component, i.e. $\mathcal{G}_{+-+}$, which has Schwinger-Keldysh indices $\a_1=+$, $\a_2=-$ and $a_3=+$. To obtain $\mathcal{G}_{+-+}$, we need to compute the integral $G_{+-+}$ given in \eqref{eq_G_integrals_a1a2a3} with $\a_1=+$, $\a_2=-$ and $a_3=+$. The quantity $G_{+-+}$ is then simply related to $\mathcal{G}_{+-+}$ using the formula \eqref{eq_G_a1a2a3}.

In this case, both internal propagators take a very simple form. Indeed, the propagator $\Pi^{\mu_1}_{+-}(\nu)$ for the left internal line is still given by the $\delta$-function \eqref{eq_propagator_-+_line1}, as written before for the partially nested component. In addition, this is also the case for the right internal line propagator $\Pi^{\mu_2}_{-+}(\mu)$:
\begin{equation}
    \Pi^{\mu_2}_{-+}(\mu)=\frac{\delta\left(\mu-\mu_2\right)}{\N_{\mu}}\,.
    \label{eq_propagator_+-_line2}
\end{equation}
Using the propagators \eqref{eq_propagator_-+_line1} and \eqref{eq_propagator_+-_line2}, from \eqref{eq_G_integrals_a1a2a3} the integral $G_{+-+}$ is
\begin{equation}
\begin{aligned}
    G_{+-+} \left(\{k_i,s_j\}\right)
    &\equiv \int\limits_{-\infty}^{+\infty}\d\nu 
    \int\limits_{-\infty}^{+\infty}\d\mu 
    \;
    \mathcal{I}_d\begin{bmatrix}
        \nu &i/2 & i/2 \\ s_1 &k_1 & k_2 \\ K & K & K
    \end{bmatrix} \delta\left(\nu-\mu_1\right)\\
    &\times
    \mathcal{I}_d\begin{bmatrix}
        \nu &\mu &i/2 & i/2 \\ s_1 & s_2 &k_3 & k_4 \\ K &K & K & K
    \end{bmatrix}
    \delta\left(\mu-\mu_2\right)
    \mathcal{I}_d\begin{bmatrix}
        \mu &i/2 & i/2 \\ s_2 &k_5 & k_6 \\ K & K & K
    \end{bmatrix}\,.
    \end{aligned}
    \label{eq_G_integrals_+-+_0}
\end{equation}
The integral over $\delta$-functions gives straightforwardly
\begin{framed}
\begin{equation}
\begin{aligned}
    G_{+-+} \left(\{k_i,s_j\}\right)=
    \mathcal{I}_d\begin{bmatrix}
        \mu_1 &i/2 & i/2 \\ s_1 &k_1 & k_2 \\ K & K & K
    \end{bmatrix}\,
    \mathcal{I}_d\begin{bmatrix}
        \mu_1 &\mu_2 &i/2 & i/2 \\ s_1 & s_2 &k_3 & k_4 \\ K &K & K & K
    \end{bmatrix}\,
    \mathcal{I}_d\begin{bmatrix}
        \mu_2 &i/2 & i/2 \\ s_2 &k_5 & k_6 \\ K & K & K
    \end{bmatrix}\,.
    \end{aligned}
    \label{eq_G_integrals_+-+}
\end{equation}
\end{framed}
The relation \eqref{eq_G_a1a2a3} then relates $G_{+-+}$ to the diagram component $\G_{+-+}$.

\subsection{Final Result}\label{subsection_full_result}

The double-exchange diagram \eqref{draw_double_exch_diagram_general} can be expressed as a sum over 8 Schwinger-Keldysh components, each having a fixed value of indices $\a_i=\pm$, $i=1,2,3$. In this work, we compute the totally nested component $\G_{+++}$ in subsection \ref{subsection_+++}, a partially nested one $\G_{-++}$ in subsection \ref{subsection_-++}, and the factorised one $\G_{+-+}$ in subsection \ref{subsection_+-+}. The other components can be obtained by complex conjugation: $\G_{---}=\G_{+++}^*$, $\G_{+--}=\G_{-++}^*$ and $\G_{-+-}=\G_{+-+}^*$. The last two can be made explicit by using the symmetry of the diagram that consists in exchanging the left and right vertices. Therefore, up to the change of kinematic parameters $k_{12}\leftrightarrow k_{56}$, $s_1\leftrightarrow s_2$ and $\mu_1 \leftrightarrow\mu_2$, the components $\G_{++-}$ and $\G_{--+}$ can be respectively obtained from $\G_{-++}$ and $\G_{+--}$. The results for $\G_{+++}$, $\G_{-++}$ and $\G_{+-+}$ are summarised in the table below.

Displaying only the quantities $G_{\a_1\a_2\a_3}$ (recall that $\G_{\a_1\a_2\a_3}$ is obtained from \eqref{eq_G_a1a2a3}), we have:
\begin{itemize}
    \item Recall that $G_{+++}=G_{+++}^{B,B}+G_{+++}^{B,S}+G_{+++}^{S,B}+G_{+++}^{S,S}$ \eqref{eq_decomposition_G+++_BorS}:

\begin{table}[H]
\centering
\renewcommand{\arraystretch}{1.5}
\begin{tabular}{|ll|l|l|l|l|l|}
\hline
\rowcolor[HTML]{C0C0C0} 
\multicolumn{2}{|l|}{\cellcolor[HTML]{C0C0C0}} & $G_{+++}^{B,B}$ & $G_{+++}^{S,B}$ &  & $G_{+++}^{B,S}$ & $G_{+++}^{S,S}$ \\ \hline
\multicolumn{1}{|c|}{} & $\kappatwo>1$ & \eqref{eq_G+++_BB_C2<1_C11p>1_full_res}
& \eqref{eq_G+++_SB_C2<1_C11p>1} & $\kappathree>1$ & \eqref{eq_G_+++_BS_C2<1_C11p>1}
& \eqref{eq_G+++_SS_C2<1_C11p>1}
\\ \cline{2-4}
\multicolumn{1}{|c|}{} & $\kappatwo<1$ and $\kappathree>1$ 
& \eqref{eq_G_+++_BB_final_res_k1<1_k2<1_k3>1} & \eqref{eq_G+++_SB_k1<1_k2<1_k3>1} & \multirow{-2}{*}{} & \multirow{-2}{*}{} & \multirow{-2}{*}{} \\ \cline{2-7} 
\multicolumn{1}{|c|}{\multirow{-3}{*}{$\kappaone<1$}} & $\kappathree<1$
& \eqref{eq_G+++_BB_C2<1_C1p<1_full_res}
& \eqref{eq_G+++_SB_C2<1_C1p<1}
& $\kappathree<1$
& \eqref{eq_G_+++_BS_C2<1_C1p<1}
& \eqref{eq_G+++_SS_C2<1_C1p<1} 
\\ \hline
\multicolumn{1}{|l|}{} & $\kappafour>1$
& \eqref{eq_G+++_BB_C2>1_C1>1_full_res}
& \eqref{eq_G_+++_SB_C2>1_C1>1}
& $\kappatwo>1$
& \eqref{eq_G_+++_BS_C2>1_C2p>1}
& \eqref{eq_G_+++_SS_C2>1_C2p>1} \\ \cline{2-7} 
\multicolumn{1}{|l|}{\multirow{-2}{*}{$\kappaone>1$}} & $\kappafour<1$
& \eqref{eq_G+++_BB_C2>1_C1<1_full_res}
& \eqref{eq_G_+++_SB_C2>1_C1<1}
& $\kappatwo<1$
& \eqref{eq_G_+++_BS_C2>1_C2p<1}
& \eqref{eq_G_+++_SS_C2>1_C2p<1}  \\ \hline
\end{tabular}
\end{table}

where the kinematic ratios are given by
\begin{equation}
\kappaone=\frac{k_{56}}{k_{34}+s_1}\;;\;\kappatwo=\frac{k_{12}}{k_{34}+s_2}\;;\;
\kappafour=\frac{k_{12}+\sqrt{k_{12}^2-s_1^2}}{k_{56}+\sqrt{k_{56}^2-s_2^2}}\;;\;
\kappathree= \frac{k_{12}}{k_{34}-s_2}\,.
\end{equation}

\item Recall that $G_{-++}=G_{-++}^B+G_{-++}^S$ \eqref{eq_G-++_BorS}:
\begin{table}[H]
\centering
\renewcommand{\arraystretch}{1.5}
\begin{tabular}{|l|l|l|}
\hline
\rowcolor[HTML]{C0C0C0}
 & $G_{-++}^B$ & $G_{-++}^S$ \\
\hline
$\kappaone<1$ & \eqref{eq_G-++_B_C2<1} & \eqref{eq_G-++_S_C2<1} \\
\hline
$\kappaone>1$ & \eqref{eq_G-++_B_C2>1} & \eqref{eq_G-++_S_C2>1} \\
\hline
\end{tabular}
\end{table}
\item $G_{+-+}$ is given by \eqref{eq_G_integrals_+-+}.
\end{itemize}
To conclude this section, we note that a simplification pattern occurs in some components summarised above: in many cases the analytic continuation to the kinematic region $k_{34}<s_1+s_2$ can be obtained in a simple way. Indeed, the ratios $(s_1/k_{34})^2$ and $(s_2/k_{34})^2$ only appear in a $\F_4$ function. For the components where the residues of the central vertex function have been collected, each term in the sum involves an Appell $\F_4$ function with at least one upper parameter being a negative integer (typically $-n/2$ or $(1-n)/2$). In this case, the series that defines $\F_4$ \eqref{eq_def_Appell_F4} terminates and $\F_4$ reduces to a polynomial of degree $\lfloor n/2\rfloor$ in $\left(s_1/k_{34}\right)^2$ and $\left(s_2/k_{34}\right)^2$. This feature allows a straightforward analytic continuation for $k_{34}<s_1+s_2$ of an initially complicated function, since $\F_4$ is entire in its kinematic arguments.
A non-terminating $\F_4$ survives then for the fully nested background-background component in the regime $\kappaone<1$, $\kappathree<1$ \eqref{eq_G+++_BB_C2<1_C1p<1_full_res} (here the residues are collected at the poles of the Legendre $Q$ factors, at these points $\I_{II}$ is regular and contributes through its non-terminating series \eqref{eq_series_rep_vertex_F4_III} instead of the terminating residue \eqref{eq_residues_III}), but in this case the combination of the two kinematic conditions imposes $k_{34}>s_1+s_2$, therefore we do not need analytic continuation. However, the latter is needed for the background-background contribution when $\kappaone<1$, $\kappatwo<1$ and $\kappathree>1$ \eqref{eq_G_+++_BB_final_res_k1<1_k2<1_k3>1}, the signal-signal contributions $G_{+++}^{S,S}$, the signal pieces $G_{-++}^{S}$ of the partially nested component, and the factorised component $G_{+-+}$. In these cases, numerical evaluation can be done using the one-fold Mellin-Barnes representation for the central vertex function \eqref{eq_one_fold_MB_N3} in the analytically continued region.

\section{Conclusions \& Future Directions}
\label{sec:conclusion}

In this work, we derived analytical results that provide the information needed to obtain a series representation of tree-level cosmological correlators from the KLF formalism. For such correlators, the outcomes of the KLF diagrammatic rules include multi-layered frequency integrals, with one layer per massive internal line.
The corresponding integrands are made of propagators and vertex functions, each coming with an interaction vertex. Unlike the propagators that carry a structure very similar to the flat-space ones up to the $i\epsilon$-prescription, the vertex functions are the key non-trivial quantities to understand in order to compute the frequency integrals, which is why these objects were at the core of this work.
The choice of complex analysis tools to carry out frequency integration required a precise understanding of the analytical properties of these vertex functions. This set of properties was obtained by using two complementary representations of the vertex functions: the defining integral one over modified Bessel functions and their series representation in terms of Lauricella functions. The first one was used for the derivation of the large-frequency asymptotic behaviours, which are needed to close the integration contours in the complex frequency plane, and the second one to determine the location of poles and their corresponding residues. Additionally, connection formulas among the different vertex functions were also worked out, which are crucial in order to perform the integrals with tractable asymptotic behaviours. Gathering these properties allows one to compute the KLF frequency integrals for a given tree-level correlator.

As an illustration of these results, we exhaustively performed the computation of a double-exchange diagram with massive internal lines. This diagrammatic computation, which was studied in the literature using methods that rely on solving differential equation systems, is based only on the new perspective developed in this work. The most involved component of the diagram was given by a two-layer frequency integral. Integration has been done layer by layer, computing at each step the residues of the integrand coming from the integration measure, propagators, and vertex functions. To find the set of collected residues at each integration layer, we determined a contour closure prescription in the complex frequency plane using the large-frequency behaviour of the relevant vertex functions. At each layer, this prescription appeared to depend on a ratio of momenta, leading to two different cases. Therefore, four different cases must be distinguished in the final solution, depending on the hierarchy between kinematic parameters.
In the end, we derived a new double series representation for this correlator, with a general term written in terms of Gauss hypergeometric and Appell functions, with the latter actually arising only through the central vertex function. This improves on previously known representations, which involved series with up to four summation layers. Moreover, the computation makes manifest the separation between the different
relevant physical pieces: a contribution to the background or to the cosmological collider signal emerges for each internal line. Although the residue computations are carried out under a given kinematic condition for each vertex function, the obtained series representation is valid in the whole physical domain up to analytic continuation.

Our work provides useful tools from the KLF formalism for the systematic study of tree-level cosmological correlators. In addition, let us highlight some limitations and openings that may lead to forthcoming research directions. An obstacle to the numerical evaluation of a tree-level correlator using a series representation arising from the formalism developed in this work is the finite radius of convergence of the Lauricella function defining series, which translates into a kinematic condition on the series representation of vertex functions. Indeed, to cover the entire kinematic domain, one needs to know analytic continuation formulas for Lauricella functions outside of the region of convergence of the defining series. Such analytic continuation formulas and series expansions are known for Gauss hypergeometric functions, however, not for higher-order Lauricella $F_C$ functions (including the Appell $F_4$ function), accounting for physical situations where a vertex has several massive lines. To tackle this issue, it would be interesting to derive an atlas of convergent series representations of these functions from the defining system of partial differential equations with the suitable analytic continuation formulas, in order to cover the whole kinematic space. It would also be worth studying whether the simplification that arises for the double-exchange diagram holds more generally, namely whether Lauricella functions appear at the end of the computations only through vertex functions. A more challenging direction would be to extend the present approach beyond tree level, where loop integrations introduce additional frequency and momentum structures that the contour techniques developed here would need to accommodate. Finally, as a natural perspective, it would also be interesting to transpose the formalism developed in this work to other inflationary models, which can include, for instance, spinning fields and chemical potentials.

\appendix

\section{Special functions}\label{appendix_def_special_functions}

In this Appendix, we gather definitions and useful relations for the different special functions used in this work. Unless otherwise specified, the formulas written here can be found or straightforwardly obtained from formulas in \cite{NISTDLMF}.

\paragraph{$\Gamma$-function.} The $\Gamma$-function is defined by the integral
\begin{equation}
    \Gamma(z)=\int_0^{+\infty}\d t\,t^{z-1}\,e^{-t}\,,
    \label{app_def_gamma}
\end{equation}
where $\Re(z)>0$. This function can be analytically continued to a meromorphic function of $\mathbb{C}$, with poles at $z=-n$, $n\in\mathbb{N}$ and residues
\begin{equation}
    \mathrm{Res}\left(\Gamma(z)\,,\,z=-n\right)=\frac{(-1)^n}{n!}\,.
    \label{eq_poles_residue_gamma}
\end{equation}
Let us also mention two important functional identities, the first one being the reflection formula:
\begin{equation}
    \frac{\pi}{\sin(\pi z)}=\Gamma(z)\Gamma(1-z)\,,
    \label{app_eq_reflection_gamma}
\end{equation}
and the duplication formula:
\begin{equation}
    \Gamma(2z)=\frac{2^{2z-1}}{\sqrt{\pi}}\Gamma\left(z+\frac{1}{2}\right)\Gamma(z)\,.
    \label{eq_gamma_duplication}
\end{equation}

\paragraph{Modified Bessel functions.}
The modified Bessel function of the first kind $I_{i\mu}$ is defined by 
\begin{equation}
    I_{i\mu}(z)=\left(\frac{z}{2}\right)^{i\mu}\sum_{k=0}^\infty \frac{\left(\frac{z}{2}\right)^{2 k}}{k!\,\Gamma(i\mu+k+1)}\,.
    \label{def_bessel_I}
\end{equation}
The other solution of the modified Bessel differential equation is the modified Bessel function of the second kind, related to $I_{i\mu}$ via the connection formula:
\begin{equation}
    K_{i\mu}(z)=\frac{i\pi}{2}\frac{1}{\sinh(\pi\mu)}\left(I_{i\mu}(z)-I_{-i\mu}(z)\right)\,.
    \label{connection_formula_bessel}
\end{equation}
The definition \eqref{def_bessel_I}, as well as the connection formula, provide the small-argument leading behaviours:
\begin{subequations}
\begin{equation}
    I_{i\mu}(z)\underset{z\rightarrow0}{\sim}\left(\frac{z}{2}\right)^{i\mu}\frac{1}{\Gamma(1+i\mu)}\;,\; i\mu\notin -\mathbb{N}\,.
    \label{small_z_bessel_I}
\end{equation}
\begin{equation}
    K_{i\mu}(z)\underset{z\rightarrow0}{\sim}\frac{1}{2}\left(\Gamma(-i\mu)\left(\frac{z}{2}\right)^{i\mu}+\Gamma(i\mu)\left(\frac{z}{2}\right)^{-i\mu}\right)\;,\; i\mu\notin -\mathbb{N}\,.
    \label{small_z_bessel_K}
\end{equation}
\end{subequations}
Also, their large-argument asymptotic behaviour is
\begin{subequations}
\begin{equation}
    I_{i\mu}(z)\underset{z\rightarrow+\infty}{\sim}\frac{e^z}{\sqrt{2\pi z}}\,,
    \label{large_z_bessel_I}
\end{equation}
\begin{equation}
    K_{i\mu}(z)\underset{z\rightarrow+\infty}{\sim}\sqrt{\frac{\pi}{2 z}}\,e^{-z}\,.
    \label{large_z_bessel_K}
\end{equation}
\label{eq_large_z_Bessels}
\end{subequations}
From the analytic continuation formula
\begin{equation}
    I_{i\mu}(z)=\frac{i}{\pi}\left(-K_{i\mu}(z e^{-i\pi})+e^{-\pi\mu}K_{i\mu}(z)\right)\,,
    \label{eq_continuation_BesselI_asymptotics}
\end{equation}
the large argument asymptotic behaviour of $I_{i\mu}$ can be refined by adding
\begin{equation}
    I_{i\mu}(z)\underset{z\rightarrow+\infty}{=}\frac{1}{\sqrt{2\pi z}}
    \left(e^z\left(1+O(z^{-1})\right)+i\,e^{-\pi\mu}e^{-z}\left(1+O(z^{-1})\right)\right)
    \,,
    \label{large_z_bessel_I_refined}
\end{equation}
up to factors cancelled by powers of $z$.
The added term is dominant when $\Re(z)<0$. 
Arising from its convergent series expansion around $z=0$, $K_{i\mu}$ has the following Mellin-Barnes integral representation:
\begin{equation}
    K_{i\mu}(z)=\frac{\left(\frac{z}{2}\right)^{i\mu}}{4\pi i}\int_{c-i\infty}^{c+i\infty}\Gamma(s)\Gamma(s-i\mu)\left(\frac{z}{2}\right)^{-2 s} \d s\,,
    \label{eq_MellinB_K}
\end{equation}
where $c>\max\left(\Re(i\mu),0\right)$ and $\arg(z)<\pi/2$. We will also use the following analytic continuation formula:
\begin{equation}
    K_{i\mu}(e^{-i\pi}z)=e^{-\pi\mu}K_{i\mu}(z)+i\pi I_{i\mu}(z)\,.
    \label{eq_analytic_continuation_BesselK}
\end{equation}

Let us finally note the simple expression of $K_{i\mu}$ when $\mu=\pm i/2$:
\begin{equation}
    K_{\frac{1}{2}}(z)=K_{-\frac{1}{2}}(z)=\sqrt{\frac{\pi}{2 z}} e^{-z}\,.
    \label{eq_besselK_cc}
\end{equation}
It leads to the following identity
\begin{equation}
    \prod_{i=1}^M K_{\frac{1}{2}}(k_i z)=\left(\frac{\pi}{2}\right)^{\frac{M-1}{2}}\sqrt{k_{1\ldots M}}\left(\prod_{i=1}^M\frac{1}{\sqrt{k_i}}\right) \frac{1}{z^{\frac{M-1}{2}}}
    K_{\frac{1}{2}}(k_{1\ldots M} z)\,,
    \label{eq_prod_K_identity}
\end{equation}
where $k_{1\ldots M}\equiv\sum_{i=1}^M k_i$. We use this last formula to simplify vertex functions involving $M\geq 2$ conformally coupled fields.

\paragraph{Gauss hypergeometric function.} The Gauss hypergeometric function $_2F_1$ is defined by the series
\begin{equation}
    _2F_1\left(
    \begin{matrix}
        a\,,\,b \\ c
    \end{matrix}\,,z\right)=\sum_{n=0}^\infty \frac{\left(a\right)_n\left(b\right)_n}{\left(c\right)_n}\frac{z^n}{n!}\;,
    \label{eq_2F1_def}
\end{equation}
that converges in the unit disk and $z\mapsto\, _2F_1\left(a,b;c;z\right)$ can be analytically continued for values of $z\in\mathbb{C}\setminus\left[1;+\infty\right]$. The Pochhammer symbol $(a)_n$ is
\begin{equation}
    \left(a\right)_n\equiv\frac{\Gamma(a+n)}{\Gamma(a)}\,.
\end{equation}
Regarding the parameters of the hypergeometric function, $a,b\mapsto\,_2F_1\left(a,b;c;z\right)$ is analytic (because $(a)_n$ stays finite $\forall a\in\mathbb{C}$)
and $c\mapsto\,_2F_1\left(a,b;c;z\right)$ is meromorphic with poles at $c=-p\,,\,p\in\mathbb{N}$. Therefore, we define the dressed Gauss hypergeometric function
\begin{equation}
    _2\F_1\left(
    \begin{matrix}
        a\,,\,b \\ c
    \end{matrix}\,,z\right)=\frac{1}{\Gamma(c)}\,
    _2F_1\left(
    \begin{matrix}
        a\,,\,b \\ c
    \end{matrix}\,,z\right)\;.
    \label{eq_2F1_dressed_def}
\end{equation}
This last function is then analytic with respect to $a,b$ and $c$.

\paragraph{Associated Legendre functions.}
The associated Legendre functions of the first and second kind are, respectively, defined using the Gauss hypergeometric function \eqref{eq_2F1_def} by
\begin{equation}
    P_{i\mu-\frac{1}{2}}^{-\frac{d-3}{2}}(x)=\frac{1}{\Gamma\left(\frac{d-1}{2}\right)}\left(\frac{x-1}{x+1} \right)^{\frac{d-3}{4}}\,
    _2F_1\left(
    \begin{matrix}
        \frac{1}{2}+i\mu\,,\,\frac{1}{2}-i\mu \\ \frac{d-1}{2}
    \end{matrix}\,;\frac{1-x}{2}\right)
    \quad,\quad x>1\,,
    \label{eq_def_Legendre_P}
\end{equation}
\begin{equation}
    Q_{i\mu-\frac{1}{2}}^{\frac{d-3}{2}}(x)=\frac{e^{i\pi\frac{d+1}{2}}\sqrt{\pi}\left(x^2-1\right)^{\frac{d-3}{4}}}{x^{\frac{d-3}{2}}(2x)^{ i\mu+\frac{1}{2}}}\frac{\Gamma\left(\frac{d-2}{2}+ i\mu\right)}{\Gamma\left(1+ i\mu\right)}\,_2F_1\left(
    \begin{matrix}
        \frac{d}{4}+ \frac{i\mu}{2}\,,\,\frac{d-2}{4}+ \frac{i\mu}{2} \\ 1+i\mu
    \end{matrix}\,,\frac{1}{x^2}\right)\,,x>1\,.
    \label{eq_def_legendre_Q}
\end{equation}
These functions satisfy the following connection formula
\begin{equation}
    P_{i\mu-\frac{1}{2}}^{-\frac{d-3}{2}}(x) =\frac{i e^{-i\pi\left(\frac{d-3}{2}\right)}}{\sinh(\pi\mu)\Gamma\left(\frac{d-2}{2}\pm i\mu\right)}
    \left(Q_{i\mu-\frac{1}{2}}^{\frac{d-3}{2}}(x)-Q_{-i\mu-\frac{1}{2}}^{\frac{d-3}{2}}(x)\right)\,,\quad x>1\,.
    \label{connection_formula_P_Q}
\end{equation}

In this work, we also use the following integral representations:
\begin{equation}
    \int_0^\infty \d y\,y^{\alpha-\frac{1}{2}}e^{-x y}K_{i\mu}(y)=\sqrt{\frac{\pi}{2}}\Gamma\left(\alpha+\frac{1}{2}\pm i\mu\right)\left(x^2-1\right)^{-\frac{\alpha}{2}}P_{i\mu-\frac{1}{2}}^{-\alpha}(x)\,,
    \label{eq_P_integral_K}
\end{equation}
where $x>1$ and $\Re\left(\alpha+1/2\pm i\mu\right)>0$, and
\begin{equation}
    \int_0^\infty \d y\,y^{\alpha-\frac{1}{2}}e^{-x y}I_{i\mu}(y)=\sqrt{\frac{2}{\pi}} e^{-i\pi\alpha}\left(x^2-1\right)^{-\frac{\alpha}{2}}Q_{i\mu-\frac{1}{2}}^{\alpha}(x)\,,
    \label{eq_Q_integral_I}
\end{equation}
where $x>1$ and $\Re\left(\alpha+1/2+ i\mu\right)>0$.

The large $\mu$ asymptotic behaviours of Legendre functions are \cite{dunster2025}:
\begin{equation}
    P_{i\mu-\frac{1}{2}}^{-\frac{d-3}{2}}(\cosh(\xi)) \underset{\mu\rightarrow+\infty}{=}
    \sqrt{\frac{2}{\pi \sinh(\xi)}}\frac{1}{\mu^{\frac{d}{2}-1}}\sin\left(\mu\,\xi+\frac{\pi}{4}\left(4-d\right)\right)\left(1+O\left(\frac{1}{\mu}\right)\right)\,.
    \label{eq_asymptotics_P}
\end{equation}
\begin{equation}
    Q_{\pm i\mu-\frac{1}{2}}^{\frac{d-3}{2}}(\cosh(\xi))\underset{\mu\rightarrow+\infty}{=}e^{i\frac{\pi}{2}(d+1)}\sqrt{\frac{\pi}{2 \sinh(\xi)}}\left(\pm i\mu\right)^{\frac{d}{2}-2}e^{\mp i\mu\xi}\left(1+O\left(\frac{1}{\mu}\right)\right)\;,\quad\xi>0\,.
    \label{eq_large_mu_Q}
\end{equation}
The function $\mu\mapsto P_{i\mu-\frac{1}{2}}^{-\frac{d-3}{2}}(x)$ is analytic, whereas $\mu\mapsto Q_{\pm i\mu-\frac{1}{2}}^{\frac{d-3}{2}}(x)$ is meromorphic with poles and residues
\begin{equation}
\begin{aligned}
&\mathrm{Res}\left(\mu\mapsto Q_{\pm i\mu-\frac{1}{2}}^{\frac{d-3}{2}}(x)\,;\,\mu=\pm i\left(n+\frac{d-2}{2}\right)\,,\,n\in\mathbb{N}\right)\\
&=\mp i\frac{e^{i\pi\frac{d-3}{2}}\sqrt{\pi}\left(x^2-1\right)^{\frac{d-3}{4}}}{2^{-\frac{d-3}{2}}(2x)^{-n}}
\frac{(-1)^n}{n!}
\frac{1}{\Gamma\left(-n-\frac{d-4}{2}\right)}\,_2F_1\left(
    \begin{matrix}
        \frac{1-n}{2}\,,\,-\frac{n}{2} \\ -n-\frac{d-4}{2}
    \end{matrix}\,,\frac{1}{x^2}\right)\,.
\end{aligned}
\label{eq_residues_legendre_Q}
\end{equation}
Finally, the associated Legendre $P$ function has the following analytic continuation at $z=e^{+i\pi}x\,,\,x>1$:
\begin{equation}
    P_{i\mu-\frac{1}{2}}^{-\frac{d-3}{2}}(e^{+i\pi}x)= \frac{e^{-i\pi\frac{d-3}{2}}}{\sinh(\pi\mu)\Gamma\left(\frac{d-2}{2}\pm i\mu\right)}
    \left(e^{\pi\mu}Q_{i\mu-\frac{1}{2}}^{\frac{d-3}{2}}(x)-e^{-\pi\mu}Q_{-i\mu-\frac{1}{2}}^{\frac{d-3}{2}}(x)\right)\,.
    \label{eq_analytic_cont_P}
\end{equation}

\paragraph{Lauricella functions.}
Lauricella functions \cite{Lauricella1893,Matsumoto_2020} are multi-variable generalisations of the hypergeometric functions, denoted $F_A$, $F_B$, $F_C$ and $F_D$. In particular, the function $F_C$ is used in this work and can be defined as
\begin{equation}
    F_C^{(N)}\Biggl(\begin{matrix}
    a\;,\; b \\ c_1\,,\;\ldots\,,\;c_{N}\end{matrix};
    x_1,\ldots,x_N\Biggr)
    =\sum_{n_1,\ldots,n_N=0}^\infty \frac{(a)_{n_1+\ldots+n_N}(b)_{n_1+\ldots+n_N}}{(c_1)_{n_1}\ldots(c_N)_{n_N}}\frac{x_1^{n_1}}{n_1!}\ldots\frac{x_N^{n_N}}{n_N!}\,.
    \label{eq_def_Lauricella_C}
\end{equation}
In the same way as the Gauss hypergeometric function $_2F_1$, this function has singularities at $c_1,\ldots,c_N=-p$, $p\in\mathbb{N}$. Then, we define the dressed Lauricella function $\F_C^{(N)}$ as
\begin{equation}
    \F_C^{(N)}\Biggl(\begin{matrix}
    a\;,\; b \\ c_1\,,\;\ldots\,,\;c_{N}\end{matrix};
    x_1,\ldots,x_N\Biggr)
    \equiv\prod_{i=1}^N\frac{1}{\Gamma(c_i)}\,F_C^{(N)}\Biggl(\begin{matrix}
    a\;,\; b \\ c_1\,,\;\ldots\,,\;c_{N}\end{matrix};
    x_1,\ldots,x_N\Biggr)\,.
    \label{eq_def_dressed_Lauricella_C}
\end{equation}
The function $\F_C^{(N)}$ is regular for every complex value of its parameters $a,b,c_1,\ldots,c_N$.
The convergence domain of the series \eqref{eq_def_Lauricella_C} is
\begin{equation}
    \sum_{i=1}^N \sqrt{|x_i|}<1\,.
    \label{eq_cvg_radius_Lauricella_FC}
\end{equation}

For $N=1$, $F_C^{(1)}$ reduces to the hypergeometric function \eqref{eq_2F1_def}, and can then be connected to the Legendre functions of the first and second kind \eqref{eq_def_Legendre_P}-\eqref{eq_def_legendre_Q}.

\paragraph{Appell $F_4$ function.}
For $N=2$, $F_C^{(2)}$ reduces to the Appell $F_4$ function:
\begin{equation}
    F_4\Biggl(\begin{matrix}
    a\;,\; b \\ c_1\,,\,c_2\end{matrix}\,;
    x,y\Biggr)=\sum_{m,n=0}^\infty \frac{(a)_{m+n}(b)_{m+n}}{(c_1)_m(c_2)_n}\frac{x^{m}}{m!}\frac{y^{n}}{n!}\,,
    \label{eq_def_Appell_F4}
\end{equation}
and $\F_C^{(2)}$ is the dressed and regularised $\F_4$:
\begin{equation}
    \F_4\Biggl(\begin{matrix}
    a\;,\; b \\ c_1\,,\,c_2\end{matrix}\,;
    x,y\Biggr)
    \equiv\frac{1}{\Gamma(c_1)\Gamma(c_2)}\,F_4\Biggl(\begin{matrix}
    a\;,\; b \\ c_1\,,\,c_2\end{matrix}\,;
    x,y\Biggr)\,.
    \label{eq_def_dressed_F4}
\end{equation}

\section{Vertex function representation with Lauricella functions}\label{appendix_derivation_vertex_fct}

In this appendix, the hypergeometric representation \eqref{eq_vertex_function_vf} of the vertex function is derived. Starting from the integral representation \eqref{def_vertex_function_K}, we write $N-1$ of the modified Bessel functions in terms of their Mellin-Barnes representation \eqref{eq_MellinB_K}. It leads to
\begin{equation}\begin{aligned}
    \mathcal{I}_d\begin{bmatrix}
        \mu_1 &\ldots &\mu_N \\ k_1 &\ldots & k_N \\ K &\ldots & K
    \end{bmatrix} = &\int \displaylimits^\infty_0\d z\; z^{\frac{d(N-2)}{2}-1}
    \prod_{j=1}^{N-1} \int_{c-i\infty}^{c+i\infty}\frac{\d s_j}{4\pi i}\\ &\times\Gamma(s_j)\Gamma(s_j-i\mu_j)\left(\frac{k_j z}{2}\right)^{i\mu_j-2 s_j}K_{i\mu_N}(k_N z) \,.
    \end{aligned}
    \label{eq_app_vertex_function_1}
\end{equation}
In this computation, we keep the frequencies $\mu_j$ real, the analytic continuation being straightforward at the end. Then, we must have $c>0$. To make progress, we can integrate over $z$, leading to
\begin{equation}
\begin{aligned}
    \int \displaylimits^\infty_0\d z\;&z^{\frac{d(N-2)}{2}-1+\sum_{j=1}^{N-1} i\mu_j-2 s_j}\,K_{i\mu_N}(k_N z)\\&=\frac{1}{4}\left(\frac{2}{k_N}\right)^{\frac{d(N-2)}{2}+\sum_{j=1}^{N-1} i\mu_j-2 s_j}\,\Gamma\left(\frac{1}{2}\left(\frac{d(N-2)}{2}+\sum_{j=1}^{N-1} i\mu_j-2 s_j\right)\pm \frac{i\mu_N}{2}\right)\,,
    \end{aligned}
    \label{eq_app_integral_z}
\end{equation}
where we used the convention \eqref{eq_convention_Gamma_fct}.
The validity of the integration over $z$ enforces the condition
\begin{equation}
    c<\frac{d(N-2)}{4(N-1)}\,,
\end{equation}
so we can take $c<d/8$.
Then,
\begin{equation}
\begin{aligned}
    \mathcal{I}_d\begin{bmatrix}
        \mu_1 &\ldots &\mu_N \\ k_1 &\ldots & k_N \\ K &\ldots & K
    \end{bmatrix} = 
    \frac{1}{4}\left(\frac{2}{k_N}\right)^{\frac{d(N-2)}{2}}\
    \int_{c-i\infty}^{c+i\infty}&\prod_{j=1}^{N-1} \frac{\d s_j}{4\pi i}\Gamma(s_j)\Gamma(s_j-i\mu_j)\left(\frac{k_j}{k_N}\right)^{i\mu_j-2 s_j}\\&\times\Gamma\left(\frac{1}{2}\left(\frac{d(N-2)}{2}+\sum_{j=1}^{N-1} i\mu_j-2 s_j\right)\pm \frac{i\mu_N}{2}\right)\,.
    \end{aligned}
    \label{eq_app_vertex_function_2}
\end{equation}
For convenience, this last integral can be rewritten using the change of variable $u_j=s_j-\frac{i\mu_j}{2}$, $j\in\{1,\ldots,N-1\}$:
\begin{equation}
\begin{aligned}
    \mathcal{I}_d\begin{bmatrix}
        \mu_1 &\ldots &\mu_N \\ k_1 &\ldots & k_N \\ K &\ldots & K
    \end{bmatrix} = 
    \frac{1}{4}\left(\frac{2}{k_N}\right)^{\frac{d(N-2)}{2}}\
    \int_{c-i\infty}^{c+i\infty}&\prod_{j=1}^{N-1} \frac{\d u_j}{4\pi i}\Gamma\left(u_j\pm\frac{i\mu_j}{2}\right)\left(\frac{k_j}{k_N}\right)^{-2 u_j}\\&\times\Gamma\left(\frac{d(N-2)}{4}-\sum_{j=1}^{N-1} u_j\pm \frac{i\mu_N}{2}\right)\,.
    \end{aligned}
    \label{eq_app_vertex_function_3}
\end{equation}
Now, we can integrate layer by layer using the residue theorem and a suitable contour prescription. Let $k\in\{1,\ldots,N-1\}$. We focus on the integral
\begin{equation}
\begin{aligned}
    I_k = 
    \int_{c-i\infty}^{c+i\infty} \frac{\d u_k}{4\pi i}\Gamma\left(u_k\pm\frac{i\mu_k}{2}\right)\Gamma\left(\frac{d(N-2)}{4}-\sum_{j=1}^{N-1} u_j\pm \frac{i\mu_N}{2}\right)\left(\frac{k_k}{k_N}\right)^{-2 u_k}\,,
    \end{aligned}
    \label{eq_app_I_k}
\end{equation}
where we kept all terms that depend on $u_k$ in \eqref{eq_app_vertex_function_3}. The integrand is meromorphic and has the following pole structure (also drawn in Fig. \ref{fig_MB_integral_vertex_fct}):
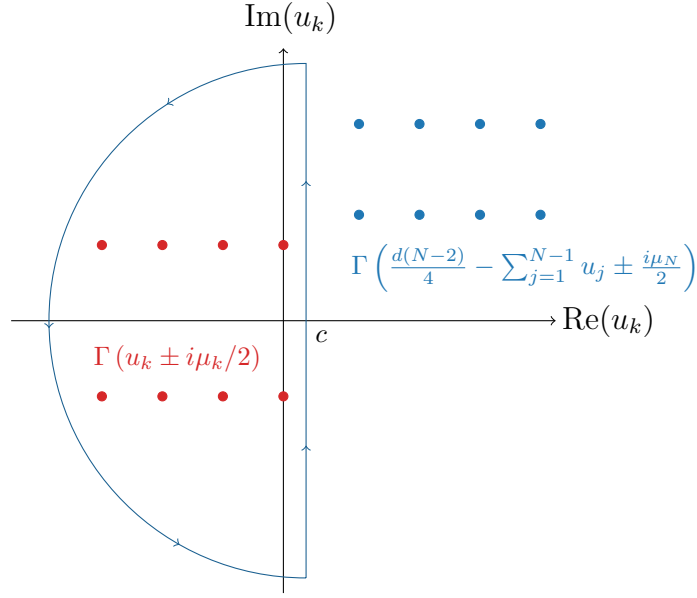
\begin{figure}[h!]\centering
\hspace{0cm}
    	\begin{tikzpicture}[scale = 2]

        \draw[black, ->] (-1.8,0) -- (1.8,0) coordinate (xaxis);
		\draw[black, ->] (0,-1.8) -- (0,1.8) coordinate (yaxis);
		\node at (2.15, 0) {$\text{Re}(u_k)$};
		\node at (0.05, 2) {$\text{Im}(u_k)$};
        
        \draw[pyred, fill = pyred] (0, 0.5) circle (.03cm);
		\draw[pyred, fill = pyred] (-.4, 0.5) circle (.03cm);
		\draw[pyred, fill = pyred] (-.8, .5) circle (.03cm);
		\draw[pyred, fill = pyred] (-1.2, .5) circle (.03cm);
        \node at (-.7, -0.25) {\textcolor{pyred}{\footnotesize$\Gamma\left(u_k\pm i\mu_k/2\right)$}};

        \draw[pyred, fill = pyred] (0, -0.5) circle (.03cm);
		\draw[pyred, fill = pyred] (-.4, -0.5) circle (.03cm);
		\draw[pyred, fill = pyred] (-.8, -.5) circle (.03cm);
		\draw[pyred, fill = pyred] (-1.2, -.5) circle (.03cm);

        \draw[pyblue, fill = pyblue] (.5, 1.3) circle (.03cm);
		\draw[pyblue, fill = pyblue] (0.9, 1.3) circle (.03cm);
		\draw[pyblue, fill = pyblue] (1.3, 1.3) circle (.03cm);
		\draw[pyblue, fill = pyblue] (1.7, 1.3) circle (.03cm);
        \draw[pyblue, fill = pyblue] (.5, .7) circle (.03cm);
		\draw[pyblue, fill = pyblue] (0.9, .7) circle (.03cm);
		\draw[pyblue, fill = pyblue] (1.3, .7) circle (.03cm);
		\draw[pyblue, fill = pyblue] (1.7, .7) circle (.03cm);
        \node at (1.6, .35) {\textcolor{pyblue}{\footnotesize$\Gamma\left(\frac{d(N-2)}{4}-\sum_{j=1}^{N-1} u_j\pm \frac{i\mu_N}{2}\right)$}};
        
        \draw[xshift=0,pyblue!80!black,decoration={markings,mark=between positions 0.1 and 1 step 0.2 with \arrow{>}},postaction={decorate}] (0.15,-1.7) -- (0.15,1.70) arc (90:270:1.7);
        \node at (.25, -.1) {\textcolor{black}{\footnotesize$c$}};
    \end{tikzpicture}
\caption{Analytic structure in the $u_k$ complex plane of the integrands of \eqref{eq_app_I_k}. We close the contour with a half-circle in the negative real part half-plane, with the assumption $k_k<k_N$.}
\label{fig_MB_integral_vertex_fct}
\end{figure} 
\begin{itemize}
    \item \textbf{\textcolor{pyred}{Poles from $\Gamma\left(u_k\pm i\mu_k/2\right)$.}} The first factor of two $\Gamma$-functions in \eqref{eq_app_I_k} has poles when the argument is a negative integer, with residues that follow from \eqref{eq_poles_residue_gamma}:
    \begin{equation}
    \mathrm{Res}\left(\Gamma\left(u_k\pm \frac{i\mu_k}{2}\right)\,,\,u_k=\mp\frac{i\mu_k}{2}-n_k\right)=\frac{(-1)^{n_k}}{n_k!}\,,
    \label{eq_poles_residue_I_k}
    \end{equation}
    where $n_k\in\mathbb{N}$.
    \item \textbf{\textcolor{pyblue}{Poles from $\Gamma\left(\frac{d(N-2)}{4}-\sum_{j=1}^{N-1} u_j\pm \frac{i\mu_N}{2}\right)$.}} The second factor of $\Gamma$-functions gives poles located at $u_k=\frac{d(N-2)}{4}-\sum_{j=1, j\neq k}^{N-1} u_j\pm \frac{i\mu_N}{2}+n_k$, $n_k\in\mathbb{N}$, in the positive real part half-plane.
\end{itemize}
Here we choose $k_j<k_N$, for every $j\in\{1,\ldots,N-1\}$, therefore the contour is closed in the negative real part half-plane. Collecting only the encircled residues \eqref{eq_poles_residue_I_k}, we obtain
\begin{equation}
\begin{aligned}
    I_k = 
    \frac{1}{2}\sum_{c_k=\pm}\sum_{n_k=0}^\infty&\frac{(-1)^{n_k}}{n_k!}\,\Gamma\left(-i c_k\mu_k-n_k\right)\\
    &\times\Gamma\left(\frac{d(N-2)}{4}-\sum_{j=1,j\neq k}^{N-1} u_j+\frac{i c_k \mu_k}{2}+n_k\pm \frac{i\mu_N}{2}\right)\left(\frac{k_k}{k_N}\right)^{i c_k\mu_k+2 n_k}\,.
    \end{aligned}
    \label{eq_app_I_k_res}
\end{equation}

Using this result, we can now iterate to integrate over each $u_j$ variable in \eqref{eq_app_vertex_function_3}, leading to
\begin{equation}
\begin{aligned}
    \mathcal{I}_d\begin{bmatrix}
        \mu_1 &\ldots &\mu_N \\ k_1 &\ldots & k_N \\ K &\ldots & K
    \end{bmatrix} &= 
    \frac{1}{2^{N+1}}\left(\frac{2}{k_N}\right)^{\frac{d(N-2)}{2}}\sum_{c_1,\dots,c_{N-1}=\pm}\,\sum_{n_1,\ldots,n_{N-1}=0}^\infty \\
    &\times\Biggl(\prod_{j=1}^{N-1}\frac{(-1)^{n_j}}{n_j!}\,\left(\frac{k_j}{k_N}\right)^{i c_j\mu_j+2 n_j}\Gamma\left(-i c_j\mu_j-n_j\right)\Biggr)\\
    &\times\Gamma\left(\frac{d(N-2)}{4}\pm \frac{i\mu_N}{2}+\sum_{j=1}^{N-1} \frac{i c_j \mu_j}{2}+n_j\right)\,.
    \end{aligned}
    \label{eq_app_vertex_function_4}
\end{equation}
Using the reflection formula \eqref{app_eq_reflection_gamma} to rewrite the factors $\Gamma\left(-i c_j\mu_j-n_j\right)$, it becomes
\begin{equation}
\begin{aligned}
    \mathcal{I}_d\begin{bmatrix}
        \mu_1 &\ldots &\mu_N \\ k_1 &\ldots & k_N \\ K &\ldots & K
    \end{bmatrix} &= 
    \frac{(i\pi)^{N-1}}{2^{N+1}}\left(\frac{2}{k_N}\right)^{\frac{d(N-2)}{2}}\sum_{c_1,\dots,c_{N-1}=\pm} \sum_{n_1,\ldots,n_{N-1}=0}^\infty\\
    &\times\prod_{j=1}^{N-1}\Biggl[\frac{c_j}{\sinh(\pi\mu_j)}\left(\frac{k_j}{k_N}\right)^{i c_j\mu_j+ 2 n_j}\frac{1}{n_j!}
    \frac{1}{\Gamma\left(1+i c_j\mu_j+n_j\right)}\Biggr]\\
    &\times
    \Gamma\left(\frac{d(N-2)}{4}\pm \frac{i\mu_N}{2}+\sum_{j=1}^{N-1} \frac{i c_j \mu_j}{2}+n_j\right)\,.
    \end{aligned}
    \label{eq_app_vertex_function_5}
\end{equation}
We can finally rewrite this expression in terms of the dressed Lauricella function $\F_C^{(N-1)}$ \eqref{eq_def_dressed_Lauricella_C} to obtain the result \eqref{eq_vertex_function_vf}.

\paragraph{Mellin-Barnes representation for analytic continuation.}
Finally, notice that, while this series representation converges only in the kinematic region \eqref{eq_cvg_radius_kinematics} where the reference leg dominates, the Mellin-Barnes representation \eqref{eq_app_vertex_function_3} underlying it, on the contrary, has a contour decay that is exponential and independent of the kinematics. Specialising to $N=3$ as relevant in this paper, this yields an explicit and rapidly convergent evaluation of the vertex function in any kinematic domain. It is obtained by performing the residue summation above for one of the two folds only, and keeping the other as a one-dimensional Mellin-Barnes integral. Explicitly:
\begin{equation}
    \mathcal{I}_d\begin{bmatrix}
        \mu_1 &\mu_2 &\mu_3 \\ k_1 &k_2 & k_3 \\ K &K & K
    \end{bmatrix}
    =\frac{1}{4}\left(\frac{2}{k_3}\right)^{\frac{d}{2}}
    \int_{c-i\infty}^{c+i\infty}\frac{\d u_1}{4\pi i}\,
    \Gamma\left(u_1\pm\frac{i\mu_1}{2}\right)\left(\frac{k_1}{k_3}\right)^{-2 u_1}
    \mathcal{T}(u_1)\,,
    \label{eq_one_fold_MB_N3}
\end{equation}
where the integral over $u_2$ has been closed in the negative-real-part half-plane, collecting the poles of $\Gamma\left(u_2\pm i\mu_2/2\right)$ \eqref{eq_poles_residue_I_k}, which resum into
\begin{equation}
    \mathcal{T}(u_1)=\frac{1}{2}\sum_{c_2=\pm}\sum_{n=0}^\infty\frac{(-1)^n}{n!}\,
    \Gamma\left(-i c_2\mu_2-n\right)\,
    \Gamma\left(\frac{d}{4}-u_1+\frac{i c_2\mu_2}{2}+n\pm\frac{i\mu_3}{2}\right)
    \left(\frac{k_2}{k_3}\right)^{i c_2\mu_2+2 n}\,.
    \label{eq_one_fold_T_N3}
\end{equation}
The sum over $n$ converges geometrically in $(k_2/k_3)^2$ as soon as $k_2<k_3$, a condition that can always be enforced by choosing the reference leg $k_3$ to be the largest momentum. The remaining one-fold integral \eqref{eq_one_fold_MB_N3} retains the exponential, kinematics-independent contour decay of \eqref{eq_app_vertex_function_3}, hence converges for any value of $k_1/k_3$.

\section{Vertex functions properties}\label{appendix_general_vertex_functions}

In this appendix additional details are given concerning the singularities of vertex functions $\mathcal{I}_d\left[\{\mu_j\},\{k_j\},\{K\}\right]$ and their asymptotic behaviour at large $\mu_k$.

\subsection{General connection formula}\label{appendix_connection_formula}

Connection formulas for vertex functions are written using the one for modified Bessel functions \eqref{connection_formula_bessel}. For instance, if we use it for the modified Bessel function with index $k\in\{1,\ldots,N-1\}$, at the level of vertex functions it gives
\begin{equation}
    \mathcal{I}_d\begin{bmatrix}
        \ldots &\mu_k&\ldots  \\ \ldots &k_k&\ldots \\ \ldots & K & \ldots
    \end{bmatrix}=\frac{i\pi}{2}\frac{1}{\sinh(\pi\mu_k)}\left(
    \mathcal{I}_d\begin{bmatrix}
        \ldots &\mu_k&\ldots  \\ \ldots &k_k&\ldots \\ \ldots & I & \ldots
    \end{bmatrix}-
    \mathcal{I}_d\begin{bmatrix}
        \ldots &-\mu_k&\ldots  \\ \ldots &k_k&\ldots \\ \ldots & I & \ldots
    \end{bmatrix}\right)\,.
    \label{eq_connection_formula_1_change_k}
\end{equation}
It reduces to \eqref{eq_connection_formula_bdy} for $k=1$.
It is then possible to iterate the use of \eqref{connection_formula_bessel} to obtain formulas involving vertex functions $\mathcal{I}_d\left[\{\mu_j\},\{k_j\},\{A_j\}\right]$ with more indices $A_j=I$. Let $F\subset\{1,\ldots,N-1\}$. The subset $F$ corresponds to the indices $j$ for which we want to turn $A_j=K$ into $A_j=I$ through the vertex function connection formula \footnote{We do not allow $N\in F$ (i.e. we keep $A_N=K$ in all the formulas) to ensure the convergence of any integral representation \eqref{def_vertex_function_general}.}. Then, applying \eqref{connection_formula_bessel} for each $j\in F$, we have
\begin{equation}
    \mathcal{I}_d\begin{bmatrix}
        \mu_1&\ldots &\mu_N  \\ k_1&\ldots &k_N \\ K&\ldots & K 
    \end{bmatrix}=\left(\frac{i\pi}{2}\right)^{\text{card}(F)}
    \prod_{j\in F}\,\sum_{c_j=\pm}\,\frac{1}{\sinh(\pi c_j\mu_j)}
    \left(
    \mathcal{I}_d\begin{bmatrix}
        \{\tilde{\mu}_l\}_{1\leq l\leq N}  \\ \{k_l\}_{1\leq l\leq N} \\ \{A_l\}_{1\leq l\leq N}
    \end{bmatrix}\right)\,,
    \label{eq_general_connection_from_vertexK}
\end{equation}
where $\text{card}(F)$ denotes the number of elements of $F$, and
\begin{subequations}
\begin{equation}
    \tilde{\mu}_l=\begin{cases}
        c_l \mu_l &\text{if}\; l\in F\\ \mu_l &\text{if}\; l\notin F
    \end{cases}\;,
\end{equation}
\begin{equation}
    A_l=\begin{cases}
        I &\text{if}\; l\in F\\ K &\text{if}\; l\notin F
    \end{cases}\;.
\end{equation}
\end{subequations}
The formula \eqref{eq_general_connection_from_vertexK} encodes all possible connection formulas written from \\ $\mathcal{I}_d\left[\{\mu_j\},\{k_j\},\{K\}\right]$. However, we can also write all connection formulas starting from a function $\mathcal{I}_d\left[\{\mu_j\},\{k_j\},\{A_j\}\right]$ with an arbitrary value of $A_j$. Let $F\subset\{1,\ldots,N-1\}$ still be the set of changing indices from $K$ to $I$. Let $G\subset\{1,\ldots,N-1\}$ be the set of indices $j$ such that $A_j=I$ in the starting vertex function $\mathcal{I}_d\left[\{\mu_j\},\{k_j\},\{A_j\}\right]$. We must have $F\cap G=\emptyset$ \footnote{Indeed, the set of indices $j\in\{1,\ldots,N-1\}$ with original label $A_j=I$ (set $G$) cannot have their label $A_j$ being changed from $K$ to $I$ in the connection formula (set $F$).}. Then, the connection formula reads
\begin{equation}
    \mathcal{I}_d\begin{bmatrix}
        \{\mu_l\}_{1\leq l\leq N}  \\ \{k_l\}_{1\leq l\leq N} \\ \{A_l\}_{1\leq l\leq N} 
    \end{bmatrix}=\left(\frac{i\pi}{2}\right)^{\text{card}(F)}
    \prod_{j\in F}\,\sum_{c_j=\pm}\,\frac{1}{\sinh(\pi c_j\mu_j)}
    \left(
    \mathcal{I}_d\begin{bmatrix}
        \{\mu'_l\}_{1\leq l\leq N}  \\ \{k_l\}_{1\leq l\leq N} \\ \{A'_l\}_{1\leq l\leq N}
    \end{bmatrix}\right)\,,
    \label{eq_full_general_connection}
\end{equation}
where
\begin{subequations}
\begin{equation}
    \mu'_l=\begin{cases}
        c_l \mu_l &\text{if}\; l\in F\\ \mu_l &\text{if}\; l\notin F
    \end{cases}\;,
\end{equation}
\begin{equation}
    A_l=\begin{cases}
        I &\text{if}\; l\in G\\ K &\text{if}\; l\notin G
    \end{cases}\;,
\end{equation}
\begin{equation}
    A'_l=\begin{cases}
        I &\text{if}\; l\in F\cup G \\ K &\text{if}\; l\notin F\cup G
    \end{cases}\;.
\end{equation}
\end{subequations}
This straightforwardly reduces to \eqref{eq_general_connection_from_vertexK} for $G=\emptyset$.

\subsection{Poles and residues of vertex functions}\label{subappendix_poles_residues_vertex_fct_general}

In this subsection, the singularity structure in the frequency complex plane of the vertex function $\mathcal{I}_d\left[\{\mu_j\},\{k_j\},\{K\}\right]$ is detailed using the series representation \eqref{eq_vertex_function_vf} that allows for explicit analytical continuation. Let us fix $k\in\{1,\ldots,N-1\}$ and then look at the properties of the function $\mu_k\mapsto\mathcal{I}_d\left[\{\mu_j\},\{k_j\},\{K\}\right]$.
First, in the series representation, since the dressed Lauricella C function is regular for every complex value of $\mu_k$ (see \eqref{eq_def_dressed_Lauricella_C}) as well as the power-dependent prefactors, the singularities of $\mu_k\mapsto\mathcal{I}_d\left[\{\mu_j\},\{k_j\},\{K\}\right]$ are simple poles and, as we will see, come only from the $\Gamma$-function factor in \eqref{eq_vertex_function_vf}.

\paragraph{Poles from the $\Gamma$-function.} Let us start by focusing on the poles and residues from the product of two $\Gamma$-functions:
\begin{equation}
\begin{aligned}
    \Gamma&\left(\frac{d(N-2)}{4}+\sum_{j=1}^{N-1} \frac{i c_j \mu_j}{2}\pm \frac{i\mu_N}{2}\right)\\&\equiv
    \Gamma\left(\frac{d(N-2)}{4}+\sum_{j=1}^{N-1} \frac{i c_j \mu_j}{2}+ \frac{i\mu_N}{2}\right)\Gamma\left(\frac{d(N-2)}{4}+\sum_{j=1}^{N-1} \frac{i c_j \mu_j}{2}- \frac{i\mu_N}{2}\right)\,.
    \end{aligned}
    \label{eq_product_gammas_vertex_fct}
\end{equation}
Knowing the singularity structure and residues of the $\Gamma$-function \eqref{eq_poles_residue_gamma}, it is straightforward to deduce that $\mu_k\mapsto\mathcal{I}_d\left[\{\mu_j\},\{k_j\},\{K\}\right]$ has poles located at
\begin{equation}
    \mu_k=c_k\left(-\sum_{j=1,j\neq k}^{N-1} c_j \mu_j -c_N \mu_N+i\left(2 p_k+\frac{d(N-2)}{2}\right)\right)\;,\; p_k\in\mathbb{N}\,,
    \label{eq_pole_vertex_fct_N}
\end{equation}
where the value of the index $c_k=\pm$ fixes one term in the sum over $c_k$ in \eqref{eq_vertex_function_vf}, and we introduce the index $c_N$, that can also take $\pm$ values and corresponds to selecting a factor in the $\Gamma$-function product \eqref{eq_product_gammas_vertex_fct}. Recall that the distribution of these poles in the complex plane is represented in Fig. \ref{fig_poles_of_IKK_N}. The corresponding residues are given by the formula \eqref{eq_general_N_residue_vertex_fct}.

A common physical situation that we encounter in particular in the computation of section \ref{section_the_double_exchange_diagram} is when the field with frequency $\mu_N$ is conformally coupled, i.e. $\mu_N=i/2$. Then, the product of $\Gamma$-functions \eqref{eq_product_gammas_vertex_fct} can be simplified using the duplication formula \eqref{eq_gamma_duplication}:
\begin{equation}
\begin{aligned}
    \Gamma&\left(\frac{d(N-2)}{4}+\sum_{j=1}^{N-1} \frac{i c_j \mu_j}{2}\pm \frac{i\mu_N}{2}\right)\\
    &=\frac{\sqrt{\pi}}{2^{\frac{d(N-2)-3}{2}+\sum_{j=1}^{N-1} i c_j \mu_j}}
    \Gamma\left(\frac{d(N-2)-1}{2}+\sum_{j=1}^{N-1} i c_j \mu_j\right)\,.
    \end{aligned}
    \label{eq_product_gammas_1_cc}
\end{equation}
Then, the poles of the vertex function $\mu_k\mapsto\mathcal{I}_d\left[\{\mu_j\},\{k_j\},\{K\}\right]$ are given by
\begin{equation}
    \mu_k=c_k\left(-\sum_{j=1,j\neq k}^{N-1} c_j \mu_j +i\left(p_k+\frac{d(N-2)-1}{2}\right)\right)\;,\;p_k\in\mathbb{N}\,,
    \label{eq_poles_when_1_cc}
\end{equation}
and then the residues are
\begin{multline}
\text{Res}\Biggl(\mu_k \mapsto\mathcal{I}_d\left[\{\mu_j\},\{k_j\},\{K\}\right], \mu_k=c_k\left(-\sum_{j=1,j\neq k}^{N-1} c_j \mu_j +i\left(p_k+\frac{d(N-2)-1}{2}\right)\right)\Biggr)\\
    =\frac{(i\pi)^{N-1}}{2^{N+1}}\left(\frac{2}{k_N}\right)^{\frac{d(N-2)}{2}}\sum_{\substack{c_1,\dots,c_{k-1},\\c_{k+1},\ldots,c_{N-1}=\pm}} \prod_{j=1,j\neq k}^{N-1} \Biggl[\frac{c_j}{\sinh(\pi\mu_j)}\left(\frac{k_j}{k_N}\right)^{i c_j\mu_j}\Biggr]\\
    \times
    \frac{\sqrt{\pi}\,2^{p_k+1}(-i c_k)}{\sinh\left(-\sum\limits_{j=1,j\neq k}^{N-1} \pi c_j \mu_j +i\pi\left(p_k+\frac{d(N-2)-1}{2}\right)\right)}\left(\frac{k_k}{k_N}\right)^{-i\sum_{j=1,j\neq k}^{N-1} c_j \mu_j -\left(p_k+\frac{d(N-2)-1}{2}\right)}
     \frac{(-1)^{p_k}}{p_k!}
    \\\times\F_C^{(N-1)}\Biggl(\begin{matrix}
    -\frac{p_k}{2}\,,\;  \frac{1-p_k}{2}\\ 1+i c_1\mu_1\,,\,\ldots\,,\, \frac{3-d(N-2)}{2}-p_k-i\sum\limits_{\substack{j=1\\j\neq k}}^{N-1} c_j \mu_j
    \end{matrix};\left(\frac{k_1}{k_N}\right)^2,\ldots,\left(\frac{k_{N-1}}{k_N}\right)^2\Biggr)\,,
    \label{eq_general_N_residue_vertex_fct_1_cc}
\end{multline}
where, as for \eqref{eq_general_N_residue_vertex_fct}, the value of $c_k=\pm$ is fixed. We then have two vertical columns of poles in the $\mu_k$ plane, one for each value of $c_k$, as represented in Fig. \ref{fig_analytic_structure_F+++B} for the example of the vertex function $\mu\mapsto\I_d \left[\{\nu,\mu, i/2, i/2\},\{s_1, s_2,k_3,k_4\},\{K,I,K,K\}\right]$.

\paragraph{No poles from the $1/\sinh(\pi\mu_k)$ factor.} As discussed previously, the only other origin of singularities in \eqref{eq_vertex_function_vf} may come from the $1/\sinh(\pi\mu_k)$ factor. In fact, let us show that $\mu_k\mapsto\mathcal{I}_d\left[\{\mu_j\},\{k_j\},\{K\}\right]$ is regular when $\mu_k\mapsto 1/\sinh(\pi\mu_k)$ diverges.
Indeed, $\mu_k\mapsto 1/\sinh(\pi\mu_k)$ has poles at $\mu_k\rightarrow ip$, $p\in\mathbb{Z}$:
\begin{equation}
    \frac{1}{\sinh(\pi\mu_k)}\underset{\mu_k\rightarrow i p}{\sim} \frac{(-1)^p}{\pi}\frac{1}{\mu_k-i p}\,.
    \label{eq_singularities_1_sh}
\end{equation}
Now, let us consider the connection formula that involves only this $1/\sinh(\pi\mu_k)$ factor, i.e. \eqref{eq_connection_formula_1_change_k}. Due to the absence of $1/\sinh(\pi\mu_k)$ in their definition (see e.g. \eqref{eq_vertex_function_IK...K} for $k=1$), the two vertex functions on the right-hand side of \eqref{eq_connection_formula_1_change_k} are regular at $\mu_k=i p$. Moreover, from the equality
\begin{equation}
    I_p(z)=I_{-p}(z)\,,
\end{equation}
we have
\begin{equation}
    \mathcal{I}_d\begin{bmatrix}
        \ldots &\mu_{k-1}&i p&\mu_{k+1}&\ldots  \\ \ldots &k_{k-1}&k_k&k_{k+1}&\ldots \\ \ldots & K & I & K & \ldots
    \end{bmatrix}=
    \mathcal{I}_d\begin{bmatrix}
        \ldots &\mu_{k-1}&-i p&\mu_{k+1}&\ldots  \\ \ldots &k_{k-1}&k_k&k_{k+1}&\ldots \\ \ldots & K & I & K & \ldots
    \end{bmatrix}\,.
    \label{eq_IKI_equality}
\end{equation}
from the integral representation \eqref{def_vertex_function_general}.
This holds in the region where the integral converges (defined by a condition over the frequencies $\{\mu_j\}_{j\neq k}$), and in fact in the whole domain where these functions are defined through the series representation, using the analytic continuation theorem \footnote{Since for every $\mu_j$, $j\neq k$, the two functions are analytic in the domain $\mathbb{C}/\mathcal{S}$, where $\mathcal{S}$ is the set of singularities defined by \eqref{eq_pole_vertex_fct_N} and are equal on the subset where the integral representation is defined (that obviously contains accumulation points), we can apply the theorem and conclude that the two functions are equal on $\mathbb{C}/\mathcal{S}$.}.
Then, since the singularities \eqref{eq_singularities_1_sh} are simple poles, taking the limit $\mu_k\rightarrow ip$ in the connection formula \eqref{eq_connection_formula_1_change_k} and using \eqref{eq_IKI_equality} proves that $\mu_k\mapsto\mathcal{I}_d\left[\{\mu_j\},\{k_j\},\{K\}\right]$ is regular and we can write
\begin{equation}
    \mathcal{I}_d\begin{bmatrix}
        \ldots &i p&\ldots  \\ \ldots &k_k&\ldots \\ \ldots & K & \ldots
    \end{bmatrix}=\frac{i(-1)^p}{2}\left(\frac{\partial}{\partial\mu_k}
    \mathcal{I}_d\begin{bmatrix}
        \ldots &\mu_k&\ldots  \\ \ldots &k_k&\ldots \\ \ldots & I & \ldots
    \end{bmatrix}\Bigg|_{\mu_k=ip}+
    \frac{\partial}{\partial\mu_k}
    \mathcal{I}_d\begin{bmatrix}
        \ldots &\mu_k&\ldots  \\ \ldots &k_k&\ldots \\ \ldots & I & \ldots
    \end{bmatrix}\Bigg|_{\mu_k=-ip}\right)\,,
    \label{eq_connection_formula_1_change_k_cont}
\end{equation}
in a way that is naturally very analogous to modified Bessel functions \cite{NISTDLMF}.

\paragraph{Analyticity condition for vertex functions.} Let us now examine under which conditions a singularity \eqref{eq_pole_vertex_fct_N_body} is cancelled by a zero of the dressed Lauricella function. Let $k\in\{1,\ldots,N-1\}$. Starting from the series representation of the vertex function $\mathcal{I}_d\left[\{\mu_j\},\{k_j\},\{K\}\right]$ \eqref{eq_vertex_function_vf}, we focus on the factor
\begin{equation}
\begin{aligned}
    &\mathcal{A}\equiv\Gamma\biggl(\frac{d(N-2)}{4}+\sum_{j=1}^{N-1} \frac{i c_j \mu_j}{2}+ \frac{i\mu_N}{2}\biggr)
    \,\Gamma\biggl(\frac{d(N-2)}{4}+\sum_{j=1}^{N-1} \frac{i c_j \mu_j}{2}- \frac{i\mu_N}{2}\biggr)
    \\&\times\F_C^{(N-1)}\Biggl(\begin{matrix}
    \frac{d(N-2)}{4}+\sum\limits_{j=1}^{N-1} \frac{i c_j \mu_j}{2}+ \frac{i\mu_N}{2}\,,\;  \frac{d(N-2)}{4}+\sum\limits_{j=1}^{N-1} \frac{i c_j \mu_j}{2}- \frac{i\mu_N}{2}\\ 1+i c_1\mu_1\,,\;\ldots\,,\;1+i c_{N-1}\mu_{N-1}
    \end{matrix};\left(\frac{k_1}{k_N}\right)^2,\ldots,\left(\frac{k_{N-1}}{k_N}\right)^2\Biggr)\,.
    \end{aligned}
    \label{eq_app_C2_only_gamma_Lauricella_factors}
\end{equation}
where we kept only the term at fixed $c_j$ made of the product of $\Gamma$-functions and the dressed Lauricella function. For compactness, we define
\begin{subequations}
\begin{equation}
    M_k\equiv\frac{d(N-2)}{4}+\sum_{j=1,j\neq k}^{N-1} \frac{i c_j \mu_j}{2}+ \frac{i\mu_N}{2}\,,
\end{equation}
\begin{equation}
    M'_k\equiv\frac{d(N-2)}{4}+\sum_{j=1,j\neq k}^{N-1} \frac{i c_j \mu_j}{2}- \frac{i\mu_N}{2}\,.
\end{equation}
\end{subequations}
The factor $\mathcal{A}$ then becomes
\begin{equation}
\begin{aligned}
    &\mathcal{A}\equiv\Gamma\biggl(M_k+ \frac{i c_k \mu_k}{2}\biggr)
    \,\Gamma\biggl(M'_k+ \frac{i c_k \mu_k}{2}\biggr)
    \\&\times\F_C^{(N-1)}\Biggl(\begin{matrix}
    M_k+ \frac{i c_k \mu_k}{2}\,,\;  M'_k+ \frac{i c_k \mu_k}{2}\\ 1+i c_1\mu_1\,,\;\ldots\,,\;1+i c_{N-1}\mu_{N-1}
    \end{matrix};\left(\frac{k_1}{k_N}\right)^2,\ldots,\left(\frac{k_{N-1}}{k_N}\right)^2\Biggr)\,.
    \end{aligned}
    \label{eq_app_C2_only_gamma_Lauricella_factors_2}
\end{equation}
We want to see under which condition on $M_k$, $M'_k$ this factor remains finite when $M_k+(i c_k \mu_k)/2\rightarrow -p$, $p\in\mathbb{N}$ (and similarly for $M'_k$). As we shall see in the following, such a condition can be met if one of the bottom arguments $1+ic_k \mu_k$ becomes a negative integer. To do so, let us first derive the following lemma:
\begin{equation}
\begin{aligned}
    \F_C^{(N)}&\Biggl(\begin{matrix}
    a\;,\; b \\ c_1\,,\;\ldots\,,\;c_{N}\end{matrix};
    x_1,\ldots,x_N\Biggr)\\
    &\underset{c_k\rightarrow -p}{=}(a)_{p+1}(b)_{p+1}x_k^{p+1}
    \F_C^{(N)}\Biggl(\begin{matrix}
    a+p+1\;,\; b+p+1 \\ c_1\,,\;\ldots\,,\,c_{k-1}\,,\,p+2\,,\,c_{k+1}\,,\,\ldots\,,\;c_{N}\end{matrix};
    x_1,\ldots,x_N\Biggr)\,.
    \end{aligned}
    \label{app_residues_lemma_FCN}
\end{equation}
This is the generalisation of a well-known identity for the Gauss hypergeometric function.\footnote{See eq. 15.2.3\_5 of \cite{NISTDLMF}.} Let us prove this result. We start from the defining series of $\mathcal{F}_C^{(N)}$ \eqref{eq_def_Lauricella_C}-\eqref{eq_def_dressed_Lauricella_C}. Let $a$,$b$,$c_1$,...,$c_{N}$ be arbitrary complex parameters, and $x_1,\ldots,x_N$ complex numbers such that the series \eqref{eq_def_Lauricella_C} converges. The dependence on $c_k$ is then given in each term of the series by a $\Gamma(c_k+n_k)$ factor in the denominator (where $n_k\in\mathbb{N}$ is the summation index). We consider the limit $c_k\rightarrow -p$, $p\in\mathbb{N}$, then the inverse of $\Gamma(c_k+n_k)$ takes the values
\begin{equation}
    \frac{1}{\Gamma(c_k+n_k)}\underset{c_k\rightarrow -p}{\rightarrow}\begin{cases}
        0 &\text{if}\; n_k \leq p\\ \frac{1}{\Gamma(n_k-p)} &\text{if}\; n_k > p
    \end{cases}\,.
\end{equation}
Then, in this limit we can rewrite the defining series of $\mathcal{F}_C^{(N)}$ as
\begin{equation}
\begin{aligned}
    \F_C^{(N)}&\Biggl(\begin{matrix}
    a\;,\; b \\ c_1\,,\;\ldots\,,\;c_{N}\end{matrix};
    x_1,\ldots,x_N\Biggr)=
    \sum_{n_1,\ldots,n_N=0}^\infty \frac{(a)_{n_1+\ldots+n_N}(b)_{n_1+\ldots+n_N}}{\Gamma(c_1+n_1)\ldots\Gamma(c_N+n_N)}\frac{x_1^{n_1}}{n_1!}\ldots\frac{x_N^{n_N}}{n_N!}\\
    &\underset{c_k\rightarrow -p}{=}
    \sum_{\substack{n_j=0\,\text{,}\, j\neq k\\ n_k=p+1}}^\infty \frac{(a)_{n_1+\ldots+n_N}(b)_{n_1+\ldots+n_N}}{\Gamma(c_1+n_1)\ldots\Gamma(n_k-p)\ldots\Gamma(c_N+n_N)}\frac{x_1^{n_1}}{n_1!}\ldots\frac{x_N^{n_N}}{n_N!}\\
    &\underset{c_k\rightarrow -p}{=}
    \sum_{\substack{n_j=0\,\text{,}\, j\neq k\\ l_k=0}}^\infty \frac{(a)_{n_1+\ldots+l_k+\ldots+n_N+p+1}(b)_{n_1+\ldots+l_k+\ldots+n_N+p+1}}{\Gamma(c_1+n_1)\ldots\Gamma(l_k+1)\ldots\Gamma(c_N+n_N)} \\
    &\qquad\qquad\times\frac{x_1^{n_1}}{n_1!}\ldots\frac{x_k^{l_k+p+1}}{\Gamma(l_k+p+2)}\ldots\frac{x_N^{n_N}}{n_N!}\\
    &\underset{c_k\rightarrow -p}{=}
    (a)_{p+1}(b)_{p+1}x_k^{p+1}
    \sum_{\substack{n_j=0\,\text{,}\, j\neq k\\ l_k=0}}^\infty \frac{(a+p+1)_{n_1+\ldots+n_N}(b+p+1)_{n_1+\ldots+n_N}}{\Gamma(c_1+n_1)\ldots\Gamma(l_k+p+2)\ldots\Gamma(c_N+n_N)} \\
    &\qquad\qquad\times\frac{x_1^{n_1}}{n_1!}\ldots\frac{x_N^{n_N}}{n_N!}\,,
    \end{aligned}
    \label{eq_app_Lauricella_C_lemma_1}
\end{equation}
and using the definition in the last line \eqref{eq_def_Lauricella_C}-\eqref{eq_def_dressed_Lauricella_C} we recover \eqref{app_residues_lemma_FCN}. In this last computation, between the second and third lines we changed the summation index to $l_k=n_k-(p+1)$ and between the third and last lines we used the following identity for Pochhammer symbols
\begin{equation}
    (a)_{b+c}=(a+b)_c \,(a)_b \,.
\end{equation}

Coming back to our initial problem, in the limit $M_k+(i c_k \mu_k)/2\rightarrow -p$, we have
\begin{equation}
    1+i c_k \mu_k \rightarrow 1- 2 p -2M_k\,.
\end{equation}
Then, this quantity is a negative integer if
\begin{equation}
    M_k=\frac{m_k}{2}\;,\;m_k\in\mathbb{N}^*\,.
    \label{eq_app_condition_Mk}
\end{equation}
Under this condition, we can apply the result \eqref{app_residues_lemma_FCN} to simplify the factor $\mathcal{A}$ in the desired limit $M_k+(i c_k \mu_k)/2\rightarrow -p$:
\begin{equation}
\begin{aligned}
&\mathcal{A}\sim\Gamma\biggl(2 M_k+ p\biggr)
\,\Gamma\biggl(M'_k+M_k+p\biggr)\,\left(\frac{k_k}{k_N}\right)^{2(M_k+p)}\\
&\times\F_C^{(N-1)}\Biggl(\begin{matrix}
2 M_k+ p\,,\;  M'_k+M_k+p \\ 1+i c_1\mu_1\,,\,\ldots\,,\,2(p+M_k)+1\,,\,\ldots\,,\,1+i c_{N-1}\mu_{N-1}
\end{matrix};\left(\frac{k_1}{k_N}\right)^2,\ldots,\left(\frac{k_{N-1}}{k_N}\right)^2\Biggr)\,,
\end{aligned}
\end{equation}
where we suitably simplified $\Gamma$-function factors. Finally, under the condition \eqref{eq_app_condition_Mk} this last expression is finite, so \eqref{eq_pole_vertex_fct_N_body} is not a singularity of the vertex function \eqref{vertex_fct_N}.

\subsection{Asymptotic behaviour of vertex functions}\label{subappendix_asymptotic_behavior_N}

\paragraph{Large $\mu_k$ behaviour.} Let $k\in\{1,\ldots,N-1\}$, and consider the vertex function $\I\left[\{\mu_j\},\{k_j\},\{A_j\}\right]$ with $A_k=I$ and $A_N=K$ ($A_j=K$ or $I$ for $j\in\{1,\ldots,N\}/\{k,N\}$) when $\mu_k$ is large. It has the following integral representation:
\begin{equation}
    \mathcal{I}_d\begin{bmatrix}
        \mu_1 &\ldots &\mu_k&\ldots&\mu_N \\ k_1 &\ldots&k_k&\ldots  & k_N \\ A_1 &\ldots & I &\ldots & K
    \end{bmatrix}
    = \int \limits^\infty_0\d z\; z^{\frac{d(N-2)}{2}-1}\prod_{j=1,j\neq k}^{N-1} \left(A_j\right)_{i\mu_j}(k_j z)\, I_{i\mu_k}(k_k z) \, K_{i\mu_N}(k_N z) \,.
    \label{eq_vertex_function_app_large_mu}
\end{equation}
Again, working under the kinematic condition \eqref{eq_cvg_radius_kinematics} ensures that this integral is convergent at infinity. The small-argument behaviour of the Bessel function $I_{i c_k \mu_k}$ \eqref{small_z_bessel_I} suggests that we need to take the argument of $\mu_k$ in the lower-half (resp. upper-half) complex plane if $c_k=+1$ (resp. $c_k=-1$). Moreover, from its series definition \eqref{def_bessel_I}, the small-argument behaviour of $I_{i c_k \mu_k}$ \eqref{small_z_bessel_I} coincides with the large-order one. Then, from \eqref{small_z_bessel_I}, at large $\mu_k$ the function $I_{i c_k \mu_k}$ increases for large values of $z$ ($i c_k \mu_k$ is large and positive to ensure the convergence of \eqref{eq_vertex_function_app_large_mu} at $z=0$), meaning that at leading order in $\mu_k$, the main contribution to the integral is at large values of $z$.
Therefore, we can approximate the other modified Bessel functions entering \eqref{eq_vertex_function_app_large_mu} by their asymptotic behaviour at $z\rightarrow +\infty$ \eqref{eq_large_z_Bessels}-\eqref{large_z_bessel_I_refined} \footnote{It is possible to show that including the next-to-leading-order term of the large-argument asymptotics of Bessel functions \eqref{eq_large_z_Bessels} indeed corrects the large $\mu_k$ behaviour with a term of magnitude $O(\mu_k^{-1})$.}. Note first that the refined large asymptotic behaviour of $I_{i\mu_l}$ \eqref{large_z_bessel_I_refined} can be rewritten in the following compact way:
\begin{equation}
    I_{i\mu_l} (k_l z) \underset{z\rightarrow+\infty}{\sim} \frac{1}{\sqrt{2 \pi k_j z}}\sum_{d_l=0,1} \left(i e^{-\pi\mu_l}\right)^{d_l}e^{ -(2 d_l-1) k_l z}\,.
    \label{eq_large_z_I_compact}
\end{equation}
Then, the integral \eqref{eq_vertex_function_app_large_mu} can be approximated by
\begin{equation}
\begin{aligned}
\mathcal{I}_d&\begin{bmatrix}
        \mu_1 &\ldots &\mu_k&\ldots&\mu_N \\ k_1 &\ldots&k_k&\ldots  & k_N \\ A_1 &\ldots & I &\ldots & K
    \end{bmatrix}\underset{\mu_k\rightarrow\infty}{\sim}
    \frac{\pi^{\frac{1}{2}\left(\lvert\K\rvert-\lvert\I\rvert\right)}}{2^{\frac{N-1}{2}} \prod\limits_{j=1,j\neq k}^N \sqrt{k_j}}\\
    &\times
    \int \limits^\infty_0\d z\; z^{\frac{(d-1)(N-2)}{2}-\frac{3}{2}}
    e^{ -\left(k_N+\sum\limits_{j\in\K} k_j\right) z}
    \sum_{d_l=0,1;l\in\I} \left(i e^{-\pi\mu_l}\right)^{d_l}e^{ -(2 d_l-1) k_l z}
    I_{i\mu_k}(k_k z)
    \,.
\end{aligned}
    \label{eq_vertex_function_app_large_mu_2}
\end{equation}
where the sets $\I$ and $\K$ are respectively defined by
\begin{subequations}
\begin{equation}
    j\in\I \;\text{if}\;A_j=I\;\text{and}\;j\neq k\,,
\end{equation}
\begin{equation}
    j\in\K \;\text{if}\;A_j=K\,.
\end{equation}
\label{eq_asymptotics_sets_IK}
\end{subequations}
Performing the change of variable $z'=k_k z$ and using the integral representation of the associated Legendre $Q$ function \eqref{eq_Q_integral_I}, we can rewrite it as
\begin{equation}
\begin{aligned}
\mathcal{I}_d&\begin{bmatrix}
        \mu_1 &\ldots &\mu_k&\ldots&\mu_N \\ k_1 &\ldots&k_k&\ldots  & k_N \\ A_1 &\ldots & I &\ldots & K
    \end{bmatrix}\underset{\mu_k\rightarrow\infty}{\sim}
    \frac{\pi^{\frac{1}{2}\left(\lvert\K\rvert-\lvert\I\rvert\right)}}{2^{\frac{N-1}{2}} k_k^{\frac{(d-1)(N-2)-1}{2}}
    \prod\limits_{j=1,j\neq k}^N \sqrt{k_j}}
    \sqrt{\frac{2}{\pi}} e^{-i\frac{\pi}{2}\left((d-1)(N-2)-2\right)}\\
    &\times  
    \sum_{d_l=0,1;l\in\I} \left(i e^{-\pi\mu_l}\right)^{d_l}\left(\left(\frac{k_t\left(\{d_l\}\right)}{k_k}\right)^2-1\right)^{\frac{1}{2}-\frac{(d-1)(N-2)}{4}}
    Q_{i\mu_k-\frac{1}{2}}^{\frac{(d-1)(N-2)}{2}-1}\left(\frac{k_t\left(\{d_l\}\right)}{k_k}\right)
    \,,
\end{aligned}
    \label{eq_vertex_function_app_large_mu_3}
\end{equation}
where
\begin{equation}
    k_t\left(\{d_l\}\right) \equiv k_N+\sum\limits_{j\in\K} k_j+\sum_{l\in\I} (2 d_l-1) k_l\,.
    \label{eq_kt_dl}
\end{equation}
Then we use the known asymptotic behaviour of the Legendre $Q$ function at large $\mu_k$ \eqref{eq_large_mu_Q} to arrive at the expression \eqref{eq_vertex_function_app_large_mu_vf}.

In this final expression, leading-order behaviours are given by the exponential terms with argument $-i\mu_k\arccosh\left(k_t\left(\{d_l\}\right)/k_k\right)$. Note that from the kinematic condition \eqref{eq_cvg_radius_kinematics}, we always have $k_t\left(\{d_l\}\right)/k_k>1$. Such an asymptotic behaviour with a single exponential behaviour is then obtained by writing the KLF integrand over $\mu_k$ in terms of vertex functions with $A_k=I$ and all other $A_j=K$ ($j=[1,\ldots,N-1]$, $j\neq k$). In that case, the comparison of the different leading behaviours of vertex functions in a KLF integrand allows us to choose a neat contour prescription to ensure a vanishing contribution from the arc at infinity.

Let us add a remark on the physical intuition that comes with the mathematical step that allows us to write \eqref{eq_vertex_function_app_large_mu_2}. The limit $\mu_k\rightarrow +\infty$ can be seen as the large-mass limit of a heavy free field in inflation. However, these fields do not survive the end of inflation, since their mode function is very small at late times. Therefore, the mode function becomes non-negligible at sufficiently early times, where the propagation of other massive fields that are interacting with this very heavy field is well-approximated by plane waves, hence the approximation made in \eqref{eq_vertex_function_app_large_mu_2}.

\paragraph{Large $\mu_j$ behaviour of residues \eqref{eq_general_N_residue_vertex_fct}.}
To conclude this Appendix, let us examine the behaviour of the residue \eqref{eq_general_N_residue_vertex_fct} when one of the frequencies $\mu_j\rightarrow+\infty$. The non-trivial part to approximate in this limit is $\F_C^{(N-1)}$. In the expression \eqref{eq_general_N_residue_vertex_fct}, the two lower arguments of $\F_C^{(N-1)}$ involving $\mu_j$ become very large, while all others remain finite. First, we observe that
\begin{equation}
    \frac{1}{(c)_n}\underset{c\rightarrow+\infty}{\sim} O(c^{-n})\,,
\end{equation}
therefore in the limit where one lower argument becomes large the leading-order term of $\F_C^{(N-1)}$ is exactly given by the first term in the series \eqref{eq_def_Lauricella_C}, and we have
\begin{equation}
    \F_C^{(N-1)}\Biggl(\begin{matrix}
    a\;,\; b \\ c_1\,,\;\ldots\,,\;c_{N-1}\end{matrix};
    x_1,\ldots,x_{N-1}\Biggr)
    \underset{c_k\rightarrow+\infty}{\sim}\prod_{i=1}^N\frac{1}{\Gamma(c_i)}\,.
    \label{eq_large_c_Lauricella}
\end{equation}
This allows us to obtain the asymptotic behaviour of \eqref{eq_general_N_residue_vertex_fct} at large $\mu_j$, $j\neq k$. Keeping only the power-law growing or decaying terms in $\mu_j$ (which give the relevant information needed for contour prescriptions), we obtain the following:
\begin{equation}
\begin{aligned}
    &\text{Res}\Biggl(\mu_k \mapsto\mathcal{I}_d\left[\{\mu_j\},\{k_j\},\{K\}\right], \mu_k=c_k\biggl(\tilde{\mu}_{k,N}+i\Bigl(2 p_k+\frac{d(N-2)}{2}\Bigl)\biggl)\;,\; p_k\in\mathbb{N}\Biggr)\\
    &\underset{\mu_j\rightarrow +\infty}{\propto}
    \sum_{c_j=\pm} \frac{1}{\sinh(\pi c_j\mu_j)\sinh\left(\pi\left(\tilde{\mu}_{k,N}+i\Bigl(2 p_k+\frac{d(N-2)}{2}\Bigl)\right)\right)}\left(\frac{k_j}{k_k}\right)^{i c_j\mu_j}\,.
    \end{aligned}
    \label{eq_general_N_residue_large_mu_j}
\end{equation}
In practical KLF computations, we rewrite the $\mu_j$-integral as a sum of two terms, each with a fixed value of $c_j$. Therefore, for the large $\mu_j$ evaluation of the integrand, one should only keep the term that has the relevant value of $c_j$ in the previous expression.

\section{Other kinematic cases in the double-exchange computation}\label{appendix_kinematic_cases_G+++}

In this Appendix, we detail the computation of the totally nested component $G_{+++}$ for the kinematic cases that are not treated in subsection \ref{subsection_+++}.

\subsection{Background-Background residues}

\paragraph{Background-Background residues, $\kappaone<1$, $\kappathree<1$.} We start by considering the immediate follow-up of the derivation made for $\kappaone<1$ and $\kappatwo>1$ in subsection \ref{subsec_second_int_layer}, which is the case where $\kappathree<1$.

In this case, to compute $G_{+++}^{1,B,B}$ \eqref{eq_G_+++_1.B_def_C2<1} we close the contour in the lower half-plane. We then collect the two following sets of residues (see the right panel of Fig. \ref{fig_analytic_structure_G+++B_C2<1}):
\begin{itemize}
\item \textbf{\textcolor{pyblue}{Poles from $\frac{\nu}{\sinh(\pi\nu)}$.}}
The first term is made of residues of the factor $\nu\mapsto\nu/\sinh(\pi\nu)$ on the negative imaginary axis. It reads
\begin{equation}
\begin{aligned}
    G_{+++}^{1,B,B}&\supset
    2\sum_{p=0}^{+\infty}(-1)^{p} \,Q_{-p-\frac{3}{2}}^{\frac{d-3}{2}}\left(\frac{k_{12}}{s_1}\right)\frac{p+1}{(p+1)^2+\mu_1^2}
    \\
    &\times 
    \mathcal{I}_d\begin{bmatrix}
        -i(p+1) &-i\left(n+\frac{d-2}{2}\right) &i/2 & i/2 \\ s_1 & s_2 & k_3 & k_4  \\ I & I & K & K
    \end{bmatrix}\,.
    \end{aligned}
    \label{eq_G+++_1BB_C2<1_C1p<1_res_sh}
\end{equation}
\item \textbf{\textcolor{pyred}{Poles from $Q_{-i\nu-\frac{1}{2}}^{\frac{d-3}{2}}\left(k_{12}/s_1\right)$.}} The other set of residues comes from the poles of the Legendre $Q$ function $Q_{-i\nu-\frac{1}{2}}^{\frac{d-3}{2}}\left(k_{12}/s_1\right)$. These poles are located at $\nu=-i\left(p+(d-2)/2\right)$ and the corresponding residues are given in \eqref{eq_residues_legendre_Q}. Their contribution to $G_{+++}^{1,B,B}$ is then
\begin{equation}
\begin{aligned}
    G_{+++}^{1,B,B}&\supset
    \frac{\pi^\frac{3}{2} e^{i\pi\frac{d-1}{2}} 2^\frac{d-1}{2}}{\sin\left(\frac{\pi d}{2}\right)}
    \left(\left(\frac{k_{12}}{s_1}\right)^2-1\right)^\frac{d-3}{4}
    \sum_{p=0}^{+\infty}
     \left(\frac{2k_{12}}{s_1}\right)^p
    \frac{1}{p!}\\
    &\times
    \,_2\F_1\left(
    \begin{matrix}
        \frac{1-p}{2}\,,\,-\frac{p}{2} \\ -p-\frac{d-4}{2}
    \end{matrix}\,,\left(\frac{s_1}{k_{12}}\right)^2\right)
    \frac{p+\frac{d-2}{2}}{\left(p+\frac{d-2}{2}\right)^2+\mu_1^2}\\
    &\times\mathcal{I}_d\begin{bmatrix}
        -i\left(p+\frac{d-2}{2}\right) &-i\left(n+\frac{d-2}{2}\right) &i/2 & i/2 \\ s_1 & s_2 & k_3 & k_4  \\ I & I & K & K
    \end{bmatrix}\,.
    \end{aligned}
    \label{eq_G+++_1BB_C2<1_C1p<1_res_III}
\end{equation}
\end{itemize}
The complete result for $G_{+++}^{B,B}$ when $\kappaone<1$ and $\kappathree<1$ is then made of the 3 contributions \eqref{eq_G+++_0BB_C2<1}, \eqref{eq_G+++_1BB_C2<1_C1p<1_res_sh} and \eqref{eq_G+++_1BB_C2<1_C1p<1_res_III}. However, we can note that the sum of the first two vanishes, using the connection formula between Legendre $P$ and $Q$ functions \eqref{connection_formula_P_Q}. Consequently, $G_{+++}^{B,B}$ is given by inserting \eqref{eq_G+++_1BB_C2<1_C1p<1_res_III} in the decomposition \eqref{eq_G_+++_.B_def_wrt_G_+++_01B_C2<1}, and we obtain
\begin{framed}
\vspace{-.5cm}
\begin{equation}
\begin{aligned}
    G_{+++}^{B,B}\left(\{k_i,s_j\}\right)
    &= \frac{e^{-i 2 \pi d} 2^{d-5} \pi^7}{\sqrt{k_1 k_2 k_3 k_4 k_5 k_6}s_1^{\frac{d-2}{2}}s_2^{\frac{d-2}{2}}k_{34}^{d-1}
    \sin^2\left(\pi\frac{d}{2}\right)}
    \left(\frac{s_1}{2 k_{34}}\right)^{\frac{d-2}{2}} \left(\frac{s_2}{2 k_{34}}\right)^{\frac{d-2}{2}}\\
    &\times \sum_{n=0}^{+\infty} \sum_{p=0}^{+\infty} \frac{\Gamma\left(2d-3+n+p\right)}{n! p!} \left(\frac{k_{56}}{k_{34}}\right)^n \left(\frac{k_{12}}{k_{34}}\right)^p
    \\
    &\times
    \,_2\F_1\left(
    \begin{matrix}
        \frac{1-p}{2}\,,\,-\frac{p}{2} \\ -p-\frac{d-4}{2}
    \end{matrix}\,,\left(\frac{s_1}{k_{12}}\right)^2\right)
    \frac{p+\frac{d-2}{2}}{\left(p+\frac{d-2}{2}\right)^2+\mu_1^2}\\
    &\times\F_4\Biggl(\begin{matrix}
    d-\frac{3}{2}+\frac{p+n}{2}\,,\, d-1+\frac{p+n}{2}\\
    p+\frac{d}{2}\,,\,n+\frac{d}{2}\end{matrix}\,;
    \left(\frac{s_1}{k_{34}}\right)^2,\left(\frac{s_2}{k_{34}}\right)^2\Biggr)\\
    &\times \frac{n+\frac{d-2}{2}}{\left(n+\frac{d-2}{2}\right)^2+\mu_2^2}
    \,_2\F_1\left(
    \begin{matrix}
        \frac{1-n}{2}\,,\,-\frac{n}{2} \\ -n-\frac{d-4}{2}
    \end{matrix}\,,\left(\frac{s_2}{k_{56}}\right)^2\right)
    \,.
    \end{aligned}
    \label{eq_G+++_BB_C2<1_C1p<1_full_res}
\end{equation}
\end{framed}
\noindent where we also used the series expression of $\I_{II}$ \eqref{eq_series_rep_vertex_F4_III}.

\paragraph{Background-Background residues, $\kappaone<1$, $\kappatwo<1$, $\kappathree>1$.}
The large frequency does not allow for a neat criterion when $\kappaone<1$, $\kappatwo<1$ and $\kappathree>1$. To tackle this issue, we decompose the vertex function $\I_{II}$ that enters the integral into the two pieces that give rise to each exponential behaviour in \eqref{eq_large_nu_integrand_C2<1}. This can be done by the use of the analytic continuation formula \eqref{eq_continuation_III_asymptotics}. It leads to
\begin{equation}
    G_{+++}^{1,.,B}=G_{+++}^{1,0,.,B}+G_{+++}^{1,1,.,B}\,,
    \label{kappa1<1_decomposition}
\end{equation}
with
\begin{subequations}
\begin{equation}
\begin{aligned}
G_{+++}^{1,0,.,B}&\equiv\frac{i}{\pi}\int\limits_{-\infty}^{+\infty}\d\nu \, \frac{\nu}{\sinh(\pi\nu)}\,
Q_{-i\nu-\frac{1}{2}}^{\frac{d-3}{2}}\left(\frac{k_{12}}{s_1}\right)
    \frac{1}{(\nu^2-\mu_{1}^2)_{i\epsilon}}
    \mathcal{I}_d\begin{bmatrix}
        \nu &-i\left(n+\frac{d-2}{2}\right) &i/2 & i/2 \\ s_1 & e^{-i\pi}s_2 & k_3 & k_4  \\ I & K & K & K
    \end{bmatrix}
    \,,
    \end{aligned}
    \label{eq_G_+++_10.B_def}
\end{equation}
\begin{equation}
\begin{aligned}
G_{+++}^{1,1,.,B}\equiv-\frac{i}{\pi}(-1)^n e^{i\pi\frac{d-2}{2}}\int\limits_{-\infty}^{+\infty}\d\nu \, &\frac{\nu}{\sinh(\pi\nu)}\,
Q_{-i\nu-\frac{1}{2}}^{\frac{d-3}{2}}\left(\frac{k_{12}}{s_1}\right)
    \frac{1}{(\nu^2-\mu_{1}^2)_{i\epsilon}}\\
    &\times\mathcal{I}_d\begin{bmatrix}
        \nu &-i\left(n+\frac{d-2}{2}\right) &i/2 & i/2 \\ s_1 & s_2 & k_3 & k_4  \\ I & K & K & K
    \end{bmatrix}
    \,.
    \end{aligned}
    \label{eq_G_+++_11.B_def}
\end{equation}
\end{subequations}
Then, using the large frequency asymptotic behaviour of the vertex functions \eqref{eq_large_mu_Q}-\eqref{eq_large_mu_IKI}, we obtain, for the integrand of $G_{+++}^{1,0,.,B}$:
\begin{equation}
Q_{-i\nu-\frac{1}{2}}^{\frac{d-3}{2}}\left(\frac{k_{12}}{s_1}\right)
    \mathcal{I}_d\begin{bmatrix}
        \nu &-i\left(n+\frac{d-2}{2}\right) &i/2 & i/2 \\ s_1 & e^{-i\pi}s_2 & k_3 & k_4  \\ I & K & K & K
    \end{bmatrix}
\underset{\nu\to\infty}{\propto}
e^{-i\nu\left(-\arccosh\left(\frac{k_{12}}{s_1}\right)+ \arccosh\left(\frac{k_{34}-s_2}{s_1}\right)\right)}\,.
\end{equation}
Since $\kappathree>1$, to collect the residues of $G_{+++}^{1,0,.,B}$, we close the contour in the upper half-plane. Similarly, for the integrand of $G_{+++}^{1,1,.,B}$:
\begin{equation}
Q_{-i\nu-\frac{1}{2}}^{\frac{d-3}{2}}\left(\frac{k_{12}}{s_1}\right)
    \mathcal{I}_d\begin{bmatrix}
        \nu &-i\left(n+\frac{d-2}{2}\right) &i/2 & i/2 \\ s_1 & s_2 & k_3 & k_4  \\ I & K & K & K
    \end{bmatrix}
\underset{\nu\to\infty}{\propto}
    e^{-i\nu\left(-\arccosh\left(\frac{k_{12}}{s_1}\right)+ \arccosh\left(\frac{k_{34}+s_2}{s_1}\right)\right)}\,.
\end{equation}
Then, the condition $\kappatwo<1$ implies a contour closure in the lower half-plane for $G_{+++}^{1,1,.,B}$.

Let us now collect the background residues of these two components, the signal residues being collected in the next subsection. We note that these two integrals have the same structure as $F_{+++}^1$ \eqref{eq_F_+++_1} up to exchanging the labels of kinematic variables, then we can use the computations done in section \ref{subsection_+++}. For $G_{+++}^{1,0,.,B}$, since the contour is closed in the upper half-plane we can deduce the result from the expression of $F_{+++}^{1,B}$ for $\kappaone>1$. It gives the contributions:
\begin{itemize}
\item \textbf{\textcolor{pyblue}{Poles from $\frac{\mu}{\sinh(\pi\mu)}$.}} The corresponding residues are:
\begin{equation}
\begin{aligned}
    G_{+++}^{1,0,.,B}\supset -\frac{2 i}{\pi} \sum_{p=0}^{+\infty} (-1)^p \,&Q_{p+\frac{1}{2}}^{\frac{d-3}{2}}\left(\frac{k_{12}}{s_1}\right)
    \frac{p+1}{(p+1)^2+\mu_1^2}\\
    &\times\mathcal{I}_d\begin{bmatrix}
        i(p+1) &-i\left(n+\frac{d-2}{2}\right) &i/2 & i/2 \\ s_1 &e^{-i\pi}s_2 & k_3 & k_4 \\ I & K & K & K
    \end{bmatrix}
    \,.
    \end{aligned}
    \label{eq_G+++_10.B_res_sh}
\end{equation}
\item \textbf{\textcolor{pyorange}{Poles from $\I_{IK}$.}} The other poles that contribute to $G_{+++}^{1,0,.,B}$ come from $I_{IK}$, with residues
\begin{equation}
\begin{aligned}
    G_{+++}^{1,0,.,B}&\supset -\frac{i 2^{d-2}\pi^2}{\sqrt{k_3 k_4} \,s_1^{d-1}}
    \sum_{c=\pm}\frac{e^{-i\pi c\frac{d}{2}}}{\sin(\pi c \frac{d}{2}) \sin(\pi c \frac{d}{2} +\pi d)}  (-1)^{n+1}\left(\frac{s_2}{s_1}\right)^{c\left(n+\frac{d-2}{2}\right)} \\
    &\times \sum_{p=0}^{+\infty} \left(\frac{2 k_{34}}{s_1}\right)^p \frac{1}{p!}Q_{c\left(n+\frac{d-2}{2}\right) +p+d-\frac{3}{2}}^{\frac{d-3}{2}}\left(\frac{k_{12}}{s_1}\right)\,
    \frac{c \left(n+\frac{d-2}{2}\right)+(p+d-1)}{\left(c \left(n+\frac{d-2}{2}\right)+p+d-1\right)^2+\mu_1^2}\\
    &\times \F_4\Biggl(\begin{matrix}
    -\frac{p}{2}\,,\, \frac{1-p}{2}\\
    1+c \left(n+\frac{d-2}{2}\right)\,,\,2- c\left(n+\frac{d-2}{2}\right)-p-d\end{matrix}\,;
    \left(\frac{s_1}{k_{34}}\right)^2,\left(\frac{s_2}{k_{34}}\right)^2\Biggr)\,.
    \end{aligned}
    \label{eq_G+++_10.B_res_IIK}
\end{equation}
\end{itemize}
For $G_{+++}^{1,1,.,B}$, the contour is closed in the lower half-plane, we obtain the result from the expression of $F_{+++}^{1,B}$ for $\kappaone<1$. The two contributions are the following:
\begin{itemize}
\item \textbf{\textcolor{pyblue}{Poles from $\frac{\mu}{\sinh(\pi\mu)}$.}}
\begin{equation}
\begin{aligned}
    G_{+++}^{1,1,.,B}\supset \frac{2}{\pi} (-1)^{n+1} e^{i\pi\frac{d+1}{2}} \sum_{p=0}^{+\infty} (-1)^p &
    Q_{-p-\frac{3}{2}}^{\frac{d-3}{2}}\left(\frac{k_{12}}{s_1}\right)
    \frac{p+1}{(p+1)^2+\mu_1^2}\\
    &\mathcal{I}_d\begin{bmatrix}
        -i(p+1) &-i\left(n+\frac{d-2}{2}\right) &i/2 & i/2 \\ s_1 &s_2 & k_3 & k_4 \\ I & K & K & K
    \end{bmatrix}
    \,.
    \end{aligned}
    \label{eq_G+++_11.B_res_sh}
\end{equation}
\item \textbf{\textcolor{pyred}{Poles from $Q_{-i\mu-\frac{1}{2}}^{\frac{d-3}{2}}\left(k_{56}/s_2\right)$.}}
\begin{equation}
\begin{aligned}
    G_{+++}^{1,1,.,B}&\supset (-1)^{n+1} \frac{2^{\frac{d-1}{2}} \pi^\frac{1}{2} e^{i\pi d}}{\sin\left(\pi\frac{d}{2}\right)}
    \left(\left(\frac{k_{12}}{s_1}\right)^2-1\right)^{\frac{d-3}{4}}
    \sum_{p=0}^{+\infty} \frac{p+\frac{d-2}{2}}{\left(p+\frac{d-2}{2}\right)^2+\mu_1^2}\frac{1}{p!}\left(\frac{2 k_{12}}{s_1}\right)^p\\
    &\times\,_2\F_1\left(
    \begin{matrix}
        \frac{1-p}{2}\,,\,-\frac{p}{2} \\ -p-\frac{d-4}{2}
    \end{matrix}\,,\left(\frac{s_1}{k_{12}}\right)^2\right)
    \mathcal{I}_d\begin{bmatrix}
        -i\left(p+\frac{d-2}{2}\right) &-i\left(n+\frac{d-2}{2}\right) &i/2 & i/2 \\ s_1 & s_2 & k_3 & k_4  \\ I & K & K & K
    \end{bmatrix}\,.
    \end{aligned}
    \label{eq_G+++_11.B_res_Q}
\end{equation}
\end{itemize}
To obtain $G_{+++}^{B,B}$, we have to sum these four components \eqref{eq_G+++_10.B_res_sh}-\eqref{eq_G+++_10.B_res_IIK}-\eqref{eq_G+++_11.B_res_sh}-\eqref{eq_G+++_11.B_res_Q} and $G_{+++}^{0,B,B}$ given by \eqref{eq_G+++_0BB_C2<1}. A significant simplification occurs: the sum of \eqref{eq_G+++_0BB_C2<1}, \eqref{eq_G+++_10.B_res_sh} and \eqref{eq_G+++_11.B_res_sh} vanishes. This can be seen by using the analytic continuation formula \eqref{eq_continuation_III_asymptotics} in the expression of $G_{+++}^{0,B,B}$ \eqref{eq_G+++_0BB_C2<1} and then the connection formula among Legendre functions \eqref{connection_formula_P_Q} to simplify the sum. The two remaining components \eqref{eq_G+++_10.B_res_IIK} and \eqref{eq_G+++_11.B_res_Q} inserted in the decompositions \eqref{eq_G_+++_.B_def_wrt_G_+++_01B_C2<1} and \eqref{kappa1<1_decomposition} give the following final result for $G_{+++}^{B,B}$ in the case $\kappaone<1$, $\kappatwo<1$ and $\kappathree>1$:
\begin{framed}
\vspace{-.5cm}
\begin{equation}
\begin{aligned}
    G_{+++}^{B,B}
    &= \frac{e^{-i\pi\frac{3}{2}(d-1)} 2^{\frac{d-7}{2}} \pi^\frac{9}{2}}{\sqrt{k_1 k_2 k_5 k_6}s_1^{\frac{d-2}{2}}s_2^{\frac{d-2}{2}}\sin\left(\pi\frac{d}{2}\right)}
    \left(\left(\frac{k_{12}}{s_1}\right)^2-1\right)^{\frac{3-d}{4}}\\
    &\times \sum_{n=0}^{+\infty} \sum_{p=0}^{+\infty} \frac{(-1)^{n+1} }{n!}\left(\frac{2 k_{56}}{s_2}\right)^n
    \frac{n+\frac{d-2}{2}}{\left(n+\frac{d-2}{2}\right)^2+\mu_2^2}
    \,_2\F_1\left(
    \begin{matrix}
        \frac{1-n}{2}\,,\,-\frac{n}{2} \\ -n-\frac{d-4}{2}
    \end{matrix}\,,\left(\frac{s_2}{k_{56}}\right)^2\right)\\
    &\times
    \Biggl[\frac{e^{i\pi\frac{3}{2}} 2^{d-2}\pi^2}{\sqrt{k_3 k_4} \,s_1^{d-1}}
    \sum_{c=\pm}\frac{e^{-i\pi c\frac{d}{2}}}{\sin(\pi c \frac{d}{2}) \sin(\pi c \frac{d}{2} +\pi d)}  \left(\frac{s_2}{s_1}\right)^{c\left(n+\frac{d-2}{2}\right)} \\
    &\times \left(\frac{2 k_{34}}{s_1}\right)^p \frac{1}{p!}
    Q_{c\left(n+\frac{d-2}{2}\right) +p+d-\frac{3}{2}}^{\frac{d-3}{2}}\left(\frac{k_{12}}{s_1}\right)\,
    \frac{c \left(n+\frac{d-2}{2}\right)+(p+d-1)}{\left(c \left(n+\frac{d-2}{2}\right)+p+d-1\right)^2+\mu_1^2}\\
    &\times \F_4\Biggl(\begin{matrix}
    -\frac{p}{2}\,,\, \frac{1-p}{2}\\
    1+c \left(n+\frac{d-2}{2}\right)\,,\,2- c\left(n+\frac{d-2}{2}\right)-p-d\end{matrix}\,;
    \left(\frac{s_1}{k_{34}}\right)^2,\left(\frac{s_2}{k_{34}}\right)^2\Biggr)\\
    &+\frac{2^{\frac{d-1}{2}} \pi^\frac{1}{2} e^{i\pi d}}{\sin\left(\pi\frac{d}{2}\right)}
    \left(\left(\frac{k_{12}}{s_1}\right)^2-1\right)^{\frac{d-3}{4}}
    \frac{p+\frac{d-2}{2}}{\left(p+\frac{d-2}{2}\right)^2+\mu_1^2}\frac{1}{p!}\left(\frac{2 k_{12}}{s_1}\right)^p\\
    &\times\,_2\F_1\left(
    \begin{matrix}
        \frac{1-p}{2}\,,\,-\frac{p}{2} \\ -p-\frac{d-4}{2}
    \end{matrix}\,,\left(\frac{s_1}{k_{12}}\right)^2\right)
    \mathcal{I}_d\begin{bmatrix}
        -i\left(p+\frac{d-2}{2}\right) &-i\left(n+\frac{d-2}{2}\right) &i/2 & i/2 \\ s_1 & s_2 & k_3 & k_4  \\ I & K & K & K
    \end{bmatrix}
    \Biggr]
    \,.
    \end{aligned}
    \label{eq_G_+++_BB_final_res_k1<1_k2<1_k3>1}
\end{equation}
\end{framed}

\paragraph{Background-Background residues, $\kappaone>1$.} Let us turn to the computation of $G_{+++}^{B,B}$ when $\kappaone>1$ \eqref{eq_condition_C2}. Then, from the expression of the first layer $F_{+++}^B$ \eqref{eq_F+++B_C2>1} in this case, the integral \eqref{eq_G_+++_.B_def_wrt_F+++B} over the second layer can be written
\begin{equation}
\begin{aligned}
    G_{+++}^{.,B}&=\frac{-e^{-i2\pi d}2^{d-4}\pi^5}{\sqrt{k_1 k_2 k_3 k_4 k_5 k_6} \,s_1^{\frac{d-2}{2}}s_2^{\frac{3d}{2}-2}}
    \left(\left(\frac{k_{12}}{s_1}\right)^2-1\right)^{\frac{3-d}{4}}
    \left(\left(\frac{k_{56}}{s_2}\right)^2-1\right)^{\frac{3-d}{4}}\\
    &\times\sum_{n=0}^{+\infty} \left(\frac{2 k_{34}}{s_2}\right)^n \frac{1}{n!}
    \Biggl[G_{+++}^{0,.,B}+G_{+++}^{1,.,B} \Biggr]\,,
    \end{aligned}
    \label{eq_G_+++_.B_def_wrt_G_+++_01B_C2>1}
\end{equation}
where
\begin{subequations}
\begin{equation}
\begin{aligned}
    G_{+++}^{0,.,B}&=\int\limits_{-\infty}^{+\infty}\d\nu \, \frac{\nu}{\sinh(\pi\nu)}\,
    Q_{i\nu-\frac{1}{2}}^{\frac{d-3}{2}}\left(\frac{k_{12}}{s_1}\right) \frac{1}{(\nu^2-\mu_{1}^2)_{i\epsilon}}\\
    &\times F(n,\nu) \,\frac{-\nu+i(n+d-1)}{\left(\nu-i(n+d-1)\right)^2-\mu_2^2}\,Q_{i \nu +n+d-\frac{3}{2}}^{\frac{d-3}{2}}\left(\frac{k_{56}}{s_2}\right)\,,
    \end{aligned}
    \label{eq_G_+++_0.B_def_C2>1}
\end{equation}
\begin{equation}
\begin{aligned}
G_{+++}^{1,.,B}&=-\int\limits_{-\infty}^{+\infty}\d\nu \, \frac{\nu}{\sinh(\pi\nu)}\,
    Q_{-i\nu-\frac{1}{2}}^{\frac{d-3}{2}}\left(\frac{k_{12}}{s_1}\right) \frac{1}{(\nu^2-\mu_{1}^2)_{i\epsilon}}\\
    &\times F(n,\nu) \,\frac{-\nu+i(n+d-1)}{\left(\nu-i(n+d-1)\right)^2-\mu_2^2}\,Q_{i \nu +n+d-\frac{3}{2}}^{\frac{d-3}{2}}\left(\frac{k_{56}}{s_2}\right)\,,
    \end{aligned}
    \label{eq_G_+++_1.B_def_C2>1}
\end{equation}
\end{subequations}
and the function $F$ is defined as
\begin{equation}
\begin{aligned}
    F(n,\nu)&\equiv \left(\frac{s_1}{s_2}\right)^{i \nu} \,\frac{1}{\sinh(\pi (\nu-i d))}\\
    &\times \,\F_4\Biggl(\begin{matrix}
    -\frac{n}{2}\,,\, \frac{1-n}{2}\\
    1+i \nu\,,\,1-i \nu-(n+d-1)\end{matrix}\,;
    \left(\frac{s_1}{k_{34}}\right)^2,\left(\frac{s_2}{k_{34}}\right)^2\Biggr)\,,
    \end{aligned}
    \label{eq_def_residue_function}
\end{equation}
where in the notation $F(n,\nu)$ we omit the dependence on kinematic variables for clarity. To write $G_{+++}^{.,B}$ as in \eqref{eq_G_+++_.B_def_wrt_G_+++_01B_C2>1}, we used the connection formula for \eqref{eq_vertex_fct_12_Q} and the change of variable $\nu'=-\nu$.

The large $\nu$ asymptotic behaviours of the integrands in $G_{+++}^{0,.,B}$ and $G_{+++}^{1,.,B}$ can be inferred from those of Legendre $Q$ functions \eqref{eq_large_mu_Q} and of the function $F$ defined just before \eqref{eq_large_nu_F} (see Appendix \ref{subappendix_G+++BB}). Then, at large $\nu$ we have
\begin{equation}
\begin{aligned}
    Q_{\pm i\nu-\frac{1}{2}}^{\frac{d-3}{2}}\left(\frac{k_{12}}{s_1}\right)F(n,\nu) \,Q_{i \nu +n+d-\frac{3}{2}}^{\frac{d-3}{2}}\left(\frac{k_{56}}{s_2}\right)
    \underset{\nu\to\infty}{\propto}
    \left(\frac{s_1}{s_2}\right)^{i\nu} e^{- i\nu\left[\pm\arccosh\left(\frac{k_{12}}{s_1}\right)+\arccosh\left(\frac{k_{56}}{s_2}\right)\right]}\,,
    \label{eq_asymptotics_integrand_nu}
\end{aligned}
\end{equation}
where we kept only the terms that allow for the determination of the contour prescription. The $\pm$ sign in the exponential is $+$ for $G_{+++}^{0,.,B}$ (resp. $-$ for $G_{+++}^{1,.,B}$). In the case of $G_{+++}^{0,.,B}$, we find that the contour can always be closed in the lower-half $\nu$-plane (details in \ref{subsubapp_contour_G+++01.B}), given that we assumed $k_{34}>s_{12}$. For $G_{+++}^{1,.,B}$, we find that the contour prescription depends on the following kinematic ratio:
\begin{equation}
    \kappafour\equiv\frac{k_{12}+\sqrt{k_{12}^2-s_1^2}}{k_{56}+\sqrt{k_{56}^2-s_2^2}}\,.
    \label{eq_condition_C1}
\end{equation}
Then, if $\kappafour>1$ (resp. $\kappafour<1$), the contour has to be closed in the upper-half (resp. lower-half) $\nu$-plane.

Figure \ref{fig_analytic_structure_G+++B_C2>1} shows the contour choice for $G_{+++}^{0,.,B}$ while for $G_{+++}^{1,.,B}$ we have two possibilities depending on the value of $\kappafour$. The singularity structure of each component that we will now describe through its computation is drawn as well. The representation of the pole structure in Fig. \ref{fig_analytic_structure_G+++B_C2>1} is valid when $d\geq 2$ is an integer. Indeed, in this case it can be shown that the function $\nu\mapsto F(n,\nu)$ \eqref{eq_def_residue_function} is analytic (see Appendix \ref{subapp_residue_fct} for a detailed proof). Otherwise, $F$ has an infinite number of poles on the imaginary axis, at $\nu=i(p+d)$, where $p$ is an integer that satisfies $p\geq n-\lfloor \frac{n}{2} \rfloor-1$ (see also Appendix \ref{subapp_residue_fct}). Since in interesting physical cases $d$ is an integer greater than $2$ (in particular $d=3$), we will show here how the computation is done under this assumption, and leave the case $d\in\mathbb{C}/\mathbb{Z}$ to Appendix \ref{subappendix_G+++BB}.

\begin{figure}[h!]
\hspace{-1cm}
\begin{subfigure}[h!]{0.4\textwidth}
        \hspace{-0.25cm}
    	\begin{tikzpicture}[scale = 2]

        \draw[black, ->] (-1.8,0) -- (1.8,0) coordinate (xaxis);
		\draw[black, ->] (0,-1.8) -- (0,1.9) coordinate (yaxis);
		\node at (2.1, 0) {$\text{Re}(\nu)$};
		\node at (0, 2) {$\text{Im}(\nu)$};
        
        \draw[pyred, fill = pyred] (0, 0.25) circle (.03cm);
		\draw[pyred, fill = pyred] (0, 0.75) circle (.03cm);
		\draw[pyred, fill = pyred] (0, 1.75) circle (.03cm);        
    \draw[pyred] (0,1.22cm) -- (0,1.28cm) arc(90:270:.03cm);
    \draw[pyorange] (0,1.28cm) -- (0,1.22cm) arc(-90:90:.03cm);
    \begin{scope}
        \clip (0,1.25cm) circle(.03cm);
        \fill[pyred] (0,1.22cm) rectangle (-.03cm,1.28cm);
        \fill[pyorange] (0,1.22cm) rectangle (.03cm,1.28cm);
    \end{scope}
    \draw[pyred] (0,1.72cm) -- (0,1.78cm) arc(90:270:.03cm);
    \draw[pyorange] (0,1.78cm) -- (0,1.72cm) arc(-90:90:.03cm);
    \begin{scope}
        \clip (0,1.75cm) circle(.03cm);
        \fill[pyred] (0,1.72cm) rectangle (-.03cm,1.78cm);
        \fill[pyorange] (0,1.72cm) rectangle (.03cm,1.78cm);
    \end{scope}
        \node at (-.7, 1.25) {\textcolor{pyred}{\footnotesize$Q_{i\nu-\frac{1}{2}}^{\frac{d-3}{2}}\left(k_{12}/s_1\right)$}};
        \node at (.9, 1.25) {\textcolor{pyorange}{\footnotesize$Q_{i\nu+n+d-\frac{3}{2}}^{\frac{d-3}{2}}\left(k_{56}/s_2\right)$}};

        \draw[pyblue, fill = pyblue] (0, -0.5) circle (.03cm);
		\draw[pyblue, fill = pyblue] (0, -1) circle (.03cm);
		\draw[pyblue, fill = pyblue] (0, -1.5) circle (.03cm);
        \draw[pyblue, fill = pyblue] (0, 0.5) circle (.03cm);
		\draw[pyblue, fill = pyblue] (0, 1) circle (.03cm);
		\draw[pyblue, fill = pyblue] (0, 1.5) circle (.03cm);
        \node at (0.4, -.53) {\textcolor{pyblue}{$\frac{\nu}{\sinh\left(\pi\nu\right)}$}};
        
        \draw[xshift=0,pyblue!80!black,decoration={markings,mark=between positions 0.1 and 1 step 0.2 with \arrow{>}},postaction={decorate}] (-1.70,0) -- (1.70,0) arc (0:-180:1.70);
    \end{tikzpicture}
\end{subfigure}
\hfill
\begin{subfigure}[h!]{0.4\textwidth}
    \hspace{-1.5cm}
    	\begin{tikzpicture}[scale = 2]

        \draw[black, ->] (-1.8,0) -- (1.8,0) coordinate (xaxis);
		\draw[black, ->] (0,-1.8) -- (0,1.9) coordinate (yaxis);
		\node at (2.1, 0) {$\text{Re}(\nu)$};
		\node at (0, 2) {$\text{Im}(\nu)$};

        \draw[pyred, fill = pyred] (0, -0.25) circle (.03cm);
		\draw[pyred, fill = pyred] (0, -0.75) circle (.03cm);
		\draw[pyred, fill = pyred] (0, -1.25) circle (.03cm);
        \node at (-.7, -.755) {\textcolor{pyred}{\footnotesize$Q_{-i\nu-\frac{1}{2}}^{\frac{d-3}{2}}\left(k_{12}/s_1\right)$}};

        \draw[pyorange, fill = pyorange] (0, 1.25) circle (.03cm);
		\draw[pyorange, fill = pyorange] (0, 1.75) circle (.03cm);
        \node at (.9, 1.25) {\textcolor{pyorange}{\footnotesize$Q_{i\nu+n+d-\frac{3}{2}}^{\frac{d-3}{2}}\left(k_{56}/s_2\right)$}};

        \draw[pyblue, fill = pyblue] (0, -0.5) circle (.03cm);
		\draw[pyblue, fill = pyblue] (0, -1) circle (.03cm);
		\draw[pyblue, fill = pyblue] (0, -1.5) circle (.03cm);
        \draw[pyblue, fill = pyblue] (0, 0.5) circle (.03cm);
		\draw[pyblue, fill = pyblue] (0, 1) circle (.03cm);
		\draw[pyblue, fill = pyblue] (0, 1.5) circle (.03cm);
        \node at (0.4, -.53) {\textcolor{pyblue}{$\frac{\nu}{\sinh\left(\pi\nu\right)}$}};
    \end{tikzpicture}
\end{subfigure}
\caption{Background residues in the $\nu$ complex plane of $G_{+++}^{0,.,B}$ \eqref{eq_G_+++_0.B_def_C2>1} (left) and $G_{+++}^{1,.,B}$ \eqref{eq_G_+++_1.B_def_C2>1} (right) for $\kappaone>1$. The figure is drawn for $d=3$ and $n=0$. Under these conditions, in the integrand of $G_{+++}^{0,.,B}$ poles coming from the two Legendre $Q$ functions are at the same location on the imaginary axis; however, we avoid them by always closing the contour with a half-circle in the lower half-plane. The function $F$ that appears in the two integrands has no singularities when $d$ is an integer (see Appendix \ref{subappendix_G+++BB}).
For $G_{+++}^{1,.,B}$, the contour is closed in the lower (resp. upper) half-plane for $\kappafour<1$ \eqref{eq_condition_C1} (resp. $\kappafour>1$).}
\label{fig_analytic_structure_G+++B_C2>1}
\end{figure} 

From the contour prescription displayed in Fig. \ref{fig_analytic_structure_G+++B_C2>1}, the only background residues in $G_{+++}^{0,.,B}$ come from the poles of the $\nu/\sinh(\pi\nu)$ factor at $\nu=-i(p+1)$, $p\in\mathbb{N}$. Their sum, which we will denote $G_{+++}^{0,B,B}$, is then
\begin{equation}
\begin{aligned}
    G_{+++}^{0,B,B}&=2i\sum_{p=0}^{+\infty}\frac{(-1)^{p+1}i(p+1)}{(p+1)^2+\mu_1^2}\frac{i(p+n+d)}{(p+n+d)^2+\mu_2^2}
    \\
    &\times Q_{p+\frac{1}{2}}^{\frac{d-3}{2}}\left(\frac{k_{12}}{s_1}\right)\,F(n,-i(p+1))\,Q_{p+n+d-\frac{1}{2}}^{\frac{d-3}{2}}\left(\frac{k_{56}}{s_2}\right)\,.
\end{aligned}
\label{eq_G+++_0BB_C2>1}
\end{equation}
For the background residues of $G_{+++}^{1,.,B}$, which we will gather in the term $G_{+++}^{1,B,B}$, we have to distinguish cases with respect to the value of $\kappafour$ \eqref{eq_condition_C1}.

\subparagraph{Background-Background residues, $\kappaone>1$ and $\kappafour>1$.} In this regime, the contour is closed in the upper-half $\nu$-plane and $G_{+++}^{1,.,B}$ is made of the following two contributions:
\begin{itemize}
\item \textbf{\textcolor{pyblue}{Poles from $\frac{\nu}{\sinh(\pi\nu)}$.}}
We first collect residues that correspond to the poles from $\nu\mapsto\nu/\sinh(\pi\nu)$ located at $\nu=i(p+1)$, $p\in\mathbb{N}$:
\begin{equation}
\begin{aligned}
    G_{+++}^{1,B,B}&\supset
    -2i\sum_{p=0}^{+\infty}\frac{(-1)^{p+1}i(p+1)}{(p+1)^2+\mu_1^2}\frac{i(2+p-n-d)}{(2+p-n-d)^2+\mu_2^2}
    \\
    &\times Q_{p+\frac{1}{2}}^{\frac{d-3}{2}}\left(\frac{k_{12}}{s_1}\right)\,F(n,i(p+1))\,Q_{-p+n+d-\frac{5}{2}}^{\frac{d-3}{2}}\left(\frac{k_{56}}{s_2}\right)\,.
    \end{aligned}
    \label{eq_G+++_1BB_C2>1_C1>1_res_sh}
\end{equation}
\item \textbf{\textcolor{pyorange}{Poles from $Q_{i\nu+n+d-3/2}^{\frac{d-3}{2}}(k_{56}/s_2)$.}} The other contribution to $G_{+++}^{1,B,B}$ comes from the function $\nu\mapsto Q_{i\nu+n+d-3/2}^{\frac{d-3}{2}}(k_{56}/s_2)$ that, from \eqref{eq_residues_legendre_Q}, has poles at
\begin{equation}
    \nu=i\left(p+n+\frac{3d-4}{2}\right)\,,\,p\in\mathbb{N}\,.
    \label{eq_G+++_1BB_C2>1_C1>1_poles_Q}
\end{equation}
The corresponding residues are
\begin{equation}
\begin{aligned}
    G_{+++}^{1,B,B}&\supset
    2^{\frac{d-1}{2}}\pi^{\frac{3}{2}} e^{i\pi\frac{d-1}{2}}
    \left(\frac{k_{56}}{s_2}\right)^{\frac{d-3}{4}}
    \left(\left(\frac{k_{56}}{s_2}\right)^2-1\right)^{\frac{d-3}{4}}\\
    &\times\sum_{p=0}^{+\infty} \frac{p+n+\frac{3d-4}{2}}{\left(p+n+\frac{3d-4}{2}\right)^2+\mu_1^2}\frac{p+\frac{d-2}{2}}{\left(p+\frac{d-2}{2}\right)^2+\mu_2^2}
    \left(\frac{2 k_{56}}{s_2}\right)^{p}
    \frac{(-1)^p}{p!}\\
    &\times
    Q_{p+n+\frac{3d-5}{2}}^{\frac{d-3}{2}}\left(\frac{k_{12}}{s_1}\right)
    F\left(n,i\left(p+n+\frac{3d-4}{2}\right)\right)
    \,_2\F_1\left(
    \begin{matrix}
        -\frac{p}{2}\,,\,\frac{1-p}{2} \\ 2-\frac{d}{2}-p
    \end{matrix}\,,\left(\frac{s_2}{k_{56}}\right)^2\right)\,.
    \end{aligned}
    \label{eq_G+++_1BB_C2>1_C1>1_res_Q}
\end{equation}
\end{itemize}
Having collected all the background residues, the full result for $G_{+++}^{B,B}$ when $\kappaone>1$ and $\kappafour>1$ is obtained by gathering the three contributions \eqref{eq_G+++_0BB_C2>1}, \eqref{eq_G+++_1BB_C2>1_C1>1_res_sh} and \eqref{eq_G+++_1BB_C2>1_C1>1_res_Q} into the decomposition \eqref{eq_G_+++_.B_def_wrt_G_+++_01B_C2>1} in terms of $G_{+++}^{0,B,B}$ and $G_{+++}^{1,B,B}$. After some algebraic simplifications, we finally have
\begin{framed}
\vspace{-.5cm}
\begin{equation}
\begin{aligned}
    G_{+++}^{B,B}&=\frac{e^{-i\pi (d+1)}\pi^6}{2^4\sqrt{k_1 k_2 k_3 k_4 k_5 k_6} \,s_1^{\frac{d-2}{2}}s_2^{\frac{3d}{2}-2}}\\
    &\times\sum_{n=0}^{+\infty} \sum_{p=0}^{+\infty} \left(\frac{2 k_{34}}{s_2}\right)^n \frac{1}{n!}
    \Biggl[i\, 2^{\frac{3}{2}}
    \left(\frac{s_1}{k_{12}}\right)^{\frac{d-3}{2}} 
    \left(\frac{s_2}{k_{56}}\right)^{\frac{3d-4}{2}}
    \left(\frac{s_1}{2 k_{12}}\right)^{p+\frac{3}{2}}
    \Gamma\left(\frac{d}{2}+ p\right)\\&
    \,_2\F_1\left(
    \begin{matrix}
        \frac{d}{4}+ \frac{p+1}{2}\,,\,\frac{d}{4}+ \frac{p}{2} \\ p+2
    \end{matrix}\,,\left(\frac{s_1}{k_{12}}\right)^2\right)
    \times\frac{(-1)^{p+1}(p+1)}{(p+1)^2+\mu_1^2}
    \\
    &\times \Biggl[ F(n,-i(p+1))
    \frac{(p+n+d)}{(p+n+d)^2+\mu_2^2}
    \left(\frac{s_2}{2 k_{56}}\right)^{p+n+1}\\
    &\times\Gamma\left(\frac{3d}{2}+p+n-1\right)
    \,_2\F_1\left(
    \begin{matrix}
        \frac{3d}{4}+ \frac{p+n}{2}\,,\,\frac{3d-2}{4}+ \frac{p+n}{2} \\ p+n+d+1
    \end{matrix}\,,\left(\frac{s_2}{k_{56}}\right)^2\right)\\
    &- F(n,i(p+1)) \frac{(2+p-n-d)}{(2+p-n-d)^2+\mu_2^2}
    \left(\frac{s_2}{2 k_{56}}\right)^{n-p-1}\\
    &\times\Gamma\left(\frac{3d}{2}+n-p-3\right)
    \,_2\F_1\left(
    \begin{matrix}
        \frac{3d-4}{4}+ \frac{n-p}{2}\,,\,\frac{3d-6}{4}+ \frac{n-p}{2} \\ n-p+d-1
    \end{matrix}\,,\left(\frac{s_2}{k_{56}}\right)^2\right)\Biggr]\\
    &+
    2\pi \left(\frac{s_1}{k_{12}}\right)^{2d-3}
    \left(\frac{k_{56}}{s_2}\right)^{\frac{d-3}{4}}
    \left(\frac{s_1 k_{56}}{s_2 k_{12}}\right)^{p}\left(\frac{s_1}{2 k_{12}}\right)^n \frac{(-1)^p}{p!}\Gamma\left(2 d+ p+n-3\right)\\
    &\times
    \,_2\F_1\left(
    \begin{matrix}
        d-\frac{1}{2}+ \frac{p+n}{2}\,,\,d-\frac{3}{2}+ \frac{p+n}{2} \\ p+n+\frac{3d-2}{2}
    \end{matrix}\,,\left(\frac{s_1}{k_{12}}\right)^2\right)
    \frac{p+n+\frac{3d-4}{2}}{\left(p+n+\frac{3d-4}{2}\right)^2+\mu_1^2}\\
    &
    F\left(n,i\left(p+n+\frac{3d-4}{2}\right)\right)
    \frac{p+\frac{d-2}{2}}{\left(p+\frac{d-2}{2}\right)^2+\mu_2^2}
    \,_2\F_1\left(
    \begin{matrix}
        -\frac{p}{2}\,,\,\frac{1-p}{2} \\ 2-\frac{d}{2}-p
    \end{matrix}\,,\left(\frac{s_2}{k_{56}}\right)^2\right)
    \Biggr]\,,
    \end{aligned}
    \label{eq_G+++_BB_C2>1_C1>1_full_res}
\end{equation}
\end{framed}
\noindent where we used the definition of the Legendre $Q$ function \eqref{eq_def_legendre_Q} in terms of a hypergeometric series.

\subparagraph{Background-Background residues, $\kappaone>1$ and $\kappafour<1$.} In this second case, to compute $G_{+++}^{1,.,B}$, the asymptotic behaviour of the integrand \eqref{eq_asymptotics_integrand_nu} requires us to close the contour in the lower-half $\nu$-plane. The background pole structure makes manifest the following two contributions:
\begin{itemize}
\item \textbf{\textcolor{pyblue}{Poles from $\frac{\nu}{\sinh(\pi\nu)}$.}}
We still have an infinite number of poles coming from the factor $\nu\mapsto\nu/\sinh(\pi\nu)$ located at $\nu=-i(p+1)$, $p\in\mathbb{N}$, with residues:
\begin{equation}
\begin{aligned}
    G_{+++}^{1,B,B}&\supset
    -2i\sum_{p=0}^{+\infty}(-1)^{p}\frac{p+1}{(p+1)^2+\mu_1^2}\frac{p+n+d}{(p+n+d)^2+\mu_2^2}
    \\
    &\times Q_{-p-\frac{3}{2}}^{\frac{d-3}{2}}\left(\frac{k_{12}}{s_1}\right)\,F(n,-i(p+1))\,Q_{p+n+d-\frac{1}{2}}^{\frac{d-3}{2}}\left(\frac{k_{56}}{s_2}\right)\,.
    \end{aligned}
    \label{eq_G+++_1BB_C2>1_C1<1_res_sh}
\end{equation}
\item \textbf{\textcolor{pyred}{Poles from $Q_{-i\nu-1/2}^{\frac{d-3}{2}}(k_{12}/s_1)$.}} We also have another contribution to $G_{+++}^{1,B,B}$ from poles of the Legendre function $\nu\mapsto Q_{-i\nu-1/2}^{\frac{d-3}{2}}(k_{12}/s_1)$ \eqref{eq_residues_legendre_Q} at $\nu=-i\left(p+(d-2)/2\right)$.
The corresponding residues give
\begin{equation}
\begin{aligned}
    G_{+++}^{1,B,B}&\supset
    2^\frac{d-1}{2} \pi^\frac{3}{2} \frac{e^{i\pi\frac{d-2}{2}}}{\sin\left(\pi\frac{d}{2}\right)}
    \left(\left(\frac{k_{12}}{s_1}\right)^2-1\right)^{\frac{d-3}{4}}\\
    &\times\sum_{p=0}^{+\infty} \frac{1}{p!} \left(\frac{2 k_{12}}{s_1}\right)^p 
    \frac{p+\frac{d-2}{2}}{\left(p+\frac{d-2}{2}\right)^2+\mu_1^2}
    \frac{p+n+\frac{3d-4}{2}}{\left(p+n+\frac{3d-4}{2}\right)^2+\mu_2^2}\\
    &\times\,_2\F_1\left(
    \begin{matrix}
        -\frac{p}{2}\,,\,\frac{1-p}{2} \\ 2-\frac{d}{2}-p
    \end{matrix}\,,\left(\frac{s_1}{k_{12}}\right)^2\right)
    F\left(n,-i\left(p+\frac{d-2}{2}\right)\right)
    Q_{p+n+\frac{3d-5}{2}}^{\frac{d-3}{2}}\left(\frac{k_{56}}{s_2}\right)\,.
    \end{aligned}
    \label{eq_G+++_1BB_C2>1_C1<1_res_Q}
\end{equation}
\end{itemize}
To write the final result for $G_{+++}^{B,B}$, we first note that the sum
of the two contributions \eqref{eq_G+++_0BB_C2>1} and \eqref{eq_G+++_1BB_C2>1_C1<1_res_sh} vanishes, due to the connection formula between Legendre functions \eqref{connection_formula_P_Q}. Then, from \eqref{eq_G_+++_.B_def_wrt_G_+++_01B_C2>1} the final result for $G_{+++}^{B,B}$ in the case $\kappaone>1$ and $\kappafour<1$ is
\begin{framed}
\vspace{-.5cm}
\begin{equation}
\begin{aligned}
    G_{+++}^{B,B}&=\frac{e^{-i\pi \left(d-\frac{1}{2}\right)}\pi^7}{2^3\sqrt{k_1 k_2 k_3 k_4 k_5 k_6} \,s_1^{\frac{d-2}{2}}s_2^{\frac{3d}{2}-2} \sin\left(\pi\frac{d}{2}\right)}
    \left(\frac{s_2}{k_{56}}\right)^{2d-3}\\
    &\times\sum_{n=0}^{+\infty} \sum_{p=0}^{+\infty}  \left(\frac{k_{34}}{s_2}\right)^n \frac{1}{n!} \left(\frac{s_2}{k_{56}}\right)^{p+n} \frac{1}{p!} \left(\frac{k_{12}}{s_1}\right)^p \\
    &\times
    \,_2\F_1\left(
    \begin{matrix}
        -\frac{p}{2}\,,\,\frac{1-p}{2} \\ 2-\frac{d}{2}-p
    \end{matrix}\,,\left(\frac{s_1}{k_{12}}\right)^2\right)
    \frac{p+\frac{d-2}{2}}{\left(p+\frac{d-2}{2}\right)^2+\mu_1^2}
    \\
    &\times
    F\left(n,-i\left(p+\frac{d-2}{2}\right)\right)
    \frac{p+n+\frac{3d-4}{2}}{\left(p+n+\frac{3d-4}{2}\right)^2+\mu_2^2}\\
    &
    \Gamma\left(2 d+ p+n-3\right)
    \,_2\F_1\left(
    \begin{matrix}
        d-\frac{1}{2}+ \frac{p+n}{2}\,,\,d-\frac{3}{2}+ \frac{p+n}{2} \\ p+n+\frac{3d-2}{2}
    \end{matrix}\,,\left(\frac{s_2}{k_{56}}\right)^2\right)\,.
    \end{aligned}
    \label{eq_G+++_BB_C2>1_C1<1_full_res}
\end{equation}
\end{framed}

Obtaining this expression ends the computation of $G_{+++}^{B,B}$ in all kinematic configurations.

\subsection{Signal-Background residues}

The signal-background contribution to $G_{+++}$ has been computed in subsection \ref{subsec_second_int_layer} for the kinematic case $\kappaone<1$ and $\kappatwo>1$. The other cases are worked out in what follows, starting with the case $\kappaone<1$ and $\kappathree<1$.

\paragraph{Signal-Background residues, $\kappaone<1$ and $\kappathree<1$.} 
We still want to compute the signal poles of $G_{+++}^{.,B}$ \eqref{eq_G_+++_.B_def_wrt_G_+++_01B_C2<1}. Recall that $G_{+++}^{0,S,B}$ has already been computed \eqref{eq_G+++_0SB_C2<1}. We now turn to $G_{+++}^{1,S,B}$. Because $\kappathree<1$ \eqref{eq_condition_C12p}, the contour closure must be done in the upper half-plane. We have
\begin{equation}
\begin{aligned}
    G_{+++}^{1,S,B}&=\frac{\pi i}{2\sinh^2(\pi\mu_1)} \sum_{c=\pm} e^{-\pi c\mu_1}
    Q_{i c\mu_1-\frac{1}{2}}^{\frac{d-3}{2}}\left(\frac{k_{12}}{s_1}\right)
    \mathcal{I}_d\begin{bmatrix}
        -c\mu_1 &-i\left(n+\frac{d-2}{2}\right) &i/2 & i/2 \\ s_1 & s_2 & k_3 & k_4  \\ I & I & K & K
    \end{bmatrix}\,.
\end{aligned}
\label{eq_G+++_0SB_C2<1_C1p<1}
\end{equation}
We now simplify the sum of the two contributions \eqref{eq_G+++_0SB_C2<1} and \eqref{eq_G+++_0SB_C2<1_C1p<1} by using the analytic continuation formula \eqref{eq_analytic_continutaion_IKI_III} for $\I_{KI}$ and the connection formula between Legendre functions \eqref{connection_formula_P_Q}. It gives
\begin{equation}
\begin{aligned}
    G_{+++}^{0,S,B}+G_{+++}^{1,S,B}&=e^{i\pi\frac{d-2}{2}}\Gamma\left(\frac{d-2}{2}\pm i\mu_1\right)
    P_{i \mu_1-\frac{1}{2}}^{-\frac{d-3}{2}}\left(\frac{k_{12}}{s_1}\right) \\
    &\quad\times\mathcal{I}_d\begin{bmatrix}
        \mu_1 &-i\left(n+\frac{d-2}{2}\right) &i/2 & i/2 \\ e^{-i\pi}s_1 & s_2 & k_3 & k_4  \\ I & I & K & K
    \end{bmatrix}\,.
\end{aligned}
\label{eq_G+++_0SB_+_G+++_1SB_C2<1_C1p<1}
\end{equation}
We finally find $G_{+++}^{S,B}$ by using the decomposition \eqref{eq_G_+++_.B_def_wrt_G_+++_01B_C2<1}:
\begin{framed}
\vspace{-.5cm}
\begin{equation}
\begin{aligned}
    G_{+++}^{S,B}
    &= \frac{e^{-i\pi\left(d+\frac{1}{2}\right)} 2^{\frac{d-7}{2}} \pi^\frac{9}{2}}{\sqrt{k_1 k_2 k_5 k_6}s_1^{\frac{d-2}{2}}s_2^{\frac{d-2}{2}}\sin\left(\pi\frac{d}{2}\right)}
    \left(\left(\frac{k_{12}}{s_1}\right)^2-1\right)^{\frac{3-d}{4}}
    \Gamma\left(\frac{d-2}{2}\pm i\mu_1\right)
    \\
    &\times 
    P_{i \mu_1-\frac{1}{2}}^{-\frac{d-3}{2}}\left(\frac{k_{12}}{s_1}\right)
    \sum_{n=0}^{+\infty} 
    \mathcal{I}_d\begin{bmatrix}
        \mu_1 &-i\left(n+\frac{d-2}{2}\right) &i/2 & i/2 \\ e^{-i\pi}s_1 & s_2 & k_3 & k_4  \\ I & I & K & K
    \end{bmatrix}\\
    &\times\frac{n+\frac{d-2}{2}}{\left(n+\frac{d-2}{2}\right)^2+\mu_2^2}\frac{1}{n!}\left(\frac{2 k_{56}}{s_2}\right)^n
    \,_2\F_1\left(
    \begin{matrix}
        \frac{1-n}{2}\,,\,-\frac{n}{2} \\ -n-\frac{d-4}{2}
    \end{matrix}\,,\left(\frac{s_2}{k_{56}}\right)^2\right)\,.
    \end{aligned}
    \label{eq_G+++_SB_C2<1_C1p<1}
\end{equation}
\end{framed}

\paragraph{Signal-Background residues, $\kappaone<1$, $\kappatwo<1$ and $\kappathree>1$.}
In this case, we have to collect the signal residues of $G_{+++}^{.,B}$ according to the decompositions \eqref{eq_G_+++_.B_def_wrt_G_+++_01B_C2<1} and \eqref{kappa1<1_decomposition}. The expression of $G_{+++}^{0,S,B}$ is already given by \eqref{eq_G+++_0SB_C2<1}, let us focus on the other two components, which are 
$G_{+++}^{1,0,S,B}$ and $G_{+++}^{1,1,S,B}$, corresponding to the signal residues of the integrals \eqref{eq_G_+++_10.B_def}-\eqref{eq_G_+++_11.B_def}. As for the case of the computation of $G_{+++}^{B,B}$ in the same kinematic condition, we can make use of the computation of $F_{+++}^{1,S}$ done in subsection \ref{subsection_+++}, because the integrals that define $G_{+++}^{1,0,S,B}$ and $G_{+++}^{1,1,S,B}$ (\eqref{eq_G_+++_10.B_def}-\eqref{eq_G_+++_11.B_def}) are similar to \eqref{eq_F_+++_1}. Therefore, for $G_{+++}^{1,0,S,B}$, we collect the residues of the poles located at $\nu=\pm\mu_1+i\epsilon$ in the upper-half $\nu$ plane:
\begin{equation}
\begin{aligned}
    G_{+++}^{1,0,S,B}=&-\frac{i}{\pi}\frac{i\pi}{2\sinh^2(\pi\mu_1)}\Biggl(e^{-\pi\mu_1}\mathcal{I}_d\begin{bmatrix}
        \mu_1 &-i\left(n+\frac{d-2}{2}\right) & i/2 & i/2 \\  s_1 & e^{-i\pi}s_2 & k_3 & k_4  \\ I & K & K & K
    \end{bmatrix}
    Q_{-i\mu_1-\frac{1}{2}}^{\frac{d-3}{2}}\left(\frac{k_{12}}{s_1}\right)
    \\&+
    e^{\pi\mu_1}\mathcal{I}_d\begin{bmatrix}
        -\mu_1 &-i\left(n+\frac{d-2}{2}\right) & i/2 & i/2 \\  s_1 &e^{-i\pi}s_2 & k_3 & k_4  \\ I & K & K & K
    \end{bmatrix}
    Q_{i\mu_1-\frac{1}{2}}^{\frac{d-3}{2}}\left(\frac{k_{12}}{s_1}\right)
    \Biggr)\,.
    \end{aligned}
    \label{eq_G+++_10S}
\end{equation}
For $G_{+++}^{1,1,S,B}$, we now collect the residues of the poles located at $\nu=\pm\mu_1-i\epsilon$ in the lower half-plane. It gives
\begin{equation}
\begin{aligned}
G_{+++}^{1,1,S,B}&=\frac{i}{\pi}(-1)^{n+1}e^{i\pi\frac{d}{2}}\frac{i\pi}{2\sinh^2(\pi\mu_1)}
\Biggl(e^{\pi\mu_1}\mathcal{I}_d\begin{bmatrix}
        \mu_1 &-i\left(n+\frac{d-2}{2}\right) & i/2 & i/2 \\ s_1 & s_2 & k_3 & k_4  \\ I & K & K & K
    \end{bmatrix}
    Q_{-i\mu_1-\frac{1}{2}}^{\frac{d-3}{2}}\left(\frac{k_{12}}{s_1}\right)
    \\&+
    e^{-\pi\mu_1}\mathcal{I}_d\begin{bmatrix}
        -\mu_1 &-i\left(n+\frac{d-2}{2}\right) & i/2 & i/2 \\ s_1 & s_2 & k_3 & k_4  \\ I & K & K & K
    \end{bmatrix}
    Q_{i\mu_1-\frac{1}{2}}^{\frac{d-3}{2}}\left(\frac{k_{12}}{s_1}\right)
    \Biggr)\,.
\end{aligned}
\end{equation}
Using the analytic continuation formula for $\I_{II}$ \eqref{eq_continuation_III_asymptotics} into $G_{+++}^{0,S,B}$ \eqref{eq_G+++_0SB_C2<1}, we can simplify the sum of the three components $G_{+++}^{0,S,B}$, $G_{+++}^{1,0,S,B}$ and $G_{+++}^{1,1,S,B}$, using the connection formulas \eqref{connection_formula_P_Q} and \eqref{eq_connection_IKK_IIK} and the analytic continuation formulas \eqref{eq_analytic_cont_P} and \eqref{eq_analytic_continuation_IKK_1}. It gives
\begin{equation}
\begin{aligned}
G_{+++}^{0,S,B}&+G_{+++}^{1,0,S,B}+G_{+++}^{1,1,S,B}=\frac{\Gamma\left(\frac{d-2}{2}\pm i\mu_1\right)}{\pi}e^{i\pi\frac{d-1}{2}}\\
&\times\Biggl(
-i
P_{i\mu_1-\frac{1}{2}}^{-\frac{d-3}{2}}\left(e^{+i\pi}\frac{k_{12}}{s_1}\right)
\mathcal{I}_d\begin{bmatrix}
        \mu_1 &-i\left(n+\frac{d-2}{2}\right) & i/2 & i/2 \\  s_1 & e^{-i\pi}s_2 & k_3 & k_4  \\ K & K & K & K
    \end{bmatrix}\\
    &+(-1)^{n+1}e^{i\pi\frac{d}{2}} P_{i\mu_1-\frac{1}{2}}^{-\frac{d-3}{2}}\left(\frac{k_{12}}{s_1}\right)
    \mathcal{I}_d\begin{bmatrix}
        \mu_1 &-i\left(n+\frac{d-2}{2}\right) &i/2 & i/2 \\ e^{-i\pi}s_1 & s_2 & k_3 & k_4  \\ K &K & K & K
    \end{bmatrix}\Biggr)\,.
\end{aligned}
\label{eq_sum_components_SB+++_k1<1_k2<1_k3>1}
\end{equation}
Inserting this result into the expression \eqref{eq_G_+++_.B_def_wrt_G_+++_01B_C2<1} with in mind the decomposition \eqref{kappa1<1_decomposition}, we finally obtain
\begin{framed}
\vspace{-.5cm}
\begin{equation}
\begin{aligned}
    G_{+++}^{S,B}
    &= \frac{e^{-i\pi(d-1)} 2^{\frac{d-7}{2}} \pi^\frac{7}{2}}{\sqrt{k_1 k_2 k_5 k_6}s_1^{\frac{d-2}{2}}s_2^{\frac{d-2}{2}}\sin\left(\pi\frac{d}{2}\right)} \Gamma\left(\frac{d-2}{2}\pm i\mu_1\right)
    \left(\left(\frac{k_{12}}{s_1}\right)^2-1\right)^{\frac{3-d}{4}}\\
    &\times \sum_{n=0}^{+\infty} \frac{1}{n!}
\Biggl(
-i
P_{i\mu_1-\frac{1}{2}}^{-\frac{d-3}{2}}\left(e^{+i\pi}\frac{k_{12}}{s_1}\right)
\mathcal{I}_d\begin{bmatrix}
        \mu_1 &-i\left(n+\frac{d-2}{2}\right) & i/2 & i/2 \\  s_1 & e^{-i\pi}s_2 & k_3 & k_4  \\ K & K & K & K
    \end{bmatrix}\\
    &+(-1)^{n+1}e^{i\pi\frac{d}{2}} P_{i\mu_1-\frac{1}{2}}^{-\frac{d-3}{2}}\left(\frac{k_{12}}{s_1}\right)
    \mathcal{I}_d\begin{bmatrix}
        \mu_1 &-i\left(n+\frac{d-2}{2}\right) &i/2 & i/2 \\ e^{-i\pi}s_1 & s_2 & k_3 & k_4  \\ K &K & K & K
    \end{bmatrix}\Biggr)\\
    &\times\frac{n+\frac{d-2}{2}}{\left(n+\frac{d-2}{2}\right)^2+\mu_2^2}\left(\frac{2 k_{56}}{s_2}\right)^n
    \,_2\F_1\left(
    \begin{matrix}
        \frac{1-n}{2}\,,\,-\frac{n}{2} \\ -n-\frac{d-4}{2}
    \end{matrix}\,,\left(\frac{s_2}{k_{56}}\right)^2\right)
    \,.
    \end{aligned}
    \label{eq_G+++_SB_k1<1_k2<1_k3>1}
\end{equation}
\end{framed}

\paragraph{Signal-Background residues, $\kappaone>1$.}
Moving to the case where $\kappaone>1$ \eqref{eq_condition_C2}, we have to use the decomposition \eqref{eq_G_+++_.B_def_wrt_G_+++_01B_C2>1} of the quantity $G_{+++}^{.,B}$ that comes from the background piece $F_{+++}^B$ \eqref{eq_F+++B_C2>1} of the first layer integral.
Here $G_{+++}^{S,B}$ is obtained by finding the signal residues of the quantities $G_{+++}^{0,.,B}$ \eqref{eq_G_+++_0.B_def_C2>1} and $G_{+++}^{1,.,B}$ \eqref{eq_G_+++_1.B_def_C2>1}.

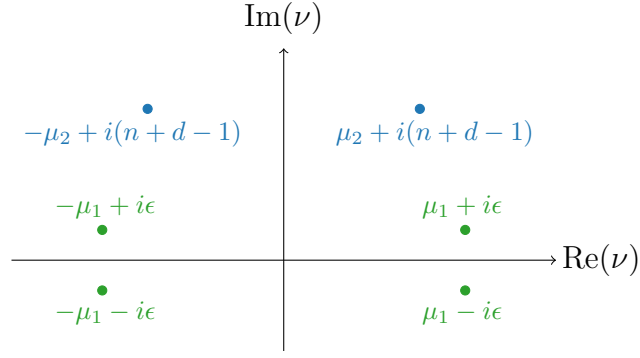
\begin{figure}[h!]\centering
    	\begin{tikzpicture}[scale = 2]

        \draw[black, ->] (-1.8,0) -- (1.8,0) coordinate (xaxis);
		\draw[black, ->] (0,-.6) -- (0,1.4) coordinate (yaxis);
		\node at (2.1, 0) {$\text{Re}(\nu)$};
		\node at (0, 1.6) {$\text{Im}(\nu)$};

        \draw[pygreen, fill = pygreen] (-1.2, -0.2) circle (.03cm);
		\draw[pygreen, fill = pygreen] (-1.2, .2) circle (.03cm);
		\draw[pygreen, fill = pygreen] (1.2, -.2) circle (.03cm);
        \draw[pygreen, fill = pygreen] (1.2, 0.2) circle (.03cm);
		\draw[pyblue, fill = pyblue] (.9, 1) circle (.03cm);
        \draw[pyblue, fill = pyblue] (-.9, 1) circle (.03cm);
        \node at (-1.18, -.35) {\footnotesize\textcolor{pygreen}{$-\mu_1-i\epsilon$}};
        \node at (-1.18, .35) {\footnotesize\textcolor{pygreen}{$-\mu_1+i\epsilon$}};
        \node at (1.18, -.35) {\footnotesize\textcolor{pygreen}{$\mu_1-i\epsilon$}};
        \node at (1.18, .35) {\footnotesize\textcolor{pygreen}{$\mu_1+i\epsilon$}};
        \node at (1, .84) {\footnotesize\textcolor{pyblue}{$\mu_2 +i(n+d-1)$}};
        \node at (-1, .84) {\footnotesize\textcolor{pyblue}{$-\mu_2 +i(n+d-1)$}};
    \end{tikzpicture}
\caption{Poles in the $\nu$ complex plane of $\color{pygreen}\left(\nu^2-\mu_1^2\right)_{i\varepsilon}^{-1}$ and $\color{pyblue}\frac{-\nu+i(n+d-1)}{\left(\nu-i(n+d-1)\right)^2-\mu_2^2}$, which are the two KLF propagators appearing in the expressions of $G_{+++}^{0,.,B}$ \eqref{eq_G_+++_0.B_def_C2>1} and $G_{+++}^{1,.,B}$ \eqref{eq_G_+++_1.B_def_C2>1} for $\kappaone>1$.}
\label{fig_signal_poles_propagators_C2>1}
\end{figure}

The signal poles of $G_{+++}^{0,.,B}$ and $G_{+++}^{1,.,B}$ originate from the two KLF propagators that appear in their expressions. The first one, $\left(\nu^2-\mu_1^2\right)_{i\epsilon}^{-1}$, has 4 poles coming from the $i\epsilon$-prescription (see Appendix \ref{subapp_details_double_exch}). The second one carries two poles that are located at the roots of the denominator. To summarise, all these poles are located at
\begin{subequations}
\begin{equation}
    \nu=\pm\mu_1\pm i\epsilon\,,
\end{equation}
\begin{equation}
    \nu=\pm\mu_2 +i(n+d-1)\,,
\end{equation}
\end{subequations}
as represented in Fig. \ref{fig_signal_poles_propagators_C2>1}.

Let us now collect the residues. For the computation of $G_{+++}^{0,S,B}$ from the integral \eqref{eq_G_+++_0.B_def_C2>1}, we always close the contour in the lower half-plane. Then $G_{+++}^{0,S,B}$ is given by the following two residues:
\begin{equation}
\begin{aligned}
    G_{+++}^{0,S,B}&=-\frac{\pi i}{2\sinh^2(\pi\mu_1)} \sum_{c=\pm} e^{\pi c\mu_1}
    Q_{i c\mu_1-\frac{1}{2}}^{\frac{d-3}{2}}\left(\frac{k_{12}}{s_1}\right)
    F(n,c\mu_1) \\
    &\times \frac{-c\mu_1+i(n+d-1)}{\left(-c\mu_1+i(n+d-1)\right)^2-\mu_2^2} Q_{i c\mu_1+n+d-\frac{3}{2}}^{\frac{d-3}{2}}\left(\frac{k_{56}}{s_2}\right)
    \,.
\end{aligned}
\label{eq_G+++_0SB_C2>1}
\end{equation}
For $G_{+++}^{1,S,B}$, we need to distinguish cases depending on the value of $\kappafour$ \eqref{eq_condition_C1}.

\subparagraph{$\kappafour>1$.} Here, the contour is closed in the upper half-plane, and $G_{+++}^{1,S,B}$ is made of the following 4 signal residues of the integral \eqref{eq_G_+++_1.B_def_C2>1}:
\begin{equation}
\begin{aligned}
    G_{+++}^{1,S,B}&=\frac{\pi i}{2\sinh^2(\pi\mu_1)} \sum_{c=\pm} e^{-\pi c\mu_1}
    Q_{-i c\mu_1-\frac{1}{2}}^{\frac{d-3}{2}}\left(\frac{k_{12}}{s_1}\right)
    F(n,c\mu_1) \\
    &\times \frac{-c\mu_1+i(n+d-1)}{\left(-c\mu_1+i(n+d-1)\right)^2-\mu_2^2} Q_{i c\mu_1+n+d-\frac{3}{2}}^{\frac{d-3}{2}}\left(\frac{k_{56}}{s_2}\right)\\
    &+\pi i \sum_{c=\pm} \frac{1}{\sinh(\pi(c\mu_2+i(n+d-1)))}
    Q_{-i c\mu_2+n+d-\frac{3}{2}}^{\frac{d-3}{2}}\left(\frac{k_{12}}{s_1}\right)
    F(n,c\mu_2+i(n+d-1)) \\
    &\times \frac{c\mu_2+i(n+d-1)}{\left(c\mu_2+i(n+d-1)\right)^2-\mu_1^2}
    Q_{i c\mu_2-\frac{1}{2}}^{\frac{d-3}{2}}\left(\frac{k_{56}}{s_2}\right)
\,.
    \end{aligned}
    \label{eq_G+++_1SB_C2>1_C1>1}
\end{equation}

Gathering the residues \eqref{eq_G+++_0SB_C2>1} and \eqref{eq_G+++_1SB_C2>1_C1>1} into \eqref{eq_G_+++_.B_def_wrt_G_+++_01B_C2>1}, and using the analytic continuation formula \eqref{eq_analytic_cont_P} for the Legendre $P$ function, we find
\begin{framed}
\vspace{-.5cm}
\begin{equation}
\begin{aligned}
    G_{+++}^{S,B}&=\frac{-e^{-i2\pi d}2^{d-4}\pi^5}{\sqrt{k_1 k_2 k_3 k_4 k_5 k_6} \,s_1^{\frac{d-2}{2}}s_2^{\frac{3d}{2}-2}}
    \left(\left(\frac{k_{12}}{s_1}\right)^2-1\right)^{\frac{3-d}{4}}
    \left(\left(\frac{k_{56}}{s_2}\right)^2-1\right)^{\frac{3-d}{4}}\\
    &\times\sum_{n=0}^{+\infty} \left(\frac{2 k_{34}}{s_2}\right)^n \frac{1}{n!}\\
    &\Biggl[
    -\frac{\pi i}{2\sinh(\pi\mu_1)} e^{i\pi\frac{d-3}{2}}\sum_{c=\pm} c\,
    \Gamma\left(\frac{d-2}{2}\pm i c\mu_1\right) P_{ic\mu_1-\frac{1}{2}}^{-\frac{d-3}{2}}\left(e^{+i\pi}\frac{k_{12}}{s_1}\right)
    F(n,c\mu_1) \\
    &\times \frac{-c\mu_1+i(n+d-1)}{\left(-c\mu_1+i(n+d-1)\right)^2-\mu_2^2} Q_{i c\mu_1+n+d-\frac{3}{2}}^{\frac{d-3}{2}}\left(\frac{k_{56}}{s_2}\right)\\
    &+\pi i \sum_{c=\pm} \frac{1}{\sinh(\pi(c\mu_2+i(n+d-1)))}
    Q_{-i c\mu_2+n+d-\frac{3}{2}}^{\frac{d-3}{2}}\left(\frac{k_{12}}{s_1}\right)
     \\
    &\times 
    F(n,c\mu_2+i(n+d-1)) \frac{c\mu_2+i(n+d-1)}{\left(c\mu_2+i(n+d-1)\right)^2-\mu_1^2}
    Q_{i c\mu_2-\frac{1}{2}}^{\frac{d-3}{2}}\left(\frac{k_{56}}{s_2}\right)
    \Biggr]\,.
    \end{aligned}
    \label{eq_G_+++_SB_C2>1_C1>1}
\end{equation}
\end{framed}

\subparagraph{$\kappafour<1$.} Here, we close the contour in the lower half-plane, so to compute $G_{+++}^{1,S,B}$ we pick only the two residues with negative imaginary part:
\begin{equation}
\begin{aligned}
    G_{+++}^{1,S,B}&=\frac{\pi i}{2\sinh^2(\pi\mu_1)} \sum_{c=\pm} e^{\pi c\mu_1}
    Q_{-i c\mu_1-\frac{1}{2}}^{\frac{d-3}{2}}\left(\frac{k_{12}}{s_1}\right) F(n, c\mu_1)
     \\
    &\times \frac{-c\mu_1+i(n+d-1)}{\left(-c\mu_1+i(n+d-1)\right)^2-\mu_2^2} Q_{i c\mu_1+n+d-\frac{3}{2}}^{\frac{d-3}{2}}\left(\frac{k_{56}}{s_2}\right)
\,.
    \end{aligned}
    \label{eq_G+++_1SB_C2>1_C1<1}
\end{equation}
Therefore, the sum of \eqref{eq_G+++_0SB_C2>1} and \eqref{eq_G+++_1SB_C2>1_C1<1} in \eqref{eq_G_+++_.B_def_wrt_G_+++_01B_C2>1} gives
\begin{framed}
\vspace{-.5cm}
\begin{equation}
\begin{aligned}
    G_{+++}^{S,B}&=\frac{e^{-i\pi \frac{3d-1}{2}}2^{d-5}\pi^6}{\sqrt{k_1 k_2 k_3 k_4 k_5 k_6} \,s_1^{\frac{d-2}{2}}s_2^{\frac{3d}{2}-2}}
    \left(\left(\frac{k_{12}}{s_1}\right)^2-1\right)^{\frac{3-d}{4}}
    \left(\left(\frac{k_{56}}{s_2}\right)^2-1\right)^{\frac{3-d}{4}}\\
    &\times\sum_{n=0}^{+\infty} \left(\frac{2 k_{34}}{s_2}\right)^n \frac{1}{n!}
    \sum_{c=\pm} \frac{c \,e^{\pi c\mu_1}}{\sinh(\pi\mu_1)}
    \Gamma\left(\frac{d-2}{2}\pm i c\mu_1\right)
    P_{ic\mu_1-\frac{1}{2}}^{-\frac{d-3}{2}}\left(\frac{k_{12}}{s_1}\right)\\
    &\times F(n,c\mu_1)\frac{-c\mu_1+i(n+d-1)}{\left(-c\mu_1+i(n+d-1)\right)^2-\mu_2^2} Q_{i c\mu_1+n+d-\frac{3}{2}}^{\frac{d-3}{2}}\left(\frac{k_{56}}{s_2}\right)
    \,.
    \end{aligned}
    \label{eq_G_+++_SB_C2>1_C1<1}
\end{equation}
\end{framed}

\subsection{Background-Signal residues}

This subsection collects results about the components $G_{+++}^{B,S}$ in all kinematic configurations.

\paragraph{Background-Signal residues, $\kappaone<1$ and $\kappathree<1$.}
We follow the computation of $G_{+++}^{B,S}$ started in subsection \ref{subsec_second_int_layer} for the case where $\kappaone<1$, using the decomposition of $G_{+++}^{B,S}$ \eqref{eq_G_+++_.S_wrt_G+++_01.S_C2<1} in terms of the two integrals $G_{+++}^{0,B,S}$ \eqref{eq_G_+++_0.S_def_C2<1}, which is given in \eqref{eq_G+++0BS_C2<1_res}, and $G_{+++}^{1,B,S}$ \eqref{eq_G_+++_1.S_def_C2<1}, which remains to be computed when $\kappathree<1$.

To this end, we can use the computation that has been done for $F_{+++}^{1,B}$ when $\kappaone<1$\footnote{Note that here we are indeed working under the condition $\kappaone<1$, but for our present computation of $G_{+++}^{1,B,S}$ we use what has been done in subsection \ref{subsubsec_1st_layer} for $\kappaone<1$ and replace $\kappaone$ with $\kappathree$ in the expressions coming from \ref{subsubsec_1st_layer}.}. The two contributions \eqref{eq_F+++1B_C2<1_res_sh} and \eqref{eq_F+++1B_C2<1_res_Q} to $F_{+++}^{1,B}$ allow us to directly write the two corresponding contributions of $G_{+++}^{1,B,S}$:
\begin{itemize}
\item \textbf{\textcolor{pyblue}{Poles from $\frac{\nu}{\sinh(\pi\nu)}$.}}
The factor $\nu/\sinh(\pi\nu)$ provides the following residues
\begin{equation}
    G_{+++}^{1,B,S}\supset 2 \sum_{n=0}^{+\infty} (-1)^n\frac{n+1}{(n+1)^2+\mu_1^2}\,
    \mathcal{I}_d\begin{bmatrix}
        -i(n+1) & \mu_2 &i/2 & i/2 \\ s_1 & e^{-i\pi}s_2 &k_3 & k_4 \\ I & K & K & K
    \end{bmatrix}
    Q_{-n-\frac{3}{2}}^{\frac{d-3}{2}}\left(\frac{k_{12}}{s_1}\right)\,.
    \label{eq_G+++1BS_C2<1_C1p<1_res_sh}
\end{equation}
\item \textbf{\textcolor{pyred}{Poles from $Q_{-i\nu-\frac{1}{2}}^{\frac{d-3}{2}}\left(k_{12}/s_1\right)$.}} The other residues come from the Legendre $Q$ function:
\begin{equation}
\begin{aligned}
    G_{+++}^{1,B,S}&\supset \frac{2^{\frac{d-1}{2}} \pi^\frac{3}{2} e^{i\pi\frac{d-1}{2}}}{\sin\left(\pi\frac{d}{2}\right)}
    \left(\left(\frac{k_{12}}{s_1}\right)^2-1\right)^{\frac{d-3}{4}}
    \sum_{n=0}^{+\infty} \frac{n+\frac{d-2}{2}}{\left(n+\frac{d-2}{2}\right)^2+\mu_1^2}\frac{1}{n!}\left(\frac{2 k_{12}}{s_1}\right)^n\\
    &\times\,_2\F_1\left(
    \begin{matrix}
        \frac{1-n}{2}\,,\,-\frac{n}{2} \\ -n-\frac{d-4}{2}
    \end{matrix}\,,\left(\frac{s_1}{k_{12}}\right)^2\right)
    \mathcal{I}_d\begin{bmatrix}
        -i\left(n+\frac{d-2}{2}\right)&\mu_2 &i/2 & i/2 \\ s_1 & e^{-i\pi}s_2 & k_3 & k_4  \\ I & K & K & K
    \end{bmatrix}\,.
    \end{aligned}
    \label{eq_G+++1BS_C2<1_C1p<1_res_Q}
\end{equation}
\end{itemize}
Again, the sum of the terms \eqref{eq_G+++0BS_C2<1_res} and \eqref{eq_G+++1BS_C2<1_C1p<1_res_sh} vanishes, using the connection formula for the Legendre $Q$ functions \eqref{connection_formula_P_Q}. Using the third term \eqref{eq_G+++1BS_C2<1_C1p<1_res_Q} in \eqref{eq_G_+++_.S_wrt_G+++_01.S_C2<1} leads to
\begin{framed}
\vspace{-.5cm}
\begin{equation}
\begin{aligned}
    G_{+++}^{B,S}
    &=\frac{e^{-i\pi(d-1)} 2^{\frac{d-9}{2}}\pi^\frac{9}{2}}{\sqrt{k_1 k_2 k_5 k_6}s_1^{\frac{d-2}{2}} s_2^{\frac{d-2}{2}}\sin\left(\pi\frac{d}{2}\right)}
    \left(\left(\frac{k_{56}}{s_2}\right)^2-1\right)^{\frac{3-d}{4}} \Gamma\left(\frac{d-2}{2}\pm i\mu_2\right)\\
    &\times
    P_{i\mu_2-\frac{1}{2}}^{-\frac{d-3}{2}}\left(\frac{k_{56}}{s_2}\right)
    \sum_{n=0}^{+\infty} \frac{1}{n!}\left(\frac{2 k_{12}}{s_1}\right)^n
    \,_2\F_1\left(
    \begin{matrix}
        \frac{1-n}{2}\,,\,-\frac{n}{2} \\ -n-\frac{d-4}{2}
    \end{matrix}\,,\left(\frac{s_1}{k_{12}}\right)^2\right)\\
    &\times
    \frac{n+\frac{d-2}{2}}{\left(n+\frac{d-2}{2}\right)^2+\mu_1^2}
    \mathcal{I}_d\begin{bmatrix}
        -i\left(n+\frac{d-2}{2}\right)&\mu_2 &i/2 & i/2 \\ s_1 & e^{-i\pi}s_2 & k_3 & k_4  \\ I & K & K & K
    \end{bmatrix}
    \,.
    \end{aligned}
    \label{eq_G_+++_BS_C2<1_C1p<1}
\end{equation}
\end{framed}

\paragraph{Background-Signal residues, $\kappaone>1$.}
Returning to the expression of $G_{+++}^{.,S}$ in terms of $F_{+++}^S$ \eqref{eq_G_+++_.S_def_wrt_F+++S}, we now use the expression of $F_{+++}^S$ found for $\kappaone>1$ \eqref{eq_F+++_S_C2>1}.
In this case, also making use of the connection formulas \eqref{eq_vertex_fct_12_Q} and \eqref{eq_connection_IKK_IIK} for the vertex functions, we can re-express $G_{+++}^{.,S}$ \eqref{eq_G_+++_.S_def_wrt_F+++S} as
\begin{equation}
\begin{aligned}
    G_{+++}^{.,S}\left(\{k_i,s_j\}\right)
    &=\frac{e^{-i\pi\frac{3}{2}(d-1)}\pi^3}{2^4 \sqrt{k_1 k_2 k_5 k_6}s_1^{\frac{d-2}{2}} s_2^{\frac{d-2}{2}}}
    \left(\left(\frac{k_{12}}{s_1}\right)^2-1\right)^{\frac{3-d}{4}}
    \left(\left(\frac{k_{56}}{s_2}\right)^2-1\right)^{\frac{3-d}{4}}\\
    &\times \,
    \Gamma\left(\frac{d-2}{2}\pm i\mu_2\right)
    P_{i\mu_2-\frac{1}{2}}^{-\frac{d-3}{2}}\left(e^{+i\pi}\frac{k_{56}}{s_2}\right)\\
    &\Biggl[G_{+++}^{0,.,S}+G_{+++}^{1,.,S}\Biggr]
    \,,
    \end{aligned}
    \label{eq_G_+++_.S_def_wrt_F+++S_C2>1}
\end{equation}
where
\begin{subequations}
\begin{equation}
\begin{aligned}
    G_{+++}^{0,.,S}&=
    \int\limits_{-\infty}^{+\infty}\d\nu \, \frac{\nu}{\sinh(\pi\nu)}\, Q_{i\nu-\frac{1}{2}}^{\frac{d-3}{2}}\left(\frac{k_{12}}{s_1}\right) \frac{1}{(\nu^2-\mu_{1}^2)_{i\epsilon}}
    \mathcal{I}_d\begin{bmatrix}
        \nu &\mu_2 &i/2 & i/2 \\ s_1 & s_2 & k_3 & k_4  \\ I &K & K & K
    \end{bmatrix}\,,
    \end{aligned}
    \label{eq_G_+++_0.S_def_C2>1}
\end{equation}
\begin{equation}
\begin{aligned}
G_{+++}^{1,.,S}&=
    -\int\limits_{-\infty}^{+\infty}\d\nu \, \frac{\nu}{\sinh(\pi\nu)}\, Q_{-i\nu-\frac{1}{2}}^{\frac{d-3}{2}}\left(\frac{k_{12}}{s_1}\right) \frac{1}{(\nu^2-\mu_{1}^2)_{i\epsilon}}
    \mathcal{I}_d\begin{bmatrix}
        \nu &\mu_2 &i/2 & i/2 \\ s_1 & s_2 & k_3 & k_4  \\ I &K & K & K
    \end{bmatrix}\,,
    \end{aligned}
    \label{eq_G_+++_1.S_def_C2>1}
\end{equation}
\label{eq_G_+++_0and1.S_def_C2>1}
\end{subequations}
using the change of variable $\nu'=-\nu$ to reduce the number of terms to two in \eqref{eq_G_+++_.S_def_wrt_F+++S_C2>1}. The two integrals $G_{+++}^{0,.,S}$ and $G_{+++}^{1,.,S}$ are identical to the same objects \eqref{eq_G_+++_0.S_def_C2<1}-\eqref{eq_G_+++_1.S_def_C2<1} in the case $\kappaone<1$, up to the rescaling $s_2\to e^{+i\pi}s_2$. Therefore, their expressions can be straightforwardly deduced from the latter. Note that the kinematic condition that arises from the leading large $\nu$ behaviour:
\begin{equation}
    \mathcal{I}_d\begin{bmatrix}
        \nu &\mu_2 &i/2 & i/2 \\ s_1 & s_2 &k_3 & k_4 \\ I & K & K & K
    \end{bmatrix}
    Q_{\pm i\nu-\frac{1}{2}}^{\frac{d-3}{2}}\left(\frac{k_{12}}{s_1}\right)
    \underset{\nu\to\infty}{\propto}e^{-i\nu\left[\arccosh\left(\frac{k_{34}+s_2}{s_1}\right)\pm\arccosh\left(\frac{k_{12}}{s_1}\right)\right]}\,,
    \label{eq_asymptotics_G+++01.S_C2>1}
\end{equation}
is, for $G_{+++}^{1,.,S}$, the kinematic criterion $\kappatwo$ \eqref{eq_condition_C11p}.
If $\kappatwo>1$ (resp. $\kappatwo<1$), the contour must be closed in the upper-half (resp. lower-half) $\nu$-plane. Up to the replacement $s_2\to e^{+i\pi}s_2$, we can compute $G_{+++}^{0,B,S}$ and $G_{+++}^{1,B,S}$ for $\kappaone>1$ using the results of the case $\kappaone<1$.

\subparagraph{$\kappatwo>1$.} In this configuration, $G_{+++}^{B,S}$ can be deduced by inserting the expression \eqref{eq_G+++1BS_C2<1_C1p>1_res_IKI} with $s_2\to e^{+i\pi}s_2$ into $G_{+++}^{.,S}$ \eqref{eq_G_+++_.S_def_wrt_F+++S_C2>1}. This leads to
\begin{framed}
\vspace{-.5cm}
\begin{equation}
\begin{aligned}
    G_{+++}^{B,S}
    &=\frac{-e^{-i\pi\frac{3d}{2}}2^{d-6}\pi^6}{\sqrt{k_1 k_2 k_3 k_4 k_5 k_6}s_1^{\frac{3d-4}{2}} s_2^{\frac{d-2}{2}}}
    \left(\left(\frac{k_{12}}{s_1}\right)^2-1\right)^{\frac{3-d}{4}}
    \left(\left(\frac{k_{56}}{s_2}\right)^2-1\right)^{\frac{3-d}{4}}\\
    &\times \,
    \Gamma\left(\frac{d-2}{2}\pm i\mu_2\right)
    P_{i\mu_2-\frac{1}{2}}^{-\frac{d-3}{2}}\left(e^{+i\pi}\frac{k_{56}}{s_2}\right)
    \sum_{c=\pm}\frac{1}{\sinh(\pi c \mu_2)} \left(\frac{s_2}{s_1}\right)^{i c\mu_2}\\
    &\times
    \sum_{n=0}^{+\infty} \left(\frac{2 k_{34}}{s_1}\right)^n \frac{1}{n!}\,
    Q_{i c \mu_2 +n+d-\frac{3}{2}}^{\frac{d-3}{2}}\left(\frac{k_{12}}{s_1}\right)\,\frac{-c\mu_2+i(n+d-1)}{\left(c\mu_2-i(n+d-1)\right)^2-\mu_1^2}\\
    &\times\frac{1}{\sinh(\pi c\mu_2-i\pi d)}\,\F_4\Biggl(\begin{matrix}
    -\frac{n}{2}\,,\, \frac{1-n}{2}\\
    2-i c \mu_2-n-d\,,\,1+i c\mu_2\end{matrix}\,;
    \left(\frac{s_1}{k_{34}}\right)^2,\left(\frac{s_2}{k_{34}}\right)^2\Biggr)    \,.
    \end{aligned}
    \label{eq_G_+++_BS_C2>1_C2p>1}
\end{equation}
\end{framed}

\subparagraph{$\kappatwo<1$.} Similarly, we insert the contribution \eqref{eq_G+++1BS_C2<1_C1p<1_res_Q} with $s_2\to e^{+i\pi}s_2$ into $G_{+++}^{.,S}$ \eqref{eq_G_+++_.S_def_wrt_F+++S_C2>1}, to obtain
\begin{framed}
\vspace{-.5cm}
\begin{equation}
\begin{aligned}
    G_{+++}^{B,S}
    &=\frac{e^{-i\pi(d-1)} 2^{\frac{d-9}{2}}\pi^\frac{9}{2}}{\sqrt{k_1 k_2 k_5 k_6}s_1^{\frac{d-2}{2}} s_2^{\frac{d-2}{2}}\sin\left(\pi\frac{d}{2}\right)}
    \left(\left(\frac{k_{56}}{s_2}\right)^2-1\right)^{\frac{3-d}{4}}
    \Gamma\left(\frac{d-2}{2}\pm i\mu_2\right)\\
    &\times 
    P_{i\mu_2-\frac{1}{2}}^{-\frac{d-3}{2}}\left(e^{+i\pi}\frac{k_{56}}{s_2}\right)
    \sum_{n=0}^{+\infty} \frac{1}{n!}\left(\frac{2 k_{12}}{s_1}\right)^n
    \,_2\F_1\left(
    \begin{matrix}
        \frac{1-n}{2}\,,\,-\frac{n}{2} \\ -n-\frac{d-4}{2}
    \end{matrix}\,,\left(\frac{s_1}{k_{12}}\right)^2\right)\\
    &\times
    \frac{n+\frac{d-2}{2}}{\left(n+\frac{d-2}{2}\right)^2+\mu_1^2}
    \mathcal{I}_d\begin{bmatrix}
        -i\left(n+\frac{d-2}{2}\right)&\mu_2 &i/2 & i/2 \\ s_1 & s_2 & k_3 & k_4  \\ I & K & K & K
    \end{bmatrix}
    \,.
    \end{aligned}
    \label{eq_G_+++_BS_C2>1_C2p<1}
\end{equation}
\end{framed}

\subsection{Signal-Signal residues}

In this subsection, we compute the signal-signal residues for the kinematic cases that are not considered in the main text.

\paragraph{Signal-Signal residues, $\kappaone<1$ and $\kappathree<1$.} 
Recall that the signal-signal residues when $\kappaone<1$ are obtained by collecting the signal residues of $G_{+++}^{.,S}$ \eqref{eq_G_+++_.S_wrt_G+++_01.S_C2<1} under this condition, through the integrals $G_{+++}^{0,S,S}$ and $G_{+++}^{1,S,S}$.
In the present case where in addition $\kappathree<1$, the expression of $G_{+++}^{0,S,S}$ \eqref{eq_G+++_0SS_C2<1} still holds, and for $G_{+++}^{1,S,S}$ the collected residues are located at $\nu=\pm\mu_1-i\epsilon$, in the lower half-plane. Then,
\begin{equation}
\begin{aligned}
    G_{+++}^{1,S,S}=&\frac{i\pi}{2\sinh^2(\pi\mu_1)}
    \sum_{c=\pm}e^{\pi c\mu_1}\mathcal{I}_d\begin{bmatrix}
        c\mu_1 &\mu_2 & i/2 & i/2 \\ s_1 &e^{-i\pi}s_2 & k_3 & k_4  \\ I & K & K & K
    \end{bmatrix}
    Q_{-i c\mu_1-\frac{1}{2}}^{\frac{d-3}{2}}\left(\frac{k_{12}}{s_1}\right)
    \,.
    \end{aligned}
    \label{eq_G+++_1SS_C2<1_C1p<1}
\end{equation}
To find $G_{+++}^{S,S}$ for $\kappaone<1$ and $\kappathree<1$, we sum the two contributions \eqref{eq_G+++_0SS_C2<1} and \eqref{eq_G+++_1SS_C2<1_C1p<1} and simplify the resulting expression by the use of the connection formula between Legendre $P$ and $Q$ functions \eqref{connection_formula_P_Q} and the analytic continuation formula of $\I_{KK}$ in terms of $\I_{IK}$ \eqref{eq_analytic_continuation_IKK_1}. In the end, we have
\begin{framed}
\vspace{-.2cm}
\begin{equation}
\begin{aligned}
    G_{+++}^{S,S} 
    &=\frac{e^{-i\pi(d+1)} \pi^3}{2^4\sqrt{k_1 k_2 k_5 k_6}s_1^{\frac{d-2}{2}}s_2^{\frac{d-2}{2}}}
    \left(\left(\frac{k_{12}}{s_1}\right)^2-1\right)^{\frac{3-d}{4}}
    \left(\left(\frac{k_{56}}{s_2}\right)^2-1\right)^{\frac{3-d}{4}}\\
    &\times\Gamma\left(\frac{d-2}{2}\pm i\mu_1\right)\Gamma\left(\frac{d-2}{2}\pm i\mu_2\right)\\
    &\times
    P_{i\mu_1-\frac{1}{2}}^{-\frac{d-3}{2}}\left(\frac{k_{12}}{s_1}\right)
    \mathcal{I}_d\begin{bmatrix}
        \mu_1 & \mu_2 &i/2 & i/2 \\ e^{-i\pi}s_1 & e^{-i\pi}s_2 & k_3 & k_4  \\ K & K & K & K
    \end{bmatrix}
    P_{i\mu_2-\frac{1}{2}}^{-\frac{d-3}{2}}\left(\frac{k_{56}}{s_2}\right)
    \,.
    \end{aligned}
    \label{eq_G+++_SS_C2<1_C1p<1}
\end{equation}
\end{framed}

\paragraph{Signal-Signal residues, $\kappaone>1$.}
To find the signal-signal residues for $\kappaone>1$, we start from $G_{+++}^{.,S}$ written in \eqref{eq_G_+++_.S_def_wrt_F+++S_C2>1} in terms of $G_{+++}^{0,.,S}$ and $G_{+++}^{1,.,S}$ \eqref{eq_G_+++_0and1.S_def_C2>1}, and compute their signal residues. In the same way as for the $\kappaone<1$ case, these residues originate from the singularities of the propagator $\left(\nu^2-\mu_1^2\right)_{i\epsilon}^{-1}$.
As noted above in the computation of background-signal residues, the integrals defining $G_{+++}^{0,.,S}$ and $G_{+++}^{1,.,S}$ \eqref{eq_G_+++_0and1.S_def_C2>1} in the case $\kappaone>1$ are the same as those in the case $\kappaone<1$ \eqref{eq_G_+++_0and1.S_def_C2<1}, up to the exchange $e^{-i\pi}s_2\to s_2$. Then, we can first compute $G_{+++}^{0,S,S}$ from its expression when $\kappaone<1$ \eqref{eq_G+++_0SS_C2<1}:
\begin{equation}
\begin{aligned}
    G_{+++}^{0,S,S}=&-\frac{i\pi}{2\sinh^2(\pi\mu_1)}\sum_{c=\pm}e^{\pi c\mu_1}\mathcal{I}_d\begin{bmatrix}
        c\mu_1 & \mu_2 &i/2 & i/2 \\ s_1 & s_2 & k_3 & k_4  \\ I & K & K & K
    \end{bmatrix}
    Q_{i c\mu_1-\frac{1}{2}}^{\frac{d-3}{2}}\left(\frac{k_{12}}{s_1}\right)
    \,.
    \end{aligned}
    \label{eq_G+++_0SS_C2>1}
\end{equation}
For the computation of $G_{+++}^{1,S,S}$, we need to distinguish cases with respect to the value of the ratio $\kappatwo$ \eqref{eq_condition_C11p}.

\subparagraph{$\kappatwo>1$.} As remarked before, we can write $G_{+++}^{1,S,S}$ from its value for $\kappaone<1$, $\kappatwo>1$ \eqref{eq_G+++_1SS_C2<1_C1p>1}:
\begin{equation}
\begin{aligned}
    G_{+++}^{1,S,S}=&\frac{i\pi}{2\sinh^2(\pi\mu_1)}
    \sum_{c=\pm}e^{-\pi c\mu_1}\mathcal{I}_d\begin{bmatrix}
        c\mu_1 &\mu_2 & i/2 & i/2 \\ s_1 & s_2 & k_3 & k_4  \\ I & K & K & K
    \end{bmatrix}
    Q_{-i c\mu_1-\frac{1}{2}}^{\frac{d-3}{2}}\left(\frac{k_{12}}{s_1}\right)
    \,.
    \end{aligned}
    \label{eq_G+++_1SS_C2>1_C2p>1}
\end{equation}
Then, summing \eqref{eq_G+++_0SS_C2>1} and \eqref{eq_G+++_1SS_C2>1_C2p>1} and using the formulas \eqref{eq_analytic_cont_P} and \eqref{eq_connection_IKK_IIK} in the expression of $G_{+++}^{.,S}$ \eqref{eq_G_+++_.S_def_wrt_F+++S_C2>1}, we find for $G_{+++}^{S,S}$:
\begin{framed}
\begin{equation}
\begin{aligned}
    G_{+++}^{S,S}
    &=\frac{e^{-i\pi(d-1)}\pi^3}{2^4 \sqrt{k_1 k_2 k_5 k_6}s_1^{\frac{d-2}{2}} s_2^{\frac{d-2}{2}}}
    \left(\left(\frac{k_{12}}{s_1}\right)^2-1\right)^{\frac{3-d}{4}}
    \left(\left(\frac{k_{56}}{s_2}\right)^2-1\right)^{\frac{3-d}{4}}\\
    &\times \,
    \Gamma\left(\frac{d-2}{2}\pm i\mu_1\right)
    \Gamma\left(\frac{d-2}{2}\pm i\mu_2\right)
    \\
    &\times
    P_{i\mu_1-\frac{1}{2}}^{-\frac{d-3}{2}}\left(e^{+i\pi}\frac{k_{12}}{s_1}\right)
    \mathcal{I}_d\begin{bmatrix}
        \mu_1 & \mu_2 &i/2 & i/2 \\ s_1 & s_2 & k_3 & k_4  \\ K & K & K & K
    \end{bmatrix}
    P_{i\mu_2-\frac{1}{2}}^{-\frac{d-3}{2}}\left(e^{+i\pi}\frac{k_{56}}{s_2}\right)
    \,,
    \end{aligned}
    \label{eq_G_+++_SS_C2>1_C2p>1}
\end{equation}
\end{framed}

\subparagraph{$\kappatwo<1$.} Here, we can use the expression of $G_{+++}^{1,S,S}$ from the case $\kappaone<1$, $\kappathree<1$ \eqref{eq_G+++_1SS_C2<1_C1p<1}:
\begin{equation}
\begin{aligned}
    G_{+++}^{1,S,S}=&\frac{i\pi}{2\sinh^2(\pi\mu_1)}
    \sum_{c=\pm}e^{\pi c\mu_1}\mathcal{I}_d\begin{bmatrix}
        c\mu_1 &\mu_2 & i/2 & i/2 \\ s_1 &s_2 & k_3 & k_4  \\ I & K & K & K
    \end{bmatrix}
    Q_{-i c\mu_1-\frac{1}{2}}^{\frac{d-3}{2}}\left(\frac{k_{12}}{s_1}\right)
    \,.
    \end{aligned}
    \label{eq_G+++_1SS_C2>1_C2p<1}
\end{equation}
Now, the sum of \eqref{eq_G+++_0SS_C2>1} and \eqref{eq_G+++_1SS_C2>1_C2p<1} simplified with the formulas \eqref{connection_formula_P_Q} and used in the expression of $G_{+++}^{.,S}$ \eqref{eq_G_+++_.S_def_wrt_F+++S_C2>1} gives
\begin{framed}
\begin{equation}
\begin{aligned}
    G_{+++}^{S,S}
    &=\frac{e^{-i\pi\left(d-\frac{1}{2}\right)}\pi^3}{2^4 \sqrt{k_1 k_2 k_5 k_6}s_1^{\frac{d-2}{2}} s_2^{\frac{d-2}{2}}}
    \left(\left(\frac{k_{12}}{s_1}\right)^2-1\right)^{\frac{3-d}{4}}
    \left(\left(\frac{k_{56}}{s_2}\right)^2-1\right)^{\frac{3-d}{4}}\\
    &\times\Gamma\left(\frac{d-2}{2}\pm i\mu_1\right)
    \Gamma\left(\frac{d-2}{2}\pm i\mu_2\right)
    \\
    &\times
    P_{i\mu_1-\frac{1}{2}}^{-\frac{d-3}{2}}\left(\frac{k_{12}}{s_1}\right)
    \mathcal{I}_d\begin{bmatrix}
        \mu_1 & \mu_2 &i/2 & i/2 \\ e^{-i\pi}s_1 & s_2 & k_3 & k_4  \\ K & K & K & K
    \end{bmatrix}
    P_{i\mu_2-\frac{1}{2}}^{-\frac{d-3}{2}}\left(e^{+i\pi}\frac{k_{56}}{s_2}\right)
    \,.
    \end{aligned}
    \label{eq_G_+++_SS_C2>1_C2p<1}
\end{equation}
\end{framed}

\section{Useful properties for the double-exchange computation}\label{appendix_double_exchange}

In this Appendix, we provide useful relations for the computation of the double-exchange diagram of section \ref{section_the_double_exchange_diagram}. In Appendix \ref{subapp_vertex_fct_double_exch} we give some analytical properties of the vertex functions that enter the diagram \eqref{draw_double_exch_diagram_general}, focusing on the function related to the central vertex. In Appendix \ref{subapp_details_double_exch} we detail, when needed, some steps of the computation of the KLF integrals, whereas Appendix \ref{subappendix_G+++BB} derives additional features of the computation of $G_{+++}^{B,B}$, in particular analytical properties of the residue function $F$ \eqref{eq_def_residue_function} that enters $G_{+++}^{B,B}$ when $\kappaone>1$.

\subsection{Vertex functions in the double-exchange}\label{subapp_vertex_fct_double_exch}

\paragraph{Connection formulas.} The function arising from the central vertex is introduced in sec. \ref{subsection_analytics_properties}, with its integral \eqref{eq_int_rep_vertex_F4} and series representation \eqref{eq_series_rep_vertex_F4_IKK}. The latter allows us to immediately write the connection formulas relating the relevant functions, which are
\begin{equation}
    \mathcal{I}_d\begin{bmatrix}
        \nu &\mu &i/2 & i/2 \\ s_1 & s_2 & k_3 & k_4  \\ K &K & K & K
    \end{bmatrix}=\frac{i\pi}{2\sinh(\pi\nu)}\left(
    \mathcal{I}_d\begin{bmatrix}
        \nu &\mu &i/2 & i/2 \\ s_1 & s_2 & k_3 & k_4  \\ I &K & K & K
    \end{bmatrix}-
    \mathcal{I}_d\begin{bmatrix}
        -\nu &\mu &i/2 & i/2 \\ s_1 & s_2 & k_3 & k_4  \\ I &K & K & K
    \end{bmatrix}\right)\,,
    \label{eq_connection_IKK_IIK}
\end{equation}
\begin{equation}
    \mathcal{I}_d\begin{bmatrix}
        \nu &\mu &i/2 & i/2 \\ s_1 & s_2 & k_3 & k_4  \\ K &K & K & K
    \end{bmatrix}=\frac{i\pi}{2\sinh(\pi\mu)}\left(
    \mathcal{I}_d\begin{bmatrix}
        \nu &\mu &i/2 & i/2 \\ s_1 & s_2 & k_3 & k_4  \\ K & I & K & K
    \end{bmatrix}-
    \mathcal{I}_d\begin{bmatrix}
        \nu &-\mu &i/2 & i/2 \\ s_1 & s_2 & k_3 & k_4  \\ K & I & K & K
    \end{bmatrix}\right)\,.
    \label{eq_connection_IKK_IKI}
\end{equation}
For the functions that appear in the right-hand side, we can also write
\begin{equation}
    \mathcal{I}_d\begin{bmatrix}
        \nu &\mu &i/2 & i/2 \\ s_1 & s_2 & k_3 & k_4  \\ I &K & K & K
    \end{bmatrix}=\frac{i\pi}{2\sinh(\pi\mu)}\left(
    \mathcal{I}_d\begin{bmatrix}
        \nu &\mu &i/2 & i/2 \\ s_1 & s_2 & k_3 & k_4  \\ I & I & K & K
    \end{bmatrix}-
    \mathcal{I}_d\begin{bmatrix}
        \nu &-\mu &i/2 & i/2 \\ s_1 & s_2 & k_3 & k_4  \\ I & I & K & K
    \end{bmatrix}\right)\,,
    \label{eq_connection_IIK_III}
\end{equation}
and
\begin{equation}
    \mathcal{I}_d\begin{bmatrix}
        \nu &\mu &i/2 & i/2 \\ s_1 & s_2 & k_3 & k_4  \\ K & I & K & K
    \end{bmatrix}=\frac{i\pi}{2\sinh(\pi\nu)}\left(
    \mathcal{I}_d\begin{bmatrix}
        \nu &\mu &i/2 & i/2 \\ s_1 & s_2 & k_3 & k_4  \\ I & I & K & K
    \end{bmatrix}-
    \mathcal{I}_d\begin{bmatrix}
        -\nu &\mu &i/2 & i/2 \\ s_1 & s_2 & k_3 & k_4  \\ I & I & K & K
    \end{bmatrix}\right)\,.
    \label{eq_connection_IKI_III}
\end{equation}
In the rest of this Appendix, to shorten the notations in the text we will denote by $\I_{KK}$ the original vertex function with all lower indices $A_j=K$, $j=1,\ldots,4$ ($\I_{KK}$ is then the function that appears in the left-hand side of \eqref{eq_connection_IKK_IIK} and \eqref{eq_connection_IKK_IKI}), by $\I_{IK}$ and $\I_{KI}$ the functions that appear in the right-hand side of respectively \eqref{eq_connection_IKK_IIK} and \eqref{eq_connection_IKK_IKI}, and finally by $\I_{II}$ the function that enters the right-hand side of \eqref{eq_connection_IIK_III} and \eqref{eq_connection_IKI_III}.

\paragraph{Poles and residues.} In the process of performing a KLF integral over a given frequency $\mu$, we use a connection formula to rewrite the integrand in terms of vertex functions that have their lower index corresponding to the frequency $\mu$ being $A_j=I$. In the first layer integration of the computation of sec. \ref{section_the_double_exchange_diagram}, we then make use of the formula \eqref{eq_connection_IKK_IKI} to write the integrand (originally expressed in terms of $\I_{KK}$) in terms of $\I_{KI}$. In the second integration over $\nu$, we use the formula \eqref{eq_connection_IKI_III} to express the integrand in terms of $\I_{II}$. Then, we are interested in the analytic properties of the two functions $\I_{KI}$ and $\I_{II}$.

Combining the series representation \eqref{eq_series_rep_vertex_F4_IKK} with the connection formula \eqref{eq_connection_IKK_IKI} leads to the following expression for $\I_{KI}$:
\begin{equation}
\begin{aligned}
    \mathcal{I}_d&\begin{bmatrix}
        \nu &\mu &i/2 & i/2 \\ s_1 & s_2 & k_3 & k_4  \\ K &I & K & K
    \end{bmatrix}
    =\frac{\pi}{2\sqrt{k_3 k_4}k_{34}^{d-1}}
    \sum_{c_1=\pm}\frac{i\pi}{2}\frac{1}{\sinh(\pi c_1\nu)}\\
    &\times\left(\frac{s_1}{2 k_{34}}\right)^{i c_1 \nu} \left(\frac{s_2}{2 k_{34}}\right)^{i \mu}
    \Gamma\left(d-1+i\left(c_1\nu+\mu\right)\right)\\
    &\times\F_4\Biggl(\begin{matrix}
    \frac{d-1}{2}+\frac{i}{2}\left(c_1\nu+\mu\right)\,,\, \frac{d}{2}+\frac{i}{2}\left(c_1\nu+\mu\right)\\
    1+i c_1\nu\,,\,1+i \mu\end{matrix}\,;
    \left(\frac{s_1}{k_{34}}\right)^2,\left(\frac{s_2}{k_{34}}\right)^2\Biggr)\,.
    \end{aligned}
    \label{eq_series_rep_vertex_F4_IKI}
\end{equation}
From this, applying \eqref{eq_connection_IKI_III} leads to
\begin{equation}
\begin{aligned}
    \mathcal{I}_d&\begin{bmatrix}
        \nu &\mu &i/2 & i/2 \\ s_1 & s_2 & k_3 & k_4  \\ I &I & K & K
    \end{bmatrix}
    =\frac{\pi}{2\sqrt{k_3 k_4}k_{34}^{d-1}}\left(\frac{s_1}{2 k_{34}}\right)^{i \nu} \left(\frac{s_2}{2 k_{34}}\right)^{i \mu}\\
    &\times
    \Gamma\left(d-1+i\left(\nu+\mu\right)\right)\\
    &\times\F_4\Biggl(\begin{matrix}
    \frac{d-1}{2}+\frac{i}{2}\left(\nu+\mu\right)\,,\, \frac{d}{2}+\frac{i}{2}\left(\nu+\mu\right)\\
    1+i \nu\,,\,1+i \mu\end{matrix}\,;
    \left(\frac{s_1}{k_{34}}\right)^2,\left(\frac{s_2}{k_{34}}\right)^2\Biggr)\,.
    \end{aligned}
    \label{eq_series_rep_vertex_F4_III}
\end{equation}
As in the general case, in these two expressions it can be seen that the singularities come from the $\Gamma$-function factor, hence the poles and corresponding residues are straightforward to find. We can also deduce them from the following simplification
\begin{equation}
\begin{aligned}
    \mathcal{I}_d\begin{bmatrix}
        \nu &\mu &i/2 & i/2 \\ s_1 & s_2 & k_3 & k_4  \\ A &B & K & K
    \end{bmatrix}
    &=\sqrt{\frac{\pi\, k_{34}}{2 k_3 k_4}}
    \mathcal{I}_{d'}\begin{bmatrix}
    \nu &\mu &i/2 \\ s_1 & s_2 & k_{34}  \\ A & B & K
    \end{bmatrix}\,,
    \end{aligned}
    \label{eq_simplification_to_F4_AB}
\end{equation}
This holds for any value of $A,B=K,I$; in particular, we recover \eqref{eq_simplification_to_F4} when $A=B=K$. Similarly, here $d'$ still satisfies $d'/2-1=d-3/2$. Poles and residues are found by applying the general formulas \eqref{eq_poles_when_1_cc} and \eqref{eq_general_N_residue_vertex_fct_1_cc} for $N=3$ to the vertex function on the right-hand side of the last expression \eqref{eq_simplification_to_F4_AB}. Since we will use $\I_{KI}$ in the first integral in subsection \ref{subsection_+++}, we look at its poles in the $\mu$ complex plane. They are located at
\begin{equation}
    \mu = -c_1\nu+i\left(n+d-1\right)\;,\; n\in\mathbb{N}\,,
    \label{eq_app_poles_IKI}
\end{equation}
and the corresponding residues are
\begin{equation}
\begin{aligned}
\text{Res}&\Biggl(\mu \mapsto\I_{KI}, \mu = -c_1\nu+i\left(n+d-1\right)\;,\; n\in\mathbb{N}\Biggr)\\
    &=\frac{\pi^2}{4\sqrt{k_3 k_4}k_{34}^{d-1}}
    \frac{1}{\sinh(\pi c_1\nu)}
    \left(\frac{s_1}{s_2}\right)^{i c_1 \nu} \left(\frac{s_2}{2 k_{34}}\right)^{-n-(d-1)}
    \frac{(-1)^n}{n!}\\
    &\times\F_4\Biggl(\begin{matrix}
    -\frac{n}{2}\,,\,\frac{1-n}{2}\\
    1+i c_1\nu\,,\,1-i c_1 \nu-n-(d-1)\end{matrix}\,;
    \left(\frac{s_1}{k_{34}}\right)^2,\left(\frac{s_2}{k_{34}}\right)^2\Biggr)\,.
    \end{aligned}
    \label{eq_residues_IKI}
\end{equation}
Since $\I_{II}$ is used in the second integral over the frequency $\nu$, we need the location of its poles in the $\nu$-plane:
\begin{equation}
    \nu = -\mu+i\left(p+d-1\right)\;,\; p\in\mathbb{N}\,,
    \label{eq_app_poles_III}
\end{equation}
with residues
\begin{equation}
\begin{aligned}
\text{Res}&\Biggl(\nu \mapsto\I_{II}, \nu = -\mu+i\left(p+d-1\right)\;,\; p\in\mathbb{N}\Biggr)\\
    &=\frac{\pi}{2\sqrt{k_3 k_4}k_{34}^{d-1}}\left(\frac{s_1}{2 k_{34}}\right)^{-\left(p+d-1\right)} \left(\frac{s_2}{ s_1}\right)^{i \mu} (-i)\frac{(-1)^p}{p!}\\
    &\times\F_4\Biggl(\begin{matrix}
    -\frac{p}{2}\,,\,\frac{1-p}{2}\\
    1 -i\mu-\left(p+d-1\right)\,,\,1+i \mu\end{matrix}\,;
    \left(\frac{s_1}{k_{34}}\right)^2,\left(\frac{s_2}{k_{34}}\right)^2\Biggr)\,.
    \end{aligned}
    \label{eq_residues_III}
\end{equation}

\paragraph{Analytic continuation.} We can derive an analytic continuation formula in the kinematic domain for $\I_{KK}$ that will help us simplify expressions in the computation of section \ref{section_the_double_exchange_diagram}. Starting from the analytic continuation formula for the modified Bessel function $K_{i\mu}$ \eqref{eq_analytic_continuation_BesselK}, at the level of vertex functions this translates into
\begin{equation}
\begin{aligned}
    \mathcal{I}_d\begin{bmatrix}
        \nu &\mu &i/2 & i/2 \\ s_1 & e^{-i\pi}s_2 & k_3 & k_4  \\ K &K & K & K
    \end{bmatrix}
    &=e^{-\pi\mu}\mathcal{I}_d\begin{bmatrix}
        \nu &\mu &i/2 & i/2 \\ s_1 & s_2 & k_3 & k_4  \\ K &K & K & K
    \end{bmatrix}+i\pi\,
    \mathcal{I}_d\begin{bmatrix}
        \nu &\mu &i/2 & i/2 \\ s_1 & s_2 & k_3 & k_4  \\ K &I & K & K
    \end{bmatrix}
    \,.
    \end{aligned}
    \label{eq_analytic_cont_IKK_0}
\end{equation}
Using the connection formula between $\I_{KK}$ and $\I_{KI}$ \eqref{eq_connection_IKK_IKI} to rewrite the right-hand side, we get
\begin{equation}
\begin{aligned}
    \mathcal{I}_d\begin{bmatrix}
        \nu &\mu &i/2 & i/2 \\ s_1 & e^{-i\pi}s_2 & k_3 & k_4  \\ K &K & K & K
    \end{bmatrix}
    &=\frac{i\pi}{2\sinh(\pi\mu)}\biggl[
    e^{\pi\mu}\mathcal{I}_d\begin{bmatrix}
        \nu &\mu &i/2 & i/2 \\ s_1 & s_2 & k_3 & k_4  \\ K &I & K & K
    \end{bmatrix} \\
    &\qquad -e^{-\pi\mu}\,
    \mathcal{I}_d\begin{bmatrix}
        \nu &-\mu &i/2 & i/2 \\ s_1 & s_2 & k_3 & k_4  \\ K &I & K & K
    \end{bmatrix}\biggr]
    \,.
    \end{aligned}
    \label{eq_analytic_continuation_IKK_2}
\end{equation}
We straightforwardly also have the following symmetric relation:
\begin{equation}
\begin{aligned}
    \mathcal{I}_d\begin{bmatrix}
        \nu &\mu &i/2 & i/2 \\ e^{-i\pi}s_1 & s_2 & k_3 & k_4  \\ K &K & K & K
    \end{bmatrix}
    &=\frac{i\pi}{2\sinh(\pi\nu)}\biggl[
    e^{\pi\nu}\mathcal{I}_d\begin{bmatrix}
        \nu &\mu &i/2 & i/2 \\ s_1 & s_2 & k_3 & k_4  \\ I &K & K & K
    \end{bmatrix} \\
    &\qquad -e^{-\pi\nu}\,
    \mathcal{I}_d\begin{bmatrix}
        -\nu &\mu &i/2 & i/2 \\ s_1 & s_2 & k_3 & k_4  \\ I &K & K & K
    \end{bmatrix}\biggr]
    \,.
    \end{aligned}
    \label{eq_analytic_continuation_IKK_1}
\end{equation}
Following the same steps, but setting the second lower index to be $I$, we can obtain an analytic continuation formula for the vertex function $\I_{KI}$ in terms of $\I_{II}$:
\begin{equation}
\begin{aligned}
    \mathcal{I}_d\begin{bmatrix}
        \nu &\mu &i/2 & i/2 \\ e^{-i\pi}s_1 & s_2 & k_3 & k_4  \\ K &I & K & K
    \end{bmatrix}
    &=\frac{i\pi}{2\sinh(\pi\nu)}\biggl[
    e^{\pi\nu}\mathcal{I}_d\begin{bmatrix}
        \nu &\mu &i/2 & i/2 \\ s_1 & s_2 & k_3 & k_4  \\ I &I & K & K
    \end{bmatrix} \\
    &\qquad -e^{-\pi\nu}\,
    \mathcal{I}_d\begin{bmatrix}
        -\nu &\mu &i/2 & i/2 \\ s_1 & s_2 & k_3 & k_4  \\ I &I & K & K
    \end{bmatrix}\biggr]
    \,.
    \end{aligned}
    \label{eq_analytic_continutaion_IKI_III}
\end{equation}
Finally, the analytic continuation formula for $I_{i\mu}$ \eqref{eq_continuation_BesselI_asymptotics} leads, in particular, to
\begin{equation}
    \mathcal{I}_d\begin{bmatrix}
        \nu &\mu &i/2 & i/2 \\ s_1 & s_2 & k_3 & k_4  \\ I &I & K & K
    \end{bmatrix}=\frac{i}{\pi}
    \left[-\mathcal{I}_d\begin{bmatrix}
        \nu &\mu &i/2 & i/2 \\ s_1 & e^{-i\pi}s_2 & k_3 & k_4  \\ I & K & K & K
    \end{bmatrix}+e^{-\pi\mu}\mathcal{I}_d\begin{bmatrix}
        \nu &\mu &i/2 & i/2 \\ s_1 & s_2 & k_3 & k_4  \\ I & K & K & K
    \end{bmatrix}
    \right]
    \label{eq_continuation_III_asymptotics}
\end{equation}

\paragraph{Asymptotic behaviour.}
The asymptotic behaviour at large frequency of $\I_{KI}$ and $\I_{II}$ can be deduced from the reduction formula \eqref{eq_simplification_to_F4_AB} and the application of the general formula \eqref{eq_vertex_function_app_large_mu_vf} for $N=3$ and $d\to d'=2d-1$ to the right-hand side of \eqref{eq_simplification_to_F4_AB}. First, we have for $\I_{KI}$ at large $\mu$:
\begin{equation}
\begin{aligned}
    \mathcal{I}_d\begin{bmatrix}
        \nu &\mu &i/2 & i/2 \\ s_1 & s_2 & k_3 & k_4  \\ K &I & K & K
    \end{bmatrix}
    &\underset{\mu\rightarrow\infty}{\sim}
    \left(\frac{\pi}{2}\right)^{\frac{3}{2}}\frac{1}{\sqrt{k_3 k_4 s_1}s_2^{d-\frac{3}{2}}}\left(\left(\frac{k_{34}+s_1}{s_2}\right)^2-1\right)^{\frac{3}{4}-\frac{d}{2}}\\
    &\times\left(i\mu\right)^{d-\frac{5}{2}} e^{-i\mu\arccosh\left(\frac{k_{34}+s_1}{s_2}\right)}
    \,.
    \end{aligned}
    \label{eq_large_mu_IKI}
\end{equation}
In this formula, the leading behaviour that will provide us with our contour prescription is $e^{-i\mu\arccosh\left(\frac{k_{34}+s_1}{s_2}\right)}$. Note that the kinematic ratio $(k_{34}+s_1)/s_2>1$ since we assumed the condition \eqref{eq_kinematic_cond_double_exchange} to be satisfied.

The same process can be applied to deduce the asymptotic behaviour of $\I_{II}$, for instance at large $\nu$:
\begin{equation}
\begin{aligned}
    \mathcal{I}_d&\begin{bmatrix}
        \nu &\mu &i/2 & i/2 \\ s_1 & s_2 & k_3 & k_4  \\ I &I & K & K
    \end{bmatrix}
    \underset{\nu\rightarrow\infty}{\sim}
    \frac{\sqrt{\pi}}{2\sqrt{2 k_3 k_4 s_2}s_1^{d-\frac{3}{2}}}
    \left(i\nu\right)^{d-\frac{5}{2}}\\
    &\times \sum_{d_0=0,1} \left(i e^{-\pi\mu}\right)^{d_0}\left(\left(\frac{k_{34}+(2 d_0 -1)s_2}{s_1}\right)^2-1\right)^{\frac{3}{4}-\frac{d}{2}}
    e^{-i\nu\arccosh\left(\frac{k_{34}+(2 d_0 -1)s_2}{s_1}\right)}
    \,.
    \end{aligned}
    \label{eq_large_nu_III}
\end{equation}
The leading behaviour is given here by two exponential factors with argument $\arccosh\left((k_{34}\pm s_2)/s_1\right)$. To disentangle the two leading factors, in the main text,  we use the analytic continuation formula \eqref{eq_continuation_III_asymptotics}.

\subsection{KLF integrand properties}\label{subapp_details_double_exch}

\paragraph{Singularities of $\mu\mapsto\mu/\sinh(\pi\mu)$.}
The KLF frequency integrands, when written in a suitable form to have correct contour prescription, carry the factor $\mu\mapsto\mu/\sinh(\pi\mu)$, which has poles at values $\mu=i n\,,\,n\in\mathbb{Z}^*$ and residues:
\begin{equation}
    \mathrm{Res}\left(\mu\mapsto\frac{\mu}{\sinh(\pi\mu)}\,;\,\mu=i n\,,\,n\in\mathbb{Z}^*\right)=\frac{(-1)^n i n}{\pi}\,.
    \label{eq_residues_sinh}
\end{equation}

\begin{figure}[h!]\centering
    	\begin{tikzpicture}[scale = 2]

        \draw[black, ->] (-1.8,0) -- (1.8,0) coordinate (xaxis);
		\draw[black, ->] (0,-.6) -- (0,.6) coordinate (yaxis);
		\node at (2.1, 0) {$\text{Re}(\mu)$};
		\node at (0, 0.75) {$\text{Im}(\mu)$};

        \draw[pygreen, fill = pygreen] (-1.2, -0.2) circle (.03cm);
		\draw[pygreen, fill = pygreen] (-1.2, .2) circle (.03cm);
		\draw[pygreen, fill = pygreen] (1.2, -.2) circle (.03cm);
        \draw[pygreen, fill = pygreen] (1.2, 0.2) circle (.03cm);
        \node at (-1.18, -.35) {\footnotesize\textcolor{pygreen}{$-\mu_\varphi-i\epsilon$}};
        \node at (-1.18, .35) {\footnotesize\textcolor{pygreen}{$-\mu_\varphi+i\epsilon$}};
        \node at (1.18, -.35) {\footnotesize\textcolor{pygreen}{$\mu_\varphi-i\epsilon$}};
        \node at (1.18, .35) {\footnotesize\textcolor{pygreen}{$\mu_\varphi+i\epsilon$}};

    \end{tikzpicture}
\caption{Analytic structure in the $\mu$ complex plane of $\mu\mapsto\left(\mu^2-\mu_\varphi^2\right)_{i\varepsilon}^{-1}$.}
\label{fig_poles_ieps}
\end{figure}
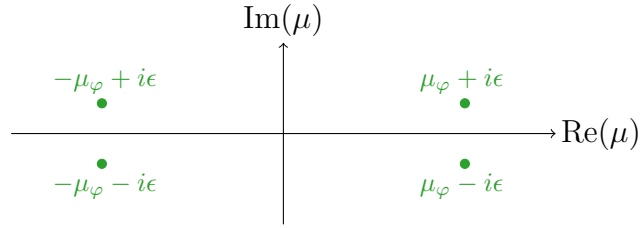

\paragraph{Singularities of $\mu\mapsto1/\left(\mu^2-\mu_\varphi^2\right)_{i\varepsilon}$.}
The KLF propagator $\mu\mapsto 1/\left(\mu^2-\mu_\varphi^2\right)_{i\varepsilon}$ that involves the $i \epsilon$-prescription expression \eqref{ieps_precription} has poles at $\mu=\pm\mu_\varphi\pm i \epsilon$, represented in Fig. \ref{fig_poles_ieps}. The corresponding residues are:
\begin{equation}
\begin{aligned}
  &  \mathrm{Res}\left(\mu\mapsto\frac{1}{\left(\mu^2-\mu_\varphi^2\right)_{i\epsilon}}\,;\,\mu=\pm \mu_\varphi-i\epsilon\right)=
    \frac{e^{\pm \pi\mu_\varphi}}{4\mu_\varphi\sinh(\pi\mu_\varphi)}\,, \\
   & \mathrm{Res}\left(\mu\mapsto\frac{1}{\left(\mu^2-\mu_\varphi^2\right)_{i\epsilon}}\,;\,\mu=\pm \mu_\varphi+i\epsilon\right)=
    -\frac{e^{\mp \pi\mu_\varphi}}{4\mu_\varphi\sinh(\pi\mu_\varphi)}\,.
\label{eq_i_eps_residues}
\end{aligned}
\end{equation}

\subsection{Additional details on the computation of $G_{+++}^{B,B}$}\label{subappendix_G+++BB}

In this sub-Appendix we provide additional details about the computation of $G_{+++}^{B,B}$, especially when $\kappaone>1$.

\subsubsection{Details on the residue function $F$}\label{subapp_residue_fct}

We gather here some details about the analytic properties of the function $F$ \eqref{eq_def_residue_function} that appears in the expression of $F_{+++}^B$ when $\kappaone>1$.

\paragraph{Asymptotic behaviour.} To perform the integral over the frequency $\nu$, we need to know the leading behaviour of $F$ when $\nu\to\infty$. From the general discussion on residues (see \eqref{eq_large_c_Lauricella} for $N=3$), we have
\begin{multline}
    \F_4\Biggl(\begin{matrix}
    -\frac{n}{2}\,,\, \frac{1-n}{2}\\
    1+i \nu\,,\,1-i \nu-(n+d-1)\end{matrix}\,;
    \left(\frac{s_1}{k_{34}}\right)^2,\left(\frac{s_2}{k_{34}}\right)^2\Biggr) \\
    \underset{\nu\to\infty}{\sim}\frac{1}{\Gamma(1+i\nu)\Gamma(1-i \nu-(n+d-1))}\,.
    \label{eq_large_nu_F4_residue}
\end{multline}
Using the reflection formula \eqref{app_eq_reflection_gamma}, we can rewrite
\begin{equation}
    \frac{1}{\Gamma(1-i \nu-(n+d-1))}=\frac{i\sinh(\pi(\nu-id))(-1)^{n+1}}{\pi}\Gamma(i \nu+n+d-1)\,.
    \label{eq_reflection_residue_F}
\end{equation}
Note also that
\begin{equation}
    \frac{\Gamma(i \nu+a)}{\Gamma(i \nu+b)}\underset{\nu\to\infty}{\sim}\left(i\nu\right)^{a-b}\,.
    \label{eq_large_nu_gamma_ratio}
\end{equation}
Combining \eqref{eq_large_nu_F4_residue}, \eqref{eq_reflection_residue_F} and \eqref{eq_large_nu_gamma_ratio} in the expression of $F$ \eqref{eq_def_residue_function}, we finally have
\begin{equation}
    F(n,\nu)\underset{\nu\to\infty}{\sim} \left(\frac{s_1}{s_2}\right)^{i \nu} \frac{i(-1)^{n+1}}{\pi}\left(i\nu\right)^{n+d-2}\,.
    \label{eq_large_nu_F}
\end{equation}
The growing or decaying behaviour in the different regions of the complex $\nu$-plane depends on the ratio $s_1/s_2$.

\paragraph{Singularities for $d\in\mathbb{C}/\mathbb{Z}$.}

Let us show that $\nu\mapsto F(n,\nu)$ has poles when $d$ is not an integer. Using the reflection formula \eqref{app_eq_reflection_gamma}, we first rewrite
\begin{equation}
    \frac{1}{\sinh\left(\pi\left(\nu-i d\right)\right)}=\frac{i}{\pi}\Gamma\left(d+i\nu\right)\Gamma\left(1-d-i\nu\right)\,.
    \label{eq_reflection_sinh_nu}
\end{equation}
Using also the series expression of $\F_4$ \eqref{eq_def_Appell_F4}, we can write $F$ in the following way:
\begin{equation}
\begin{aligned}
    F(n,\nu)&=\left(\frac{s_1}{s_2}\right)^{i \nu} \,\frac{i}{\pi}
    \sum_{m,l=0}^{+\infty} \left(-\frac{n}{2}\right)_{m+l} \left(\frac{1-n}{2}\right)_{m+l}
    \\
    &\times \frac{\Gamma\left(d+i\nu\right)\Gamma\left(1-d-i\nu\right)}{\Gamma\left(1+i\nu+l\right)\Gamma\left(1-i\nu-(n+d-1)+m\right)}
    \frac{\left(\frac{s_1}{k_{34}}\right)^{2m}}{m!}
    \frac{\left(\frac{s_2}{k_{34}}\right)^{2l}}{l!}\,,
    \end{aligned}
    \label{eq_residue_fct_series}
\end{equation}
where we grouped the $\Gamma$-function factors, since they make the possible singularities manifest. Note first that the series is truncated, because if $n$ is even, we have $(-n/2)_{l+m}=0$ for $l+m\geq n/2+1$, and if $n$ is odd, it is the factor $(\frac{1-n}{2})_{l+m}$ that vanishes when $l+m\geq (n+1)/2$. In all cases, the sum is truncated up to $l+m\leq \lfloor \frac{n}{2} \rfloor$ and then reduced to a finite sum.

The possible singularities of $F$ would come from the $1/\sinh$ factor in \eqref{eq_def_residue_function}, therefore, let us examine the limit $\nu\to i(p+d)$, with $p\in\mathbb{Z}$. In this limit, such a divergence can be cancelled if one of the $\Gamma$-functions in the denominator of \eqref{eq_residue_fct_series} becomes infinite. We have
\begin{equation}
    \Gamma\left(1+i\nu+l\right)\Gamma\left(1-i\nu-(n+d-1)+m\right)
    \underset{\nu\to i(p+d)}{\sim}
    \Gamma\left(1-p-d+l\right)\Gamma\left(2-n+p+m\right)\,.
\end{equation}
Since in this case $d$ is not an integer, the factor $\Gamma\left(1-p-d+l\right)$ stays finite for any value of $p\in\mathbb{Z}$. However, the second factor can be divergent. It will cancel the singularity of $1/\sinh$ if for every value of the summation index $m\in \left[0,\ldots,\lfloor \frac{n}{2} \rfloor\right]$ the argument $2-n+p+m$ is a negative integer. Otherwise, the function $F$ has a pole if for at least one value of $m\in \left[0,\ldots,\lfloor \frac{n}{2} \rfloor\right]$, we have $2-n+p+m\geq 1$, i.e. if $2-n+p+\lfloor \frac{n}{2} \rfloor\geq 1$. This translates into the following condition for the integer $p$, that is related to the pole location:
\begin{equation}
    p\geq n-\lfloor \frac{n}{2} \rfloor-1\,.
    \label{eq_conditon_pole_res_fct_d_complex}
\end{equation}
The residue of $F$ that corresponds to the pole at $\nu= i(p+d)$ where $p\in\mathbb{Z}$ satisfies \eqref{eq_conditon_pole_res_fct_d_complex} is then
\begin{equation}
\begin{aligned}
    \text{Res}\biggl(\nu\mapsto F(n,\nu)\,&;\, \nu=i(p+d)\biggr)\\
    &= \left(\frac{s_2}{s_1}\right)^{p+d} \frac{(-1)^p}{\pi}
    \,\F_4\Biggl(\begin{matrix}
    -\frac{n}{2}\,,\, \frac{1-n}{2}\\
    1-p-d\,,\,2-n+p\end{matrix}\,;
    \left(\frac{s_1}{k_{34}}\right)^2,\left(\frac{s_2}{k_{34}}\right)^2\Biggr)\,.
    \end{aligned}
    \label{eq_residue_F_d_complex}
\end{equation}

\paragraph{Analyticity for $d\in\mathbb{Z}$, $d\geq 2$.}
We now show that if $d\in\mathbb{Z}$ and $d\geq 2$, the function $F$ is analytic. Let us return to the series expression \eqref{eq_residue_fct_series} for $F$: the possible singularities still come from the $\Gamma$-functions written in the second line. Moreover, in the present case, the factor $\Gamma(1+i\nu+l)$ in the denominator can be divergent in the limit $\nu\to i(p+d)$, where $p\in\mathbb{Z}$.

Let us show that for every value of $p\in\mathbb{Z}$, the fraction
\begin{equation}
    \frac{\Gamma\left(d+i\nu\right)\Gamma\left(1-d-i\nu\right)}{\Gamma\left(1+i\nu+l\right)\Gamma\left(1-i\nu-(n+d-1)+m\right)}\,,
    \label{eq_gamma_4_factor}
\end{equation}
stays finite. As before, in the denominator exactly one of the two $\Gamma$ factors is divergent. Then, we must have at least one $\Gamma$ factor in the denominator that diverges for any value of $p$. Recalling that $l+m\leq \lfloor \frac{n}{2}\rfloor$, we can see in the denominator that the first $\Gamma$ function diverges for every value of $l\in[0,\ldots,\lfloor \frac{n}{2}\rfloor]$ if $p$ satisfies
\begin{equation}
    p\geq \lfloor \frac{n}{2}\rfloor+1-d\,,
\end{equation}
and similarly the second one is divergent for
\begin{equation}
    p\leq n-\lfloor \frac{n}{2}\rfloor-2\,.
\end{equation}
We see that for $d\geq 2$, these two conditions cover all integer values of $p$ (possibly overlapping). Therefore, under the condition $d\in\mathbb{Z}$, $d\geq 2$ the factor \eqref{eq_gamma_4_factor} is finite in the limit $\nu\to i(d+p)$ and then $F$ is analytic.

\subsubsection{Details on the contour prescription for $G_{+++}^{0,.,B}$ and $G_{+++}^{1,.,B}$}\label{subsubapp_contour_G+++01.B}

This part contains the details needed to give the contour prescriptions used in subsection \ref{subsec_second_int_layer}, by finding the kinematic ratios $\kappafour$ \eqref{eq_condition_C1} (in the case where $\kappaone>1$), $\kappatwo$ and $\kappathree$ \eqref{eq_condition_C11p}-\eqref{eq_condition_C12p} (for $\kappaone<1$).

\paragraph{$\kappaone>1$.}
To find the kinematic condition \eqref{eq_condition_C1} that results from the asymptotic behaviour at large $\nu$ given by \eqref{eq_asymptotics_integrand_nu}, we need to rewrite the hyperbolic cosine as
\begin{equation}
    \arccosh(x)=\log\left(x+\sqrt{x^2-1}\right)\;,\; x>1\,.
    \label{eq_arccosh_log}
\end{equation}
Then, for $G_{+++}^{0,.,B}$ \eqref{eq_G_+++_0.B_def_C2>1}, we can rewrite the right-hand side of \eqref{eq_asymptotics_integrand_nu} as
\begin{equation}
    \left(\frac{s_1}{s_2}\right)^{i\nu} e^{- i\nu\left[\arccosh\left(\frac{k_{12}}{s_1}\right)+\arccosh\left(\frac{k_{56}}{s_2}\right)\right]}
    =\left(\frac{k_{56}+\sqrt{k_{56}^2-s_2^2}}{s_1}\right)^{-i\nu}
    e^{- i\nu\arccosh\left(\frac{k_{12}}{s_1}\right)}\,.
\end{equation}
Since $\kappaone>1$, we have $k_{56}>k_{34}+s_1$, which implies $k_{56}>k_{34}$ and $k_{56}>s_1$. Since we also have $k_{56}>s_2$ from momentum conservation at the right vertex, the argument in the parentheses above is always greater than $1$, and since $k_{12}>s_1$, the previous expression is exponentially decreasing for $\Im(\nu)<0$. That is why for $G_{+++}^{0,.,B}$ we close the integration contour in the lower half-plane.

For $G_{+++}^{1,.,B}$ \eqref{eq_G_+++_1.B_def_C2>1}, the right-hand side of \eqref{eq_asymptotics_integrand_nu} gives, using \eqref{eq_arccosh_log},
\begin{equation}
\begin{aligned}
    \left(\frac{s_1}{s_2}\right)^{i\nu} &e^{- i\nu\left[\arccosh\left(\frac{k_{56}}{s_2}\right)-\arccosh\left(\frac{k_{12}}{s_1}\right)\right]}
    \\&=\left(\frac{s_1}{s_2}\right)^{i\nu}\left(\left(\frac{k_{56}}{s_2}\right)^2+\sqrt{\left(\frac{k_{56}}{s_2}\right)^2-1}\right)^{-i\nu}
    \left(\left(\frac{k_{12}}{s_1}\right)^2+\sqrt{\left(\frac{k_{12}}{s_1}\right)^2-1}\right)^{i\nu}\\
    &=\left(\frac{k_{12}+\sqrt{k_{12}^2-s_1^2}}{k_{56}+\sqrt{k_{56}^2-s_2^2}}\right)^{i\nu}
    =\kappafour^{i\nu}
    \,.
    \end{aligned}
\end{equation}
We find the condition $\kappafour$ \eqref{eq_condition_C1}. Depending on whether $\kappafour>1$ or $\kappafour<1$, the contour is closed in the upper or lower half-plane.

\paragraph{$\kappaone<1$.} Here, we start from the large $\nu$ expansion \eqref{eq_large_nu_integrand_C2<1}. For the component $G_{+++}^{0,.,B}$ \eqref{eq_G_+++_0.B_def_C2<1}, we fix the $\pm$ sign in the right-hand side to be $+$. In the convergence domain \eqref{eq_kinematic_cond_double_exchange} we have $(k_{34}\pm s_2)/s_1>1$, so that, together with $k_{12}/s_1>1$, the bracketed argument is real and positive and the integrand decays for $\Im(\nu)<0$: we close the contour in the lower half-plane. For the component $G_{+++}^{1,.,B}$ \eqref{eq_G_+++_1.B_def_C2<1}, we use the fact that $\arccosh$ is an increasing function on $[1,+\infty)$. It implies that we have to compare the magnitudes of the two arguments, i.e. $k_{12}/s_1$ and $(k_{34}\pm s_2)/s_1$, hence the conditions on the numbers $\kappatwo$ \eqref{eq_condition_C11p} and $\kappathree$ \eqref{eq_condition_C12p}. 

\subsubsection{Computation of $G_{+++}^{B,B}$ for $\kappaone>1$ when $d\in\mathbb{C}/\mathbb{Z}$}

When $d\in\mathbb{C}/\mathbb{Z}$, $G_{+++}^{B,B}$ receives an additional contribution from the residues of the function $F$ \eqref{eq_residue_F_d_complex}. In the cases of interest, these residues are located in the upper-half complex $\nu$-plane (more precisely if $\Re(d)>1$). Therefore, these residues do not contribute to the term $G_{+++}^{0,B,B}$ \eqref{eq_G_+++_0.B_def_C2>1}, since contour closure is always done in the lower half-plane, but they contribute to $G_{+++}^{1,B,B}$ \eqref{eq_G_+++_1.B_def_C2>1} in the case where $\kappafour>1$, as:
\begin{equation}
\begin{aligned}
    G_{+++}^{1,B,B}&\supset \frac{2}{\sin(\pi d)} \sum_{p\geq n-\lfloor n/2\rfloor-1} \left(\frac{s_2}{s_1}\right)^{p+d} 
    Q_{p+d-\frac{1}{2}}^{\frac{d-3}{2}}\left(\frac{k_{12}}{s_1}\right) \frac{p+d}{\left(p+d\right)^2+\mu_1^2}\\
    &\times
    \F_4\left(
    \begin{matrix}
        -\frac{n}{2}\,,\,\frac{1-n}{2} \\ 1-p-d\,,\, 2-n+p
    \end{matrix}\,,\left(\frac{s_1}{k_{34}}\right)^2\,,\,\left(\frac{s_2}{k_{34}}\right)^2\right)
    \frac{n-p-1}{\left(n-p-1\right)^2+\mu_2^2}
    Q_{n-p-\frac{3}{2}}^{\frac{d-3}{2}}\left(\frac{k_{56}}{s_2}\right)\,.
    \end{aligned}
\end{equation}
In the present case, this term has to be added to the result \eqref{eq_G+++_BB_C2>1_C1>1_full_res} ($\kappaone>1$, $\kappafour>1$) for $G_{+++}^{B,B}$, using \eqref{eq_G_+++_.B_def_wrt_G_+++_01B_C2>1}. In the case $\kappafour<1$, no contribution is added due to the contour prescription.

\bibliographystyle{JHEP}
\bibliography{references.bib}

@misc{NISTDLMF,
         key = "{\relax DLMF}",
       title = "{\it NIST Digital Library of Mathematical Functions}",
howpublished = "\url{https://dlmf.nist.gov/}, Release 1.2.6 of 2026-03-15",
         url = "https://dlmf.nist.gov/",
        note = "F.~W.~J. Olver, A.~B. {Olde Daalhuis}, D.~W. Lozier, B.~I. Schneider,
                R.~F. Boisvert, C.~W. Clark, B.~R. Miller, B.~V. Saunders,
                H.~S. Cohl, and M.~A. McClain, eds."}

@inbook{Matsumoto_2020, place={Cambridge}, title={Appell and Lauricella Hypergeometric Functions}, booktitle={Encyclopedia of Special Functions: The Askey-Bateman Project}, publisher={Cambridge University Press}, author={Matsumoto, K.}, editor={Koornwinder, Tom H. and Stokman, Jasper V.Editors}, year={2020}, pages={79?100}}

@article{Lauricella1893,
  title={Sulle funzioni ipergeometriche a piu variabili},
  author={Giuseppe Lauricella},
  journal={Rendiconti del Circolo Matematico di Palermo},
  year={1893},
  volume={7},
  pages={111-158},
  url={https://api.semanticscholar.org/CorpusID:122316343}
}

@article{Bzowski:2012ih,
    author = "Bzowski, Adam and McFadden, Paul and Skenderis, Kostas",
    title = "{Holography for inflation using conformal perturbation theory}",
    eprint = "1211.4550",
    archivePrefix = "arXiv",
    primaryClass = "hep-th",
    doi = "10.1007/JHEP04(2013)047",
    journal = "JHEP",
    volume = "04",
    pages = "047",
    year = "2013"
}

@article{Bzowski:2013sza,
    author = "Bzowski, Adam and McFadden, Paul and Skenderis, Kostas",
    title = "{Implications of conformal invariance in momentum space}",
    eprint = "1304.7760",
    archivePrefix = "arXiv",
    primaryClass = "hep-th",
    doi = "10.1007/JHEP03(2014)111",
    journal = "JHEP",
    volume = "03",
    pages = "111",
    year = "2014"
}

@article{Bzowski:2015yxv,
    author = "Bzowski, Adam and McFadden, Paul and Skenderis, Kostas",
    title = "{Evaluation of conformal integrals}",
    eprint = "1511.02357",
    archivePrefix = "arXiv",
    primaryClass = "hep-th",
    reportNumber = "IMPERIAL-TP-2015-PM-02",
    doi = "10.1007/JHEP02(2016)068",
    journal = "JHEP",
    volume = "02",
    pages = "068",
    year = "2016"
}

@article{Sleight:2019hfp,
    author = "Sleight, Charlotte and Taronna, Massimo",
    title = "{Bootstrapping Inflationary Correlators in Mellin Space}",
    eprint = "1907.01143",
    archivePrefix = "arXiv",
    primaryClass = "hep-th",
    reportNumber = "PUPT-2590",
    doi = "10.1007/JHEP02(2020)098",
    journal = "JHEP",
    volume = "02",
    pages = "098",
    year = "2020"
}

@article{Sleight:2019mgd,
    author = "Sleight, Charlotte",
    title = "{A Mellin Space Approach to Cosmological Correlators}",
    eprint = "1906.12302",
    archivePrefix = "arXiv",
    primaryClass = "hep-th",
    doi = "10.1007/JHEP01(2020)090",
    journal = "JHEP",
    volume = "01",
    pages = "090",
    year = "2020"
}

@article{Sleight:2020obc,
    author = "Sleight, Charlotte and Taronna, Massimo",
    title = "{From AdS to dS exchanges: Spectral representation, Mellin amplitudes, and crossing}",
    eprint = "2007.09993",
    archivePrefix = "arXiv",
    primaryClass = "hep-th",
    doi = "10.1103/PhysRevD.104.L081902",
    journal = "Phys. Rev. D",
    volume = "104",
    number = "8",
    pages = "L081902",
    year = "2021"
}

@article{Sleight:2021plv,
    author = "Sleight, Charlotte and Taronna, Massimo",
    title = "{From dS to AdS and back}",
    eprint = "2109.02725",
    archivePrefix = "arXiv",
    primaryClass = "hep-th",
    doi = "10.1007/JHEP12(2021)074",
    journal = "JHEP",
    volume = "12",
    pages = "074",
    year = "2021"
}

@article{Qin:2023bjk,
    author = "Qin, Zhehan and Xianyu, Zhong-Zhi",
    title = "{Inflation correlators at the one-loop order: nonanalyticity, factorization, cutting rule, and OPE}",
    eprint = "2304.13295",
    archivePrefix = "arXiv",
    primaryClass = "hep-th",
    doi = "10.1007/JHEP09(2023)116",
    journal = "JHEP",
    volume = "09",
    pages = "116",
    year = "2023"
}

@article{Weinberg:2005vy,
    author = "Weinberg, Steven",
    title = "{Quantum contributions to cosmological correlations}",
    eprint = "hep-th/0506236",
    archivePrefix = "arXiv",
    reportNumber = "UTTG-01-05",
    doi = "10.1103/PhysRevD.72.043514",
    journal = "Phys. Rev. D",
    volume = "72",
    pages = "043514",
    year = "2005"
}

@article{Chen:2017ryl,
    author = "Chen, Xingang and Wang, Yi and Xianyu, Zhong-Zhi",
    title = "{Schwinger-Keldysh Diagrammatics for Primordial Perturbations}",
    eprint = "1703.10166",
    archivePrefix = "arXiv",
    primaryClass = "hep-th",
    doi = "10.1088/1475-7516/2017/12/006",
    journal = "JCAP",
    volume = "12",
    pages = "006",
    year = "2017"
}

@article{Hogervorst:2021uvp,
    author = "Hogervorst, Matthijs and Penedones, Jo{\~a}o and Vaziri, Kamran Salehi",
    title = "{Towards the non-perturbative cosmological bootstrap}",
    eprint = "2107.13871",
    archivePrefix = "arXiv",
    primaryClass = "hep-th",
    doi = "10.1007/JHEP02(2023)162",
    journal = "JHEP",
    volume = "02",
    pages = "162",
    year = "2023"
}

@misc{dunster2025,
      title={Simplified uniform asymptotic expansions for associated Legendre and conical functions}, 
      author={T. M. Dunster},
      year={2025},
      eprint={2410.03002},
      archivePrefix={arXiv},
      primaryClass={math.CA},
      url={https://arxiv.org/abs/2410.03002}, 
}

@article{Achucarro:2022qrl,
    author = "Ach{\'u}carro, Ana and others",
    title = "{Inflation: Theory and Observations}",
    eprint = "2203.08128",
    archivePrefix = "arXiv",
    primaryClass = "astro-ph.CO",
    month = "3",
    year = "2022"
}

@article{Hang:2024xas,
    author = "Hang, Yanfeng and Shen, Cong",
    title = "{A note on kinematic flow and differential equations for two-site one-loop graph in FRW spacetime}",
    eprint = "2410.17192",
    archivePrefix = "arXiv",
    primaryClass = "hep-th",
    doi = "10.1007/JHEP09(2025)209",
    journal = "JHEP",
    volume = "09",
    pages = "209",
    year = "2025"
}

@article{Arkani-Hamed:2018kmz,
    author = "Arkani-Hamed, Nima and Baumann, Daniel and Lee, Hayden and Pimentel, Guilherme L.",
    title = "{The Cosmological Bootstrap: Inflationary Correlators from Symmetries and Singularities}",
    eprint = "1811.00024",
    archivePrefix = "arXiv",
    primaryClass = "hep-th",
    doi = "10.1007/JHEP04(2020)105",
    journal = "JHEP",
    volume = "04",
    pages = "105",
    year = "2020"
}

@article{Baumann:2019oyu,
    author = "Baumann, Daniel and Duaso Pueyo, Carlos and Joyce, Austin and Lee, Hayden and Pimentel, Guilherme L.",
    title = "{The cosmological bootstrap: weight-shifting operators and scalar seeds}",
    eprint = "1910.14051",
    archivePrefix = "arXiv",
    primaryClass = "hep-th",
    doi = "10.1007/JHEP12(2020)204",
    journal = "JHEP",
    volume = "12",
    pages = "204",
    year = "2020"
}

@article{Baumann:2020dch,
    author = "Baumann, Daniel and Duaso Pueyo, Carlos and Joyce, Austin and Lee, Hayden and Pimentel, Guilherme L.",
    title = "{The Cosmological Bootstrap: Spinning Correlators from Symmetries and Factorization}",
    eprint = "2005.04234",
    archivePrefix = "arXiv",
    primaryClass = "hep-th",
    doi = "10.21468/SciPostPhys.11.3.071",
    journal = "SciPost Phys.",
    volume = "11",
    pages = "071",
    year = "2021"
}

@article{Pajer:2020wxk,
    author = "Pajer, Enrico",
    title = "{Building a Boostless Bootstrap for the Bispectrum}",
    eprint = "2010.12818",
    archivePrefix = "arXiv",
    primaryClass = "hep-th",
    doi = "10.1088/1475-7516/2021/01/023",
    journal = "JCAP",
    volume = "01",
    pages = "023",
    year = "2021"
}

@article{Pimentel:2022fsc,
    author = "Pimentel, Guilherme L. and Wang, Dong-Gang",
    title = "{Boostless cosmological collider bootstrap}",
    eprint = "2205.00013",
    archivePrefix = "arXiv",
    primaryClass = "hep-th",
    doi = "10.1007/JHEP10(2022)177",
    journal = "JHEP",
    volume = "10",
    pages = "177",
    year = "2022"
}

@article{Wang:2022eop,
    author = "Wang, Dong-Gang and Pimentel, Guilherme L. and Ach{\'u}carro, Ana",
    title = "{Bootstrapping multi-field inflation: non-Gaussianities from light scalars revisited}",
    eprint = "2212.14035",
    archivePrefix = "arXiv",
    primaryClass = "astro-ph.CO",
    doi = "10.1088/1475-7516/2023/05/043",
    journal = "JCAP",
    volume = "05",
    pages = "043",
    year = "2023"
}

@article{Jazayeri:2022kjy,
    author = "Jazayeri, Sadra and Renaux-Petel, S{\'e}bastien",
    title = "{Cosmological bootstrap in slow motion}",
    eprint = "2205.10340",
    archivePrefix = "arXiv",
    primaryClass = "hep-th",
    doi = "10.1007/JHEP12(2022)137",
    journal = "JHEP",
    volume = "12",
    pages = "137",
    year = "2022"
}

@article{Qin:2022fbv,
    author = "Qin, Zhehan and Xianyu, Zhong-Zhi",
    title = "{Helical inflation correlators: partial Mellin-Barnes and bootstrap equations}",
    eprint = "2208.13790",
    archivePrefix = "arXiv",
    primaryClass = "hep-th",
    doi = "10.1007/JHEP04(2023)059",
    journal = "JHEP",
    volume = "04",
    pages = "059",
    year = "2023"
}

@article{Qin:2023ejc,
    author = "Qin, Zhehan and Xianyu, Zhong-Zhi",
    title = "{Closed-form formulae for inflation correlators}",
    eprint = "2301.07047",
    archivePrefix = "arXiv",
    primaryClass = "hep-th",
    doi = "10.1007/JHEP07(2023)001",
    journal = "JHEP",
    volume = "07",
    pages = "001",
    year = "2023"
}

@article{Arkani-Hamed:2023kig,
    author = "Arkani-Hamed, Nima and Baumann, Daniel and Hillman, Aaron and Joyce, Austin and Lee, Hayden and Pimentel, Guilherme L.",
    title = "{Differential equations for cosmological correlators}",
    eprint = "2312.05303",
    archivePrefix = "arXiv",
    primaryClass = "hep-th",
    doi = "10.1007/JHEP09(2025)009",
    journal = "JHEP",
    volume = "09",
    pages = "009",
    year = "2025"
}

@article{Arkani-Hamed:2023bsv,
    author = "Arkani-Hamed, Nima and Baumann, Daniel and Hillman, Aaron and Joyce, Austin and Lee, Hayden and Pimentel, Guilherme L.",
    title = "{Kinematic Flow and the Emergence of Time}",
    eprint = "2312.05300",
    archivePrefix = "arXiv",
    primaryClass = "hep-th",
    doi = "10.1103/dsjm-tckw",
    journal = "Phys. Rev. Lett.",
    volume = "135",
    number = "3",
    pages = "031602",
    year = "2025"
}

@article{De:2023xue,
    author = "De, Shounak and Pokraka, Andrzej",
    title = "{Cosmology meets cohomology}",
    eprint = "2308.03753",
    archivePrefix = "arXiv",
    primaryClass = "hep-th",
    doi = "10.1007/JHEP03(2024)156",
    journal = "JHEP",
    volume = "03",
    pages = "156",
    year = "2024"
}

@article{Baumann:2024mvm,
    author = "Baumann, Daniel and Goodhew, Harry and Lee, Hayden",
    title = "{Kinematic flow for cosmological loop integrands}",
    eprint = "2410.17994",
    archivePrefix = "arXiv",
    primaryClass = "hep-th",
    doi = "10.1007/JHEP07(2025)131",
    journal = "JHEP",
    volume = "07",
    pages = "131",
    year = "2025"
}

@article{Westerdijk:2026msm,
    author = "Westerdijk, Tom and Yang, Chen",
    title = "{Kinematic Flow for Banana Loops and Unparticles}",
    eprint = "2604.22918",
    archivePrefix = "arXiv",
    primaryClass = "hep-th",
    month = "4",
    year = "2026"
}

@article{Ke:2026laa,
    author = "Ke, Ji-Yuan and He, Ping",
    title = "{An Alternative Viewpoint on Kinematic Flow from Tubing Splitting}",
    eprint = "2605.17751",
    archivePrefix = "arXiv",
    primaryClass = "hep-th",
    month = "5",
    year = "2026"
}

@article{Baumann:2026atn,
    author = "Baumann, Daniel and Joyce, Austin and Lee, Hayden and Salehi Vaziri, Kamran",
    title = "{Differential Equations for Massive Correlators}",
    eprint = "2604.08658",
    archivePrefix = "arXiv",
    primaryClass = "hep-th",
    month = "4",
    year = "2026"
}

@article{Goodhew:2020hob,
    author = "Goodhew, Harry and Jazayeri, Sadra and Pajer, Enrico",
    title = "{The Cosmological Optical Theorem}",
    eprint = "2009.02898",
    archivePrefix = "arXiv",
    primaryClass = "hep-th",
    doi = "10.1088/1475-7516/2021/04/021",
    journal = "JCAP",
    volume = "04",
    pages = "021",
    year = "2021"
}

@article{Melville:2021lst,
    author = "Melville, Scott and Pajer, Enrico",
    title = "{Cosmological Cutting Rules}",
    eprint = "2103.09832",
    archivePrefix = "arXiv",
    primaryClass = "hep-th",
    doi = "10.1007/JHEP05(2021)249",
    journal = "JHEP",
    volume = "05",
    pages = "249",
    year = "2021"
}

@article{Maldacena:2011nz,
    author = "Maldacena, Juan M. and Pimentel, Guilherme L.",
    title = "{On graviton non-Gaussianities during inflation}",
    eprint = "1104.2846",
    archivePrefix = "arXiv",
    primaryClass = "hep-th",
    reportNumber = "PUPT-2371",
    doi = "10.1007/JHEP09(2011)045",
    journal = "JHEP",
    volume = "09",
    pages = "045",
    year = "2011"
}

@article{Raju:2012zr,
    author = "Raju, Suvrat",
    title = "{New Recursion Relations and a Flat Space Limit for AdS/CFT Correlators}",
    eprint = "1201.6449",
    archivePrefix = "arXiv",
    primaryClass = "hep-th",
    reportNumber = "HRI-ST-1201",
    doi = "10.1103/PhysRevD.85.126009",
    journal = "Phys. Rev. D",
    volume = "85",
    pages = "126009",
    year = "2012"
}

@article{Goodhew:2021oqg,
    author = "Goodhew, Harry and Jazayeri, Sadra and Lee, Mang Hei Gordon and Pajer, Enrico",
    title = "{Cutting cosmological correlators}",
    eprint = "2104.06587",
    archivePrefix = "arXiv",
    primaryClass = "hep-th",
    doi = "10.1088/1475-7516/2021/08/003",
    journal = "JCAP",
    volume = "08",
    pages = "003",
    year = "2021"
}

@article{Jazayeri:2021fvk,
    author = "Jazayeri, Sadra and Pajer, Enrico and Stefanyszyn, David",
    title = "{From locality and unitarity to cosmological correlators}",
    eprint = "2103.08649",
    archivePrefix = "arXiv",
    primaryClass = "hep-th",
    doi = "10.1007/JHEP10(2021)065",
    journal = "JHEP",
    volume = "10",
    pages = "065",
    year = "2021"
}

@article{Belrhali:2026ygh,
    author = "Belrhali, Nathan and Poisson, Arthur and Renaux-Petel, S{\'e}bastien",
    title = "{Laplace Space for Cosmological Correlators}",
    eprint = "2606.27309",
    archivePrefix = "arXiv",
    primaryClass = "hep-th",
    month = "6",
    year = "2026"
}

@article{Belrhali:2026jqe,
    author = "Belrhali, Nathan and Poisson, Arthur and Renaux-Petel, S{\'e}bastien",
    title = "{Massive Cosmological Correlators from Flat Space: a Laplace-Space Approach}",
    eprint = "2606.27311",
    archivePrefix = "arXiv",
    primaryClass = "hep-th",
    month = "6",
    year = "2026"
}

@article{Belrhali:2026ktb,
    author = "Belrhali, Nathan and Poisson, Arthur and Renaux-Petel, S{\'e}bastien and Werth, Denis",
    title = "{De Sitter Momentum Space}",
    eprint = "2601.15228",
    archivePrefix = "arXiv",
    primaryClass = "hep-th",
    month = "1",
    year = "2026"
}

@article{Belrhali:2026rkn,
    author = "Belrhali, Nathan and Poisson, Arthur and Renaux-Petel, S{\'e}bastien and Werth, Denis",
    title = "{Kontorovich-Lebedev-Fourier Space for de Sitter Correlators}",
    eprint = "2604.15251",
    archivePrefix = "arXiv",
    primaryClass = "hep-th",
    month = "4",
    year = "2026"
}

@article{Werth:2024mjg,
    author = "Werth, Denis",
    title = "{Spectral representation of cosmological correlators}",
    eprint = "2409.02072",
    archivePrefix = "arXiv",
    primaryClass = "hep-th",
    doi = "10.1007/JHEP12(2024)017",
    journal = "JHEP",
    volume = "12",
    pages = "017",
    year = "2024"
}

@article{Melville:2024ove,
    author = "Melville, Scott and Pimentel, Guilherme L.",
    title = "{A de Sitter S-matrix from amputated cosmological correlators}",
    eprint = "2404.05712",
    archivePrefix = "arXiv",
    primaryClass = "hep-th",
    doi = "10.1007/JHEP08(2024)211",
    journal = "JHEP",
    volume = "08",
    pages = "211",
    year = "2024"
}

@article{Fan:2025scu,
    author = "Fan, Bingchu and Xianyu, Zhong-Zhi",
    title = "{Anatomy of family trees in cosmological correlators}",
    eprint = "2509.02684",
    archivePrefix = "arXiv",
    primaryClass = "hep-th",
    doi = "10.1007/JHEP12(2025)179",
    journal = "JHEP",
    volume = "12",
    pages = "179",
    year = "2025"
}

@article{Xianyu:2023ytd,
    author = "Xianyu, Zhong-Zhi and Zang, Jiaju",
    title = "{Inflation correlators with multiple massive exchanges}",
    eprint = "2309.10849",
    archivePrefix = "arXiv",
    primaryClass = "hep-th",
    doi = "10.1007/JHEP03(2024)070",
    journal = "JHEP",
    volume = "03",
    pages = "070",
    year = "2024"
}

@article{Liu:2024str,
    author = "Liu, Haoyuan and Xianyu, Zhong-Zhi",
    title = "{Massive inflationary amplitudes: differential equations and complete solutions for general trees}",
    eprint = "2412.07843",
    archivePrefix = "arXiv",
    primaryClass = "hep-th",
    reportNumber = "USTC-ICTS/PCFT-24-56",
    doi = "10.1007/JHEP09(2025)183",
    journal = "JHEP",
    volume = "09",
    pages = "183",
    year = "2025"
}

@article{Aoki:2024uyi,
    author = "Aoki, Shuntaro and Pinol, Lucas and Sano, Fumiya and Yamaguchi, Masahide and Zhu, Yuhang",
    title = "{Cosmological correlators with double massive exchanges: bootstrap equation and phenomenology}",
    eprint = "2404.09547",
    archivePrefix = "arXiv",
    primaryClass = "hep-th",
    doi = "10.1007/JHEP09(2024)176",
    journal = "JHEP",
    volume = "09",
    pages = "176",
    year = "2024"
}

@article{Xianyu:2025lbk,
    author = "Xianyu, Zhong-Zhi and Zang, Jiaju",
    title = "{Massive inflationary amplitudes: new representations and degenerate limits}",
    eprint = "2511.08677",
    archivePrefix = "arXiv",
    primaryClass = "hep-th",
    doi = "10.1007/JHEP03(2026)122",
    journal = "JHEP",
    volume = "03",
    pages = "122",
    year = "2026"
}
\end{document}